\documentclass[a4paper,12pt]{article}
\usepackage{macros}

\title{An introduction to 2d conformal field theory}
\date{}

\author{Satoshi Nawata, Runkai Tao, Daisuke Yokoyama}

\begin{document}
\maketitle
\setcounter{tocdepth}{2}
\abstract{These lecture notes provide a comprehensive introduction to 2d conformal field theory, covering foundational topics such as free fields, minimal models, Zamolodchikov's $c$-theorem, Wess-Zumino-Witten models, modular invariant partition functions, and the connections to entanglement entropy. The material presented is based on lectures at the Southeast University Summer School and the course at Fudan University.}
\tableofcontents

\section{Introduction}

Conformal field theories play a central role in the study of quantum field theory due to their profound relevance in multiple areas of modern theoretical physics: quantum field theories (QFTs) at an infrared fixed point, two-dimensional critical phenomena, and string theory. Moreover, two-dimensional conformal field theories (2d CFTs) are grounded on a mathematically rigorous foundation, facilitating new developments in mathematics. Remarkably, these trends continue to advance, underscoring the lasting impact of CFTs on both physics and mathematics.

In this course, we will focus on conformal field theories of \textbf{two dimensions}. However,  the subject is so rich that we can only cover the basics. To begin with, let us begin by introducing the concept of conformal field theories in the context of physics.

\subsection{Renormalization group flow and Wilson-Fisher fixed points}

In quantum field theory with a Lagrangian description, a partition function is conceptually defined by using Feynman path integral. The basic integration variables are the Fourier modes $\phi_k$ of a field $\phi$. With a cutoff $\Lambda$, we can schematically express it as
\[
\cZ= \prod_{|k|<\Lambda} \int \mathcal{D}\phi_k\exp\left[-\cS_\Lambda[\phi_k]\right]~.
\]
To relate the coupling constants in a theory with an energy cutoff $\Lambda$ to those in a theory with a lower energy cutoff $\Lambda'<\Lambda$, we change the variable $\phi \rightarrow \phi+\phi'$, where $\phi'$ only has non-trivial Fourier components in $\Lambda'<|k|<\Lambda$ and $\phi$ only has non-trivial Fourier components in $|k|<\Lambda'$. After integrating out the field $\phi'$, we obtain a new action in terms of $\phi$
\[
\exp\left(-\cS_{\Lambda'}[\phi]\right)\ \stackrel{\mathrm{def}}{=}\  \int_{\Lambda'  \leq |k| \leq \Lambda} \mathcal{D}\phi   \exp\left[-\cS_\Lambda[\phi]\right].
\]

Integrating out the high-energy quantum modes results in changes to the coupling constants in the action, leading to a new, effective action for our low-energy theory. Repeatedly integrating out thin shells in momentum space will lead to a continuous trajectory through the space of couplings: this is \textbf{renormalization group flow}.
\begin{figure}[ht]\centering
\includegraphics[width=10cm]{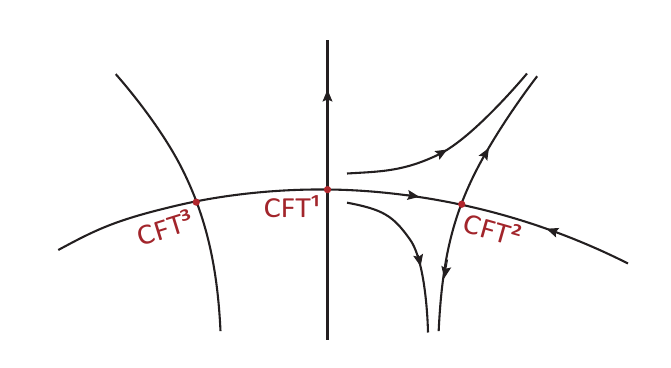}
\end{figure}

By employing naive dimensional analysis, we find that the possible parameter space becomes finite-dimensional. Although one can consider infinitely many interaction terms:
\[
\cS_{int}=\int d^d x\sum g_i \mathcal{O}_i(\phi)~,
\]
the terms $\cO_i$ with $d<\Delta^i$ are suppressed at low energy, which are called \textbf{irrelevant}. Therefore, we just need to consider finitely many terms $\cO_i$ with $d\ge\Delta^i$ at low energy. To understand the behavior of a theory for these terms, we analyze the beta function $\beta(g)$, which describes how the coupling parameter $g$ changes with respect to the energy scale $\Lambda$
\begin{equation}
\beta(g)=\frac{\partial g}{\partial \log(\Lambda)}=\Lambda \frac{\partial g}{\partial \Lambda}.
\end{equation}
The non-trivial beta function will result in
the breaking of conformal invariance quantum mechanically. For example, the one-loop  $\beta$ function  for the $\phi^4$ theory in $d=4$ dimensions is
\[
\beta(\lambda)=\frac{3}{16\pi^2}\lambda^2.
\]
Since it is positive, the coupling increases with the energy scale. One the other hand, the one-loop $\beta$-function of QCD is 
\[
\beta(g)=-\frac{9g^3}{16\pi^2}.
\]
This beta-function shows the coupling constant decreases as we increase the energy. Such theory is \textbf{asymptotic free}. The above two examples show how the theory depends on the
length scale through quantum effects. However,  if $\beta=0$, the theory is scale-invariant, \textit{i.e.} the couplings are fixed under RG flow. The coupling constant $g^*$ with $\beta(g^*)=0$ is called \textbf{Wilson-Fisher fixed point}, and the theory at the fixed point is believed to have conformal symmetry. Therefore, conformal field theories appear at special points in the parameter spaces of quantum field theories, and they play an important role in understanding families of quantum field theories.
\begin{figure}[ht]\centering
\includegraphics[width=7cm]{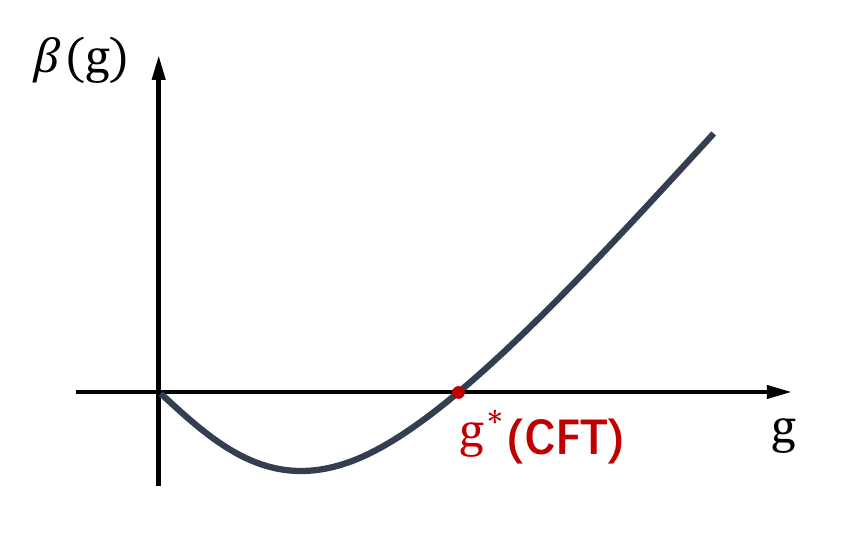}
\includegraphics[width=9cm]{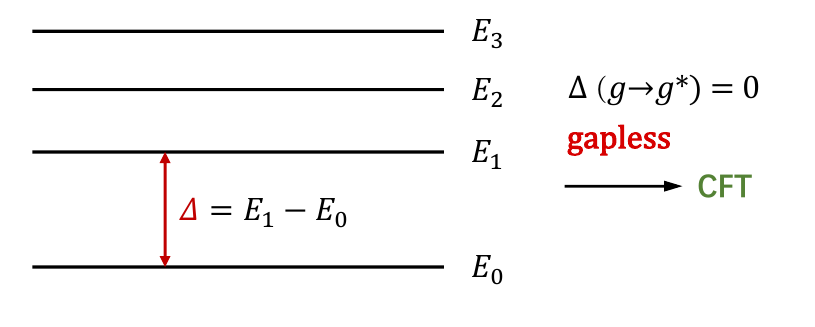}
\caption{Conformal field theory at infra-red fixed point $\beta(g^*)=0$ and illustration of spectrum of states in quantum field theory}
\label{fig:beta-function}
\end{figure}

As we will see below, conformal field theories have more enhanced symmetries than general quantum field theories. Moreover, 2d CFTs are special because conformal symmetry becomes infinite-dimensional. In this lecture, we will study 2d CFTs by making full use of an infinite-dimensional symmetry. Nonetheless, even in higher dimensions, conformal symmetries are incredibly powerful, offering profound insights into QFTs. Furthermore, if a CFT is endowed with sufficient supersymmetry, it becomes tractable, so recent studies have revealed aspects of QFTs lacking a Lagrangian description.

\subsection{Critical phenomena}

Let us recall the basics of statistical mechanics. The partition function is defined by
\[
\mathcal{Z}=e^{-F/T}=\sum_{i} \exp \left(-\frac{\cE_i}{T} \right)~,
\]
where $F$ is called the free energy, and we set the Boltzmann constant $k_B=1$. Then, the probability $P_i$ that the system is in a state with energy $\cE_i$ is given by the Boltzmann distribution
\[
P_i=\frac1\cZ \exp(-\frac{\cE_i}{T})~.
\]
Therefore, the expectation value of an operator $\cO$ is given by
\[
\langle \cO \rangle=\frac{1}{\mathcal{Z}} \sum_{i}  \cO_i  ~ e^{- \cE_i / T}
\]
For instance, the average energy is
\[
E=\langle\cE\rangle=-T^2\frac{\partial}{\partial T}\left(\frac{F}{T}\right)= \frac{1}{\mathcal{Z}} \sum_{i}  \cE_i  ~ e^{- \cE_i / T}
\]

\begin{figure}[ht]\centering
\includegraphics[width=8cm]{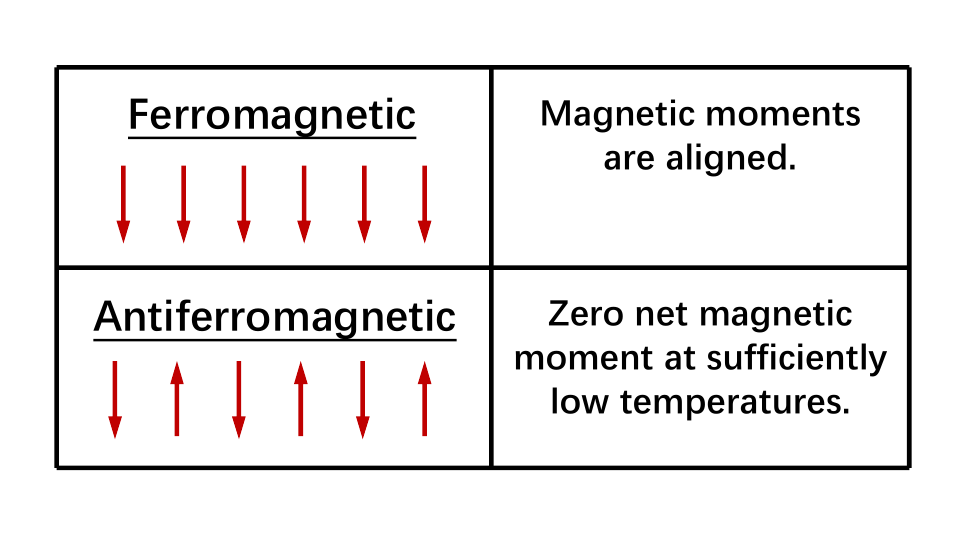}
\end{figure}

Now we consider the \textbf{Ising model}  of a $d$-dimensional lattice $\Lambda$ with  $N$ lattice sites. For each lattice site $k \in \Lambda$, there is a discrete variable $\sigma_k\in  \{+1, -1\}$, representing the site's spin configuration.
The Hamiltonian of the Ising model is given by
\be\label{Ising-Hamiltonian}
\mathcal{H}=- J \sum_{\left\langle i , i^{\prime} \right\rangle} \sigma_{i} \sigma_{i^{\prime}}-B \sum_{i} \sigma_{i}
\ee
where $B$  is the external magnetic field. If $J > 0$, the alignment of spins is energetically favored and the configuration is called \textbf{ferromagnetic}. Conversely, for $J < 0$, the system is in an \textbf{antiferromagnetic} phase. In the following discussion, we focus on the case $J > 0$.

At low temperatures ($T$), the system settles into one of two ground states, leading to a jump discontinuity in the magnetization $M$ across the line of \textbf{first-order} transitions at $B=0$. A first-order phase transition is characterized by a discontinuity in some first derivative of the free energy $F$. In this example, the average energy $E$ is discontinuous.

At high temperatures, thermal fluctuations dominate, and the spins are randomly oriented, resulting in a \textbf{paramagnetic} state. A phase transition occurs at the critical temperature $T = T_c$, known as the \textbf{Curie temperature}. The magnetization, which serves as the \textbf{order parameter} of the Ising model, is discontinuous across this phase transition.

\begin{figure}[ht]\centering
\includegraphics[width=5cm]{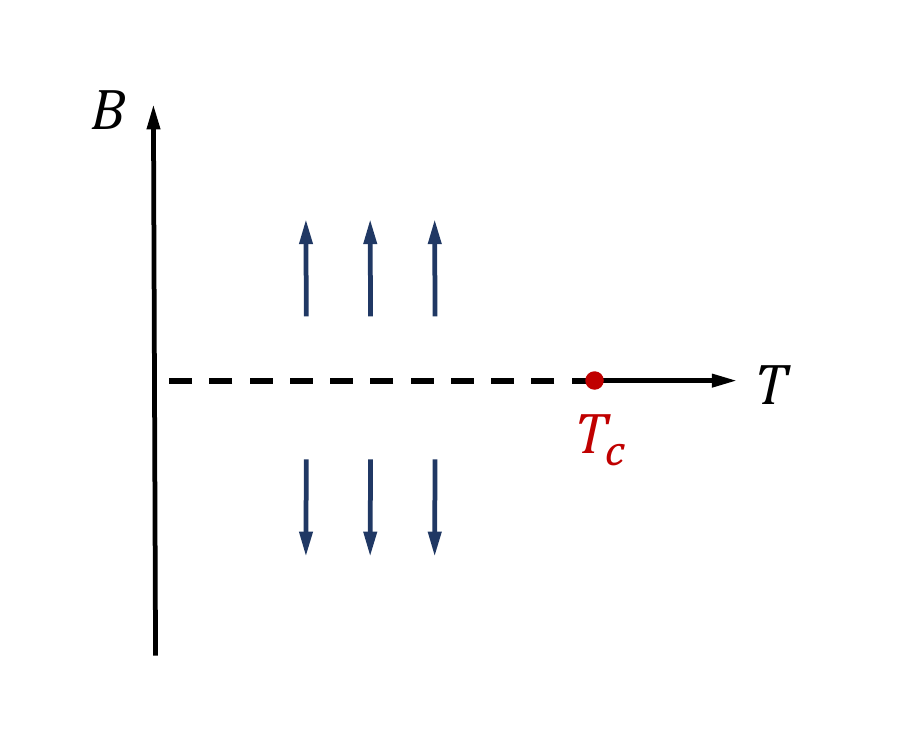}
\includegraphics[width=11cm]{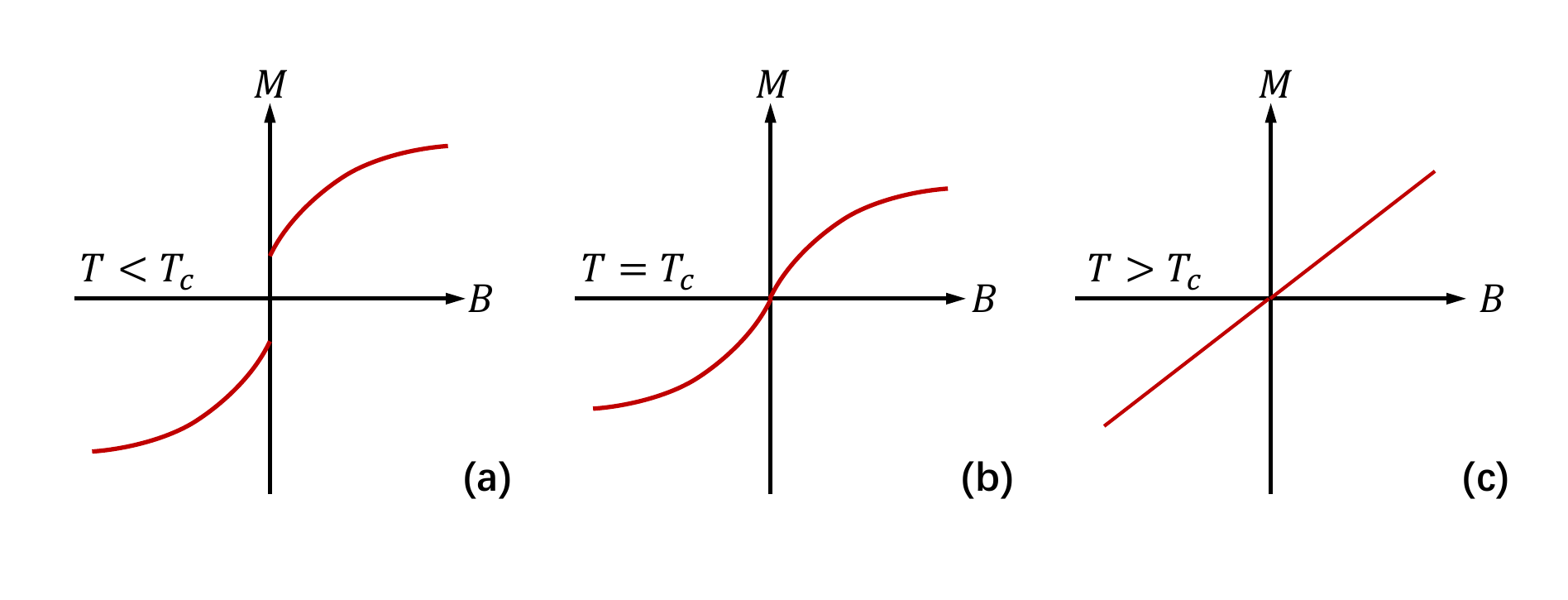}
\end{figure}

\begin{figure}[ht]
\[
\begin{array}{ccc}
\includegraphics[width=.25\textwidth]{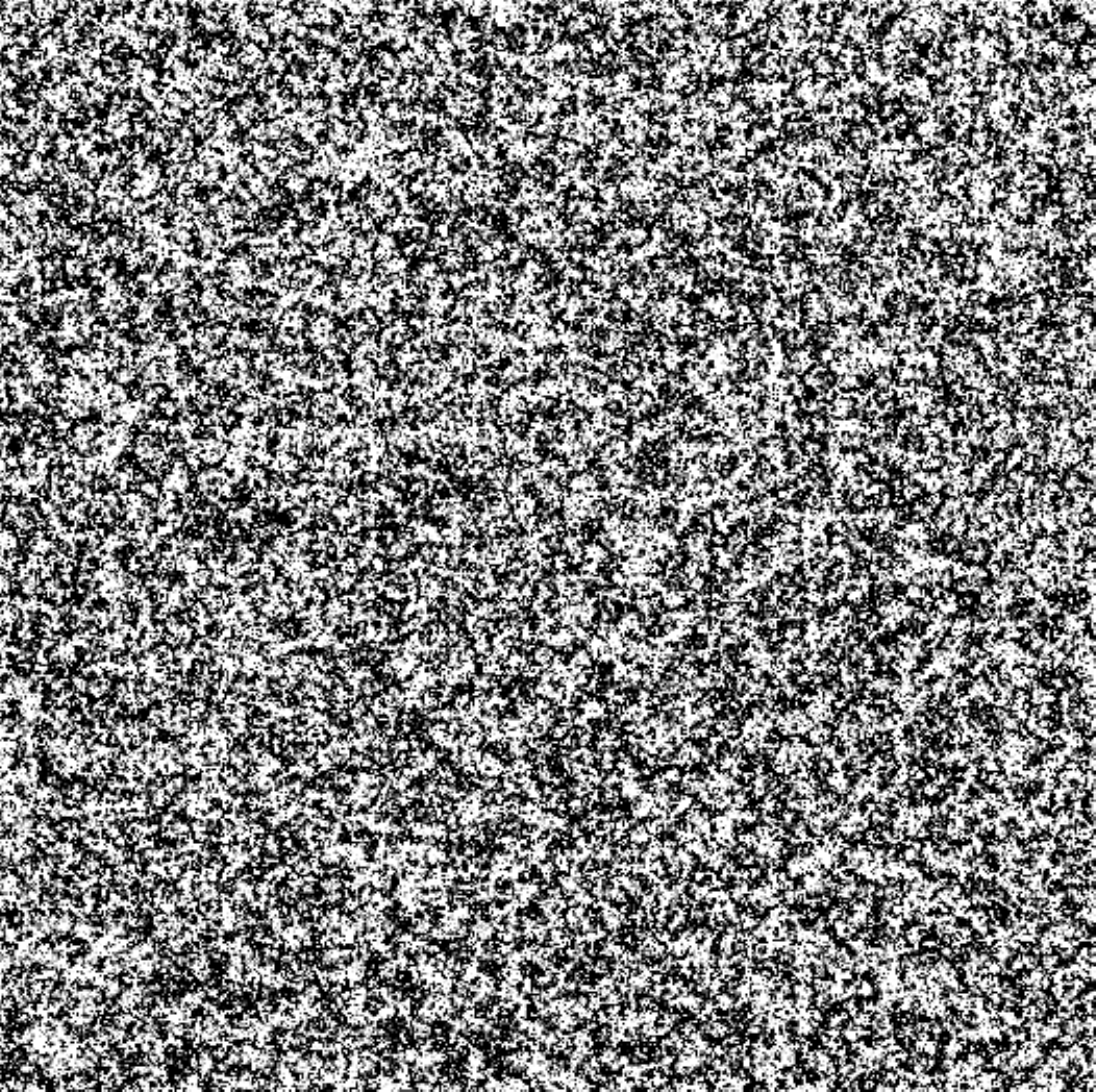} &
\includegraphics[width=.25\textwidth]{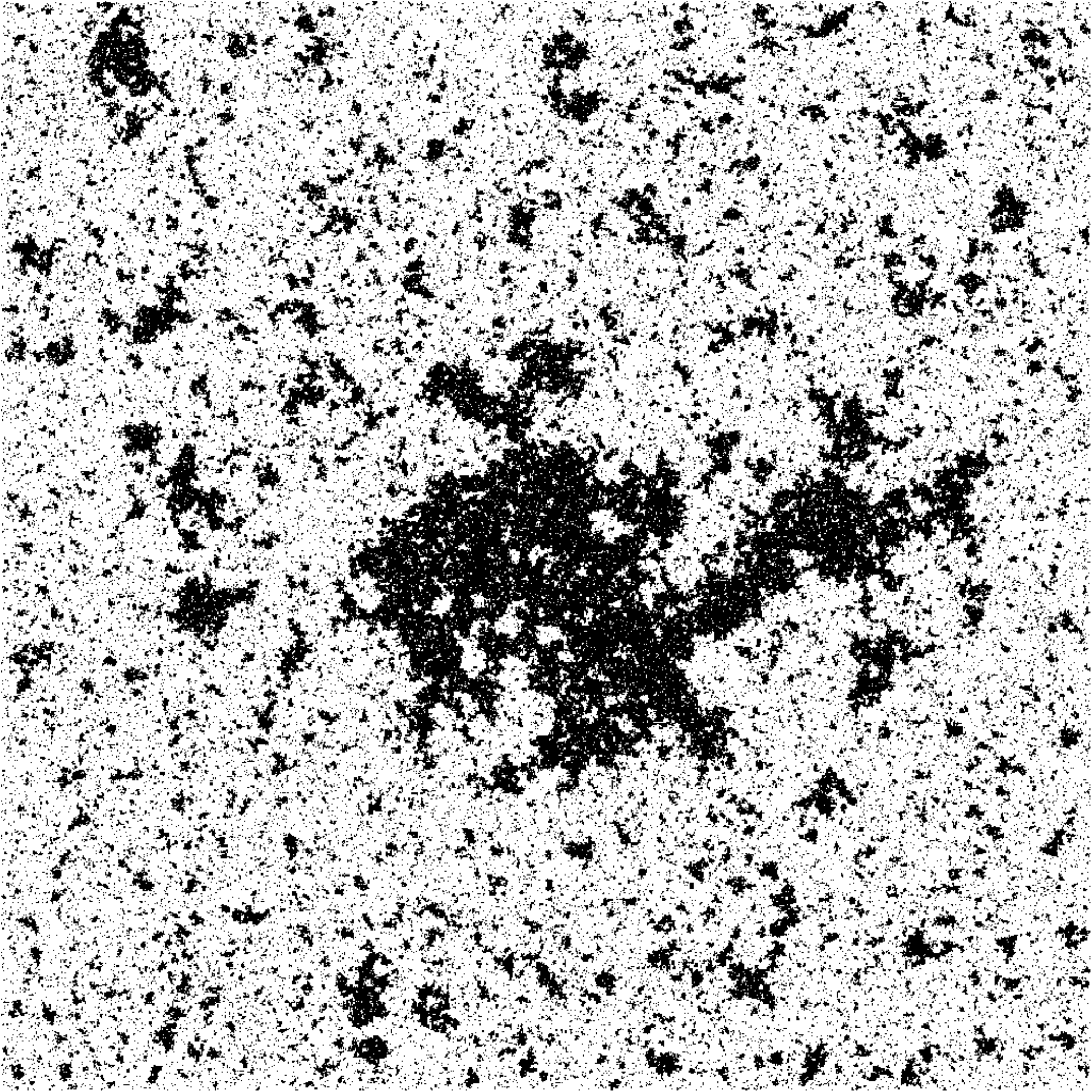} &
\includegraphics[width=.25\textwidth]{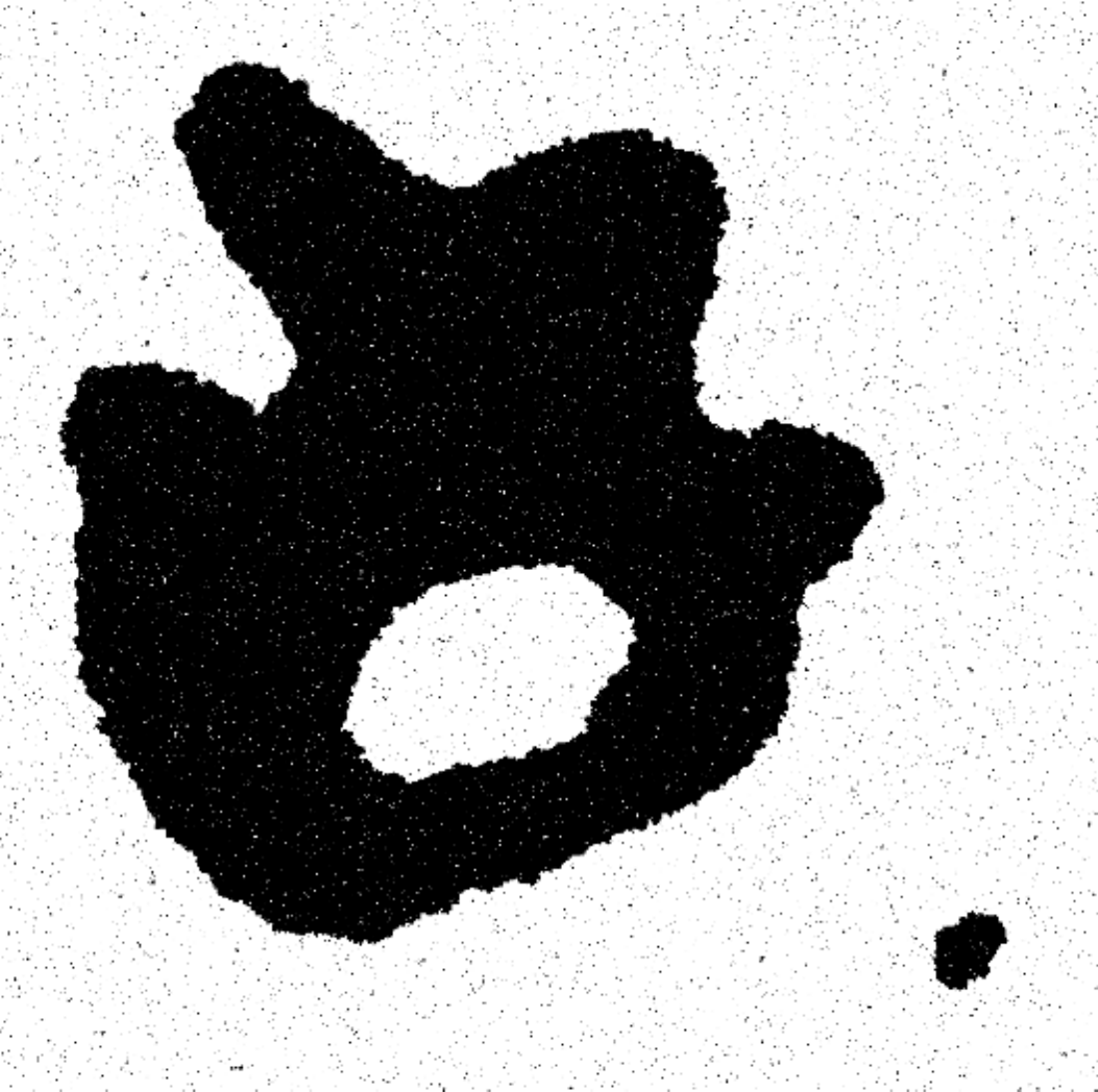}  \\
\text{High temperature} &
\text{Curie temperature} &
\text{Low temperature}
\end{array}
\]
\caption{Temperature dependence of Ising model.}

\[
\begin{array}{ccccc}
\includegraphics[width=.25\textwidth]{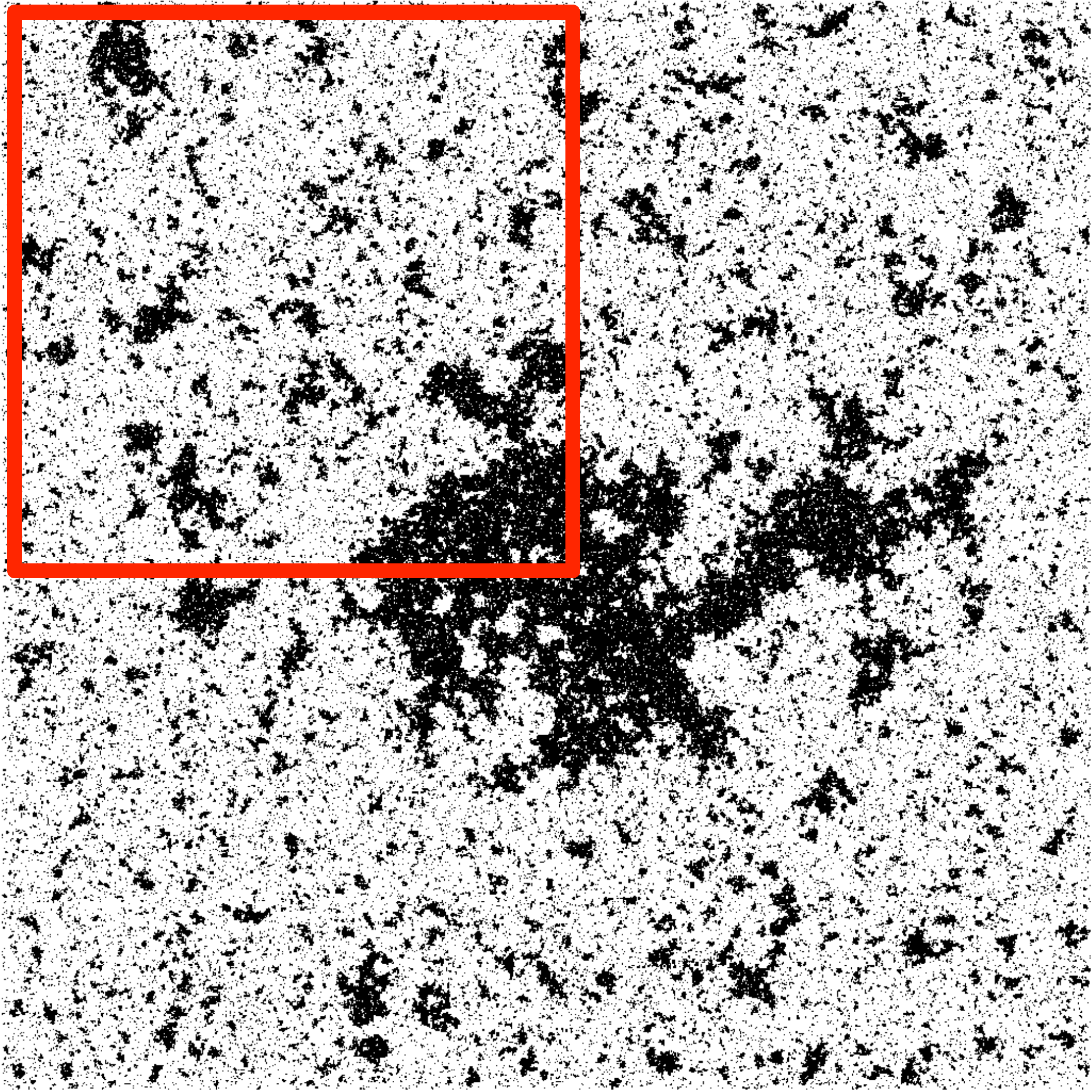} &{\raisebox{2cm}{$\to$}}&
\includegraphics[width=.25\textwidth]{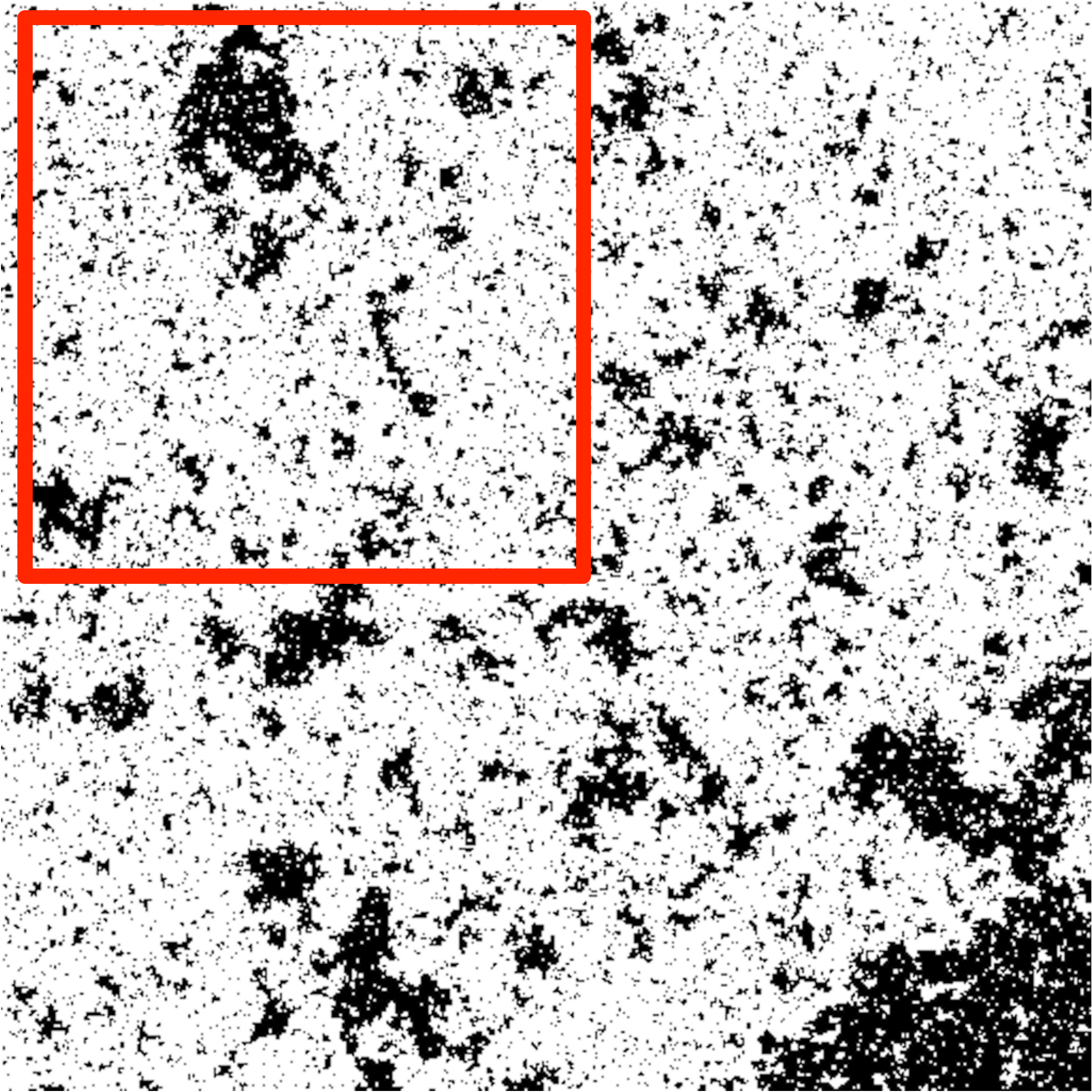} &{\raisebox{2cm}{$\to$}}&
\includegraphics[width=.25\textwidth]{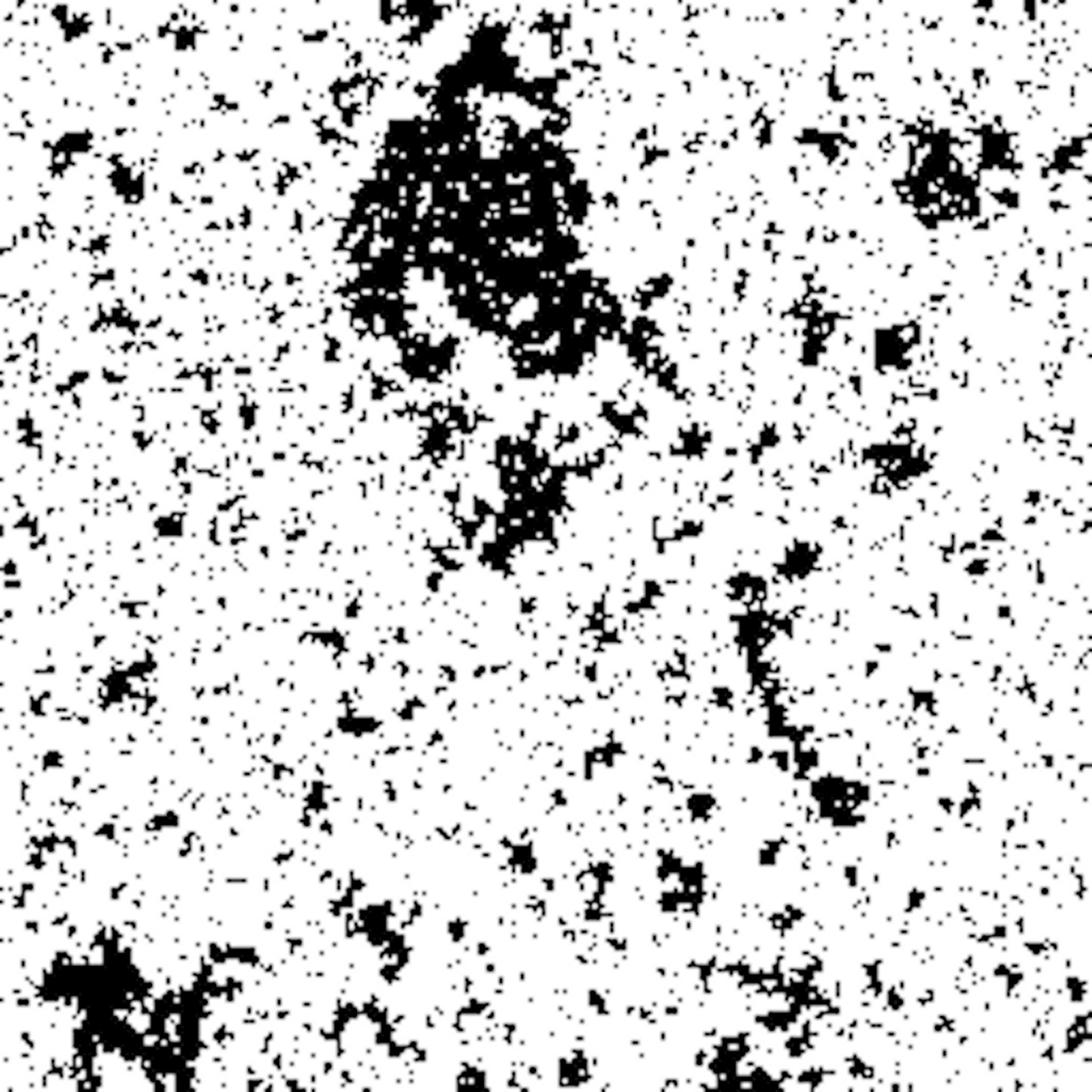}
\end{array}
\]
\caption{At the critical temperature, zooming into the red squares, the pattern does not change so much. Figures are taken from \cite{Tachikawa:2011rp}.}\label{fig:scaling}
\end{figure}

To describe this critical point, we need to introduce a notion of correlation functions.  A two-point correlation function is defined as
\[
G(r)=\left\langle \sigma_{i} \sigma_{j} \right\rangle-\left\langle \sigma_{i} \right\rangle \left\langle \sigma_{j} \right\rangle \approx r^{- \tau} \mathrm{e}^{- r / \xi}
\]
where $r=|i-j|$ is the distance between the sites $i$ and $j$. In the 2d Ising model, one has $\tau =1/2$ for $T>T_c$ and $\tau =2$ for $T<T_c$.
For $r<\xi$, the spins are correlated, indicating a high probability of having the same value. 
On the other hand, for sufficiently large $r\gg \xi$, the probability decreases exponentially with $r$. Hence, the mean cluster size of correlated spins is  $\xi$, which is called the \textbf{correlation length}. As a temperature $T$ approaches $T_c$, the correlation length $\xi$ diverges, and such a critical point is called  \textbf{second-order}. At the critical point,  fluctuations on all length scales become important, resulting in a fractal-like spin configuration  (Figure \ref{fig:scaling})\footnote{Thanks to Yuji Tachikawa for the permission of figure use.}. Consequently, the corresponding mass scale $1/\xi$ disappears, and physics is described by conformal field theory at the critical point \cite[ISZ88-No.2]{Cardy:1984bb}. (See Youtube video \cite{Thucydides411}.)

Near the critical point,  the correlation lengths and the thermodynamic observables obey power laws, and the behavior is characterized by the values of the \textbf{critical exponents}.
Introducing the reduced temperature $t$ and the reduced magnetic field $h$
\begin{equation}\label{t-h}
t :=\frac{T-T_{c}}{T_{c}} \quad \text{and} \quad h :=\frac{B}{T_{c}}~,
\end{equation}
the conventional critical exponents $\a,\b,\g,\d,\nu,\eta$ are now defined by
\begin{equation}\label{critical-exponents}
\begin{matrix}
C \sim t^{-\alpha}&(h=0)&\textrm{specific heat}\\
M \sim (-t)^\beta &(t < 0, h=0)&\textrm{spontaneous magnetization}\\
\chi \sim t^{-\gamma}&(h=0)&\textrm{zero field susceptibility}\\
M=h^{1/\delta} &(t=0)&\textrm{magnetization}\\
\xi \sim t^{-\nu}&(h=0)&\textrm{correlation length}\\
G(r) \sim  |r|^{2-d-\eta}  &(t=0,h=0)&\textrm{2pt-function}\end{matrix}
\end{equation}
where
\[
C :=-\frac{T}{N} \frac{\partial^{2} F}{\partial T^{2}} , \quad M :=-\frac{1}{N} \frac{\partial F}{\partial B} \quad , \quad \chi:=\frac{\partial M}{\partial B}~.
\]
In fact, there are relationships among these critical exponents, leaving only two independent exponents, $\nu$ and $\eta$. For the 2d Ising model, the exponents are
\[
\a=0~,\quad \b=\frac18~,\quad \g=\frac74~, \quad \d=15~, \quad \nu=1~, \quad \eta=\frac14
\]
Remarkably, these critical exponents remain unchanged even if the lattice shape is altered or next-nearest-neighbor interactions are included. This independence from microscopic details is known as \textbf{universality}.

At the critical point, the two-point function can be written as
\[
\langle \sigma(x)\sigma(0)\rangle=\frac{C}{|x|^{2\Delta_\sigma}}
\]
where $\Delta_\sigma$ is called the \textbf{scaling dimension} of $\sigma$.
One can similarly define a local energy density operator
\[
\varepsilon_{i}=\sum_{i^{\prime}} J \left(i , i^{\prime} \right) \sigma_{i} \sigma_{i^{\prime}}~.
\]
Then, in the 2d Ising model, their scaling dimensions are
\be\label{Ising-conf-dim}\Delta_\sigma =\frac18~,\qquad \Delta_\e =1~,\ee
which we will see in the subsequent lectures.

\subsection{String theory}
Conformal field theories (CFTs) play a central role in string theory, serving as the mathematical foundation that underpins the formulation and analysis of string dynamics \cite{GSW,Polchinski}. In string theory, the propagation of strings on a manifold $M$ is described by a nonlinear sigma model $\Sigma \to M$, where $\Sigma$ is a two-dimensional Riemann surface representing the string worldsheet, and $M$ is the target space manifold. For instance, in bosonic string theory, the action is expressed as
\bea\nonumber
 &\cS=\frac{1}{4\pi \alpha'} \int d^2x \left(\sqrt h h^{ab} \partial_a X^\mu \partial_b X^\nu G_{\mu\nu}(X)
 +i \varepsilon^{ab} \partial_a X^\mu \partial_b X^\nu B_{\mu\nu}(X)
 +\alpha' \sqrt h R^{(2)} \Phi(X)
 \right) \ .
\eea
where $h$ is the metric on the worldsheet $\Sigma$, $G_{\mu\nu}(X)$ is the metric on the target space $M$, $B_{\mu\nu}(X)$ is an antisymmetric B-field, and $\Phi(X)$ is the dilaton field. To describe a consistent string theory, the model must satisfy conformal invariance at the quantum level. Specifically, the dimensionality of $M$ is constrained to be 26 for bosonic string theory and 10 for superstring theories with supersymmetry. Furthermore, maintaining conformal invariance at the quantum level requires the vanishing of the one-loop renormalization group beta function, which implies that the target manifold $M$ must be Ricci-flat.

In particular, Calabi-Yau sigma models are described by $\mathcal{N}=2$ superconformal field theories. It was discovered that $\mathcal{N}=2$ superconformal field theories associated with pairs of Calabi-Yau manifolds are equivalent, a phenomenon known in mathematics as \textbf{mirror symmetry}. This equivalence indicates that two Calabi-Yau manifolds, although topologically distinct, yield the same physical theory, leading to significant advances in both mathematics and theoretical physics.

Moreover, the relationship between string theory on anti-de Sitter (AdS) spaces and conformal field theories led to the proposal of the \textbf{AdS/CFT correspondence}. This duality posits that a string theory defined on an AdS space is equivalent to a conformal field theory on the boundary of that space. For instance, the correspondence suggests that the non-perturbative formulation of type IIB superstring theory on AdS${}_5 \times S^5$ is captured by the four-dimensional $\mathcal{N}=4$ superconformal field theory. The AdS/CFT correspondence has been extensively studied over the past decades, significantly impacting various fields of physics, including quantum gravity, particle physics, and condensed matter theory.

\subsection*{References}

There is a vast array of literature on 2d conformal field theories (CFTs), with each resource focusing on specific aspects of this expansive subject. The classical lecture notes include \cite{cardy1988conformal,ginsparg1988applied,zamolodchikov1989conformal}. Additionally, the well-known "yellow book" \cite{francesco2012conformal} serves as an invaluable reference, effectively functioning as a comprehensive dictionary for CFT topics. This lecture note is primarily based on these foundational works, with Sections \ref{sec:conf-inv} and \ref{sec:basics} serving as concise summaries of the corresponding parts in \cite{francesco2012conformal}.

In compiling this note, we have also included several standard references relevant to various related fields. For statistical mechanics, see \cite{cardy1996scaling} and \cite{henkel2013conformal}. String theory is covered in \cite{Blumenhagen:2009zz}, \cite{schellekens1996introduction}, and \cite{ketov1995conformal}. For insights into integrable systems, consider \cite{gomez2005quantum} and \cite{mussardo2010statistical}. 

In the realm of mathematics, the representation-theoretic aspects of 2d CFTs have been extensively explored, notably by Kac \cite{kac1994infinite}. The mathematical foundation for 2d CFTs was established in \cite{segal1988definition,tsuchiya1989conformal} (also included in \cite{jimbo2014integrable}), leading to significant insights into various modern mathematical fields. These areas include vertex operator algebras, moduli spaces, low-dimensional topology, and geometric representation theory. Key references in these domains include \cite{frenkel1989vertex,kohno2002conformal,frenkel2004vertex,beilinson2004chiral}.

For those interested in conformal field theories in higher dimensions, the following references provide comprehensive insights: \cite{Nakayama:2013is,Qualls:2015qjb,Rychkov:2016iqz,Simmons-Duffin:2016gjk}.

Upon completion of this course, it is highly recommended that readers engage with some of the seminal papers collected in \cite{itzykson1988conformal}. These papers are frequently referenced throughout this lecture note. For related yet more mathematically oriented papers, see the collections in \cite{jimbo2012conformal} and \cite{jimbo2014integrable}. 

Even in this fast-moving community, the importance of reading classical works cannot be overemphasized, as they provide the foundational insights and methodologies that continue to influence contemporary research.

\subsection*{Acknowledgments and Disclaimer}

We express our gratitude to the organizers of the Southeast University Summer School for their kind invitation and to the students there for their valuable feedback. We are also deeply thankful to the students of Fudan University who attended the lecture series in 2018 and 2021, whose insightful comments greatly contributed to the improvement of these notes. Special thanks go to Chang Shu for assistance in typing the text and creating the figures.

The literature on two-dimensional conformal field theory (2d CFT) is vast, and it is beyond our scope to provide a comprehensive citation of all relevant works or to accurately trace the historical development of the field. We encourage readers to explore the original literature for a deeper understanding. As noted earlier, many excellent books on 2d CFT already exist, and this note covers only a small subset of this expansive subject. Nonetheless, we hope these notes serve as a concise and useful introduction to the field for students.

\section{Conformal invariance}\label{sec:conf-inv}

This section studies the concept of conformal symmetry in detail. For additional information and in-depth explanations, we recommend consulting \cite[\S4]{francesco2012conformal}, \cite[\S1, \S2]{ginsparg1988applied}, and \cite[\S2]{Blumenhagen:2009zz}.

\subsection{Conformal transformation}

Roughly speaking, a conformal transformation is a mapping between two spaces that preserves angles but not necessarily lengths. Formally, consider a differentiable map \(\phi: U \rightarrow V\), where \(U\) and \(V\) are open subsets of manifolds \(M\) and \(M'\) respectively. Let \(g\) and \(g'\) denote the metric tensors of \(M\) and \(M'\). In differential geometry, we can pull back the metric \(g'\) using the map \(\phi\). A transformation is said to be conformal if this pullback metric \(\phi^{*}g'\) satisfies
\begin{equation}
  \phi^{*}g'=\Lambda g \, .
\end{equation}
If we denote \(x' = \phi(x)\) for \(x \in U\), this condition can be expressed in a covariant form as
\begin{equation}
\label{general-def-ct}
g'_{\rho\sigma}(x')
\pdv{x'^{\rho}}{x^\mu}
\pdv{x'^\sigma}{x^\nu}
=
\Lambda(x)g_{\mu\nu}(x)
\,,
\end{equation}
where \(\Lambda(x)\) ensures that angles between vectors are preserved under the transformation. Notably, in one dimension, every smooth transformation is inherently conformal due to the nature of scalar metrics.

Additionally, from \eqref{general-def-ct}, we can derive that the inner product of two arbitrary vectors \(V'^{\mu}_{1}(x')\) and \(V'^{\mu}_{2}(x')\) in \(M'\) scales under a conformal map:
\begin{equation}
  \label{inner-product-relation}
  g_{\mu\nu}(x)V_1^\mu(x)V_2^\nu(x)
  =
  g'_{\mu\nu}(x')V'^\mu_1(x')V'^\nu_2(x')
  \Lambda
  \, ,
\end{equation}
where \(V^{\mu}_{1}(x)\) and \(V^{\mu}_{2}(x)\) are the pullback vectors. This equation demonstrates that the angle between two vectors in \(M'\) remains unchanged when mapped to \(M\) via a conformal transformation.

In this lecture, we focus on the case where \(M' = M\), implying \(g' = g\). We will consider \(M\) as a flat space with a constant metric \(\eta_{\mu\nu} = \text{diag}(-1, +1, \dots, +1)\). Thus, the condition for a conformal transformation simplifies to
\begin{equation} \label{def-flat-ct}
\eta_{\rho\sigma}
\pdv{x'^\rho}{x^\mu}
\pdv{x'^\sigma}{x^\nu}
=
\Lambda(x)\eta_{\mu\nu}
\, .
\end{equation}
Under a conformal transformation, the line element transforms as follows
\begin{equation}
  \dd s^2
  =
  \eta_{\mu\nu}
  \dd x^\mu \dd x^\nu
  \rightarrow
  \eta_{\rho\sigma}
  \dd x'^\rho \dd x'^\sigma
  =
  \eta_{\rho\sigma}
  \pdv{x'^\rho}{x^\mu}
  \pdv{x'^\sigma}{x^\nu}
  \dd x^\mu \dd x^\nu
  =\Lambda(x)\dd s^2
  \, .
\end{equation}
Two remarks should be stated here. The composition of two conformal transformations is also conformal, suggesting that conformal transformations form a group. Secondly, when \(\Lambda(x) = 1\), the conformal transformations reduce to the Poincaré group transformations, which include translations, rotations, and Lorentz boosts, preserving both angles and lengths.

\subsection{Conditions for conformal invariance}
In this part, we need to figure out some basic conformal transformations in flat space. To begin with, we study the infinitesimal transformations
\begin{equation}  x'^\rho=x^\rho+\e^\rho(x)+\mathcal{O}(\e^2)
 \, ,
\end{equation}
with an infinitesimal variable
$\e(x)\ll1$.
Plugging into the conformal condition
\eqref{def-flat-ct},
up to first order in $\e$, the infinitesimal form of conformal transformation may be derived as:
\begin{equation} \label{inf-constraint-K}   \partial_{\mu}\e_\nu   +   \partial_{\nu}\e_\mu   =   K(x) \eta_{\mu\nu}   \, ,
\end{equation}
with $K(x)=\Lambda(x)-1$, some infinitesimal function. However, since we want to find out the explicit form for $\e(x)$, $K(x)$ needs to be cancelled in the constraint formula. By tracing both sides of \eqref{inf-constraint-K}, we have
\begin{equation}   K(x)   =   \frac{2 \partial^{\mu}\e_\mu}  {d}   \, .
\end{equation}
Plugging back to \eqref{inf-constraint-K}, we find the restriction depending on $\e(x)$ to make the transformation conformal:
\begin{equation} \label{inf-cft-e-condition}   \partial_{\mu}\e_\nu   +   \partial_{\nu}\e_\mu   =   \frac{2}{d}   (\partial\cdot\e)   \eta_{\mu\nu}   \, .
\end{equation}


Let us derive two useful equations for later purposes. First of all, by taking $\partial^{\nu}$ to both sides of \eqref{inf-cft-e-condition}, we can obtain
\begin{equation}   \partial_{\mu}   (\partial\cdot\e)   +   \square\e_\mu   =   \frac{2}{d}   \partial_{\mu}   (\partial\cdot\e)   \, .
\end{equation}
Therefore, only when $d=2$, we will have
\begin{equation}\label{Laplacian-2d}
\square\e_\mu=0~,
\end{equation}
which makes $d=2$ conformal transformations special.
We then take $\partial_{\nu}$ to both sides:
  \begin{equation} \label{inf-cft-e-condition-2}   \partial_{\mu}\partial_{\nu}   (\partial\cdot\e)   +   \square   \partial_{\nu}\e_\mu   =   \frac{2}{d}   \partial_\nu\partial_{\mu}   (\partial\cdot\e)   \, .
\end{equation}
    Interchanging $\mu \leftrightarrow \nu$, adding back to \eqref{inf-cft-e-condition-2} and using \eqref{inf-cft-e-condition}, we get
\begin{equation}
 \label{e-useful-relation-1} ( \eta_{\mu\nu}\square + (d-2) \partial_{\mu}\partial_{\nu} ) (\partial\cdot\e) = 0 \, .
\end{equation}
    Finally, tracing spacetime indices gives
\begin{equation}
 \label{e-useful-relation-2} (d-1)\square (\partial\cdot\e) = 0 \, .
\end{equation}

The second useful expression is obtained by taking $\partial_{\rho}$ of \eqref{inf-cft-e-condition} and permuting indices:
\bea
 \partial_{\rho} \partial_{\mu} \e_\nu + \partial_{\rho} \partial_{\nu} \e_\mu &= \frac{2}{d} \eta_{\mu\nu} \partial_{\rho} (\partial\cdot\e) \, , \notag \\
\partial_{\nu} \partial_{\rho} \e_\mu +  \partial_\mu \partial_\nu \epsilon_\rho &= \frac{2}{d} \eta_{\rho\mu} \partial_{\nu} (\partial\cdot\e) \, , \notag \\
\partial_{\mu} \partial_{\nu} \e_\rho +  \partial_\rho \partial_\mu \epsilon_\nu &= \frac{2}{d} \eta_{\nu\rho} \partial_{\mu} (\partial\cdot\e) \, . \notag
\eea
     Subtracting the first line from the sum of the last two leads to
\begin{equation}
 \label{inf-cft-e-relation-3} 2\partial_{\mu} \partial_{\nu} \e_\rho = \frac{2}{d} ( - \eta_{\mu\nu} \partial_\rho + \eta_{\rho\mu} \partial_{\nu} + \eta_{\nu\rho} \partial_{\mu} ) (\partial\cdot\e) \, .
\end{equation}

    \subsection{Conformal transformation in \texorpdfstring{$d \geq 3$}{d>2}}\label{sec:conf-gen}  For the case $d=1$, the relation \eqref{e-useful-relation-2} imposes no constraint to $\e(x)$, which accords with what we have claimed, that any smooth transformation in one dimension is conformal. The case $d=2$ will be studied later in detail. In this subsection, let us focus on the case $d\ge 3 $.

    According to \eqref{e-useful-relation-2} for $d \geq 3$, we have $\square (\partial\cdot\e)=0$. Substituting back to condition \eqref{e-useful-relation-1} gives $  \partial_{\mu}  \partial_{\nu}  (\partial\cdot\e)  =  0 $ ,implying that $(\partial\cdot\e)$ is at most linear in $x^\mu$, i.e. $  (\partial\cdot\e)  =  A+B_{\mu}x^\mu $ .When plugging this linear expression back to \eqref{inf-cft-e-relation-3}, we immediately obtain that $  \partial_{\mu}  \partial_{\nu}  \e_\rho $ equals to a constant. This follows that $\e_\mu$ is at most quadratic in $ x^{\nu}$ so that we can make the ansatz:
\begin{equation}
 \label{e-ansatz} \e_\mu = a_\mu + b_{\mu\nu} x^\nu + c_{\mu\nu\rho}x^\nu x^\rho \, ,
\end{equation}
    where $ c_{\mu\nu\rho} = c_{\mu\rho\nu} $. For the infinitesimal nature of $\e_\mu$, all the constants: $  a_\mu,b_{\mu\nu},c_{\mu\rho\nu}  \ll 1 $.

    Noticing that all the terms in \eqref{inf-cft-e-condition} have only one derivative and that the conformal constraints for these constants should be independent of $x^\mu$, we can study the various terms in \eqref{e-ansatz} separately.  The constant term $a_\mu$ in \eqref{e-ansatz} is not constrained by \eqref{inf-cft-e-condition}. It describes infinitesimal translation $x'^\mu=x^\mu+a^\mu$, for which the generator is the momentum operator $P_\mu=-i\partial_{\mu}$ \footnote{ 	All the generators of this form can be calculated by definition:
\begin{equation}
 	  \label{def-generator} 	  iG_a\Phi 	  = 	  \frac 	 {\delta x^\mu} 	 {\delta \omega_a} 	  \partial_{\mu} 	  \Phi 	  \, , 	\end{equation}
    	if we suppose that the fields are not affected by the transformation with $\omega_a$ the corresponding parameter. }.  Inserting the linear term into \eqref{inf-cft-e-condition} , we find
\begin{equation}
 \label{linear-condition}   b_{\mu\nu} + b_{\nu\mu} = \frac{2}{d} ( \eta^{\rho\sigma} b_{\sigma\rho} ) \eta_{\mu\nu} \, .
\end{equation}
    We then split $b_{\mu\nu}$ into symmetric and antisymmetric parts, i.e. $b_{\mu\nu}=n_{\mu\nu}+m_{\mu\nu}$, with $n_{\mu\nu}=n_{\nu\mu}$ and $m_{\mu\nu}=-m_{\nu\mu}$. The antisymmetric part $m_{\mu\nu}$ satisfies \eqref{linear-condition} automatically, while the symmetric part gives constraint: $  n_{\mu\nu}  =  \frac{1}{d}  (   \eta^{\rho\sigma}   n_{\sigma\rho} )  \eta_{\mu\nu} $ , telling that the symmetric part should be diagonal. Therefore, we can split $b_{\mu\nu}$ as
\begin{equation}
 	b_{\mu\nu} 	= 	\alpha\eta_{\mu\nu} 	+ 	m_{\mu\nu} 	\, .
\end{equation}
    The symmetric term $\alpha\eta_{\mu\nu}$ describes infinitesimal scale transformation $  x'^\mu  =  (1+\alpha)x^\mu $ with generator $   D= -ix^\mu\partial_\mu $. The antisymmetric part $m_{\mu\nu} $ corresponds to infinitesimal rotations $   x'^\mu   =   (    \delta^\mu_\nu    +   {m^\mu}_\nu  )   x^\nu $ with generator being the angular momentum operator $   L_{\mu\nu}   =   i(   x_\mu\partial_\nu   -   x_\nu\partial_\mu  ) $.

  Now let us consider the quadratic term. By inserting the quadratic term into \eqref{inf-cft-e-relation-3}, after straight-forward calculation, we find that
\begin{equation}
   c_{\mu\nu\rho}   =   \eta_{\rho\mu}f_\nu   +   \eta_{\mu\nu}f_\rho   -   \eta_{\nu\rho}f_\mu   \, ,
\end{equation}
    with $f_\mu=\frac{1}{d}{c^\rho}_{\rho\mu}$. The resulting transformations are
\begin{equation}
 \label{inf-SCT}   x'^\mu = x^\mu + 2(x\cdot f)x^\mu -(x\cdot x)f^\mu \, ,
\end{equation}
    which are called \textbf{special conformal transformations} (SCT) and the corresponding generator is $  K_\mu =  -i(      2 x_\mu x^\nu\partial_\nu      -      (x\cdot x)\partial_{\mu} ) $ .

    \subsection{Finite conformal transformation} 
\begin{figure}[ht] \centering \includegraphics [width=0.7\linewidth] {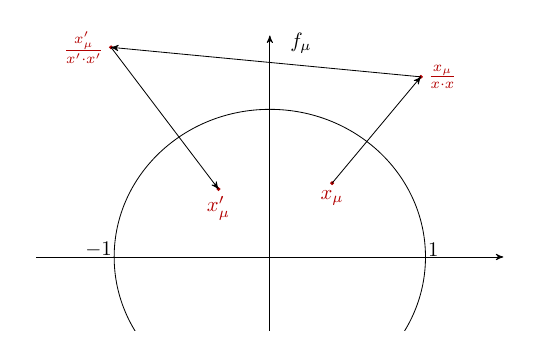} \caption { Illustration of a finite Special Conformal Transformation } \label{fig:SCT-ill}
\end{figure}

So far, we have identified all basic infinitesimal conformal transformations and their generators. However, in order to determine the conformal group, we also need the corresponding finite transformations. All the basic finite conformal transformations are shown in Table \ref{Tab:fin-ct}.
\begin{table}[htbp] 	\centering
  \begin{tabular}{lll} \toprule   Transformations & & Generators \\
\midrule   translation & $x'^\mu=x^\mu+a^\mu$ & $P_\mu=-i \partial_\mu$ \\
  dilatation & $x'^\mu= \alpha x^\mu$ & $D=-ix^\mu    \partial_\mu   $ \\
  rotation & $x'^\mu    ={M^\mu}_\nu x^\nu   $ & $L_{\mu\nu}=    i(    x_\mu\partial_\nu    -    x_\nu\partial_\mu   )   $ \\
  SCT & $x'^{\mu}=    \frac   {x^\mu-(x\cdot x)f^\mu}   {     1-2(f\cdot x)     +     (f\cdot f)(x\cdot x)}   $ & $K_\mu=-i(    2 x_\mu x^\nu\partial_{\nu}    -    (x\cdot x)\partial_\mu   )   $ \\
\bottomrule
\end{tabular}
	 \caption{Finite conformal transformations 	 and corresponding generators} \label{Tab:fin-ct}
\end{table}
 The finite transformations for translation and rotation are Poincare transformations we are familiar with. The finite transformation for dilatation can be derived from its infinitesimal form easily. Thus, what we focus on is the finite transformation for SCT.  In fact, we can guess this finite form of transformation if we realize that an inversion $x'^\mu=\frac{x^\mu}{x\cdot x}$ with respect to the unit sphere is also a conformal map. However, it has neither infinitesimal form nor parameters. Using the transitive property for a conformal map, we can thus first make an inversion of $x^\mu$, then translate by $f^\mu$, and inverse the result again to construct a new conformal transformation. An illustration in two dimensions is shown in Figure \ref{fig:SCT-ill}.
\begin{equation}
 \label{fin-SCT} x'^\mu =\frac {   \frac{x^\mu}{x\cdot x}   -f^\mu } {   (\frac{x^\mu}{x\cdot x}   -f^\mu)   (\frac{x^\mu}{x\cdot x}   -f^\mu) } =\frac {x^\mu-(x\cdot x)f^\mu} {  1-2(f\cdot x)  +  (f\cdot f)(x\cdot x) } \, .
\end{equation}
  It is easy to verify that when $f^\mu \rightarrow 0$, \eqref{fin-SCT} reduces to the infinitesimal form of SCT \eqref{inf-SCT}. Furthermore, by using \eqref{def-flat-ct},  the conformal factor for SCT can be computed as
\begin{equation}
   \Lambda(x)   =   (1-2(f\cdot x)+ (f\cdot f)(x\cdot x))^{-2}
\end{equation}
Finally, we observe from \eqref{fin-SCT} that a point $x^\mu=\frac{1}{f\cdot f}f^\mu$ is mapped to infinity. Therefore, in order to define the arbitrary conformal transformation, the point at infinity should be included.

    \subsection{Conformal groups and algebras}  The conformal transformations form a group, called the conformal group of dimension $d\ge3$. Actually, the algebra generated by infinitesimal conformal transformations is the Lie algebra of the conformal group, called the conformal algebra.

    Let us identify the conformal algebra  of dimension $d\ge3$. The dimension of the algebra is $N=d+1   +\frac{d(d-1)}{2}   +d   =\frac{(d+2)(d+1)}{2} $ by counting the generators in Table \ref{Tab:fin-ct}. The commutation rules with no central charge can be obtained by direct calculations, and non-trivial commutation relations are
\bea
 \label{com-rela-conf} \comm{D}{P_\mu} &= i P_\mu ~,\cr \comm{D}{K_\mu} &= -i K_\mu ~,\cr \comm{K_\mu}{P_\nu} &= 2i( \eta_{\mu\nu}D - L_{\mu\nu} ) ~,\cr \comm{K_\rho}{L_{\mu\nu}} &= i( \eta_{\rho\mu}K_\nu - \eta_{\rho\nu}K_\mu ) \, ,\\
\comm{P_\rho}{L_{\mu\nu}} &= i( \eta_{\rho\mu}P_\nu - \eta_{\rho\nu}P_\mu ) ~,\cr \comm{L_{\mu\nu}}{L_{\rho\sigma}} &= i( \eta_{\rho\nu}L_{\mu\sigma} + \eta_{\mu\sigma}L_{\nu\rho} - \eta_{\rho\mu}L_{\nu\sigma} - \eta_{\nu\sigma}L_{\mu\rho} )\, .
\eea
     The others commute. To put the above commutation relations into a simpler form, we define the following generators.
\bea
    &J_{\ \mu,\nu}    =    L_{\mu,\nu}    \, ,\cr
&J_{-1,0} = D \, ,\cr
&J_{-1,\mu} = \frac{1}{2} (P_\mu-K_\mu) \, ,\\
&J_{\ 0,\mu} = \frac{1}{2} (P_\mu+K_\mu) \, .\notag
\eea
     By using \eqref{com-rela-conf}, we can verify that $J_{m,n}$ with $m,n=-1,0,1,\dots d$, satisfy the following commutation relations:
\begin{equation}
 \label{simper-conf-algebra} \comm{J_{mn}}{J_{rs}} = i( \eta_{ms}J_{nr} + \eta_{nr}J_{ms} - \eta_{mr}J_{ns} - \eta_{ns}J_{mr} ) \, .
\end{equation}
    For Euclidean $d$-dimensional space $\mathbb{R}^{d}$, we see that the metric $\eta_{mn}$ here is $\eta_{mn} =$ $\text{diag}(-1,1,\dots,1)$ so that we identify \eqref{simper-conf-algebra} with the commutation relations of the Lie algebra $\mathfrak{so}(1,d+1)$. We can generalize the result to the space $\mathbb{R}^{p,q}$, with $d=p+q \geq 3$. The corresponding conformal group is $SO(p+1,q+1)$.

    \subsection{2d conformal transformation}\label{sec:2d-conf-trans} In this section, we will focus on the special case of two dimensions with the Euclidean metric. From \eqref{inf-cft-e-condition}, the infinitesimal condition for conformal transformation in two dimensions reads as follows:
\begin{equation}
 \label{C-R equation} \partial_1\e_1 = \partial_2\e_2 \, , \qquad \partial_1\e_2 = -\partial_2\e_1 \, .
\end{equation}
These are indeed the Cauchy-Riemann equations in complex analysis. Thus, if we introduce complex coordinates
\bea
 \label{coordinate transformation} z= x^1+i x^2 \,  ,\qquad \e=\e^1+i \e^2 \,  ,\cr \overline{z}= x^1-ix^2 \,  ,\qquad \overline{\e}= \e^1-i \e^2 \, ,
\eea
\eqref{C-R equation} tells us that $\e(z)$ is a holomorphic function (in some open set). Equivalently any infinitesimal holomorphic transformation $z'=z+ \e(z)$ gives rise to a two-dimensional conformal transformation.
This can be seen from \eqref{Laplacian-2d} because the Laplacian is $\partial \overline\partial \e_\mu=0$ in the complex coordinates.

\begin{figure}[ht] \centering \subfigure[$z$]{ \includegraphics[width=0.3\textwidth]{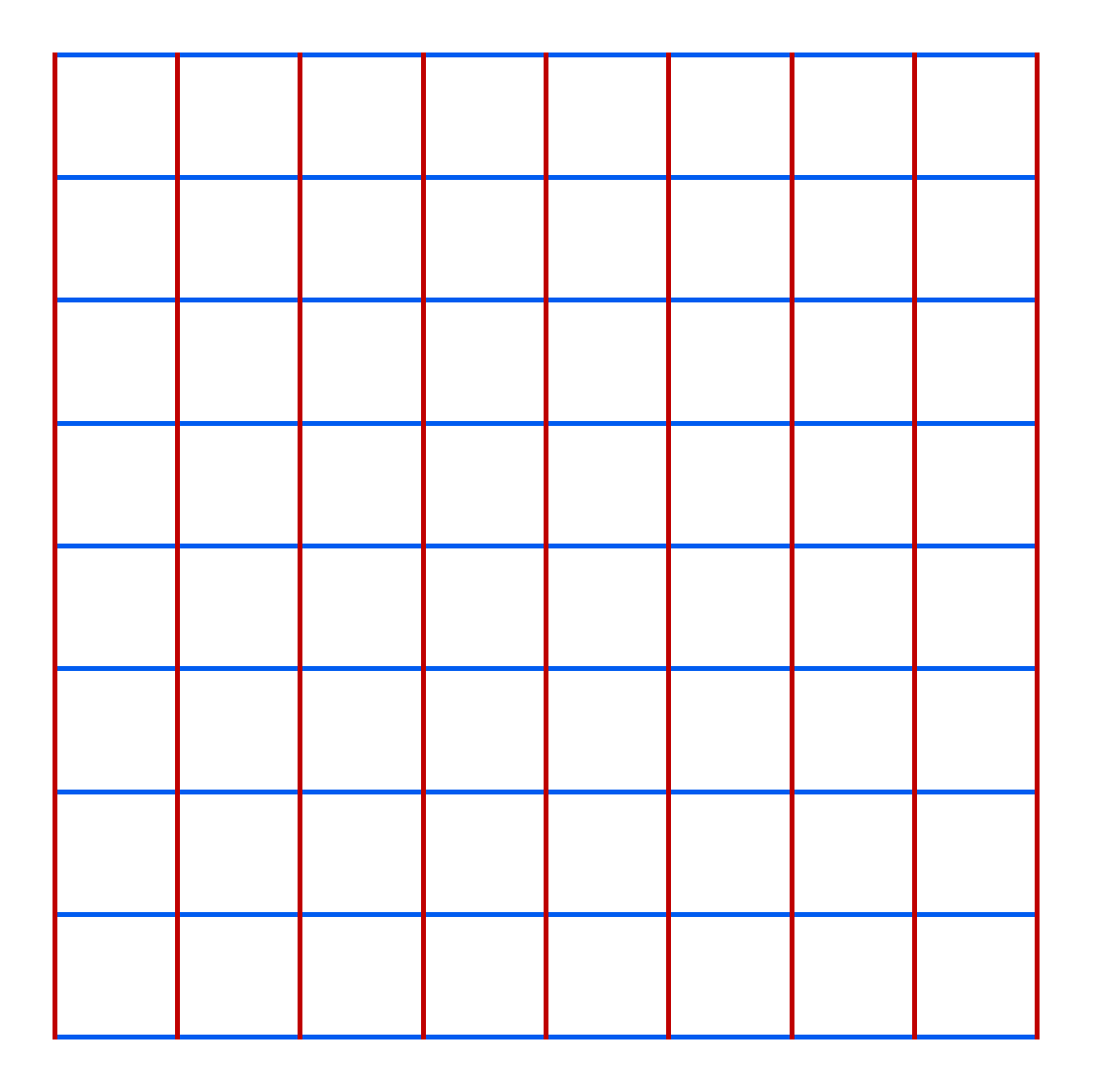}} \subfigure[$z'=z^2$]{ \includegraphics[width=0.3\textwidth]{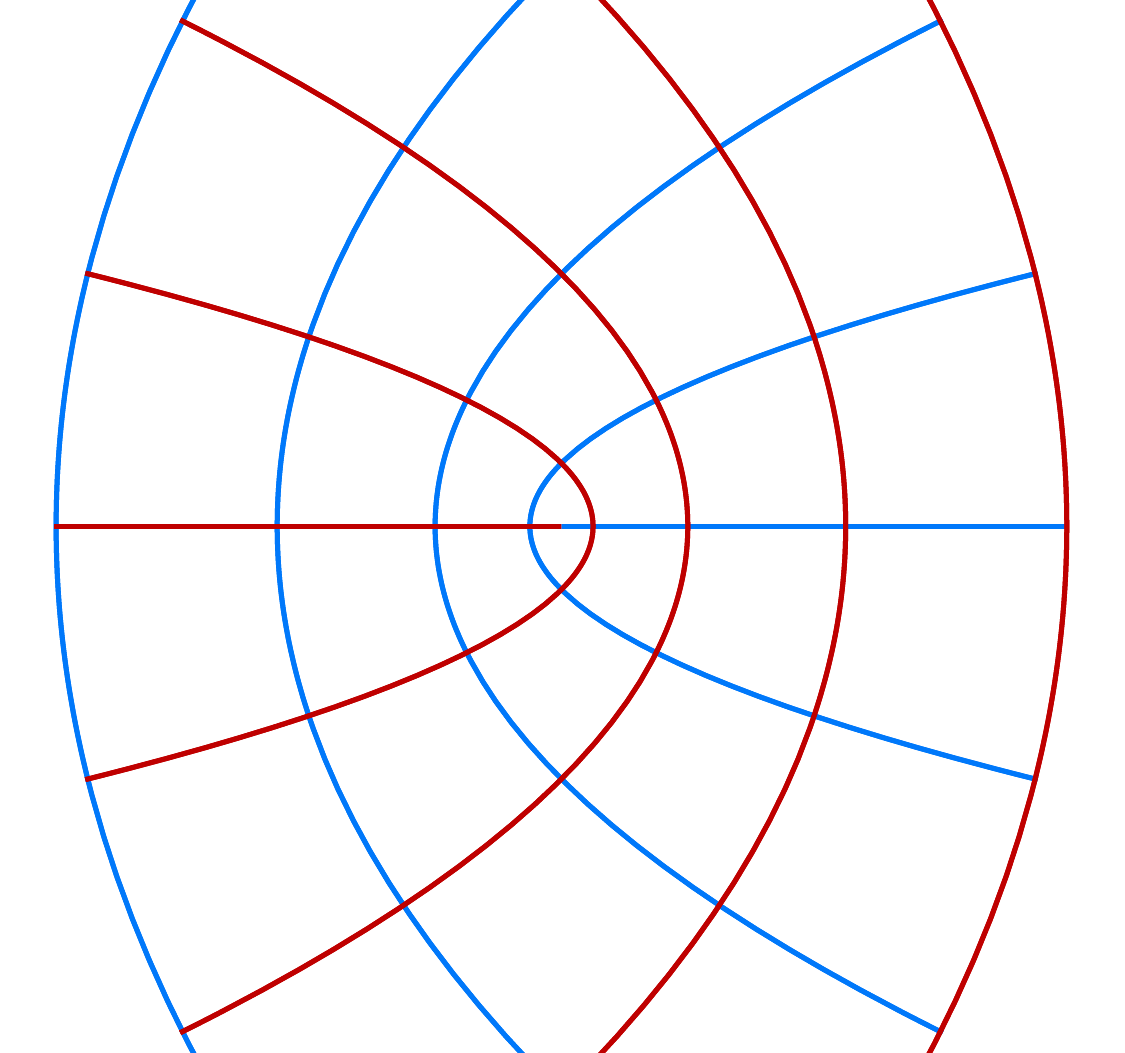}} \subfigure[$z'=1/z$]{ \includegraphics[width=0.3\textwidth]{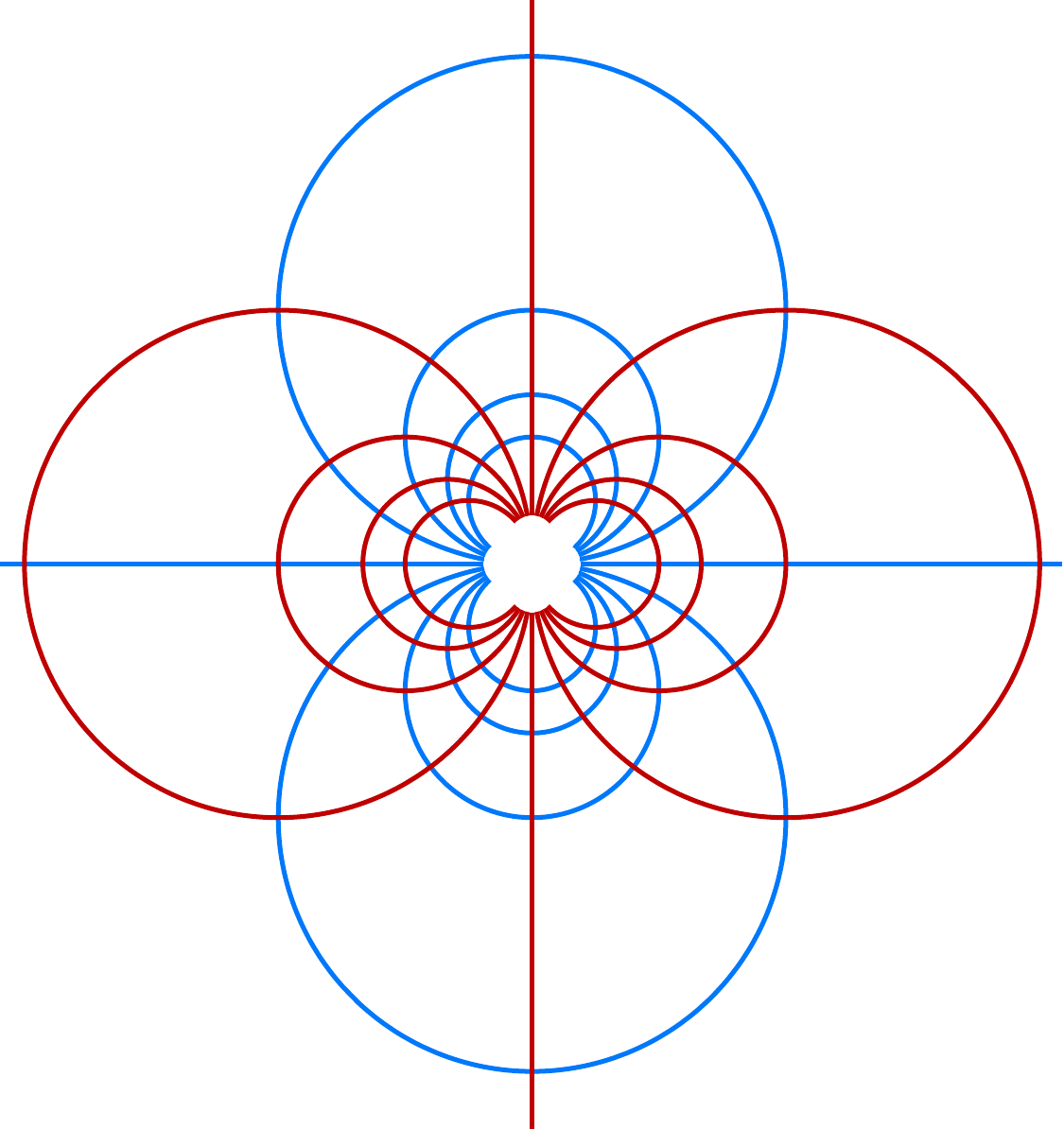}} \subfigure[$z'=z\abs{z}$]{ \includegraphics[width=0.3\textwidth]{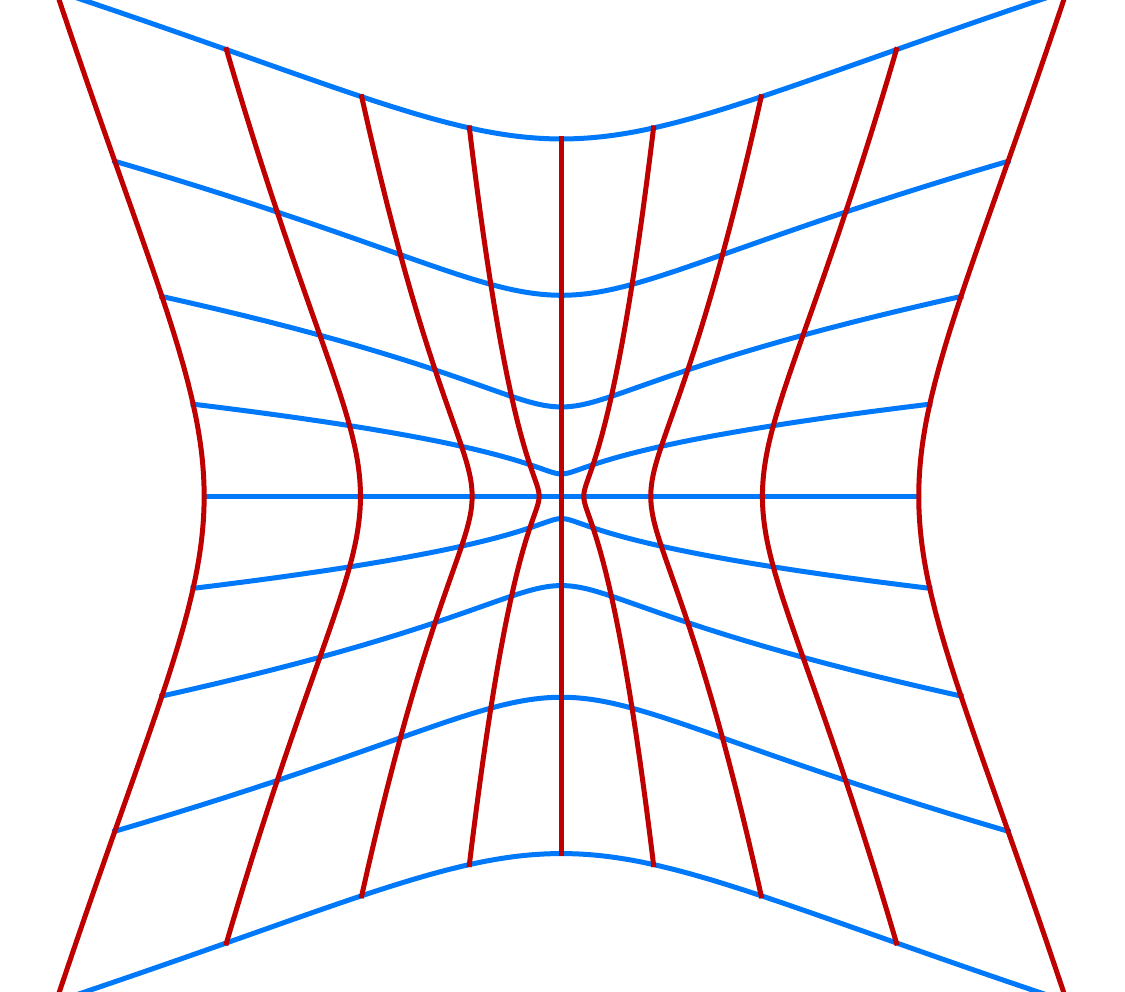}}
    \caption{Coordinate transformation: The transformation from square lattice (a), onto the lattice in (b) and (c) is conformal, while the transformation onto the lattice in (d) is not.} \label{fig:2dcft-ill}
\end{figure}

     In fact, any holomorphic function on the complex plane is conformal (i.e, preserves angles). To prove this, we should rewrite the definition of a conformal transformation \eqref{def-flat-ct} with complex variables. Let us first, rewrite the line element:
\begin{equation}
 ds^2=(\dd x^1)^2+(\dd x^2)^2   =\dd z \dd \overline{z}   \, .
\end{equation}
Therefore, in terms of complex coordinates, the metric can be written as
\begin{equation}
 \label{metric-complex-space} g_{\alpha\beta} = \mqty(    0 & \frac{1}{2}\\
   \frac{1}{2} & 0   ) \, , \qquad g^{\alpha\beta} = \mqty(    0 & 2 \\
   2 & 0   ) \, .
\end{equation}
    The condition \eqref{def-flat-ct} now becomes:
\begin{equation}
 \label{def-flat-ct-complex} g_{\alpha\beta} \pdv{z'^\alpha}{z^\gamma} \pdv{z'^\beta}{z^\delta} = \Lambda(z,\overline{z})g_{\gamma\delta} \, ,
\end{equation}
    which is satisfied, if $z'=f(z)$ is a holomorphic function. The conformal factor is then $\Lambda(z,\overline{z}) = \abs{\pdv{f(z)}{z}}^2$. Figure \ref{fig:2dcft-ill} shows some examples on conformal transformations. We can see that an angle at any point is preserved when $z'=f(z)$ is holomorphic.

\subsubsection*{Witt algebra}  As we have seen above, for an infinitesimal conformal transformation in two dimensions, $\e(z)$ has to be holomorphic in some open set. We can perform a Laurent expansion of $\e(z)$ locally around $z=0$. A general infinitesimal conformal transformation thus can be written as
\bea
 \label{2d-inf-Laurant} z'&=z+\e(z)  =z +\sum_{n\in\mathbb{Z}}  \e_n(-z^{n+1})  \notag \, , \\
\overline{z}' &=z+ \overline{\e}(\overline{z})  =\overline{z} + \sum_{n\in\mathbb{Z}}  \overline{\e}_n  (-\overline{z}^{n+1})  \, ,
\eea
     where $\e_n$ and $\overline{\e}_n$ are infinitesimal parameters. Supposing a scalar and dimensionless field $\phi(z,\overline{z})$ living on the plane, the effect of such a mapping would be:
\begin{equation}
 \label{tran-scalar-field} \phi(z,\overline{z}) \rightarrow \phi'(z',\overline{z}') = \phi(z,\overline{z}) \, ,
\end{equation}
    or
\bea
 \delta{\phi} &\equiv \phi'(z,\overline{z}) -\phi(z,\overline{z}) =\phi'(z'-\e(z),\overline{z}'-\bar\e(\overline{z})) -\phi(z,\overline{z}) \cr
&= -\e(z)\partial\phi -\bar\e(\overline{z})\overline{\partial}\phi +\mathcal{O}(\e^2,\bar\e^2) = -\sum_{n\in\mathbb{Z}} (\e_n l_n \phi(z,\overline{z}) + \bar\e_n \overline{l}_n \phi(z,\overline{z}) ) \, ,
\eea
     where we have introduced the generators
\begin{equation}
 l_n=- z^{n+1}\partial\,, \qquad \overline{l}_n=-\overline{z}^{n+1} \bar\partial \, .
\end{equation}
    We find that the number of independent infinitesimal conformal transformations is infinite. The commutators for these local generators can be calculated straightforwardly:
\bea
 \label{Witt-algebra} \comm{l_n}{l_m} &= (n-m) l_{n+m} \, ,\cr \comm{\overline{l}_n}{\overline{l}_m} &= (n-m)\overline{l}_{n+m} \, , \cr
\comm{l_n}{\overline{l}_m} &=0 \, .
\eea
     The first commutation relations define one copy of the so-called Witt algebra. We can observe that the local conformal algebra \eqref{Witt-algebra} is the direct sum of two Witt algebras.

\subsubsection*{M\"obius transformation} A global conformal transformation in two dimensions is defined to be invertible and well-defined on the Riemann sphere $S^2 \simeq \mathbb{C}\cup \{\infty\}$. Consider the Laurent expansion:
\[
l_n = -z^{n+1}\partial_z
\]
where \( z \) is a complex coordinate. The operator \( l_n \) is nonsingular at \( z = 0 \) only for \( n \geq -1 \). To analyze behavior at infinity, we perform a change of variable \( z = -{1}/{w} \). Under this transformation, the generator \( l_n \) becomes:
\[
l_n = \left(-\frac{1}{w}\right)^{n-1}\partial_w
\]
which is nonsingular at \( w = 0 \) only for \( n \leq 1 \).
Thus, the globally defined conformal transformations on the Riemann sphere are generated by \( l_{-1} \), \( l_0 \), and \( l_1 \). These three generators form a closed subalgebra of the Witt algebra \eqref{Witt-algebra}, defined globally on the Riemann sphere. The discussion of the antiholomorphic part is analogous.

Let us examine the global conformal transformation of the sphere. Using the Laurent expansion \( l_n = -z^{n+1}\partial_z \), we observe the following:
\begin{itemize} 
    \item \textbf{Translation:} The operator \( l_{-1} = -\partial_z \) generates the infinitesimal transformation \( z' = z - \epsilon \), where \( \epsilon \) is an infinitesimal constant. This corresponds to a translation \( z \mapsto z + b \), with \( b \) being an arbitrary complex constant.
\item\textbf{Dilatation and Rotation:} For \( n = 0 \), the operator \( l_0 = -z\partial_z \) generates the linear transformation \( z' = az \), where \( a \) is an arbitrary complex number. The modulus of \( a \) corresponds to a dilatation, while the argument (phase) of \( a \) corresponds to a rotation. To gain geometric intuition, we perform a change of variables \( z = re^{i\theta} \), leading to:
  \be \label{l0}
  l_0 = -\frac{1}{2}r\partial_r + \frac{i}{2}\partial_\theta \quad \text{and} \quad \bar{l}_0 = -\frac{1}{2}r\partial_r - \frac{i}{2}\partial_\theta
  \ee 
  Consequently, the linear combinations
  \be \label{l0combinations}
  l_0 + \bar{l}_0 = -r\partial_r \quad \text{and} \quad i(l_0 - \bar{l}_0) = -\partial_\theta
  \ee
  reveal that \( l_0 + \bar{l}_0 \) generates dilatations, and \( i(l_0 - \bar{l}_0) \) generates rotations.
  \item\textbf{Special Conformal Transformation:} The operator \( l_1 \) generates the infinitesimal quadratic transformation. By changing the variable \( w = -\frac{1}{z} \), we find \( l_1 = -\partial_w \), corresponding to a translation in \( w \): \( w \mapsto w - c \). Thus, \( l_1 \) generates a special conformal transformation given by \( z \mapsto \frac{z}{cz + 1} \).
\end{itemize}

The complete set of mappings discussed above is of the form
\begin{equation}
 z\mapsto \frac{a z+b}{cz+d} \, ,
\end{equation}
    with $a,b,c,d \in \mathbb{C} $. For this transformation to be invertible, we have to require that $ad-bc \neq 0$. If this is the case, we can scale the constants $a,b,c,d$, such that $ad-bc =1$. Furthermore, the expression is unaffected by taking all of $a,b,c,d$ to minus themselves. These mappings are called projective transformations. We see each global conformal transformation in two dimensions can be associated with a matrix $\mqty(a & b \\
c & d)$. For instance:
\bea
 &\text{translation}:\qquad z\mapsto z+b \quad \Leftrightarrow \quad \mqty(1 & b \\
0 & 1) \, ,\cr
&\text{dilatation}:\quad z\mapsto a z\, , a\in \mathbb{R} \, \Leftrightarrow \, \mqty(\sqrt{a} & 0\\
    0 & 1/\sqrt{a}   ) \, ,\cr
&\text{rotation}: \qquad z\mapsto e^{i\theta}z\, , \theta\in \mathbb{R} \, \Leftrightarrow \, \mqty(e^{i\theta/2} & 0\\
    0 & e^{-i\theta/2}   ) \, ,\cr
&\text{SCT}: \qquad\qquad z\mapsto \frac{z}{c z +1} \quad \Leftrightarrow \quad \mqty(1 & 0 \\
    c & 1) \, .\notag
\eea
     We can infer that the conformal group of the Riemann sphere $S^2=\mathbb{C} \cup \{\infty \}$ is the  M\"obius group $SL(2,\mathbb{C})/\mathbb{Z}_2$.

\section{Basics in conformal field theory}\label{sec:basics}

Conformal field theories are quantum field theories invariant under conformal transformations. This fundamental property leads to powerful constraints on the dynamics and structure of the theory, especially in two dimensions where the conformal symmetry is infinite-dimensional. In this section, we explore how these constraints manifest and significantly enhance our understanding of 2d systems. For a comprehensive treatment, refer to \cite[\S5]{francesco2012conformal}, \cite[\S3]{ginsparg1988applied}, and \cite[\S2]{Blumenhagen:2009zz}.

    \subsection{Noether theorem and Ward-Takahashi identity} 

In this subsection, we review the importance of symmetry in quantum field theory, emphasizing the implications of continuous symmetries as articulated by Noether's theorem and the Ward-Takahashi identity.

\subsubsection*{Continuous symmetry transformation} 

\begin{figure}[ht] \centering \caption{Rotation transformation of a vector field.} \includegraphics[width=0.5\linewidth]{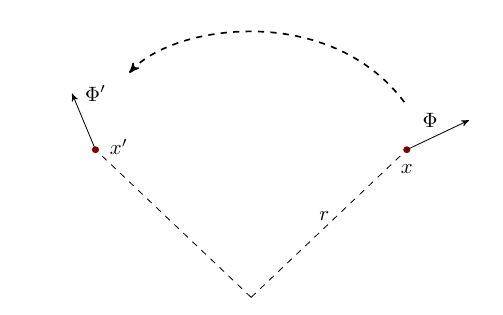} \label{fig:symmetry}
\end{figure}

We begin by examining the transformation properties of fields under coordinate changes. In earlier sections, we discussed how scalar fields transform and identified the corresponding symmetry generators. However, fields can have more complex transformation properties than those of simple scalars. For instance, Figure \ref{fig:symmetry} illustrates the rotation transformation of a vector field, demonstrating how the field $\Phi$ transforms under coordinate changes.
\begin{equation}
 \label{trans-field-general} x\rightarrow x'\, , \qquad \Phi(x)\rightarrow\Phi'(x') =\mathcal{F}(\Phi(x)) \, .
\end{equation}
    Functional change $\mathcal{F}$ can be obtained by studying the representation theory of the coordinate transformation group. We will not illustrate more on this, but give some examples for later use. 

    Let us start with a rather trivial one, a translation:
\begin{equation}
 \label{translation-field} x'=x+a\, , \qquad \Phi'(x+a)=\Phi(x)\, ,
\end{equation}
    which implies that functional change $\mathcal{F}$ is trivial here.  Next, we consider a Lorentz transformation (rigid rotation). In general, it takes the following form:
\begin{equation}
 \label{lorentz-field} x'^\mu ={\Lambda^\mu}_\nu x^\nu\, , \qquad \Phi'(\Lambda x) =U(\Lambda)\Phi(x)\, ,
\end{equation}
    where $U(\Lambda)$ is a transformation matrix acting on $\Phi$.\footnote{ In quantum field theory, $\Phi$ is more likely to be treated as an operator with transformation rule:
\begin{equation}
 \Phi'(\Lambda x) =U^{-1}(\Lambda)\Phi(x)U(\Lambda)\, . \notag
\end{equation}
    } The infinitesimal form of $U(\Lambda)$ is
\begin{equation}
 \label{U-Lambda-inf} U(\Lambda)=1-\frac{1}{2}i \omega_{\rho\nu} S^{\rho\nu}\, ,
\end{equation}
    where $\omega_{\rho\nu}$ are infinitesimal parameters for Lorentz transformation and $S^{\rho\nu}$ is a representation of the Lorentz algebra.  Finally, under a scaling transformation, we have
\begin{equation}
 \label{scale-field} x'=\lambda x \, , \qquad \Phi'(\lambda x)=\lambda^{-\Delta} \Phi(x) \, ,
\end{equation}
    where $\Delta$ is the scaling dimension of the field.

\subsubsection*{Generator} In a field theory, the generator of a symmetry depends on transformations of fields as well as coordinates. The infinitesimal form of \eqref{trans-field-general} can be written as:
\bea
 \label{trans-field-general-inf} x'^{\mu} &= x^\mu+\omega_a \frac{\delta x^\mu}{\delta \omega_a} \, , \cr \Phi'(x') &= \Phi(x) + \omega_a \frac{\delta\mathcal{F}} {\delta \omega_a}(x) \, .
\eea
     Here $\{\omega_a\}$ is a set of infinitesimal parameters. Now it is customary to define the generator $G_a$ of a symmetry transformation by the following expression:
\begin{equation}
 \label{def-generater-general} \delta_\omega \Phi(x) \equiv \Phi'(x)-\Phi(x) \equiv -i\omega_a G_a \Phi(x)\, ,
\end{equation}
which represents the infinitesimal transformation of the field at the same point. Together with \eqref{trans-field-general-inf}, to first order, we obtain the explicit expression:
\begin{equation}
 \label{def-generator-general-explicit} iG_a \Phi = \frac{\delta x^\mu}{\delta \omega_a} \partial_\mu \Phi - \frac{\delta\mathcal{F}}{\delta \omega_a} \, .
\end{equation}
    For a field satisfying $\Phi'(x')=\Phi(x)$, the expression reduces to \eqref{def-generator}. Using this definition, and combining it with the infinitesimal transformations in the cases of translation, rotation, and scaling, we can write the corresponding generators in field space as follows
\bea
 \label{generator-three-field} &\text{translation}: \qquad  P_\nu=-i\partial_\nu \, , \cr
&\text{rotation} : \qquad L^{\rho\nu} =i(x^\rho\partial^\nu -x^\nu\partial^\rho) +S^{\rho\nu} \, ,\\
&\text{dilatation}: \qquad D=-i x^\nu \partial_\nu -i\Delta \, . \notag
\eea
We observe that when the field transformation $\mathcal{F}(\Phi)$ is trivial, three expressions reduce to the first three generators in Table \ref{Tab:fin-ct}.

\subsubsection*{Noether's Theorem} 
Noether's theorem is a fundamental result in classical field theory, establishing a deep connection between symmetries and conserved quantities. We will now derive Noether's theorem, which asserts that for every continuous symmetry of the action, there is a corresponding conserved current at the classical level. Consider an action in $d$ dimensions
\begin{equation}
 \cS=\int \dd^d x \mathcal{L}(\Phi,\partial_\mu \Phi)
\end{equation}
    which is endowed with symmetry generated by $\omega_aG_a$. Under the continuous transformation \eqref{trans-field-general}, the action becomes
\bea
 \cS' &= \int \dd^d x' \mathcal{L}( \Phi'(x'),\partial_\mu'\Phi'(x')) \cr
&=\int \dd^d x \abs{\pdv{x'}{x}} \mathcal{L} (\mathcal{F}(\Phi(x)),(\pdv{x^\nu}{x'^\mu}) \partial_\nu \mathcal{F}(\Phi(x)))\, .
\eea
     Using the infinitesimal form for transformation \eqref{trans-field-general-inf}, we get
\bea
 \cS'=\int \dd^d x \bigg(1+\partial_\mu  (\omega_a \frac{\delta x^\mu} {\delta \omega_a})\bigg)  \mathcal{L}  \Bigg(  \Phi+\omega_a  \frac{\delta\mathcal{F}} {\delta\omega_a},  \bigg(  \delta^\nu_\mu  -  \partial_\mu  (  \omega_a\frac{\delta x^\nu} {\delta \omega_a} )  \bigg)  \bigg(  \partial_\nu \Phi+  \partial_\nu  (  \omega_a\frac{\delta\mathcal{F}} {\delta\omega_a} )  \bigg) \Bigg)\, .
\eea
     Since $\omega_aG_a$ is symmetry of the action, $\delta \cS=\cS'-\cS$  vanishes if $\omega_a$ is a constant. Thus, the contribution from all terms with no derivatives of $\omega_a$ in the variation must be zero. Therefore, to first order, the variation involves only the first derivatives of $\omega_a$, obtained by expanding the Lagrangian so that we can write
\begin{equation}
 \delta \cS=- \int \dd^d x ~ j^\mu_a \partial_\mu \omega_a\, ,
\end{equation}
    where $j^\mu_a$ is called the current associated with the infinitesimal transformation:
\begin{equation}
 \label{cano-current} j^\mu_a =\bigg( \pdv{\mathcal{L}}{ (\partial_\mu\Phi)} \partial_\nu \Phi - \delta^\mu_\nu \mathcal{L} \bigg) \frac{\delta x^\nu}{\delta\omega_a} - \pdv{\mathcal{L}}{(\partial_\mu \Phi)} \frac{\delta\mathcal{F}}{\delta\omega_a} \, .
\end{equation}
    The integral by parts yields
\begin{equation}
 \label{variation-action-locally} \delta \cS=\int \dd^d x \,\omega_a \partial_\mu j^\mu_a \, .
\end{equation}
    If the field configurations satisfy the classical equations of motion, the action is invariant under any variation of the fields. Thus $\delta \cS$ should vanish for any parameters $\omega_a(x)$. This implies the conservation law:
\begin{equation}
 \partial_\mu j^\mu_a=0 \, .
\end{equation}
    The associated  charge
\begin{equation}
 \label{def-conserved-charge} Q_a=\int \dd^{d-1} x \, j^0_a \, ,
\end{equation}
    is conserved because its time derivative vanishes
\begin{equation}
 \frac{d}{dt}{Q}_a=\int \dd^{d-1} x ~ \partial_0 j_a^0 =-\int \dd^{d-1}x ~ \partial_i j^i =0\, .
\end{equation}
    One remark  to be mentioned here is that we may freely add to the expression \eqref{cano-current} the divergence of an antisymmetric tensor without affecting the conservation law:
\begin{equation}
 \label{current-freedom} j^\mu_a \rightarrow j^\mu_a +\partial_\nu B^{\nu\mu}_a \, , \qquad B^{\nu\mu}_a=- B^{\mu\nu}_a \, .
\end{equation}

In a field theory, the energy-momentum tensor $T^{\mu v}$ is defined from the change of the action under an arbitrary coordinate transformation $x^{'\mu} = x^{\mu}+\omega^{\mu}(x)$ as
\begin{equation}
 \label{ener-mon-tensor}\delta \mathcal{S}=\int d^{d} x T^{\mu \nu} \partial_{\mu} \omega_{\nu}~.
\end{equation}
In particular, under the translation $\omega^\mu=\textrm{const}$, we obtain the form of the canonical energy-momentum tensor by using \eqref{cano-current}:
\begin{equation}
 \label{cano-ener-mon-tensor} T^{\mu\nu}_{\text{c}}= -\eta^{\mu\nu}\mathcal{L} +\pdv{\mathcal{L}}{(\partial_\mu\Phi)} \partial^\nu \Phi \, ,
\end{equation}
Note that the energy-momentum tensor defined in this way \eqref{ener-mon-tensor} is generally not symmetric under $\mu$ and $\nu$. Nevertheless, we have freedom to modify this tensor, as illustrated in \eqref{current-freedom}, to make it symmetric.
\begin{equation}
 \label{freedom-energy-momentum-tensor} T^{\mu\nu}_{\text{B}}= T^{\mu\nu} +\partial_\rho B^{\rho\mu\nu}\, ,\quad B^{\rho\mu\nu}=-B^{\mu\rho\nu}\, .
\end{equation}
Additionally, we often need to further adjust the form of the energy-momentum tensor due to the last term in \eqref{cano-current}. For more detail, the reader can refer to \cite[\S2.5,\S4.2.2]{francesco2012conformal}

Now, let us consider the property of the energy-momentum tensor in conformal field theory. Under a conformal transformation $x^{'\mu} = x^{\mu}+\epsilon^{\mu}(x)$, we have a conserved current
\begin{equation}\label{conformal-current}
  j_{\mu}=T_{\mu \nu} \epsilon^{v}~.
\end{equation}
Since a translation is a conformal transformation, the current conservation implies
\begin{equation}
0=\partial^{\mu} j_{\mu}=\partial^{\mu}\left(T_{\mu \nu} \epsilon^{\nu}\right)=\left(\partial^{\mu} T_{\mu \nu}\right) \epsilon^{\nu}
\end{equation}
yielding $\partial^{\mu} T_{\mu \nu}=0$. For more general conformal transformations $x^{'\mu} = x^{\mu}+\epsilon^{\mu}(x)$, the current conservation imposes
\begin{align}
0=\partial^{\mu} j_{\mu} &=\left(\partial^{\mu} T_{\mu \nu}\right) \epsilon^{\nu}+T_{\mu \nu}\left(\partial^{\mu} \epsilon^{\nu}\right) \cr
&=0+\frac{1}{2} T_{\mu \nu}\left(\partial^{\mu} \epsilon^{\nu}+\partial^{\nu} \epsilon^{\mu}\right) \cr
&=\frac{1}{d} T_{\mu}{ }^{\mu}(\partial \cdot \epsilon)
\end{align}
where we use \eqref{inf-cft-e-condition} in the last line. Thus, the energy-momentum tensor is traceless $ T_{\mu}{ }^{\mu}=0$ in conformal field theory.

\subsubsection*{Correlation functions} In the previous sections, we focused on the effect of a continuous symmetry classically. At the quantum level, the primary objects of study are correlation functions. A general correlation function is defined as
\begin{equation}
 \label{def-correlation-function} \expval{\Phi(x_1)\cdots \Phi(x_n)}= \frac{1}{\cZ}\int [\cD \Phi] \Phi(x_1)\cdots \Phi(x_n) \exp{-\cS[\Phi]}\, ,
\end{equation}
    where $\cZ$ is the vacuum functional:
\begin{equation}
 \cZ=\int [\cD \Phi] \exp(-\cS[\Phi]) \, .
\end{equation}
    The path integral notation $[\cD \Phi]$ indicates integration over all possible field configurations of $\Phi$. The operators in the correlation function $\Phi(x_1)\cdots \Phi(x_n)$ are automatically time-ordered when using the path integral method, i.e.
\begin{equation}
 \expval{\mathcal{T}(\Phi(x_1) \cdots \Phi(x_n))} = \expval{\Phi(x_1)\cdots\Phi(x_n)}\, ,
\end{equation}
    where $\mathcal{T}$ is the time ordering operator which makes the field operator in time order:
\begin{equation}
 \mathcal{T}(\Phi(x_1) \cdots \Phi(x_n)) =\Phi(x_1)\cdots\Phi(x_n)\, , \qquad \text{if} \quad t_1 > t_2 > \cdots > t_n \, .
\end{equation}
    In Euclidean space, time direction can be arbitrary. We will specify the time direction in the two-dimensional case later.

    A continuous symmetry leads to constraints on the correlation function:
\begin{equation}
 \label{tran-correlation-function} \expval{\Phi(x_1')\cdots\Phi(x_2')} = \expval{\mathcal{F}(\Phi(x_1))\cdots      \mathcal{F}(\Phi(x_n)) }\, .
\end{equation}
This identity is a consequence of the invariance of the action and the path integral measure under the transformation:
\bea
 \expval{\Phi(x_1')\cdots \Phi(x_n')} &= \frac{1}{\cZ} \int [\cD \Phi] \Phi(x_1')\cdots \Phi(x_n') \exp(-\cS[\Phi]) \cr
&= \frac{1}{\cZ} \int [\cD \Phi'] \Phi'(x'_1)\cdots \Phi'(x'_n) \exp(-S[\Phi']) \cr
&= \frac{1}{\cZ} \int [\cD \Phi] \mathcal{F}(\Phi(x_1))\cdots \mathcal{F}(\Phi(x_n)) \exp(-\cS[\Phi]) \cr
&= \expval{\mathcal{F}(\Phi(x_1))\cdots \mathcal{F}(\Phi(x_n))} \notag\, ,
\eea
where in the second line, we rename the integration variable $\Phi \rightarrow \Phi'$. On the third line, the symmetry assumptions for the action and the functional integral measure are applied. \eqref{tran-correlation-function} immediately implies that the correlation function is invariant under translation.

\subsubsection*{Ward-Takahashi identity} The quantum version of Noether's theorem is known as the Ward-Takahashi identity. We first change the functional field variable in \eqref{def-correlation-function} by an infinitesimal transformation \eqref{def-generater-general}
\begin{equation}
 \Phi'(x)=\Phi(x)- i \omega_a G_a \Phi(x) \, ,
\end{equation}
    The correlation function is invariant under the symmetry so that we can write
\begin{equation}
 \expval{X}=\frac{1}{\cZ} \int [\cD \Phi'] (X+\delta X) \exp{-\cS[\Phi]-\int \dd^d x\omega_a (x) \partial_\mu j^\mu_a} \, ,
\end{equation}
    where $X$ denotes the product $\Phi(x_1)\cdots \Phi(x_n)$ of fields in \eqref{def-correlation-function}. Expanding this to first order in $\omega_a(x)$ yields
\begin{equation}
 \label{Ward-identity-ver2} \expval{\delta X}=\int \dd^d x \partial_\mu \expval{j^\mu_a (x)X} \omega_a(x)\, .
\end{equation}

     The variation $\delta X$ is explicitly given by
\bea
 \delta X &= -i \sum_{i=1}^n ( \Phi(x_1)\cdots G_a\Phi(x_i) \cdots \Phi(x_n) ) \omega_a(x_i) \cr
&=-i \int \dd^d x \, \omega_a(x) \sum_{i=1}^{n} ( \Phi(x_1)\cdots G_a \Phi(x_i) \cdots \Phi(x_n) ) \delta(x-x_i)
\eea
Since the above two expressions hold for any infinitesimal function \( \omega^a(x) \), we can express the following local relation:
\bea
 \label{Ward-Takahashi-Identity} \pdv{x^\mu} &\expval{ j^\mu_a(x)\Phi(x_1)\cdots \Phi(x_n) }\cr
&=-i\sum_{i=1}^{n} \delta(x-x_i) \expval{ \Phi(x_1)\cdots G_a\Phi(x_i) \cdots \Phi(x_n) }\, .
\eea
     This is the Ward-Takahashi identity for the current $j^\mu_a$.

     The Ward-Takahashi identity allows us to identify the conserved charge \( Q_a \) defined in \eqref{def-conserved-charge} as the generator of the symmetry transformation in the Hilbert space of quantum states. Let $Y= \Phi(x_2)\cdots \Phi(x_n)$ and suppose that the time $t=x^0_1$  is larger than all the times in \( Y \). We integrate the Ward-Takahashi identity in a very thin box bounded by \( t_- < t \), \( t_+ > t \), and by spatial infinity, which excludes all other points \( x_2, \dots, x_n \). This gives
\begin{equation}
 \expval{Q_a(t_+)\Phi(x_1)Y} - \expval{Q_a(t_-)\Phi(x_1)Y} =-i \expval{G_a \Phi(x_1)Y}\, .
\end{equation}
    In the limit $t_-\rightarrow t_+$, for an arbitrary set of fields $Y$, we obtain
\begin{equation}
 \label{charge-com-generator} \comm{Q_a}{\Phi}=-i G_a \Phi\, .
\end{equation}
This shows that the conserved charge \( Q_a \) is the generator in the operator formalism.

    \subsection{Primary fields} 
    In the previous section, we reviewed some concepts of field theory in \( d \) dimensions and derived the current expressions for three important currents in conformal symmetry. Now, we will focus specifically on two dimensions. First, we need to define how fields transform under a conformal transformation. From \eqref{Witt-algebra}, we see that \( l_0 \) and \( \bar{l}_0 \) are the Cartan subalgebra of the conformal algebra. Therefore, we can assume that physical states are eigenstates of \( l_0 \) and \( \bar{l}_0 \) with eigenvalues \( \{h, \bar{h}\} \), known as the \textbf{conformal dimensions}. Under a scaling transformation \( z \to \lambda z \), \( \bar{z} \to \lambda \bar{z} \), the field operator transforms as
\begin{equation}
 \label{def-conformal-dimension} \phi'(\lambda z,\overline\lambda \overline{z}) = (\lambda)^{-h} (\overline\lambda)^{-\overline{h}} \phi(z,\overline{z})\, .
\end{equation}
    If a field transforms under an arbitrary global conformal map $SL(2,\bC)$, $z\rightarrow w(z), \overline{z} \rightarrow \overline{w}(\overline{z})$,
\begin{equation}
 \label{def-quasi-primary-field} \phi'(w,\overline{w}) =\bigg(\frac{\dd w}{\dd z}\bigg)^{-h} \bigg(\frac{\dd \overline{w}}{\dd \overline{z}}\bigg)^{-\overline{h}} \phi(z,\overline{z})\, ,
\end{equation}
    then we call it a \textbf{quasi-primary field}. Under infinitesimal transformation $z \rightarrow z'=z+\e(z)$, we obtain the infinitesimal transformation of quasi-primary field:
\begin{equation}
 \label{trans-inf-quasi-primary} \delta_{\e,\overline\e}\phi \equiv \phi'(z',\overline{z}') -\phi(z, \overline{z}) =-(h\phi\partial \e+\e \partial \phi) - (\overline{h}\phi\overline{\partial} \overline{\e} + \overline{\e}\overline{\partial}\phi )\, .
\end{equation}
    A field whose transformation under any conformal transformation given by \eqref{def-quasi-primary-field} or equivalently \eqref{trans-inf-quasi-primary} is called a \textbf{primary field}. Therefore all primary fields are also quasi-primary, but the reverse is not true. If an operator is not primary, it is generally called secondary.

    The definition \eqref{def-quasi-primary-field} imposes strong constraints on their correlation functions. For the two-point function of chiral quasi-primary fields, translation invariance implies that the two-point function is of the form
\begin{equation}
 \expval{\phi_1(z)\phi_2(w)} =g(z-w) \, .
\end{equation}
    Then under conformal transformation $z \rightarrow \lambda z $, we have
\begin{equation}
 \lambda^{-(h_1+h_2)} g(z-w)=g(\lambda(z-w))\, ,
\end{equation}
    where $h_1$ and $h_2$ are the conformal dimensions of $\phi_1$  and $\phi_2$, respectively. This leads to a constraint for the two-point function. Since the relation above satisfies for arbitrary $\lambda$, we obtain that $g(z-w)$ is of the form
\begin{equation}
 g(z-w)=\frac{C_{12}}{(z-w)^{h_1+h_2}}\, ,
\end{equation}
where $C_{12}$ is a constant. Additionally, the two-point function should be consistent with \eqref{def-quasi-primary-field} when we apply the conformal transformation $z \rightarrow -1/z$. This constraint requires that conformal dimensions of the two fields must be equal, i.e.
\begin{equation}
 \label{two-point-func-form} \expval{\phi_1(z)\phi_2(w)} =\frac{C_{12}}{(z-w)^{2h}}\, , \quad \text{if} \quad h_1=h_2=h\, .
\end{equation}
    The same treatment can be used to obtain the form of the three-point function for quasi-primary operators. For the holomorphic part, we have
\begin{equation} \label{three-point-func-form}
 \expval{\phi_1(z_1)\phi_2(z_2)\phi_3(z_3)} =\frac{C_{123}}{z_{12}^{h_1+h_2-h_3} z_{23}^{h_2+h_3-h_1} z_{13}^{h_3+ h_1- h_2}} \, ,
\end{equation}
    where $z_{ij}=z_i-z_j $. (Exercise)  Although these results are obtained by Polyakov in \cite{Polyakov:1970xd}, determining the four-point function and beyond requires the use of an infinite-dimensional symmetry \cite[ISZ88-No.1]{belavin1984infinite}, which will be discussed in \S\ref{sec:bootstrap}.  

    \subsection{Conformal Ward-Takahashi identity}  Now, let us apply Ward-Takahashi identities to general conformal invariance, known as \textbf{conformal Ward-Takahashi identity}. Conformal invariance gives rise to the current \eqref{conformal-current}. Substituting it into \eqref{Ward-identity-ver2}, we obtain the conformal Ward-Takahashi identity
\begin{equation}
 \label{conformal-ward-identity-st} \delta_\e\expval{X} =\int_D \dd^2 x~ \partial_\mu \expval{T^{\mu\nu}(x)\e_\nu(x)X}\, .
\end{equation}
    Here, the integral is taken over the domain $D$ containing the positions of all the fields in the $X$. In two dimensions, it is convenient to use the (anti-)holomorphic coordinate $(z,\overline{z})$ as introduced in \eqref{coordinate transformation}.
    Writing each component explicitly, we have
\bea\label{T-barT}
 T_{zz} &=  \frac{1}{4}(T_{00}-2iT_{10}- T_{11}) \, , \\
T_{\overline{z}\overline{z}} &= \frac{1}{4}(T_{00}+2iT_{10}-T_{11}) \, , \\
T_{z\overline{z}} &= T_{\overline{z}z} =\frac{1}{4} (T_{00}+T_{11})
\eea
The traceless property of the energy-momentum tensor is $T_{z\overline{z}}=T_{\overline{z}z}=0$. From the conservation law $\partial_\mu T^{\mu \nu} =0$, we see that $T_{zz}$ and $T_{\overline{z}\overline{z}}$ are holomorphic and antiholomorphic respectively. Furthermore we introduce the notation \begin{equation}\label{holomorphic-EM} T=-2\pi T_{zz}\, , \quad \overline{T}=- 2\pi T_{\overline{z}\overline{z}} \, .
\end{equation}

      Using Gauss's theorem and in two dimensions, we find
\begin{equation}
 \delta_{\e,\overline\e}\expval{X} =\frac{i}{2}  \int_C \{ -\dd z\expval{T^{\overline{z}\overline{z}}\e_{\overline{z}}X} +\dd \overline{z}\expval{T^{zz}\e_z X} \} \, ,
\end{equation}
    where we have defined $\e(z)= \e^z$ and $\overline{\e}(\overline{z})=\e^{\overline{z}}$, respectively.
    Using \eqref{holomorphic-EM}, we get
\begin{equation}
 \label{Conformal-Ward-Identity-2d} \delta_{\e,\overline\e}\expval{X} =- \frac{1}{2\pi i} \oint_C \dd z \,\e(z)\expval{T(z)X} + \frac{1}{2\pi i} \oint_C \dd \overline{z}\, \overline{\e}(\overline{z}) \expval{\overline{T}(\overline{z})X}\, ,
\end{equation}
    where the contour $C=\partial D$ should contain all the positions of fields in $X$. Notice that the validity of \eqref{Conformal-Ward-Identity-2d} extends beyond primary fields, and thus it may be taken as a definition of the effect of conformal transformations on an arbitrary local field within a correlation function.  Suppose that $X$ is a product of $n$ primary fields $\phi_i(w_i)$,
    \[
    X(w_1,\ldots,w_n):=\phi_1(w_1)\cdots \phi_n(w_n)~.
    \] Using simple complex analysis, we write the variation  \eqref{trans-inf-quasi-primary} of a primary field
\bea
 (\partial_{w_i} \e({w_i}))\phi_i({w_i},\overline{w}_i)&=\frac1{2\pi i}\oint_{C} dz~ \frac{\e(z) \phi_i({w_i},\overline{w}_i)}{(z-{w_i})^2}\cr \e({w_i})(\partial_{w_i}\phi_i({w_i},\overline{w}_i))&=\frac1{2\pi i}\oint_{C} dz ~\frac{\e(z)\partial_{w_i} \phi_i({w_i},\overline{w}_i)}{z-{w_i}}~,
\eea

Therefore, the conformal Ward identity \eqref{Conformal-Ward-Identity-2d} can be written as
\begin{equation}
 \expval{T(z)X}=\sum_i
 \bigg\{
   \frac{1}{z-w_i}\partial_{w_i}
   \expval{X}
   +
   \frac{h_i}{(z-w_i)^2}\expval{X}
 \bigg\}
 +\text{reg.}\, ,
\end{equation}
where "reg." refers to a holomorphic function of \( z \) that is regular at \( z = w_i \). The anti-holomorphic counterpart is similar. This expression provides the singular behavior of the correlator of the energy-momentum tensor \( T(z) \) with a set of primary fields \( X \) as \( z \) approaches the point \( w_i \). This implies the operator product expansion (OPE) of the energy-momentum tensor with primary fields by removing the bracket \( \langle \cdots \rangle \):
\begin{equation}
\label{Virasoro-primary}
 T(z)X \sim \sum_i
\bigg\{
\frac{1}{z-w_i}\partial_{w_i}
 X
+
\frac{h_i}{(z-w_i)^2} X
\bigg\}
\, .
\end{equation}
The symbol \( \sim \) indicates that regular terms are ignored. The OPE also contains infinitely many regular terms which, for the energy-momentum tensor, cannot be derived from the conformal Ward identity. The \textbf{operator product expansion} (OPE) represents a product of operators and describes the behavior as two operators approach each other. In general, we would write the OPE of two operators \( A(z) \) and \( B(w) \) as
\begin{equation}
  A(z)B(w)=\sum_{n=-\infty}^{\infty}
  \frac{\{AB\}_n(w)}{(z-w)^n} \, ,
\end{equation}
where the composite operators $\{AB\}_n(w)$ are nonsingular at $w=z$. In particular for a single primary field $\phi$ with conformal dimensions $h$ and $\overline{h}$ , we have:
\begin{align}
\label{OPE-T-primaryfield-z}
  T(z)\phi(w,\overline{w})&\sim   \frac{h}{(z-w)^2}\phi(w,\overline{w})   +   \frac{1}{z-w}\partial \phi(w,\overline{w})   \, ,\\
\label{OPE-T-primaryfield-barz}
  \overline{T}(\overline{z})   \phi(w,\overline{w})&\sim   \frac{\overline{h}}  {(\overline{z}-\overline{w})^2}   \phi(w,\overline{w})   +   \frac{1}{\overline{z}-\overline{w}}   \overline\partial \phi(w,\overline{w})
  \, .
\end{align}
The expression above can be treated as an alternative definition for a primary field. For a secondary field, but with conformal dimension $h$ and $\overline{h}$, its OPE with energy-momentum tensor has more singular terms, i.e.
\begin{align} \label{OPE-T-seconfield-z} T(z)\phi(w,\overline{w})&\sim \cdots+
\frac{h}{(z-w)^2}\phi(w,\overline{w})
+
\frac{1}{z-w}\partial \phi(w,\overline{w})
\, ,\\
\label{OPE-T-seconfield-barz}
\overline{T}(\overline{z})
\phi(w,\overline{w})&\sim \cdots +
\frac{\overline{h}}
{(\overline{z}-\overline{w})^2}
\phi(w,\overline{w})
+
\frac{1}{\overline{z}-\overline{w}}
\overline\partial \phi(w,\overline{w})
\, .
\end{align}

\subsection{Example: Free boson}
Now we will study the free boson field as an example. We first review some basic concepts in field theory.

\subsubsection*{Calculation of two-point function}

A two-point function is a correlation function of two operators. Therefore, from the definition of correlation function \eqref{def-correlation-function}, we can easily write the definition of a two-point function for the bosonic field as
\begin{equation}
\label{def-two-point-function}
  \expval{\mathcal{T}(\phi(x_1)\phi(x_2))}=
  \frac{1}{\cZ}\int [\cD \Phi]
  \phi(x_1)\phi(x_2) \exp{-\cS[\Phi]}\, .
\end{equation}
The time-ordering operator $\mathcal{T}$ is automatically satisfied in a path integral. Then, we can define a generating functional for the theory:
\begin{equation}
\label{def-generating-functional}
  \cZ[J]=\int [\cD \Phi] \exp{-(S-\int \dd^d x \phi(x) J(x)}\, ,
\end{equation}
where $J(x)$ is an auxiliary "current". Then the two-point function can be generated by the functional derivative of $J$ :
\begin{equation}
  \label{two-point-correlation-functional-derivative}
  \expval{\phi(x_1)\phi(x_2)}=
 \frac{1}{\cZ[0]} \frac{\delta}{\delta J(x_1)}
  \frac{\delta}{\delta J(x_2)}\cZ[J]\Bigg|_{J=0}\, .
\end{equation}
To calculate the two-point function more generally, we first assume that the action $S$ takes the following form:
\begin{equation}
  \label{general-form-action}
  \cS=\frac{1}{2}
  \int \dd^d x \dd^d y\,
  \phi(x)A(x,y)\phi(y)\, .
\end{equation}
Let us introduce a variable transformation $\phi\rightarrow \phi+\phi_0$ without changing the path integral measure and result as we have illustrated before. The term in the exponential of the path integral becomes:
\bea
  -\cS+\int \dd^d x \phi(x)J(x)
  \rightarrow&
  -\frac{1}{2}\int \dd^d x \dd^d y
  \phi(x)A(x,y)\phi(y)
  -\int \dd^d x \dd^d y
  \phi(x)A(x,y) \phi_0(y)-
  \cr
  &- \frac{1}{2}\int \dd^d x \dd^d y
  \phi_0(x)A(x,y)\phi_0(y)
  +\int \dd^d x\phi_0(x)J(x)
  +\int \dd^d x \phi(x)J(x)\, ,
  \notag
\eea
Now we choose $\phi_0$ to satisfy the equation:
\begin{equation}  \int \dd^d y A(x,y) \phi_0(y)=J(x)\,   \rightarrow
 \phi_0(x)=\int \dd^d y A^{-1}(x,y)J(y)\, .
\end{equation}
Then, the generating functional can be
written as
\begin{equation}
  \cZ[J]=N \exp{\frac{1}{2}
  	\int \dd^d x \dd^d y \,
  	J(x) A^{-1}(x,y) J(y)
}\, ,
\end{equation}
where the factor $N$ is independent of $J$
\begin{equation}
  N=\int [\cD \Phi]
  \exp{-\frac{1}{2} \int \dd^d x \dd^d y \,
  \phi(x) A(x,y) \phi(y)}\, .
\end{equation}
Therefore, by using
\eqref{two-point-correlation-functional-derivative},
we have
\begin{equation}
\label{two-point-function-calculation}
  \expval{\mathcal{T}(\phi(x)\phi(y))}
  =A^{-1}(x,y)
\end{equation}
We will apply the formula above to calculate the two-point function of specific theories.

\subsubsection*{Two-Point function in Free boson}
In two dimensions, the massless free boson has the following Euclidean action:
\begin{equation}
\label{def-Free-boson action}
  \cS=\frac{1}{8\pi}\int \dd^2 x
~
     \partial_\mu \varphi
     \partial^\mu \varphi
\, ,
\end{equation}
which is conformal. Comparing with \eqref{general-form-action}, we have
\begin{equation}
  A(x,y)=- \frac{1}{4\pi} \delta^{(2)}(x-y)\square \, ,
\end{equation}
From \eqref{two-point-function-calculation}, we can calculate the two-point function $K(x,y) \equiv \expval{\varphi(x_1)\varphi(x_2)}$ by solving the following equation:
\begin{equation}
 -\frac{1}{4\pi}\square K(x,y)=\delta^{(2)}(x-y)\, ,
\end{equation}
From \eqref{tran-correlation-function} and the trivial nature of $\mathcal{F}$ for a scalar field, $K(x,y)$ is invariant under translation and rotation. Thus, we can write $K(x,y)\equiv K(\rho)$ with $\rho=\abs{x-y}$, and integrate over $x$ within a disk of radius $\rho$ around y. We find
\bea
  1 &= \frac{1}{2}  \int^{r}_0 \dd \rho \rho
  \bigg(
    -\frac{1}{\rho}\pdv{\rho}(\rho K'(\rho))
  \bigg)
  \cr
    &= \frac12 (-rK'(r))\, .
\eea
The solution of the two-point function for massless free boson can be obtained up to an additive constant,
\begin{equation}
  \label{two-point-function-free-boson}
  \expval{\varphi(x)\varphi(y)}=-
  \ln{\abs{x-y}}^2\, .
\end{equation}

\subsubsection*{Wick's Theorem}
In the last part, we introduced the concept of the two-point function for a theory. Now, we will briefly review Wick’s Theorem, which highlights the fundamental role of the two-point function in quantum field theory.

In quantum field theory, a free (elementary) field operator can be decomposed into two parts:
\begin{equation}
  \phi(x)=\phi^{+}(x)+\phi^{-}(x)\, ,
\end{equation}
where \( \phi^-(x) \) contains only annihilation operators, and \( \phi^+(x) \) contains only creation operators. The normal ordering \( :\ : \) is defined to place all annihilation operators to the right of the creation operators. For instance, 
\begin{equation}
  :\phi(x_1)\phi(x_2):=\phi(x_1)^+\phi(x_2)^+
 +\phi(x_1)^-\phi(x_2)^-+\phi(x_1)^+\phi(x_2)^-
 +\phi(x_2)^+\phi(x_1)^- \, .\notag
\end{equation}
The time-ordered product for \( x^0_1 > x^0_2 \) gives
\begin{equation}
  \mathcal{T}(\phi(x_1)\phi(x_2))
  =\phi(x_1)^+\phi(x_2)^+
 +\phi(x_1)^-\phi(x_2)^-+\phi(x_1)^+\phi(x_2)^-
 +\phi(x_1)^-\phi(x_2)^+ \, .\notag
\end{equation}
Thus, we obtain the relation between the time-ordered product and the normal-ordered product for \( x^0_1 > x^0_2 \) as
\begin{equation}
  \mathcal{T}(\phi(x_1)\phi(x_2))
  =:\phi(x_1)\phi(x_2):
 +\comm{\phi(x_1)^-}{\phi(x_2)^+}\, .
\end{equation}
The commutator on the right side is a constant. By taking the expectation value of both sides, we find that this constant is simply the two-point function
\begin{equation}
  \mathcal{T}(\phi(x_1)\phi(x_2))=
  :\phi(x_1)\phi(x_2): +
  \expval{\phi(x_1)\phi(x_2)}\, .
\end{equation}
Note that the same expression holds for \( x^0_1 < x^0_2 \).  It is convenient to introduce \textbf{Wick contraction} for the computation of two-point functions and OPE
\begin{equation}
  \contraction{}{\phi_1}{(x_1)}{\phi_2}
  \phi_1(x_1)\phi_2(x_2)
  \equiv
  \expval{\phi_1(x_1)\phi_2(x_2)}\, .
\end{equation}
Wick’s Theorem states that the time-ordered product of operators at different points in space-time is equal to the normal-ordered product, plus all possible ways of contracting pairs of fields within it.  For instance,
\bea
  \mathcal{T}(\phi_1\phi_2\phi_3\phi_4)   &=   :\phi_1\phi_2\phi_3\phi_4:   +:   \contraction{}{\phi_1}{}{\phi_2}   \phi_1\phi_2\phi_3\phi_4   :   +:   \contraction{}{\phi_1}{\phi_2}{\phi_3}   \phi_1\phi_2\phi_3\phi_4   :   +:   \contraction{}{\phi_1}{\phi_2\phi_3}{\phi_4}   \phi_1\phi_2\phi_3\phi_4   :\cr
  &+:   \contraction{\phi_1}{\phi_2}{}{\phi_3}   \phi_1\phi_2\phi_3\phi_4   :   +:   \contraction{\phi_1}{\phi_2}{\phi_3}{\phi_4}   \phi_1\phi_2\phi_3\phi_4   :   +:   \contraction{\phi_1\phi_2}{\phi_3}{}{\phi_4}   \phi_1\phi_2\phi_3\phi_4   :   +:   \contraction{}{\phi_1}{}{\phi_2}   \contraction{\phi_1\phi_2}{\phi_3}{}{\phi_4}   \phi_1\phi_2\phi_3\phi_4   :   \cr
  &+:   \contraction{}{\phi_1}{\phi_2}{\phi_3}   \contraction[2ex]{\phi_1}{\phi_2}{\phi_3}{\phi_4}   \phi_1\phi_2\phi_3\phi_4   :   +:   \contraction{}{\phi_1}{\phi_2\phi_3}{\phi_4}   \contraction[2ex]{\phi_1}{\phi_2}{}{\phi_3}   \phi_1\phi_2\phi_3\phi_4   :   \notag \, .
\eea
The normal-ordering can be applied to the product of two operators at the same point in space-time. In this case, we can apply Wick's theorem in the following way
\bea
  &\mathcal{T}(   :\phi_1(x)\phi_2(x)::\phi_3(y)\phi_4(y):)   \cr   &=   :\phi_1(x)\phi_2(x)\phi_3(y)\phi_4(y):   +:   \contraction{}{\phi_1}{(x)\phi_2(x)}{\phi_3}   \phi_1(x)\phi_2(x)\phi_3(y)\phi_4(y)   :   +:   \contraction{}{\phi_1}{(x)\phi_2(x)\phi_3(y)}{\phi_4}   \phi_1(x)\phi_2(x)\phi_3(y)\phi_4(y)   :   \cr
  &+:   \contraction{\phi_1(x)}{\phi_2}{(x)}{\phi_3}   \phi_1(x)\phi_2(x)\phi_3(y)\phi_4(y)   :   +:   \contraction{\phi_1(x)}{\phi_2}{(x)\phi_3(y)}{\phi_4}   \phi_1(x)\phi_2(x)\phi_3(y)\phi_4(y)   :   +:   \contraction{}{\phi_1}{(x)\phi_2(x)}{\phi_3}   \contraction[2ex]{\phi_1(x)}{\phi_2}{(x)\phi_3(y)}{\phi_4}   \phi_1(x)\phi_2(x)\phi_3(y)\phi_4(y)   :   \cr
  &+:   \contraction{}{\phi_1}{(x)\phi_2(x)\phi_3(y)}{\phi_4}   \contraction[2ex]{\phi_1(x)}{\phi_2}{(x)}{\phi_3}   \phi_1(x)\phi_2(x)\phi_3(y)\phi_4(y)   :\notag\, . \eea
For fermions, Wick’s Theorem is almost identical, except that exchanging two fermions introduces a minus sign.

\subsubsection*{OPEs in Free boson}
Now we are ready to calculate OPE for the massless free boson. The  equation of motion for the action \eqref{def-Free-boson action} is
\begin{equation}
  \square \varphi=0\, .
\end{equation}
In complex coordinates, we have
\begin{equation}\label{eom-boson}
  \overline{\partial}\partial \varphi(z,\overline z)=0\, ,
\end{equation}
implying that $\partial\varphi$ and $\overline{\partial}\varphi$ are holomorphic and antiholomorphic respectively. The canonical energy-momentum tensor defined by \eqref{cano-ener-mon-tensor} is
\begin{equation}
  T_{\mu\nu}=\frac{1}{4\pi}\big[
  \partial_\mu \varphi \partial_\nu \varphi
 -\frac{1}{2}\eta_{\mu\nu}
  \partial_\rho\varphi\partial^\rho\varphi\big]\, .
\end{equation}
It is automatically symmetric and traceless, and its quantum version in complex coordinates is
\begin{equation}
  \label{EM-tensor-holo}
  T(z)=- \frac12 :\partial\varphi \partial\varphi:\, .
\end{equation}
In order for its vacuum expectation value to vanish, the energy-momentum tensor has to be normal ordered.  The antiholomorphic part is similar. We can write two-point function \eqref{two-point-function-free-boson} in terms of the complex coordinates, this is
\begin{equation}
  \expval{\varphi(z,\overline{z})\varphi(w,\overline{w})}  =-   \{     \ln(z-w)+\ln(\overline z-\overline w)   \}  +\text{const}\, . \end{equation}
We shall concentrate more on the holomorphic field $\partial_z \varphi$. By taking derivatives $\partial_z$ and $\partial_w$ from above, we have
\begin{equation}   \expval{\partial \varphi(z,\overline z)   	\partial \varphi(w, \overline w) }   =   -   \frac{1}{(z-w)^2}\, .
\end{equation}
Hence, the OPE of $\partial\varphi$ with itself is
\begin{equation}
  \partial \varphi(z) \partial \varphi(w)
  \sim
  -
  \frac{1}{(z-w)^2}\, .
\end{equation}
The OPE of $T(z)$ with $\partial\varphi$ may be calculated from Wick's theorem:
\bea
  T(z)\partial\varphi(w) &= -\frac12 :\partial \varphi(z)
  \partial \varphi(z):\partial\varphi(w)\cr
  &\sim- : \partial\varphi(z)\partial
  \contraction{}{\varphi}{(z):\partial}{\varphi}
  \varphi(z):\partial \varphi(w)
  \cr
  &\sim
  \frac{\partial\varphi(z)}{(z-w)^2}\, .
\eea
By expanding $\partial\varphi(z)$ around $w$, we arrive at the OPE
\begin{equation}
  T(z)\partial \varphi(w)\sim
  \frac{\partial\varphi(w)}{(z-w)^2}
  +\frac{\partial^2\varphi(w)}{z-w}\, .
\end{equation}
Comparing with \eqref{OPE-T-primaryfield-z}, we see $\partial\varphi(z)$ is a primary field with conformal dimension h=1. Wick's theorem also allows us to calculate the OPE of the
energy-momentum tensor with itself.
\begin{equation}
  T(z)T(w) \sim
  \frac{1/2}{(z-w)^4}
  +\frac{2T(w)}{(z-w)^2}
  +\frac{\partial T(w)}{(z-w)}\, .
\end{equation}

\subsubsection*{Transformation of the energy-momentum tensor}

It is easy to see that the energy-momentum tensor is not a primary field because of the term \[\frac{1/2}{(z-w)^4}~.\] From \eqref{OPE-T-seconfield-z}, we see that the energy-momentum tensor for free massless boson is a secondary field, with conformal dimension $h=2$. For a general 2d conformal field theory, $TT$ OPE takes the form
\begin{equation}
\label{OPE-TT}
T(z)T(w) \sim \frac{c/2}{(z-w)^4} +\frac{2T(w)}{(z-w)^2} +\frac{\partial T(w)}{(z-w)}\, ,
\end{equation}
with $c$ a constant called the \textbf{central charge}. The central charge depends on a 2d CFT, and it is one of the most important information of a 2d CFT. For the free boson, it is equal to one.

By using conformal Ward-Takahashi identity
\eqref{Conformal-Ward-Identity-2d}, we get an infinitesimal variation of $T$ under a local conformal
transformation.
\bea
  \delta_\e T(w)   &=-\frac{1}{2\pi i}   \oint_C \dd z \, \e(z)T(z) T(w)   \cr
  &=-\frac{c}{12}\partial^3\e(w)  -2 T(w)\partial \e(w) -\e(w)\partial   T(w)\, . \eea
Its finite transformation can be written as $z \rightarrow w(z)$ is
\begin{equation}
  \label{trans-T-finite}
\widetilde  T(w)=\bigg(\dv{w}{z}\bigg)^{-2}
  \big[
       T(z) -\frac{c}{12}\{w;z\}
  \big]\, ,
\end{equation}
where $\{w;z\}$ is the additional term called the \textbf{Schwarzian derivative}:
\begin{equation}
  \{w;z\}=\frac{w^{\prime \prime \prime}}{w^{\prime}}-\frac{3}{2}\left(\frac{w^{\prime \prime}}{w^{\prime}}\right)^{2}~.
\end{equation}

\begin{figure}[ht]
	\centering
	\includegraphics[width=0.7\linewidth]{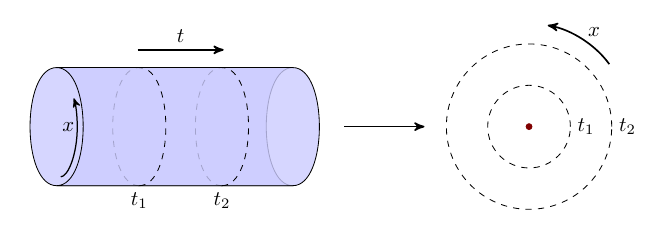}
		\caption{Mapping from the cylinder to the complex plane.}
    \label{fig:cylinder-plane-map}
\end{figure}

\subsection{Virasoro algebra}

\subsubsection*{Radial quantization}
In the previous discussion, we observed that the operator formalism distinguishes between time and space directions. However, in a 2D CFT, it is most convenient to work with complex coordinates in the Euclidean signature. In fact, there is a natural choice in the context of string theory by using radial quantization. We may first define our theory on an infinite space-time cylinder, with time \( t \) going from \(-\infty\) to \(+\infty\) along the “flat” direction of the cylinder, and space being compactified with a coordinate \( x \) going from 0 to \( \ell \), where the points \( (0, t) \) and \( (\ell, t) \) are identified. Here, \( t \) is still the Euclidean time, and the cylinder can be described by a complex coordinate \( x = t + ix \). We then map the cylinder onto a complex plane by a conformal transformation
\begin{equation}     z=e^{2\pi \xi /\ell}\, \end{equation}
   as illustrated in Figure \ref{fig:cylinder-plane-map}  \cite[ISZ88-No.18]{Cardy:1984rp}. The remote past \( (t \to -\infty) \) is mapped to the origin \( z = 0 \), while the remote future \( (t \to +\infty) \) corresponds to infinity. Now, the theory lives on the complex plane where the radial ray and the circle around the origin correspond to the \( t \)-direction and the cylinder's circle, respectively. From quantum mechanics, we know that the generator of time translations is the Hamiltonian, which in this case corresponds to the dilatation operator. Similarly, the generator for space translations is the momentum operator, corresponding to rotations.  Recalling \eqref{l0combinations},   we can write
\begin{equation} \label{H-P}  	H=L_0+\overline{L}_0\, ,   	\qquad   	P=i(L_0-\overline{L}_0)\, \end{equation}
   where $L_0$ and $\overline{L}_0$ are the zero modes of the energy-momentum tensors, as we will see below.

\begin{figure}[ht] 	\centering 	\caption[]{Subtraction of contours} 	\label{fig:ope-com-relation} 	\includegraphics[width=0.7\linewidth]{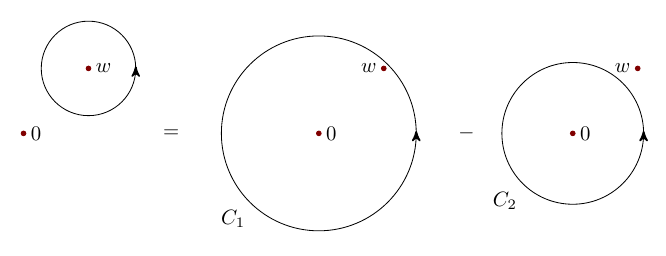} \end{figure}

A time-ordering product now becomes the radial ordering product:
\begin{equation} \mathcal{R}(\phi_1(z)\phi_2(w))= \begin{cases} \phi_1(z)\phi_2(w) & \abs{z}>\abs{w}\, , \\ \phi_2(w)\phi_1(z) & \abs{z}<\abs{w}\, . \end{cases} \end{equation}
 As usual, we will always omit the radial-ordered operator in the correlation function as well as in the OPE expansion. One consequence after specifying time direction is that we can relate OPE to commutation relations. For this, let consider the contour integral around $w$ for  two holomorphic field $a(z)$ and $b(w)$
\begin{equation} \label{relation-OPE-com-1}   \oint_w \dd z\, a(z)b(w) =   \oint_{C_1} \dd z \,a(z)b(w)-   \oint_{C_2} \dd z \,b(w)a(z)  =\comm{A}{b(w)}\, , \end{equation}
 where  the operator $A$ is the contour integral of $a(z)$ at a fixed time
\begin{equation}   A=\oint \dd z \,a(z)\, . \end{equation}
 Here we take the contours $C_1$ and $C_2$ at fixed-radius $\abs{w}+\e$ and $\abs{w}-\e$ with a small positive number $\e$  as illustrated in Figure \ref{fig:ope-com-relation} . Then, with $B=\oint \dd z \,b(z) $, we can generalize the relation \eqref{relation-OPE-com-1} to
\begin{equation} \label{rel-comm-OPE} \comm{A}{B}=\oint_0 \dd w \comm{A}{b(w)} = \oint_0 \dd  w \oint_w \dd z a(z)b(w)\, . \end{equation}

\subsubsection*{State/Operator correspondence}
   After the radial quantization, states living on the circle can be generated by inserting operators. The vacuum state $\ket{0}$ corresponds to inserting nothing. The initial state $\ket{\phi_{\text{in}}}$ corresponds to inserting an operator at the origin since the infinite past on the cylinder is mapped to $z=\overline{z}=0$,
\begin{equation}\label{state-operator}   \ket{\phi_{\text{in}}}  =\lim_{z,\overline{z} \rightarrow 0}   \phi(z,\overline z)\ket{0}\, . \end{equation}
As we have seen, a 2D CFT has scaling transformation as a symmetry. The state defined on a circle with a fixed radius can be scaled down to the origin by the scaling transformation as in \eqref{state-operator}, which results in an operator. Conversely, an operator at the origin can be scaled up to a state on a circle around the origin.  Therefore, there is a one-to-one correspondence between a state and an operator in 2d CFT.  This is called the \textbf{state-operator correspondence}.

In this Hilbert space, we must also define a bilinear product, by defining an asymptotic out state. In Minkowski space, Hermitian conjugation does not affect the space-time coordinates. However, since the Euclidean time $t_{\text{E}}= i t_{\text{M}}$, the Euclidean time must be reversed upon Hermitian conjugation. In radial quantization, this corresponds to the mapping $z\rightarrow 1/\overline{z}$. Therefore, the definition of Hermitian conjugation is
\begin{equation}   \label{def-hemi-phi}   [\phi(z,\overline{z})]^\dagger  =\overline{z}^{-2h}z^{-2\overline{h}}   \phi(1/\overline{z},1/z)\, , \end{equation}
 where $h$ and $\overline{h}$ are conformal dimensions of quasi-primary field $\phi$. The prefactors on the r.h.s. can be justified by demanding that the hermitian conjugation exchanges the conformal dimensions for holomorphic and antiholomorphic parts. We then define the asymptotic out state:
\begin{equation}   \bra{\phi_\text{out}}=   \ket{\phi_\text{in}}^\dagger\, . \end{equation}
 Then, the inner product is well defined:
\bea \braket{\phi_{\text{out}}}{\phi_{\text{in}}} &= \lim_{z,\overline{z}, w , \overline{w} \rightarrow 0} \overline{z}^{-2h}z^{-2\overline{h}} \bra{0}\phi(1/\overline{z},1/z) \phi(w,\overline{w})\ket{0}\cr &= \lim_{\xi,\overline{\xi}\rightarrow \infty} \overline{\xi}^{2h} \xi^{2\overline h} \bra{0}\phi(\overline \xi, \xi)\phi(0,0)\ket{0}\, . \eea
 According to the expression $\eqref{two-point-func-form}$, the last expression is independent of $\xi$, which also justifies the prefactors appearing in \eqref{def-hemi-phi}.

The mode expansion of a  field $\phi(z,\overline{z})$ with conformal dimension $(h,\overline h)$ is conventionally given by
\begin{equation}   \label{mode-expanson-conf-field}   \phi(z,\overline{z}) =   \sum_{m\in\mathbb{Z}}   \sum_{n\in\mathbb{Z}}   z^{-m-h}\overline{z}^{-n-\overline{h}}   \phi_{m,n}\, . \end{equation}
 A straightforward Hermitian conjugation yields
\begin{equation}   \phi(z,\overline{z})^\dagger =   \sum_{m\in\mathbb{Z}}   \sum_{n\in \mathbb{Z}}   \overline{z}^{-m-h}   z^{-n-\overline{h}}   \phi_{m,n}^\dagger \, , \end{equation}
 while the definition \eqref{def-hemi-phi} gives
\begin{equation}   \phi(z,\overline{z})^\dagger =   \overline{z}^{-2h}z^{-2\overline{h}}   \phi(1/\overline{z}, 1/z)   =   \sum_{m \in \mathbb{Z}}   \sum_{n \in \mathbb{Z}}   \phi_{-m,-n}   \overline{z}^{-m-h}z^{-n-\overline{h}}\, . \end{equation}
 Comparing the above two expressions for mode expansions, we get
\begin{equation}   \phi^\dagger_{m,n}=\phi_{-m,-n} \end{equation}
 This is the usual expression for the Hermitian conjugate of modes, not only for primary fields.

\subsubsection*{Conformal generators and Virasoro algebra}
In \S\ref{sec:2d-conf-trans}, we have derived the conformal generators $l_n$ and $\overline{l}_n$ in coordinate space, which form two copies of the Witt algebra \eqref{Witt-algebra}. Now, we shall derive the conformal generators in Hilbert space and its corresponding algebra. We have already seen in \eqref{charge-com-generator} that the conserved charge defined by \eqref{def-conserved-charge} acts as a generator of the corresponding symmetry in Hilbert space. Since the current for conformal symmetry has the form of $T^{\mu\nu}\e_\nu$, writing in complex coordinate and dropping the antiholomorphic part for convenience, we arrive at the conserved charge by performing the contour integral of equal time as
\begin{equation} \label{charge-conformal-Hilbert} Q_\e=\frac{1}{2\pi i} \oint_C \dd z \, \e(z) T(z)\, . \end{equation}
 Then, the conformal Ward-Takahashi identities \eqref{Conformal-Ward-Identity-2d} becomes the same as \eqref{charge-com-generator}
\begin{equation}   \delta_\e \Phi(z)=- \comm{Q_\e}{\Phi(z)}\, , \end{equation}
which verifies that the conserved charge in \eqref{charge-conformal-Hilbert} acts like a conformal generator as well. Now we expand the holomorphic part of the energy-momentum as follows
\begin{equation}   T(z) =\sum_{n \in \mathbb{Z}}   z^{-n-2} L_n\, , \end{equation}
 with $L_n$ equals to
\begin{equation}   L_n=\frac{1}{2 \pi i} \oint \dd z\, z^{n+1} T(z)\, . \end{equation}
 If we expand $\e(z)$ as
\begin{equation}  \e(z)=\sum_{n \in \mathbb{Z}} z^{n+1} \e_n\, , \end{equation}
 we find the conserved charge
\begin{equation}   Q_\e=\sum_{n\in\mathbb{Z}} \e_n L_n\, . \end{equation}
 Therefore, $L_n$ and its antiholomorphic part $\overline{L}_n$ become a set of conformal generators in Hilbert space. The commutator of $L_n$ can be obtained by using \eqref{rel-comm-OPE}
\begin{equation}   \comm{L_n}{L_m}=\frac{1}{(2\pi i)^2}   \oint_0 \dd w \, w^{m+1} \int_w \dd z \, z^{n+1}   T(z)T(w)\, . \end{equation}
 Plugging $TT$ OPE \eqref{OPE-TT}, we finally obtain the \textbf{Virasoro algebra}
\begin{equation} \label{Virasoro-algebra}   \comm{L_n}{L_m}=(n-m)L_{n+m}  +\frac{c}{12}n(n^2-1)\delta_{n+m,0}\, . \end{equation}
 The antiholomorphic part is similar. When central charge $c$ vanishes, the Virasoro algebra reduces to the Witt algebra.

 \subsection{Verma module}
 This subsection will briefly introduce the Hilbert space of a 2d CFT constructed by primary field operators and generators in Virasoro algebra. We have previously mentioned that the vacuum state $\ket{0}$ corresponds to a state with no operator inserted. Since $T(z)\ket{0}$ and $\overline{T}(\overline{z})\ket{0}$ is supposed to be well-defined as $z$,$\overline{z}$ goes to zero, a vacuum state should satisfy
 \begin{equation}   L_n \ket{0}=0\, \qquad   \overline{L}_n\ket{0}=0 \, . \qquad   (n \geq -1) \end{equation}
We know that asymptotic states can be obtained by acting an operator at the origin on the vacuum state. By inserting a primary field, the obtained state becomes the eigenstate of $L_0$.
 \begin{equation}   \label{def-asymptotic}   \ket{h,\overline{h}} \equiv \phi(0,0)\ket{0}\, . \end{equation}
  A simple demonstration follows from \eqref{rel-comm-OPE} and OPE for primary field \eqref{OPE-T-primaryfield-z}:
 \bea   \comm{L_n}{\phi(w,\overline{w})}   &= \frac{1}{2\pi i}   \oint_w  \dd z\, z^{n+1} T(z)\phi(w,\overline{w})   \cr   &= h(n+1) w^n \phi(w,\overline{w})  +w^{n+1}\partial\phi(w,\overline{w})\,. \eea
  With $w \rightarrow 0$ above, we immediately reach the conclusion that $\ket{h,\overline{h}}$ is an eigenstate of $L_0$.
 \begin{equation}    L_0 \ket{h,\overline{h}}=h \ket{h,\overline{h}}\, ,    \qquad    \overline{L}_0 \ket{h,\overline{h}}=\overline{h} \ket{h,\overline{h}}\, . \end{equation}
  Likewise, we have for $n > 0$
 \begin{equation}   L_n\ket{h,\overline{h}}=0 \, ,   \qquad   \overline{L}_n \ket{h, \overline{h}}=0 \, .   \qquad (n > 0) \end{equation}
  The generators $L_{-m}(m>0)$ also increase the conformal dimension, by virtue of the Virasoro algebra \eqref{Virasoro-algebra}:
 \begin{equation}    \comm{L_0}{L_{-m}}=m L_{-m}\, . \end{equation}
  This means the excited state with higher conformal dimensions can be obtained by successively applying of $L_{-m}$ on $\ket{h}$
 \begin{equation}     \label{def-excited-state} L_{-\lambda} \ket{h}	\equiv L_{-\lambda_1}L_{-\lambda_2}\cdots L_{-\lambda_n}\ket{h}\, , 	\qquad (\lambda=\lambda_1 \geq \lambda_2 \cdots \geq \lambda_n\geq 1) \end{equation}
  where $L_{-\lambda_i}$ appear in decreasing order by convention. The states defined in \eqref{def-excited-state} are called descendants of $\ket{h}$, with eigenvalue of $L_0$
 \begin{equation} 	h'=h+|\lambda| 	\equiv h+ N\, , \end{equation}
  where $N=|\lambda|=\sum_i \lambda_i$ is called the level of the descendant.  The subset of the full Hilbert space, generated by the asymptotic state \( |h\rangle \) and its descendants, is closed under the action of the Virasoro algebra. This subspace forms a representation of the Virasoro algebra and is referred to as a Verma module.

 \begin{table}[htbp]
 	\centering
 	\begin{tabular}{l c ll}
 		\toprule
 		  $l$
 		& $p(l)$
 		&
 		\\
 		\midrule
 		  $0$
 		& $1$
 		& $\ket{h}$
 		\\
 		  $1$
 		& $1$
 		& $L_{-1}\ket{h}$
 		\\
 		  $2$
 		& $2$
 		& $L^2_{-1}\ket{h}, L_{-2}\ket{h}$
 		\\
 		  $3$
 		& $3$
 		& $L^3_{-1}\ket{h},
 		  L_{-2}L_{-1}\ket{h},
 		  L_{-3}\ket{h}
 		$
 		\\
 		  $4$
 		& $5$
 		& $L^4_{-1}\ket{h},
 		L_{-2}L^2_{-1}\ket{h},
 		L_{-3}L_{-1}\ket{h}.
 		L^2_{-2}\ket{h},
 		L_{-4}\ket{h}
 		$
 		\\
 		\bottomrule
 	\end{tabular} 	\caption{States of a Verma module}\label{Tab:Verma module}
 \end{table}

 Table \ref{Tab:Verma module}  displays the states within a Verma module up to level four, where \( p(l) \) represents the number of linearly independent states at level \( l \). This number corresponds to the partitions of the integer \( l \). We can also define the generating function of the Verma module as
 \begin{equation}\label{eq:character} \textrm{Tr}(q^{L_0-\frac{c}{24}})=\frac{1}{\eta(q)}\equiv q^{-\frac{1}{24}} \prod_{n=1}^{\infty}   \frac{1}{1-q^n}  =\sum_{l=0}^{\infty}p(l)  q^{l-\frac{1}{24}}\, , \end{equation}
  which can be also interpreted as the generating function of the number $p(l)$ of the partition of $l$. The function \[ \eta(q):=q^{1/24} \prod_{n=1}^{\infty}(1-q^n) \] is called the \textbf{Dedekind $\eta$-function}.

The inner product is defined following the previous definition of the Hermitian conjugate \eqref{def-hemi-phi}: $L_m^\dagger=L_{-m}$. Therefore the dual state $\ket{h}$ satisfies
 \begin{equation} 	\bra{h} L_j=0 \, .\qquad (j<0) \end{equation}
  Then the inner product of two states $L_{-\mu}\ket{h}$ and $L_{-\lambda}\ket{h}$ is
 \begin{equation} 	\bra{h} L_{\mu_m}\cdots L_{\mu_1} 	L_{-\lambda_1}\cdots L_{-\lambda_n}\ket{h}\, . \end{equation}
This product can be evaluated using the Virasoro algebra. The inner product between two states vanishes unless they belong to the same level, and it depends solely on the highest weight \( h \) and the central charge \( c \). For the representation theory to be unitary, meaning it contains no negative-norm states, there must be strict constraints on \( (h, c) \). For example, consider the norm of the state \( L_{-n} |h\rangle \)
 \bea 	\bra{h}L_n L_{-n}\ket{h} = 	\bigg[2n h+\frac{1}{12}cn(n^2-1)\bigg]\bra{h}\ket{h}. \eea
  It is easy to see that the norm becomes negative for $n$ sufficiently large if $c<0$. Moreover, for $n=1$, we find that the theory with negative conformal dimensions is not unitary.

 \section{Free boson and free fermion}\label{sec:free}
 In this section, we are going to demonstrate more simple examples in CFT. \subsection{Free boson}\label{sec:free-boson}     In the previous sections, we have learned the calculation of some basic OPEs in the massless free boson theory. We will further study this theory. We also refer the reader to \cite[\S6]{francesco2012conformal}, \cite[\S6,\S7]{ginsparg1988applied} and \cite[\S2]{Blumenhagen:2009zz}.

 \subsubsection*{Mode Expansion}
  First, let us recall that the action of free boson  \eqref{def-Free-boson action}
 \begin{equation}    \label{def-free-boson-nol}    \cS=\frac{1}{8\pi} \int \dd^2 x    \partial_\mu \varphi \partial^\mu \varphi    \, . \end{equation}
  First, we consider the free boson on the cylinder, in  Minkowski space-time. The fact that the fields  $\varphi$ live on the cylinder implies that they  may satisfy periodic boundary conditions:
 \begin{equation}    \label{periodic-boundary-condition}    \varphi(x+\ell,t)=\varphi(x,t)\, \end{equation}
   where $\ell$ is the circumference of the cylinder.  The classical equation of motion for $\varphi$ is
 \begin{equation}    [\partial_t^2-\partial_x^2]    \varphi(x,t) =0 \, . \end{equation}
   It is a good exercise to verify that $\varphi$  of the following form  satisfies the above two conditions:
 \begin{equation}    \label{mode-expansion-cylinder}    \varphi_{\text{cyl}}=\varphi_0+\frac{4\pi}{\ell}    \pi_0 t+i\sum_{n\neq 0} \frac{1}{n}    \bigg(      a_n e^{-2\pi i n(t-x)/\ell}      +      \overline{a}_n e^{-2\pi i n(t+x)/\ell}      \bigg)\, \end{equation}
  where 'cyl' means the field defined on the cylinder. The canonical momentum for the theory is
 \begin{equation} \Pi=\frac{\delta \cS}{\delta (\partial_t\varphi)}=\frac{1}{4\pi}\partial_t\varphi\, , \end{equation}
  For quantum fields, we impose the following canonical commutation relations
 \bea   \label{comm-rel-cano-variables}   &\comm{\varphi(x,t)}{\Pi(y,t)}= i \delta(x -y)   \, ,\cr   &\comm{\varphi(x,t)}{\varphi(y,t)}=0   \, ,\cr   &\comm{\Pi(x,t)}{\Pi(y,t)} =0   \, , \eea
  In terms of the modes in \eqref{mode-expansion-cylinder}, these are equivalent to
 \bea   \label{com-rel-mode-expan-boson}   &\comm{a_k}{a_l}   =\comm{\overline{a}_k}{\overline{a}_l}   =k \delta_{k+l,0} \, ,   \cr   &\comm{a_k}{\overline{a}_l}=0 \, , \eea
  for $k,l\neq 0$, and
 \begin{equation} 	\comm{\varphi_0}{\pi_0}=i\, . \end{equation}

 The Hamiltonian for free boson theory is
 \begin{equation}   H=\frac{1}{8\pi}   \int \dd x [:(\partial_t \varphi)^2 :   +:(\partial_x\varphi)^2:]\, , \end{equation}
  Plugging the mode expansion \eqref{mode-expansion-cylinder}, the Hamiltonian is then expressed as
 \begin{equation}   \label{Hamiltonian-mode-expansion}   H=\frac{2 \pi}{\ell} \pi^2_0 +   \frac{\pi}{\ell} \sum_{n\neq 0}   (:a_{-n} a_n:+:\overline{a}_{-n} \overline{a}_n:)\, , \end{equation}
  The commutation relations \eqref{com-rel-mode-expan-boson} lead to the relation
 \begin{equation}   \comm{H}{a_{-m}}=\frac{2\pi}{\ell} m a_{-m}\, , \end{equation}
  which means $a_{-m}(m>0)$ act as creation operators. When applied to an eigenstate $|E\rangle$ of $H$ of energy $E|E\rangle$, the state $a_{-m}|E\rangle$ is another eigenstate with energy $(E+2m\pi/\ell)a_{-m}|E\rangle$. For the same reason, $a_{m}(m>0)$ is an annihilation operator.

\subsubsection*{Relation between Cylinder and Plane}
Now, we move to Euclidean space-time ($t \rightarrow -i t$) by taking $w=t-i x$ and $\overline{w}=t+i x$. The mode expansion of boson field on the cylinder now is then
\begin{equation}\label{mode-expansion-boson} \varphi_{\text{cyl}}(w,\overline{w}) = \varphi_0-i\frac{2\pi}{\ell} \pi_0 (w+\overline{w})+i\sum_{n\neq 0} \frac{1}{n} \bigg( a_n e^{-2\pi n w /\ell} + \overline{a}_n e^{-2\pi n\overline{w})/\ell}
\bigg)\, ,
\end{equation}
with now $w\sim w+i\ell$. Using a conformal transformation as in Figure \ref{fig:cylinder-plane-map}, we map all the operators from the cylinder  to the complex plane:
\begin{equation}   z=e^{2\pi w /\ell}\, , \qquad   \overline{z}=e^{2\pi \overline{w}/\ell}\, . \end{equation}
 Since  $\pi$ is invariant under conformal transformations, we finally obtain the expansion by simply replacing $w$ by $z$.
\begin{equation}   \varphi_{\text{pl}}(z, \overline{z})  =\varphi_0-i \pi_0 \ln(z\overline{z})  +i \sum_{n\neq 0} \frac{1}{n}   (     a_n z^{-n}+\overline{a}_n \overline{z}^{-n}  )\, . \end{equation}
 Now we concentrate on the holomorphic field $\partial\varphi$
\begin{equation}   i\partial\varphi(z)=\frac{\pi_0}{z}+   \sum_{n \neq 0} a_n z^{-n-1}\, . \end{equation}
 We may write the zero modes by $a_0$ and $\overline{a}_0$:
\begin{equation}   a_0 \equiv \overline{a}_0 \equiv \pi_0 \end{equation}
 Now the commutation relations of \eqref{com-rel-mode-expan-boson} can be extended to include the zero mode operator without changing the form of the algebra. The mode expansion of $\partial\varphi$ is  consistent with the expansion of a primary field with $h=1$ \eqref{mode-expanson-conf-field}:
\begin{equation} 	i \partial\varphi(z)=\sum_n a_n z^{-n-1} \, . \end{equation}
 We can calculate a two-point function by using mode expansion. For $(\abs{z}>\abs{w})$:
\begin{equation}    \expval{\varphi(z)\partial\varphi(w)}   =\sum_{m,n\neq 0} \frac{1}{n}    \expval{a_n a_m} z^{-n} w^{-m-1} \, . \end{equation}
 According to the commutation relation \eqref{com-rel-mode-expan-boson} and the fact that $a_m$ and $a_{-m}$ are annihilation and creation operators respectively, it follows that
\begin{equation}   \expval{\varphi(z)\partial\varphi(w)}  =\frac{1}{w}\frac{w/z}{1-w/z}  =\frac{1}{z-w} \, . \end{equation}
 Its differentiation with respect to $z$ provides the two-point function.
\begin{equation} 	\expval{\partial\varphi(z)\partial\varphi(w)} 	=-\frac{1}{(z-w)^2} \, . \end{equation}

The holomorphic energy-momentum tensor is given by
\begin{equation} 	T(z)=- \frac{1}{2} :\partial\varphi(z) 	\partial\varphi(z): 	= \frac{1}{2} 	\sum_{n,m\in \mathbb{Z}} 	z^{-n-m-2} : a_n a_m :\, , \end{equation}
 which implies
\bea    &L_n=\frac{1}{2}    \sum_{m\in \mathbb{Z}} : a_{n-m} a_m:\,    \qquad (n\neq 0) \cr    &L_0=\sum_{n>0} a_{-n}a_n+\frac{1}{2} a_0^2~. \eea
 The Hamiltonian \eqref{Hamiltonian-mode-expansion} now can be written as,
\begin{equation}   H=\frac{2\pi}{\ell}(L_0+\overline{L}_0)\, . \end{equation}
 This confirms the role of $L_0$ and $\overline{L}_0$ as a Hamiltonian. Since we place the free boson on a cylinder of size $\ell$, energy is proportional to $\frac{2\pi}{\ell}$, which is called the \textbf{finite-size scaling} \cite[ISZ88-No.18,19,25]{Cardy:1984rp,Cardy:1984mun,cardy1986operator}.  The mode operator $a_m$ plays a similar role  to $L_m$ with respect to $L_0$, because of the commutation relation $\comm{L_0}{a_{-m}}=m a_{-m}$. Therefore its effect on the conformal dimension is the same as that of $L_m$.

By definition, the normal ordering prescription implies that $\expval{T(z)}=0$.\footnote{However, as we will see in the next section, this is not the case for the fermionic theory with anti-periodic boundary condition.} We can always define the normal ordering product by subtracting all the singular terms from the OPE. In fact the general expectation value for energy-momentum tensor can be written as
\begin{equation}   \label{def-EM-tensor 2}   \expval{T(z)} =- \frac{1}{2} \lim_{\e\rightarrow 0}   \bigg(       \expval{\partial\varphi(z+\e) \partial\varphi(z)} +      \frac{1}{\e^2}   \bigg)\, . \end{equation}
 We will see that this relation is useful when we consider the anti-periodic boundary condition later. By plugging in to the two-point function, we see that $\expval{T(z)}=0$, which implies that $(L_0)_{\text{pl}}$ vanishes on the vacuum. We now map the theory back to a cylinder $z\rightarrow w=\frac{\ell}{2\pi}\ln z$, by using \eqref{trans-T-finite}:
\begin{equation} \label{cyl-plane}  T_{\text{cyl}} (w)=   \bigg(         \frac{2\pi}{\ell}   \bigg)^2   \bigg[   T_{\text{pl}}(z)z^2-\frac{1}{24}   \bigg]\, . \end{equation}
 where we use the central charge $c=1$ for the free boson. Taking the expectation value on both sides, we have
\begin{equation}   \expval{T_{\text{cyl}}}=- \frac{1}{24}   \bigg(       \frac{2\pi}{\ell}   \bigg)^2 \, . \end{equation}
 This implies
\begin{equation}   L_{0,\text{cyl}}  =\frac12 a_0^2+\sum_{n>0} a_{-n} a_n-\frac{1}{24}\, . \end{equation}
The Hamiltonian is now written as
\begin{equation}   H=\frac{2\pi}{\ell}   (L_{0,\text{cyl}}+\overline{L}_{0,\text{cyl}})\, . \end{equation}
 Actually, in the general case, Hamiltonian of a theory defined on a cylinder with central charge $c$ can be written as
\begin{equation}\label{hamiltonian-plane} 	H=\frac{2\pi}{\ell}(L_{0,\text{pl}}+\overline{L}_{0,\text{pl}}-\frac{c}{12})\, . \end{equation}
 Thus, we can infer that the central charge $c$ shows up as the vacuum energy of a theory on a cylinder, and this is one instance of finite-size effects \cite[ISZ88-No.20]{Bloete:1986qm}.

\subsection{Free fermion}

\subsubsection*{OPEs in Free fermion}
Now we consider another simple model: free fermion. In two dimensions, the action of a free Majorana fermion is
\begin{equation} \label{action-free-fermion} \cS=\frac{1}{4\pi} \int \dd^2 x \Psi^\dagger \gamma^0\gamma^\mu \partial_\mu \Psi \, , \end{equation}
 where the gamma matrices $\gamma^\mu$ satisfy the so-called \textbf{Clifford algebra}:
\begin{equation} \label{Dirac-algebra} \{\gamma^\mu,\gamma^\nu\}=2 \eta^{\mu\nu}\, , \end{equation}
 and we impose the \textbf{Majorana condition} $(\Psi^*=\Psi)$ to the fermionic field to remove a half of the degrees of freedom. In the Euclidean space $\eta^{\mu\nu}=\text{diag}(1,1)$, we take a basis of Dirac matrices as
\begin{equation} \gamma^0=\mqty(0 & 1\\ 1 & 0)\, , \qquad \gamma^1=i \mqty(0 & -1 \\ 1 & 0) \, . \end{equation}
 Using this basis, we can express the action as
\begin{equation} \label{fermion-action-complwx} \cS= \frac{1}{2\pi}  \int \dd^2 x (\overline \psi \partial \overline{\psi} + \psi \overline{\partial}\psi) \, , \end{equation}
 where we write the two-component spinor $\Psi$ as $(\psi, \overline{\psi})$.  Since the equations of motion are $\partial\bar\psi=0$ and $\overline{\partial}\psi=0$,  $\psi(z)$ and $\overline{\psi}(\overline{z})$ holomorphic and  antiholomorphic field, respectively.

Now let us calculate the two-point function as in the free fermion
\begin{equation} K_{ij}=\expval{\Psi_i(x)\Psi_j(y)} \quad (i,j=1,2) \, . \end{equation}
 The action can be expressed by
\begin{equation} \cS= \frac{1}{4\pi} \int \dd^2 x \dd^2 y \Psi_i(x) A_{ij}(x,y) \Psi_j(y) \, , \end{equation}
 where the kernel is
\begin{equation} A_{ij}(x,y)= \delta(x-y)(\gamma^0 \gamma^\mu)_{ij} \partial_\mu\, . \end{equation}
 Recalling \eqref{two-point-function-calculation}, the propagator $K_{ij}$ is the inverse of $A_{ij}$. Therefore, we can write the equation for $K$ as
\begin{equation} (\gamma^0\gamma^\mu)_{ik} \pdv{x^\mu} K_{kj}(x,y) = \delta(x-y)\delta_{ij} \, . \end{equation}
 In terms of complex coordinates, this becomes
\begin{equation} \frac{1}{\pi}\mqty(\overline{\partial} & 0 \\ 0           &\partial) \left(\begin{array}{l l}{\langle \psi (z , \overline{z}) \psi (w , \overline{w}) \rangle} &{\langle \psi (z , \overline{z}) \overline{\psi} (w , \overline{w}) \rangle} \\{\langle \overline{\psi} (z , \overline{z}) \psi (w , \overline{w}))} &{\langle \overline{\psi} (z , \overline{z}) \overline{\psi} (w , \overline{w}) \rangle} \end{array} \right)= \frac{1}{\pi}\mqty( \overline{\partial}\frac{1}{z-w} & 0 \\ 0 & \partial \frac{1}{\overline{z}-\overline{w}} ) \, , \end{equation}
 where we have used the complex form of the $\delta$-function: \[\delta (x)=\frac{1}{\pi} \overline{\partial} \frac{1}{z}=\frac{1}{\pi} \partial \frac{1}{\overline{z}}~.\] Therefore, we obtain the two-point functions for the fermionic fields
\bea \langle \psi (z , \overline{z}) \psi (w , \overline{w}) \rangle &=\frac{1}{z-w}\, ,\\ \langle \overline{\psi} (z , \overline{z}) \overline{\psi} (w , \overline{w}) \rangle &= \frac{1}{\overline{z}-\overline{w}}\, ,\\ \langle \psi (z , \overline{z}) \overline{\psi} (w , \overline{w}) \rangle &= 0 \, . \eea
 Thus the OPE of two holomorphic fields can be written as:
\begin{equation} \psi (z) \psi (w) \sim  \frac{1}{z-w} \end{equation}
 In order to see whether the fermion field is a primary field or not, we can calculate its OPE with the energy-momentum tensor. By using \eqref{cano-ener-mon-tensor} in complex-coordinate form, we can calculate all the components of the energy-momentum tensor.
\bea T^{\overline z \overline{z}} &= 2 \frac{\partial \mathcal{L}}{\partial (\overline{\partial} \Phi)} \partial \Phi= \frac{1}{\pi}  \psi \partial \psi \, , \\ T^{z z} &= 2 \frac{\partial \mathcal{L}}{\partial (\partial \Phi)} \overline{\partial} \Phi = \frac{1}{\pi}  \overline{\psi} \overline{\partial} \overline{\psi} \, , \\ T^{z \overline{z}} &= 2 \frac{\partial \mathcal{L}}{\partial (\partial \Phi)} \partial \Phi-2 \mathcal{L}=-  \frac{1}{\pi}  \psi \overline{\partial} \psi\, . \eea
 The traceless condition $T^{z\overline{z}}$ is preserved when taking into account the equation of motion, as we have discussed. The holomorphic part is defined as:
\begin{equation} \label{energy-momentum tensor: fermion} T(z)=-\frac12 : \psi (z) \partial \psi (z):\, . \end{equation}
 The normal-ordering product can be written in an equivalent way for the free field as follow:
\begin{equation} : \psi \partial \psi : (z)=\lim_{w \rightarrow z} (\psi (z) \partial \psi (w)-\langle \psi (z) \partial \psi (w) \rangle)\, , \end{equation}
which is the same expression \eqref{def-EM-tensor 2} as in bosonic field theory. Then we can calculate the OPE between the fermion field and energy-momentum tensor directly.
\begin{equation} T (z) \psi (w) \sim \frac{\frac{1}{2} \psi (w)}{(z-w)^{2}}+\frac{\partial \psi (w)}{z-w}\, . \end{equation}
 This immediately tells us that in the free fermion model, $\psi$ is a primary field with conformal dimension $1/2$. $TT$ OPE can also be obtained directly by calculation.
\begin{equation} T (z) T (w) \sim \frac{1 / 4}{(z-w)^{4}}+\frac{2 T (w)}{(z-w)^{2}}+\frac{\partial T (w)}{(z-w)}\, . \end{equation}
 This expression verifies the general form of $TT$ OPE \eqref{OPE-TT}, with $1/2$ the central charge in this model.

\subsubsection*{Mode Expansion}
Next, we examine the free fermion field on a cylinder, following an approach similar to that used for the boson field. We consider a cylinder with circumference \( L \), applying two distinct types of boundary conditions:
\bea \psi (x+\ell) \equiv \psi (x) &\quad \text{Ramond} (\mathrm{R})\, ,\\ \psi (x+\ell) \equiv-\psi (x)&\quad \text{Neveu-Schwarz (NS)}\, . \eea
Analogous to the case of free bosons, the mode expansion for the free fermion is given by:
\begin{equation} \psi (x , t)=\sqrt{\frac{2 \pi}{\ell}} \sum_{k\in \bZ+\nu} b_{k} e^{- 2 \pi k w / \ell}\, . \end{equation}
In the periodic case (R), \( k \) takes integer values (\( \n = 0 \)), while in the anti-periodic case (NS), \( k \) assumes half-integer values (\( \n = \frac{1}{2} \)). As before, we introduce the complex coordinate \( w = t_E - ix \), where \( t_E \) represents Euclidean time. For quantum fermionic fields, the canonical anti-commutation relation is introduced as follows
\begin{equation} \label{commutation-relation-bk} \left\{b_{k} , b_{q} \right\}=\delta_{k+q , 0}\, , \end{equation}
 where $b_{-k}$ for $k>0$ can be considered as a creation operator and $b_{k}$ for $k>0$ as an annihilation one. Note that, in the R sector, there exists a zero mode $b_0$ which leads to a degeneracy of the vacuum.

\subsubsection*{Mapping to Complex Plane}
By applying the coordinate transformation \( z = e^{{2\pi w}/{L}} \), the cylinder is mapped onto the plane. Unlike in the free boson theory, the field \( y \) possesses a conformal dimension of \(1/2 \), which introduces a prefactor in accordance with the conformal transformation properties of a primary field \eqref{def-quasi-primary-field}:
\begin{equation} \psi_{\text{cyl}}(w) \rightarrow \psi_{\text{pl}}(z) = \sqrt{\frac{\ell}{2\pi z}}\psi_{\text{cyl}}(z)\, . \end{equation}
 Therefore, on the plane, the mode expansion changes to
\begin{equation} \label{mode-expansion-fermion} \psi(z)=\sum_{k\in \bZ+\nu} b_k z^{-k-1/2} \, . \end{equation}
 This coincides with the general expansion formula for a primary field \eqref{mode-expanson-conf-field}. Now the prefactor $z^{-1/2}$ has interchanged the role of two boundary conditions: The NS condition corresponds to a periodic field ($\nu=\frac{1}{2})$ and the R condition to an anti-periodic field ($\nu=0$)
\bea \psi (e^{2 \pi i} z ) &=-\psi (z) \quad \text{Ramond (R)}\, ,\\ \psi (e^{2 \pi i} z ) &= \psi (z) \quad \text{Neveu-Schwarz (NS)} \, . \eea
 We now calculate the two-point function in the NS sector by using the mode expansion:
\begin{equation} \langle \psi (z) \psi (w) \rangle=\sum_{k ,{q} \in \mathbb{Z}+1 / 2} z^{- k-1 / 2} w^{- q-1 / 2} \left\langle b_{k} b_{q} \right\rangle\, . \end{equation}
 Applying the commutation relations \eqref{commutation-relation-bk}, this can be calculated:
\begin{equation} \langle \psi (z) \psi (w) \rangle = \sum_{k \in \mathbb{Z}+1 / 2 , k > 0} z^{- k-1 / 2} w^{k-1 / 2} =\frac{1}{z-w}\, . \end{equation}
 The final result is the same as the two-point function for fermions we have derived before. However, in the Ramond sector, the two-point function is different.
\begin{equation} \langle \psi (z) \psi (w) \rangle = \frac{1}{2 \sqrt{z w}}+\sum_{k=1}^{\infty} z^{- k-1 / 2} w^{k-1 / 2} = \frac{1}{2} \frac{\sqrt{z / w}+\sqrt{w / z}}{z-w}\, . \end{equation}
 When  $w\rightarrow z$, this result coincides with the previous one. So, the boundary condition does not affect the two-point function locally. However, globally when $z$ or $w$ is taken around the origin, there arises a minus sign due to the boundary condition. We find that different boundary conditions lead to different two-point functions on the plane.

\subsubsection*{Vacuum Expectation Value}
In this section, we will see the boundary condition will affect the vacuum expectation of the energy-momentum tensor and finally, affect the vacuum energy. As we have mentioned before, the energy-momentum tensor can be calculated by the following normal-ordering prescription:
\begin{equation} \langle T (z) \rangle=\frac{1}{2} \lim_{\e \rightarrow 0} \left(-\langle \psi (z+\varepsilon) \partial \psi (z) \rangle+\frac{1}{\varepsilon^{2}} \right)\, . \end{equation}
 In the NS sector, the vacuum expectation value of the energy-momentum tensor vanishes as before.
\begin{equation} \expval{T(z)}=0 \, . \end{equation}
 However, in the R sector, the calculation yields:
\begin{equation} \langle T (z) \rangle=- \frac{1}{4} \lim_{w \rightarrow z} \partial_{w} \left(\frac{\sqrt{z / w}+\sqrt{w / z}}{z-w} \right)+\frac{1}{2 (z-w)^{2}} = \frac{1}{16 z^{2}}\, . \end{equation}
 This means boundary conditions do affect the expectation value of the energy-momentum tensor. Now plugging the mode expansion expression \eqref{mode-expansion-fermion} into the energy-momentum tensor \eqref{energy-momentum tensor: fermion}:
\bea T_{\mathrm{pl.}} (z) &= \frac{1}{2} \sum_{k , q\in \bZ+\nu} \left(k+\frac{1}{2} \right) z^{- q-1 / 2} z^{- k-3 / 2} : b_{q} b_{k}:\cr &= \frac{1}{2} \sum_{n \in \bZ, k\in \bZ+\nu} \left(k+\frac{1}{2} \right) z^{- n-2} : b_{n-k} b_{k}:\, . \eea
 We can easily extract the conformal generator from the expression above:
\begin{equation} L_{n}=\frac{1}{2} \sum_{k\in \bZ+\nu} \left(k+\frac{1}{2} \right) : b_{n-k} b_{k}: ~.\end{equation}
 We can write $L_0$ as follows by using the commutation relations \eqref{commutation-relation-bk}.
\bea\label{L0-fermion} L_{0} &= \sum_{k > 0} k b_{- k} b_{k} \quad \left(\mathrm{NS} : k \in \mathbb{Z}+\frac{1}{2} \right) \, , \\ L_{0} &= \sum_{k > 0} k b_{- k} b_{k}+\frac{1}{16} \quad (\mathrm{R} : k \in \mathbb{Z})\, , \eea
 where the constant term is fixed from the vacuum expectation value of energy-momentum tensor for NS and R sector, since $L_0$ is the coefficient of $1/z^2$ in the expansion. As mentioned before, the Hamiltonian of such a theory defined on the cylinder can be determined by the following expression with $L_0$ on the complex plane \eqref{hamiltonian-plane}  with $c$ the central charge which is $1/2$, in this theory. From the relation above, we can immediately obtain the vacuum energy in the holomorphic part as follows:
\begin{equation} \frac{\ell}{2 \pi} E_{0}=\left\{\begin{array}{l l}{- \frac{1}{48}} &{\text{NS sector}} \\{+ \frac{1}{24}} &{\text{R sector}} \end{array} \right. \end{equation}
 The same result can be obtained by mapping the energy-momentum tensor from the complex plane to the cylinder, by using \eqref{trans-T-finite}:
\begin{equation} T_{\text{cyl}} (w)= \bigg( \frac{2\pi}{\ell} \bigg)^2 \bigg[ T_{\text{pl}}(z)z^2-\frac{c}{24} \bigg]\, , \end{equation}
 with $c=1/2$. Taking expectation value from both sides of above, and then treating NS sector and R sector respectively, we find
\begin{equation} \left\langle T_{\mathrm{cyl.}} \right\rangle=\left\{\begin{aligned}-\frac{1}{48} \left(\frac{2 \pi}{\ell}\right)^{2} & \text{NS sector} \\ \frac{1}{24} \left(\frac{2 \pi}{\ell}\right)^{2} & \text{R sector} \end{aligned} \right. \end{equation}
 This implies the same vacuum energy as before.

\subsection{Torus partition functions}\label{sec:free-torus}

\subsubsection*{Moduli  of a two-torus}
\begin{figure}
	\centering
	\includegraphics[width=0.7\linewidth]{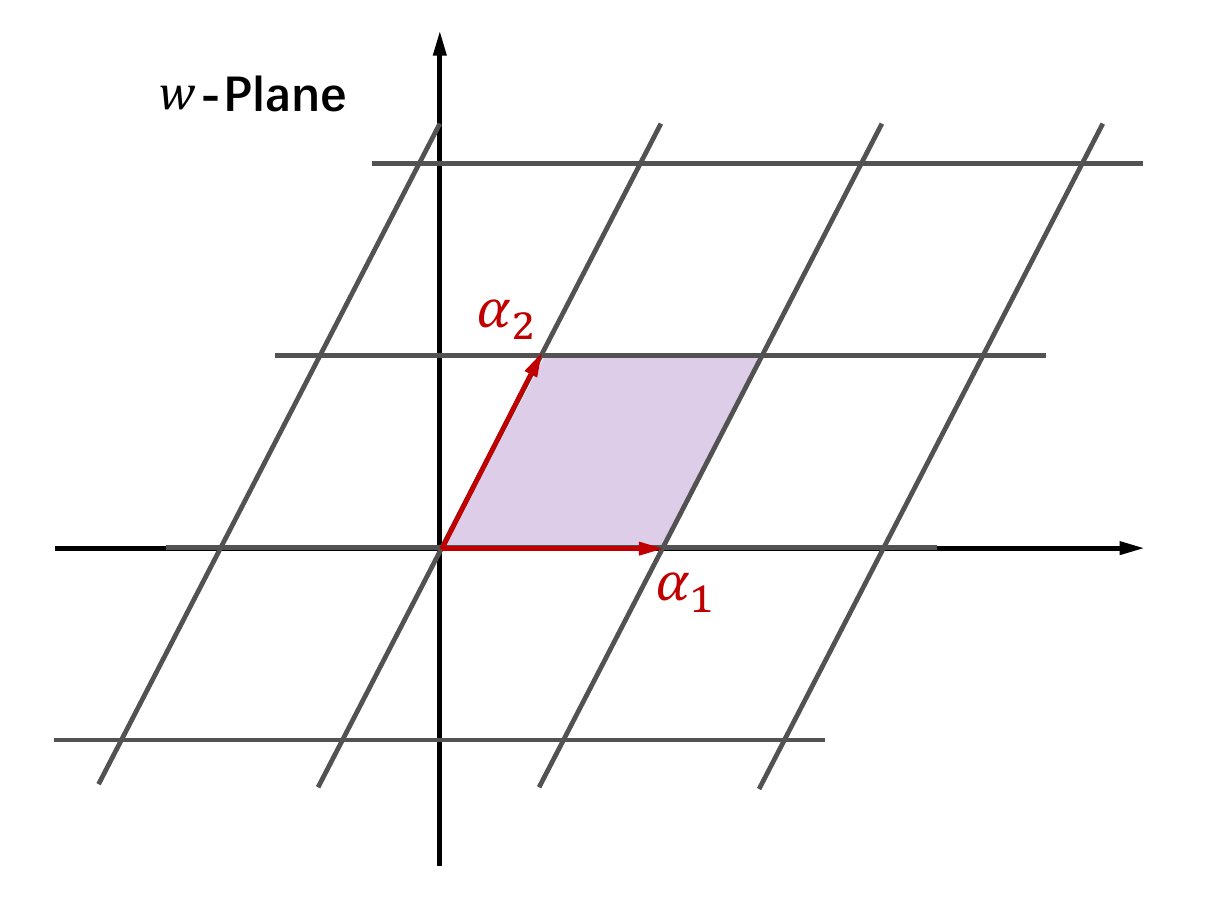}
	\caption{Lattice of a torus}
	\label{fig:torus-lattice}
\end{figure}

In this chapter, we are going to study CFTs defined on a torus $T^2$. The easiest way to obtain such a theory while using the previous result is to
cut out a finite piece of the infinite cylinder described by
$w=x^0+i x^1 $, where $x^0$ is the Euclidean time, and we identify
the circumference of cylinder $l \equiv 2\pi$ in this section.
Thus, the conformal mapping $z= e^w$  defines a theory
from the cylinder to a complex plane.
We will now perform the compactification to the torus. A
torus can be defined by identifying points $w$ in the complex plane
$\mathbb{C}$ as
\begin{equation}
 w \sim w+m \alpha_1+n \alpha_2 \, , \qquad  m, n \in \mathbb{Z}
\end{equation}
where $(\alpha_1, \alpha_2)$ is a pair of complex numbers. This pair
spans a lattice whose smallest cell illustrates the shape of a torus as in Figure \ref{fig:torus-lattice}. Thus, we can introduce the \textbf{complex structure}
\begin{equation}\label{complex-str}
 \tau =\frac{\alpha_2}{\alpha_1}=\tau_1+i \tau_2 \, ,
\end{equation}
which parametrizes the shape of a torus.
However, there are various choices of pairs $(\alpha_1,\alpha_2)$ giving the
same lattice and thus the same torus. If the two pairs $(\beta_1,\beta_2)$ and $(\alpha_1,\alpha_2)$ describe the same lattice, then they are related by an invertible $2\times2$ matrix with integral entries
\begin{equation}
\mqty(\beta_2 \\
\beta_1)
= \mqty(a & b \\
c & d)
\mqty(\alpha_2 \\
\alpha_1)\, ,
\qquad
a,b,c,d \in \mathbb{Z} \, .
\end{equation}
Moreover, because the area is preserved, we have $ad-bc= 1$ so that the matrix is an element of $SL(2,\bZ)$. Furthermore, the torus illustrated by $(\alpha_1, \alpha_2)$ is the same as the one by $(-\alpha_1, -\alpha_2)$, so that the transformation by the $\bZ_2$ subgroup of $SL(2,\bZ)$ is trivial. All in all, the symmetry of the lattice with fixed shape is  $SL(2, \mathbb{Z})/\mathbb{Z}_2$, which is called the \textbf{modular group}  or the \textbf{mapping class group} of the torus. Finally, by choosing
$(\alpha_1,\alpha_2)=(1,\tau)$, a modular group acts on the complex structure of the torus as
$\tau$ as
\begin{equation}
\label{modular-group}
\tau \mapsto \frac{a \tau+b}{c \tau+d}\quad \text{with}\quad \left(\begin{array}{l l}{a}&{b}\\{c}&{d}\end{array} \right) \in S L (2 , \mathbb{Z}) / \mathbb{Z}_{2 }~.
\end{equation}

Now let us look at some particular cases. $T$-transformation:
\begin{equation}
 T: \tau \to \tau+1\, ,
\end{equation}
which makes the lattice $(\alpha_1, \alpha_2)=(1,\tau)$ to $(1, \tau+1)$, thus defining the same lattice. And $U$-transformation:
\begin{equation}
U: \tau \to \frac{\tau}{\tau+1}\, ,
\end{equation}
which makes the lattice $(\alpha_1, \alpha_2)=(1,\tau)$ to $(1+\tau, \tau)$, thus defining the same lattice. However, it seems more convenient to work with $S$- transformation instead of $U$-transformation.
\begin{equation}
S: \tau \to-\frac{1}{\tau}
\end{equation}
which interchanges the lattice $(\alpha_1,\alpha_2)\leftrightarrow(- \alpha_2, \alpha_1)$\, . One can easily show that
\begin{equation}
 S= U T^{-1} U \,, \qquad
 S^2=\mathbb{1}\, \qquad
 (ST)^3=\mathbb{1}\, .
\end{equation}
In fact, $S$ and $T$ can prove to be the generators of the modular group $SL(2,\mathbb{Z})/\mathbb{Z}_2$. Therefore, it is sufficient to check whether a theory is modular invariant by studying its behavior under $T$ and $S$ transformations.
\begin{figure}
	\centering
	\begin{minipage}[c]{0.4\textwidth}
		\centering
		\includegraphics[width=7.5cm]{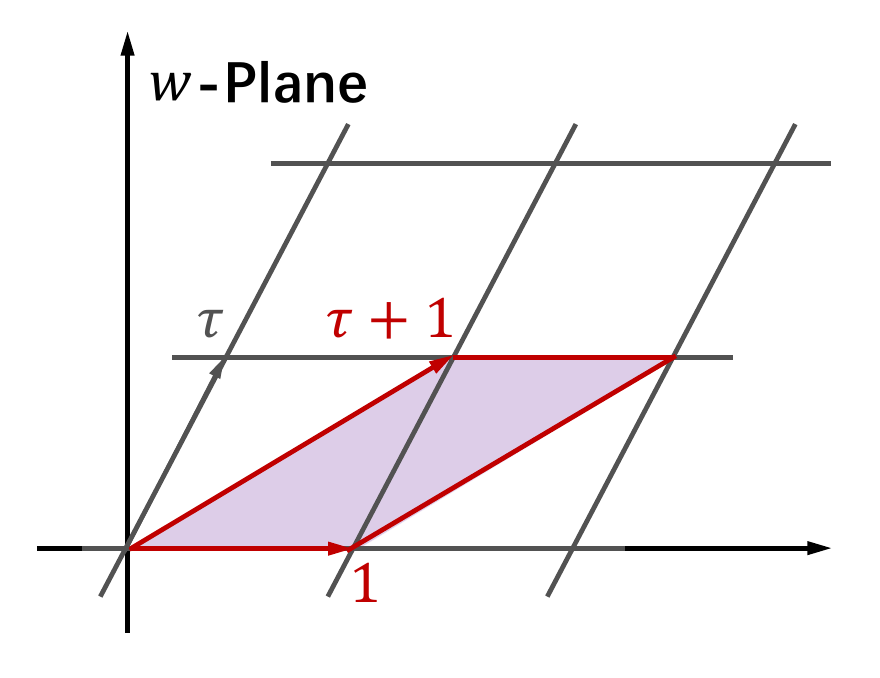}
		\caption{$T$-transformation}\label{fig:T-transformation}
	\end{minipage}
	\begin{minipage}[c]{0.4\textwidth}
		\centering
		\includegraphics[width=7.5cm]{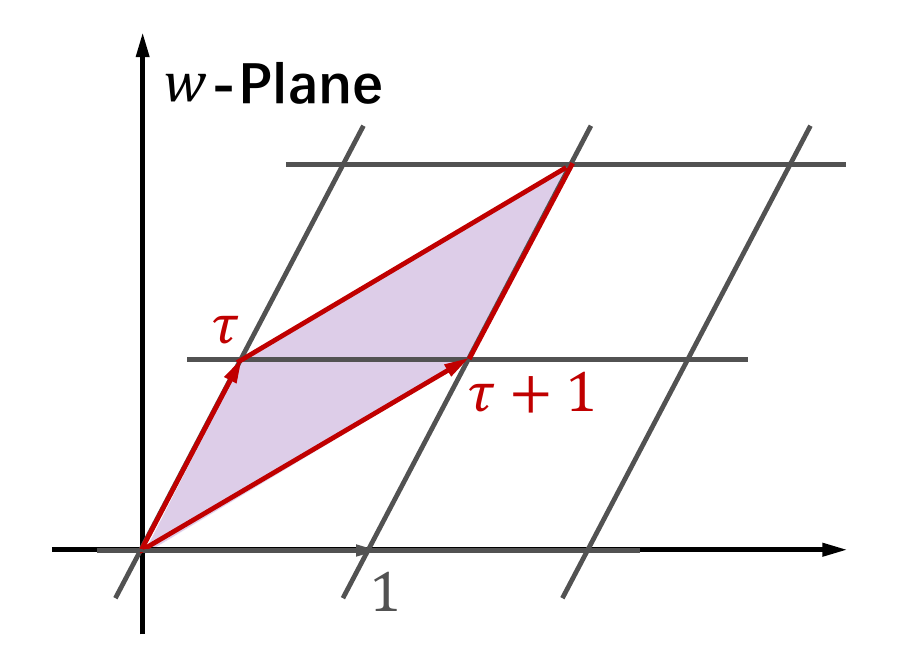}
		\caption{$U$-transformation}
	\end{minipage}
\end{figure}

\subsubsection*{Torus partition function}

Now we will define a torus partition function $\cZ$ in terms of Virasoro generators. The story is analogous to the partition function in statistical mechanics
\be\label{Boltzmann}
\cZ=\sum_{\textrm{all states}} \exp(-\beta H)~.
\ee
The sum over all states can be understood as identifying the Hilbert space at $t=0$ and $t=\beta=1/T$. This is equivalent to compactifying the Euclidean time on a  circle of radius $\beta=1/T$.

We can similarly define a partition function on a torus with complex structure $\tau=\tau_1+i \tau_2$ \eqref{complex-str}. First, we make the time translation of length $\tau_2$ and the spatial translation of length $\tau_1$, and then we identify the Hilbert space. This can be defined by the Hamiltonian formalism as
\begin{equation}
\mathcal{Z}=\operatorname{Tr}_{\mathcal{H}}\left(e^{-2 \pi \tau_{2} H} e^{2 \pi \tau_{1} P}\right)~.
\end{equation}
Here the trace is taken over all states in the Hilbert space $\mathcal{H}$ of the theory. Since we can write
the Hamiltonian and momentum are written as the zero modes of Virasoro generator \eqref{H-P},
\begin{equation}
H_{\textrm{cyl}}=L_{0, \textrm{cyl}}+\overline{L}_{0, \textrm{cyl}}, \quad P_{\textrm{cyl}}=i\left(L_{0, \textrm{cyl}}-\overline{L}_{0, \textrm{cyl}}\right) .
\end{equation}
the partition function can be expressed as
\begin{equation}
\cZ=\mbox{Tr}_\mathcal{H}\left(e^{2\pi i \tau L_{0,\textrm{cyl}}} e^{-2\pi i \overline{\tau} \overline{L}_{0,\textrm{cyl}}} \right)~.
\end{equation}
The conformal transformation \eqref{cyl-plane} tells us that the Virasoro generators on a cylinder and on the plane are related by $L_{0,\textrm{cyl}}=L_{0,\textrm{pl}}-\frac{c}{24}$.  (The same applies to the anti-holomorphic part.)
By writing defining $q=\exp(2\pi i \tau)$, the torus partition function can be written as
\begin{equation}
\mathcal{Z}=\operatorname{Tr}_{\mathcal{H}}\left(q^{L_{0}-\frac{c}{24}} \bar{q}^{\bar{L}_{0}-\frac{\bar{c}}{24}}\right)~.
\end{equation}
Here we omit the subscript $\textrm{pl}$.
Note that (the holomorphic part of) this expression is equivalent to the generating function  \eqref{eq:character} of the Verma module. As we will see later, the partition function is indeed expressed in terms of the characters of irreducible representations
\be \label{character-decomp}
\mathcal{Z}(\tau)=\sum_{(h, \bar{h})} \mathcal{N}_{h \bar{h}} \chi_{h}(\tau) \bar{\chi}_{\bar{h}}(\bar{\tau})~,
\ee
where $\cN_{\overline{h}h}$ is the multiplicity of the irreducible representation labelled by $(h,\overline{h})$ in the Hilbert space $\cH$.

\subsubsection*{Free boson}

Let us compute the torus partition function in the simplest example: free boson. The computation is similar to  In free boson, it follows from \eqref{Hamiltonian-mode-expansion} that $L_0$ is expressed as the modes $a_m$
\begin{equation}
L_{0}=\frac{1}{2} a_{0} a_{0}+\sum_{k=1}^{\infty} a_{-k} a_{k}~.
\end{equation}
In addition, the Hilbert space $\cH$ of the free boson is the Fock space which can be obtained by acting with creation operators $a_{-k}$ for $k>0$ on the vacuum $|0\rangle$. From \eqref{com-rel-mode-expan-boson}, we can derive
\be \label{L-a-com}
[L_0,a_{-k}]=k a_{-k}~.
\ee
Thus, using the geometric progression $(1-x)^{-1}=1+x+x^2+\ldots$, the contribution from states $\bigoplus_\ell a_{-k}^\ell|0\rangle$ to the partition function can be read off as
\[\Tr_{\bigoplus_\ell a_{-k}^\ell|0\rangle} q^{L_0}=\sum_{\ell=0}^\infty q^{\ell k}=\frac1{1-q^k}~.\]
Since a general state in the Fock space can be written as
\begin{equation}
|n_1,n_2,n_3,\cdots\rangle=a_{-1}^{n_1}a_{-2}^{n_2}\cdots|0\rangle, \quad \mbox{with}\quad  n_i\in \mathbb{Z}_{\ge0}~,
\end{equation}
the holomorphic part of the torus partition function is
\begin{equation}\label{free-boson-process}
\begin{aligned}
\operatorname{Tr}_{\mathcal{H}}\left(q^{L_{0}-\frac{c}{24}}\right) &=q^{-\frac{1}{24}} \sum_{n_{1}=0}^{\infty} \sum_{n_{2}=0}^{\infty} \cdots\left\langle n_{1}, n_{2}, \cdots\left|q^{L_{0}}\right| n_{1}, n_{2}, \cdots\right\rangle \\
&=q^{-\frac{1}{24}} \prod_{k=1}^{\infty} \frac{1}{1-q^{k}} .
\end{aligned}
\end{equation}
This is exactly the Dedekind $\eta$-function \eqref{eq:eta}, and including the anti-holomorphic contribution, we have therefore found the partition function
\begin{equation}\label{ZBprime}
\cZ_{\textrm{B}}'(\tau,\overline{\tau})=\frac{1}{|\eta(\tau)|^2}.
\end{equation}

However, as the modular transformations of the Dedekind $\eta$-function are given in \eqref{ST-eta-theta}, this is \emph{not} invariant under the $S$-transformation. To make it modular-invariant, we need to consider the contribution from the zero mode $a_0$.
As in \eqref{L-a-com}, $a_0$ commutes with $L_0$ so that this can be regarded as a free parameter $a_0|0\rangle :=|a_0\rangle$. The trace over the Hilbert space means that we need to integrate over all the zero modes:
\begin{equation}
\int d a_{0}(q \bar{q})^{\frac{a_{0}^{2}}{2}}=\int d a_{0} e^{2 \pi i \tau_{2} a_{0}^{2}}=\frac{1}{\sqrt{2 \tau_{2}}}
\end{equation}
Multiplying this factor to \eqref{ZBprime}, we finally obtain the modular invariant partition function of the free boson
\begin{equation}
\cZ_{\textrm{B}}(\tau,\overline{\tau})=\frac{1}{\sqrt{2\tau_2}|\eta(\tau)|^2}~.
\end{equation}

\subsubsection*{Free boson on a circle}\label{sec:compactified-boson}

\begin{figure}[ht]
	\centering
	\includegraphics[width=0.45\linewidth]{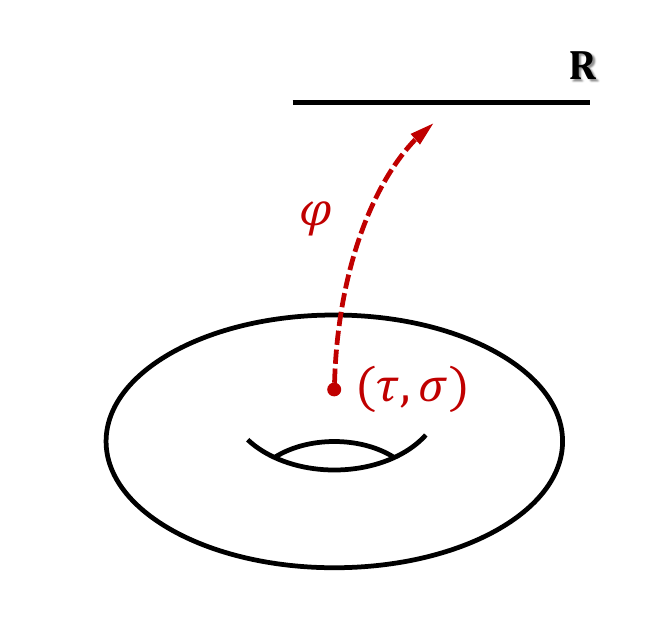}
	\includegraphics[width=0.45\linewidth]{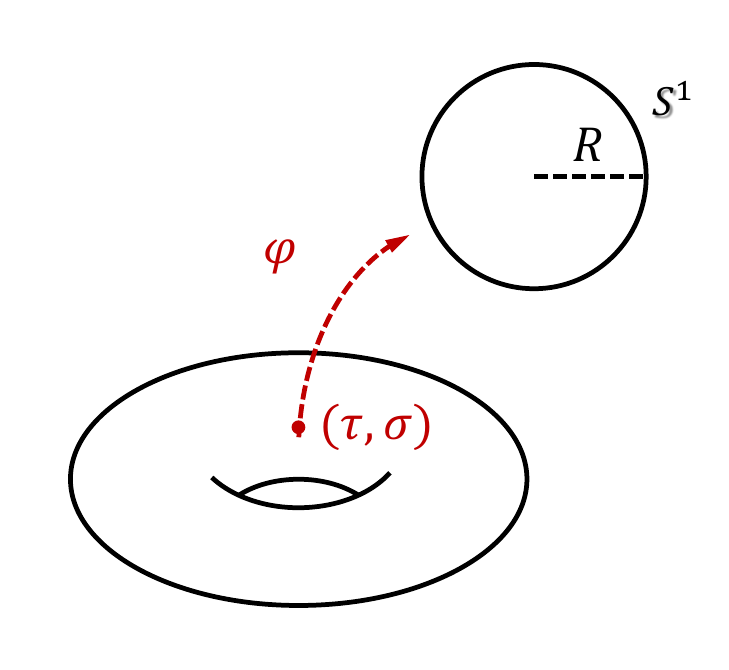}
	\caption{Left: free boson. Right: free boson on a circle.} \label{fig:LSM}
\end{figure}

The free boson considered above can be understood as a linear sigma model where a bosonic field $\varphi:T^2\to \bR$ is a map from a torus $T^2$ to a line $\bR$.
As another example, we consider the free boson compactified on a circle $S^1$ of radius $R$. This model can be interpreted as a linear sigma model $\varphi:T^2\to S^1$ from a torus $T^2$ to a circle $S^1$ of radius $R$. (See Figure \ref{fig:LSM}.) This is equivalent to imposing the periodicity on the bosonic field
\begin{equation}\label{circboson}
\varphi(z, \bar{z}) \sim \varphi(z, \bar{z})+2 \pi R m, \qquad m \in \mathbb{Z}
\end{equation}
To see how this compactification changes the partition function, let us recall the mode expansion \eqref{mode-expansion-boson} of the bosonic field $\varphi$:
\begin{equation}\label{mode-expansion-boson2}
\varphi(w,\overline{w})
= \varphi_0-i\frac{2\pi}{\ell}
\left[(a_0+\overline a_0)t -i(a_0-\overline a_0)x \right]+ i\sum_{n\neq 0} \frac{1}{n}
\bigg(
a_n e^{2\pi n w /\ell}
+
\overline{a}_n e^{2\pi n\overline{w})/\ell}
\bigg)\, .
\end{equation}
Since we map $T^2$ to $S^1$, we can impose the following periodicity to the field $\varphi$ under the rotation $x\to x+\ell$ on a spatial circle of the torus $T^2$
\begin{equation}\label{rotation}
\varphi(t,x+\ell )=\varphi(t,x)+2\pi R w \quad w\in \bZ~,\
\end{equation}
Since $w$ counts how many times the spatial circle of the torus wraps on the target $S^1$, it is called the \textbf{winding number}. (Note that $w$ here is no longer the coordinate of the torus, so do not mix them up.)
Substituting this condition into \eqref{mode-expansion-boson2}, we obtain the constraint
\begin{equation}
a_0-\overline{a}_0=R w~\quad w\in \bZ.
\end{equation}

In addition, it follows from \eqref{mode-expansion-boson2} that the center of mass momentum of the string is $p=(a_0+\overline a_0)/2$. Since a boson field is compactified on a circle, the momentum is quantized as
\[
a_0+\overline a_0=\frac{2n}{R}~\quad n\in \bZ~,
\]
where $n$ is called the \textbf{Kaluza-Klein momentum}.

Thus, a vacuum state is labeled by the winding number and Kaluza-Klein momentum as $|n,w\rangle$ where
\[
a_0|n,w\rangle=\left(\frac{n}{R}+\frac{Rw}{2}  \right) |n,w\rangle, \;\;\;\;\;\; \overline{a}_0 |n,w\rangle =\left(\frac{n}{R}-\frac{Rw}{2}  \right) |n,w\rangle~.
\]
The contributions from descendants are as before \eqref{ZBprime}. However, we now sum over all the zero modes  $a_0$ and $\overline{a}_0$, which is \cite[ISZ88-No.40]{Yang:1987mf} \cite[ISZ88-No.43]{di1987relations}
\begin{equation}
  \mathcal{Z}_{R}(\tau, \overline{\tau})=\frac{1}{|\eta(\tau)|^{2}} \sum_{n, w} q^{\frac{1}{2}\left(\frac{n}{R}+\frac{R w}{2}\right)^{2}} \overline{q}^{\frac{1}{2}\left(\frac{n}{R}-\frac{R w}{2}\right)^{2}} ~.\label{eq:bosonZ}
\end{equation}
It is straightforward to check that it is invariant under the $T$-transformation.  The invariance under the  $S$-transformation requires the \textbf{Poisson resummation formula}
\begin{equation}
  \sum_{n \in \mathbb{Z}} \exp \left(-\pi a n^{2}+b n\right)=\frac{1}{\sqrt{a}} \sum_{k \in \mathbb{Z}} \exp \left(-\frac{\pi}{a}\left(k+\frac{b}{2 \pi i}\right)^{2}\right)~.
\end{equation}
The derivation of the invariance under the modular $S$-transformation is left as an exercise.

The astute student notices that the partition function \eqref{eq:bosonZ} enjoys the identity
\begin{equation}\label{T-duality}
\cZ_{2/R}(\tau,\overline{\tau})=\cZ_{R}(\tau, \overline{\tau})~,
\end{equation}
which is called $T$-\textbf{duality}. T-duality plays an important role in string theory. This means that closed strings cannot distinguish circles of radii $R$ and $2/R$.
This shows that closed strings see geometry very differently from the way that we see it conventionally. The self-dual radius $R=\sqrt{2}$ can be interpreted as a minimal length scale of a circle in closed strings where the symmetry is enhanced from U(1) to SU(2) as we will see in \S\ref{sec:SU(2)k}.

\subsubsection*{Free fermions}

Next, let us consider the torus partition function of the free fermion. The mode expansion for a free fermion $\psi(z)$ is given in \eqref{mode-expansion-fermion}, and for Neveu-Schwarz boundary conditions, the modes $b_k$ take purely half-integer values $k\in\mathbb{Z}+\frac12$. On a torus with the coordinate $w$, this expansion corresponds to anti-periodic boundary condition along the spatial circle.
A general state in the Hilbert space $\mathcal{H}_{NS}$ of this theory is obtained by acting creation operators $b_{-s}$ on the vacuum $|0\rangle$
\begin{equation}
|n_{\frac12}, n_{\frac32}\cdots\rangle =(b_{-\frac12})^{n_{\frac12}} (b_{-\frac32} )^{n_{\frac32}}\cdots |0\rangle,
\end{equation}
where each occupation number must be $n_s=0,1$ due to the fermion-statistics.

As before, we need to evaluate the eigenvalues of $L_0=\sum_{s=\frac12}^\infty s b_{-s}b_s$ in \eqref{L0-fermion}. Using  the anti-commutation relation \eqref{commutation-relation-bk} of the fermionic modes, we have
\[
[L_0,b_k]=-k b_k~.
\]
Thus, it is straightforward to compute the character
\bea\label{AA}
\chi_{A A}(\tau) &=\operatorname{Tr}_{\mathcal{H}_{N S}}\left(q^{L_{0}-\frac{c}{24}}\right) \\
&=q^{-\frac{1}{48}} \sum_{n_{1}=0}^{1} \sum_{\frac{3}{2}=0}^{1} \cdots\left\langle n_{\frac{1}{2}}, n_{\frac{3}{2}} \cdots\left|q^{L_{0}}\right| n_{\frac{1}{2}}, n_{\frac{3}{2}} \cdots\right\rangle \\
&=q^{-\frac{1}{48}} \prod_{r=0}^{\infty}\left(1+q^{r+\frac{1}{2}}\right)=\sqrt{\frac{\vartheta_{3}(\tau)}{\eta(\tau)}}
\eea
For the fermionic character, we impose the anti-periodic boundary condition on the time circle of the torus.  The ($AA$) notation indicates that the anti-periodic boundary condition is imposed on both the time and spatial circle of the torus.

Let us investigate the property of \eqref{AA} under the modular transformation. Using the properties \eqref{ST-eta-theta} of $\eta$ and $\vartheta$ under the modular transformations, we have
\begin{equation}
\begin{aligned}
&S: \chi_{A A}(-1 / \tau)=\chi_{A A}(\tau) \\
&T: \chi_{A A}(\tau+1)=e^{-\frac{i \pi}{24}} \sqrt{\frac{\vartheta_{4}(\tau)}{\eta(\tau)}} .
\end{aligned}
\end{equation}
Although the $S$-transformation exchanges the time and spatial circle of the torus, Since we impose the same anti-periodic boundary condition on both circles, the $S$-transformation does not change the boundary conditions. On the other hand, the  $T$-transformation maps the time circle to the (1,1)-cycle of the torus as illustrated in Figure \ref{fig:T-transformation}. The periodic boundary condition is imposed on the (1,1)-cycle because imposing the anti-periodic boundary conditions twice ends up with the periodic boundary condition, \textit{i.e.} $A\times A=P$. Hence,  the $T$-transformation changes the boundary condition, which results in a different $\vartheta$-function.

In order to construct a modular invariant partition function, we need to take into account all the boundary conditions. To this end, we introduce the fermion number operator $F$ which assigns $F=0$ for bosons and $F=1$ for fermions.
Then, we can define a new character (or it is called an index)
\begin{equation}
\chi_{P A}=\operatorname{Tr}_{\mathcal{H}_{N S}}\left((-1)^{F} q^{L_{0}-\frac{c}{24}}\right)=q^{-\frac{1}{48}} \prod_{r=0}^{\infty}\left(1-q^{r+\frac{1}{2}}\right)=\sqrt{\frac{\vartheta_{4}(\tau)}{\eta(\tau)}}~.
\end{equation}
The additional term $(-1)^F$ in the argument of this trace imposes the periodicity condition in the time direction.

Again let us check the modular properties of this new character. The modular $T$-transformation takes the boundary condition back to the ($AA$) sector because the anti-periodic boundary condition is imposed on the (1,1)-cycle due to $P\times A=A$. Subsequently, we have
\[
T: \chi_{P A}(\tau+1)=e^{-\frac{i \pi}{24}} \chi_{A A}(\tau)~.
\]
On the other hand, the modular $S$-transformation leads to
\begin{equation}\label{AP}
S: \chi_{PA}(-1/\tau)=\sqrt{2}q^{\frac{1}{24}}\prod_{r\geq1}(1+q^r)=\sqrt{\frac{\vartheta_2(\tau)}{\eta(\tau)}}~,
\end{equation}
because the different boundary conditions are imposed on the time and spatial circle. Since the $S$-transformation exchanges them, we end up with $(PA)$ boundary condition after the $S$-transformation, \textit{i.e.} the anti-periodic boundary condition on the time circle and the periodic boundary condition on the spatial circle. This is the character of the Ramond sector
\[
\chi_{A P}:=\operatorname{Tr}_{\mathcal{H}_{R}}\left(q^{L_{0}-\frac{c}{24}}\right)=\sqrt{\frac{\vartheta_2(\tau)}{\eta(\tau)}}
\]
where the modes $b_k$ take integer values of $k\in\bZ$. Note that the prefactor $\sqrt{2}$ in \eqref{AP} stems from the fact that there are two ground states $|0\rangle$, $\psi_0|0\rangle$ due to the fermion zero mode $\psi_0$. Multiplying the anti-holomorphic part, this prefactor takes into account the contribution from the two ground states.

Similarly, we can read off the modular properties of this character:
\begin{equation}
T(\chi_{AP}(\tau))=e^{\frac{i\pi}{12}}\chi_{AP}(\tau), \;\;\;\;\;\; S(\chi_{AP}(\tau))=\chi_{PA}(\tau)
\end{equation}
One may wonder what $\chi_{PP}(\tau)$ is. In this case,  we need to impose periodic boundary conditions on both time and spatial circles. Due to the presence of $(-1)^F$, the contributions from the two ground states $|0\rangle$, $\psi_0|0\rangle$ cancel each other. The same argument can be applied to the excited states so that the partition function $\chi_{PP}(\tau)$  just vanishes.

We start with the partition function in NS sector, and perform modular transformations. Eventually, we obtain different boundary conditions. Hence, to obtain the modular-invariant partition function of the free fermion, we just need to sum over all of them
\begin{equation}\label{free-fermion-PF}
\cZ_{\textrm{F}}(\tau,\overline{\tau})=\frac12 \left( \left| \frac{\vartheta_3}{\eta} \right|+\left| \frac{\vartheta_4}{\eta} \right|+\left| \frac{\vartheta_2}{\eta} \right|   \right).
\end{equation}
We multiply an overall factor of $1/2$ so that we count the NS ground state only once.

\subsubsection*{Special functions}
\begin{figure}[ht]
	\centering
	\includegraphics[width=0.8\linewidth]{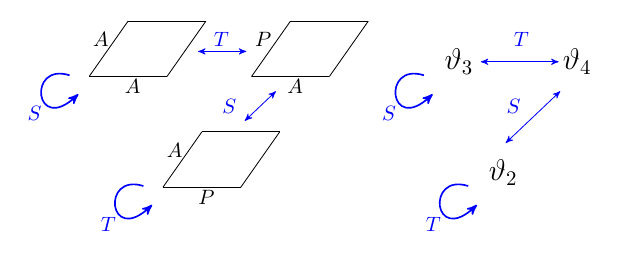}
	\caption{$T$ and $S$ transformation of $\vartheta$ function}
	\label{fig:modular-theta}
\end{figure}
The special functions we have used in the text are summarized below. You can also consult with Polchinski \cite[\S 7.2]{Polchinski} or Blumenhagen \& Plauschinn \cite[\S 4.2]{Blumenhagen:2009zz}.

Their infinite product forms are
\bea
 \label{eq:eta}
 &\eta(\tau)=q^{\frac{1}{24}} \prod_{n=1}^\infty (1-q^n) \ , \\
 &\vartheta_2(\tau)=2 q^{\frac{1}{8}} \prod_{n=1}^\infty (1-q^n) (1+q^n)^2 \ , \\
 &\vartheta_3(\tau)=\prod_{n=1}^\infty (1-q^n) (1+q^{n-\frac{1}{2}})^2 \ , \\
 &\vartheta_4(\tau)=\prod_{n=1}^\infty (1-q^n) (1-q^{n-\frac{1}{2}})^2 \ ,
\eea
where $q=e^{2\pi i \tau}$.
The first one is called the Dedekind eta function,
the others are called (Jacobi or elliptic or simply without these) theta functions.

Their $T$ and $S$-transformations are
\bea\label{ST-eta-theta}
 &\eta(\tau+1)=e^{i\pi/12} \eta(\tau) \ ,   &\eta(-1/\tau)=\sqrt{-i\tau} \eta(\tau) \ ,\\
 &\vartheta_2(\tau+1)=e^{i\pi/4} \vartheta_2(\tau) \ ,  &\vartheta_2(-1/\tau)=\sqrt{-i\tau} \vartheta_4(\tau) \\
 &\vartheta_3(\tau+1)=\vartheta_4(\tau) \ , &\vartheta_3(-1/\tau)=\sqrt{-i\tau} \vartheta_3(\tau)  \\
 &\vartheta_4(\tau+1)=\vartheta_3(\tau) \  , &  \vartheta_4(-1/\tau)=\sqrt{-i\tau} \vartheta_2(\tau) \ .
\eea
In addition, there are two important identities. One is
called Jacobi identity:
\begin{align*}
 (\vartheta_3(\tau))^4=(\vartheta_2(\tau))^4 +(\vartheta_4 (\tau))^4 \ .
\end{align*}
The other is called Jacobi triple product identity:
\begin{align*}
 \vartheta_2(\tau) \vartheta_3(\tau) \vartheta_4(\tau)=2 \eta^3(\tau) \ ,
\end{align*}
from which you can show the modular transformation of the Dedekind $\eta$-function.

\section{Minimal models}
This chapter is devoted to the most important class of conformal field theories, called \textbf{minimal models} first studied in the seminal paper \cite[ISZ88-No.1]{belavin1984infinite}. These theories
are characterized by a Hilbert space made of finite numbers of representations of the Verma modules. We refer the reader to \cite[\S7,\S8]{francesco2012conformal} and  \cite[\S4,\S5]{ginsparg1988applied} for more details.

\subsection{Conformal family}
From the previous study, we know that primary fields play a fundamental role in conformal field theory. The asymptotic state is given by $\ket{h}=\phi(0)\ket{0}$, which satisfies $L_0 \ket{h}=h\ket{h}$ and $L_n \ket{h}=0$ for  $n>0$. In the Verma module section, we also defined descendant states leading by primary $\ket{h}$, by acting a series of $L_{-n}$ on $\ket{h}$. We first study the field operator corresponding to the descendant states. The natural definition of the descendant field associated with the state $L_{-n}\ket{h}$ is
\begin{equation}
\label{def-descendant-field}
\phi^{(- n)} (w) \equiv \left(L _{- n} \phi \right) (w)=\frac{1}{2 \pi i} \oint _{w} d z \frac{1}{(z-w)^{n-1}} T (z) \phi (w)\, .
\end{equation}
By using the OPE \eqref{OPE-T-primaryfield-z}, the above expression can be calculated explicitly.
In particular, we have
\begin{equation}
 \phi^{(0)} (w)=h \phi (w)\, , \quad \text{and} \quad \phi^{(- 1)} (w)=\partial \phi (w)\, .
\end{equation}
Now consider the correlation function with one descendant field in it, i.e.
\begin{equation}
  \left\langle \left(L _{- n} \phi \right) (w) X \right\rangle\, ,
\end{equation}
where $X=\phi _{1} \left(w _{1} \right) \cdots \phi _{N} \left(w _{N} \right)$ is a product of primary fields with conformal dimensions $h_i$.
This correlation function can be calculated by substituting the definition \eqref{def-descendant-field}.
\bea
\label{differential-equation}
 \left\langle \phi^{(- n)} (w) X \right\rangle
 &= \frac{1}{2 \pi i} \oint _{w} d z (z-w)^{1-n} \langle T (z) \phi (w) X \rangle\notag \\
 &=- \frac{1}{2 \pi i} \oint _{\left(w _{i} \right)} d z (z-w)^{1-n} \sum _{i} \{\frac{1}{z-w _{i}} \partial _{w _{i}} \langle \phi (w) X \rangle+ \frac{h _{i}}{\left(z-w _{i} \right)^{2}} \langle \phi (w) X \rangle \}\notag \\
 &\equiv \mathcal{L} _{- n} \langle \phi (w) X \rangle \quad (n \geq 1)\, .
\eea
In the second step, the residue theorem is applied by
reversing the contour including $w$ only and summing
the contributions from all the poles at $w_i$. With
the help of the relevant OPEs, the definition operator $\mathcal{L}_{-n}$ is given by
\begin{equation}\label{L-diff}
 \mathcal{L} _{- n}=\sum _{i} \left\{\frac{(n-1) h _{i}}{\left(w _{i}-w \right)^{n}}-\frac{1}{\left(w _{i}-w \right)^{n-1}} \partial _{w _{i}} \right\}\, .
\end{equation}
This result tells us that the evaluation of a correlation function containing one descendant field
$\phi^{(-n)}$ is controlled by its primary field. $\mathcal{L}_{-1}$ is equivalent to $\partial/\partial{w}$, since the operator
\begin{equation}
\frac{\partial}{\partial w} + \sum _{i} \frac{\partial}{\partial w _{i}}\, ,
\end{equation}
annihilates any correlation functions because of the translation invariance.

The descendant states corresponding to $L_{-k}L_{-n}\ket{h}$ can be defined recursively.
\begin{equation}
\phi^{(- k ,-n)} (w) \equiv \left(L _{- k} L _{- n} \phi \right) (w)
=  \frac{1}{2 \pi i} \oint _{w} d z (z-w)^{1-k} T (z) \left(L _{- n} \phi \right) (w)\, .
\end{equation}
In particular, we have
\begin{equation}
\phi^{(0 ,-n)} (w)=(h + n) \phi^{(- n)} (w) \quad \text{and} \quad
\phi^{(- 1 ,-n)} (w)=\partial _{w} \phi^{(- n)} (w)\, .
\end{equation}
And it can be shown directly that
\begin{equation}\label{descendant-eq}
\left\langle \phi^{\left(- \lambda_{1}, \ldots,-\lambda_l \right)} (w) X \right\rangle=\mathcal{L} _{- \lambda_{1}} \cdots \mathcal{L} _{- \lambda_l} \langle \phi (w) X \rangle\, .
\end{equation}
That is nothing but to apply the differential operators in succession. Consequently, correlation
functions of descendant fields are determined from those of the primary field.  If we obtain all the correlation functions of primary fields in a theory, then the correlation functions, including descendant fields, are thus all determined.

Moreover, all the OPEs with $T(z)$ with descendant states can be calculated in a similar fashion. For example, for
$\phi^{(-1)}=\partial\phi$
\bea
T(z)\partial\phi(w)&\sim \frac{2h\phi(w)}{(z-w)^3}
+\frac{(h+1)\partial\phi(w)}{(z-w)^2}
+\frac{\partial^2\phi(w)}{z-w}\notag\\
&\sim\frac{2\phi^{(0)}(w)}{(z-w)^3}
+\frac{(h+1)\phi^{(-1)}(w)}{(z-w)^2}
+\frac{\phi^{(-1,-1)}(w)}{z-w}\, ,
\eea
where we neglect the regular terms as usual.
The set of a primary field $\phi$
and all of its descendants is called the \textbf{conformal family} of $\phi$, denoted by $[\phi]$. The OPEs with $T(z)$ are closed in the conformal family.

\subsection{Minimal models}\label{sec:MM}
In a conformal field theory (or more generally quantum field theories),
we want to understand correlation functions in the
theory. For free bosons and fermions,
all correlation functions can be calculated by directly using the partition function.
However, it is not so straightforward to determine all the correlation functions for the interacting CFTs.
If there are only finitely many highest weight representations of Virasoro algebra (or finitely many primary fields), one can perform exact computations of correlation functions \cite[ISZ88-No.1]{belavin1984infinite}. Such CFTs are called \textbf{minimal models}.  Let us consider a theory with the set of finitely many highest
weights $\{h_1,\cdots h_p\}$ so that OPEs take the following form:
\begin{equation}\label{op-algebra}
\phi _{1} (z , \overline{z}) \phi _{2} (0,0)
=\sum_p \sum_{\{\lambda,\overline{\lambda}\}} C^{\{\lambda,\overline{\lambda}\}}_{12p}
z^{h _{p}-h _{1}-h _{2} + |\lambda|} \overline{z} ^{\overline{h} _{p}-\overline{h} _{1}-\overline{h} _{2} + |\overline{\lambda}|}
\phi^{\{\lambda,\overline{\lambda}\}}_p(0,0)\, ,
\end{equation}
where $|\lambda|=\sum_i \lambda_i$ and $|\overline{\lambda}|=\sum_i \overline{\lambda}_i$, and $\phi_{i}$ is the primary field with
highest weight $h_i$. As we have seen in the last subsection,  all the correlation functions can be
reduced to the correlation function constructed by
primary fields. Hence, OPEs of primary fields and conformal symmetry completely determine correlation functions if there are only finitely many $p$. In this section, we will learn the method to solve correlation functions in such a model.

\subsubsection*{Kac Determinant}

A representation of the Virasoro algebra is said to be unitary if it contains no negative-norm states. In the section of the Verma module, we have mentioned that the negative of the central charge and highest weight will make the states non-unitary. As we see below, the unitarity imposes strong constraints on $(c,h)$.

In addition to unitarity, one has to pay attention to zero-norm states, indeed. If a state $|\chi\rangle$ in a Verma module $V(c,h)$ which is not a highest weight state satisfies the condition
\[
L_n|\chi\rangle=0~, \quad \textrm{for} ~ n>0~,
\]
it is called a \textbf{singular vector}. In fact, a singular vector $|\chi\rangle$ is orthogonal to any state $L_{-\lambda}|h\rangle$ \eqref{def-excited-state} in the Verma module $V(c,h)$
\[
\langle \chi|L_{-\lambda}|h\rangle=0~.
\]
Therefore, the norm of a singular vector is zero; $\langle\chi|\chi\rangle=0$ so that it is also called a \textrm{null vector} or a \textrm{null state}.

Suppose that there exists a singular vector $|\chi\rangle$ at level $N$ in the Verma module, namely $L_0|\chi\rangle=(h+N)|\chi\rangle$. Then 	its descendants
\[
L_{-\lambda}\ket{\chi}\equiv L_{-\lambda_1}L_{-\lambda_2}\cdots L_{-\lambda_l}\ket{\chi}\, ,
	\qquad (\lambda_1 \geq \lambda_2 \cdots \geq \lambda_l \geq 1)
\]
are all null, and moreover they form a submodule in the Verma module $V(c,h)$. Therefore, in this situation, the Verma module is \textbf{reducible}. In fact, the Verma module $V(c,h)$ is irreducible if and only if it contains no singular vector. Otherwise, an irreducible representation of the Virasoro algebra is
\[
L(c,h)=V(c,h)/J(c,h)
\]
where $J(c,h)$ consists of all null states and their descendants
\be 
J(c,h)=\bigoplus_{\substack{ \chi: \textrm{null}\\ \lambda : \textrm{partitions}}}~ \{L_{-\lambda}\ket{\chi}\}	
~.
\ee

To see if a Verma module has a null state, let us define a Gram matrix as
$M_{ij}=\bra{i}\ket{j}$ where $\ket{i}$ are basis vectors.
For instance, at level two, we have 2 basis
$L_{-2}\ket{h}$, $L_{-1}^2\ket{h}$.
Therefore, the Gram matrix is
\bea
G_{h}^{(2)}\equiv \begin{pmatrix}{\left\langle h \left| L _{2} L _{- 2} \right| h \right\rangle} &{\left\langle h \left| L _{1}^{2} L _{- 2} \right| h \right\rangle} \\{\left\langle h \left| L _{2} L _{- 1}^{2} \right| h \right\rangle} &{\left\langle h \left| L _{1}^{2} L _{- 1}^{2} \right| h \right\rangle} \end{pmatrix}=\begin{pmatrix}{4 h + \frac{c}2} &{6 h} \\{6 h} &{4 h (1 + 2 h)} \end{pmatrix} \, .
\eea
We focus on a zero eigenvector of this matrix, which gives a linear combination with zero norm. We write the determinant of this matrix as
\begin{equation}
2(16 h^{3}-10 h^{2} + 2 h^{2} c + h c )=32 (h-h _{1,1} (c) ) (h-h _{1,2} (c) ) (h-h _{2,1} (c) )\, ,
\end{equation}
and
\bea\label{level-2}
h _{1,1} (c) &= 0 \cr
h _{1,2}(c)&= \frac{1}{16} (5-c) + \frac{1}{16} \sqrt{(1-c) (25-c)}\cr
h _{2,1} (c)&= \frac{1}{16} (5-c)-\frac{1}{16} \sqrt{(1-c) (25-c)}~.
\eea
The $h=0$ root originates from the null state at level 1, $L_{-1}\ket{0}=0$, which directly implies the vanishing of $L_{-1}(L_{-1}\ket{0})=0$. This is a general phenomenon: if a null state (a state with zero norm) $\ket{h+n}=0$ appears at level $n$, then at level $N > n$, there are $P(N-n)$ null states of the form $L_{-\lambda_1}\cdots L_{-\lambda_k}\ket{h+n}=0$ (with $\sum_i \lambda_i=N-n$). Consequently, the existence of a null state for a specific value of $h$ at level $n$ implies that the determinant at level $N$ will have a $[P(N-n)]$-th order zero for that value of $h$.

At level $N$, the Hilbert space is spanned by all states of the form:
\begin{equation}
\sum _{|\lambda|=N} a_{\lambda} L_{- \lambda}  | h \rangle\, .
\end{equation}
From these, we can select $P(N)$ basis states to construct the Gram matrix at level $N$, referred to as the \textbf{Kac-Shapovalov} matrix, with matrix elements defined as follows:
\begin{equation}\label{KS-matrix}
G_{h}^{(N)}(\mu,\lambda)\equiv\left\langle h \left| L _{\mu_{\ell}} \cdots L _{\mu_{1}} L _{- \lambda_{1}} \cdots L _{- \lambda_{k}} \right| h \right\rangle
\end{equation}
where $\sum _{i=1}^{\ell} \mu_{i}=\sum _{j=1}^{k} \lambda_{j}=N$. We denote this matrix by $M_N(c,h)$. If $\det M_N(c,h)$ vanishes, it indicates the existence of a linear combination of states with zero norm for that particular $c$ and $h$. If the determinant is negative, $M_N(c,h)$ has an odd number of negative eigenvalues, implying that the representation of the Virasoro algebra for those values of $c$ and $h$ includes states of negative norm, and is therefore non-unitary.

\begin{figure}[ht]
	\centering
	\includegraphics[width=.8\linewidth]{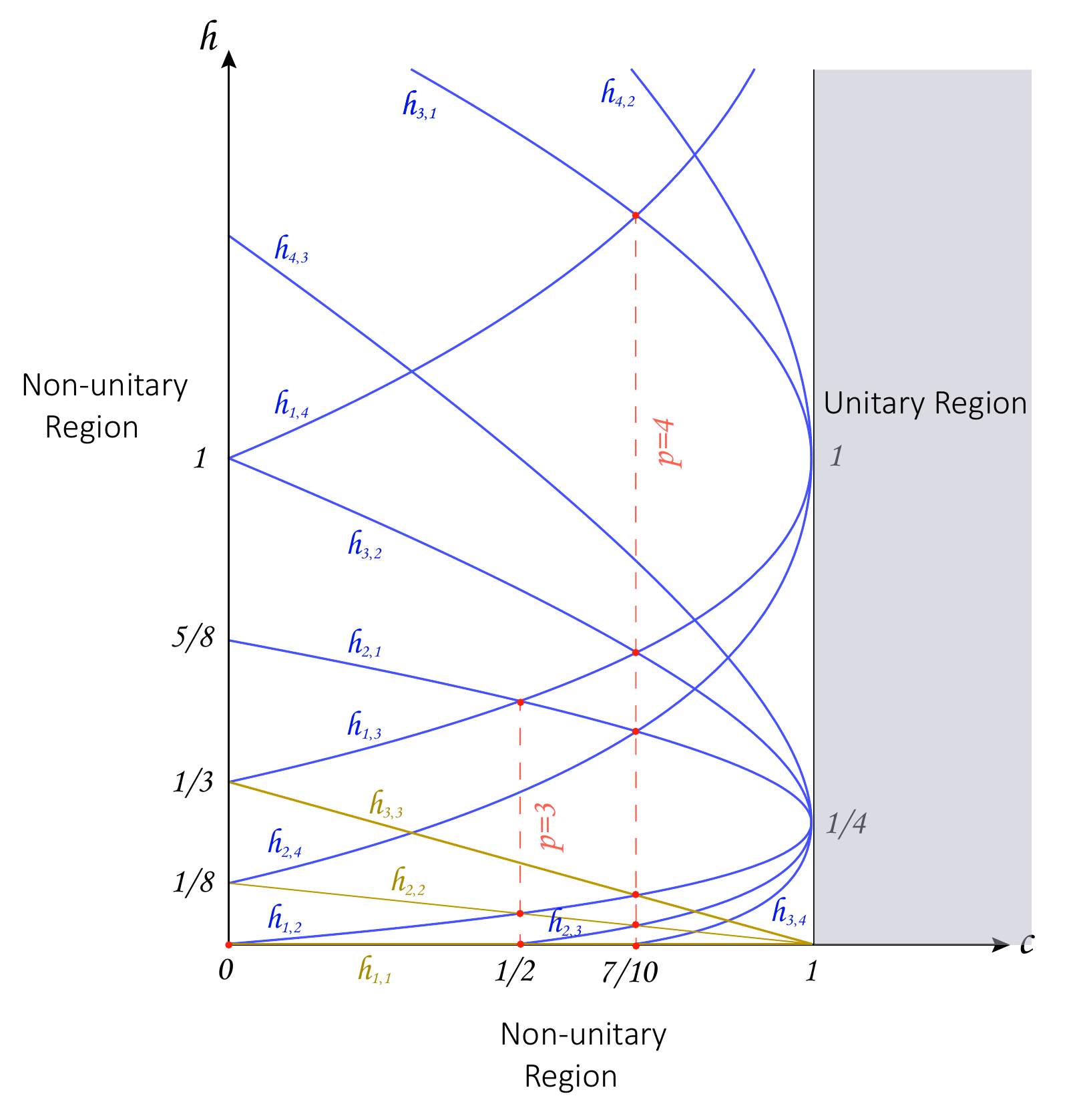}
	\caption{Kac determinant}
	\label{fig:kac-determinant}
\end{figure}

Kac has found a closed form expression of the determinant of $N$ level Gram matrix, which is called \textbf{Kac determinant}  \cite[ISZ88-No.7]{kac1979contravariant}
\begin{equation}
\label{Kac-determinant}
\det G_h^{(N)} =\alpha _{N} \prod _{r,s  \leq N} \left(h-h _{r,s} (c) \right)^{P (N-rs)}\, ,
\end{equation}
where $\alpha_N$ is a positive constant independent of $c$
and $h$. The Kac determinant provides complete information about the null states in a Verma module.
From this expression, one can see that a null state first appears at level $n=rs$ if $h=h_{r,s}$. Thus at level $N>rs$, the
zero order for $h_{r,s}$ is $P(N-rs)$. In fact, there are two useful equivalent ways to write $h_{r,s}$; the first expression is
\bea
\label{rel-hrs-alpha}
 h _{r,s} (c) &= h _{0} + \frac{1}{4} \left(r \alpha _{+} + s \alpha _{-} \right)^{2}\, ,\notag\\
 h _{0} &= \frac{1}{24} (c-1)\, ,\\
 \alpha _{\pm} &= \frac{\sqrt{1-c} \pm \sqrt{25-c}}{\sqrt{24}}\, ,\notag
\eea
and the second is
\bea
\label{rel-hrs-m}
h _{r,s} (p) &= \frac{[ (p + 1) r-p s ]^{2}-1}{4 p (p + 1)}\, ,\notag\\
c &= 1-\frac{6}{p (p + 1)}\, ,
\eea
where $p$ has two branches
\begin{equation}
\label{m branches}
p=- \frac{1}{2} \pm \frac{1}{2} \sqrt{\frac{25-c}{1-c}}
\end{equation}
From this expression, one can see that when $c=1$, a Verma module is irreducible if $h\neq n^2/4$ for $n\in\bZ$. In addition, when $c=0$, a Verma module is irreducible if $h\neq (n^2-1)/24$ for $n\in\bZ$.

Both of the above expressions provide the following
result for $h_{r,s}(c)$, which is illustrated  in Figure \ref{fig:kac-determinant}
\begin{equation}
\label{rel-hrs}
  h _{r,s} (c)=\frac{1-c}{96} \left[ \left((r + s) \pm (r-s) \sqrt{\frac{25-c}{1-c}} \right)^{2}-4 \right]\, .
\end{equation}
One can convince oneself that $h_{1,2}$ and $h_{2,1}$ are as in \eqref{level-2}.
The Kac determinant changes sign when the values of $(c,h^2)$ cross
each curve $h=h_{r,s}(c)$.

\subsubsection*{Minimal models $\cM_{p,p'}$}
Belavin-Polyakov-Zamolodchikov have noticed that all the correlation functions can be determined by differential equations when the Kac determinant becomes zero.
For example, let us first consider the situation in which there are null states at level two. In general, a null state $\ket{\chi}$ at level two takes the form \begin{equation}
\ket{\chi}=L _{- 2} | h \rangle + a L _{- 1}^{2} | h \rangle \, .
\end{equation}
By applying the $L_1$ to the above equation, we have
\bea
\left[ L _{1} , L _{- 2} \right] | h \rangle + a [ L _{1} , L _{- 1}^{2} ] | h \rangle &= 3 L _{- 1} | h \rangle + a \left(L _{- 1} 2 L _{0} + 2 L _{0} L _{- 1} \right) | h \rangle\notag \\
&= (3 + 2 a (2 h + 1)) L _{- 1} | h \rangle=0
\, ,
\eea
which requires that
\begin{equation}
a=- \frac{3}{2 (2 h + 1)}\, .
\end{equation}
By applying $L_2$, we find that
\bea
\left[ L _{2} , L _{- 2} \right] | h \rangle + a[ L _{2} , L _{- 1}^{2}] | h \rangle &= \left(4 L _{0} + \frac{c}{2}\right) | h \rangle + 3 a L _{1} L _{- 1} | h \rangle \notag \\
&= (4 h + \frac{c}2 + 6 a h) | h \rangle=0\, .
\eea
Therefore, the central charge must satisfy
\begin{equation}
c=2(-6ah-4h)=2h\frac{5-8h}{2h+1}
\end{equation}
Writing $h$ in terms of $c$, we have
\begin{equation}
h=\frac{1}{16}\left\{
5-c\pm \sqrt{(c-1)(c-25)}
\right\}\, ,
\end{equation}
which are equal to $h_{1,2}$ and $h_{2,1}$ in \eqref{level-2}.

Since the null state $\ket{\chi}$ is orthogonal to any states in the Verma module, we have
\begin{equation}
  \label{BPZ}
   \langle \chi (z) X \rangle =0
\end{equation}
where $X$ is a product of arbitrary primary fields as usual. Since $\chi(z)$ can be obtained by acting a linear combination of Virasoro generators \eqref{L-diff} on a primary operator, this equation becomes a linear differential equation for a correlation function of primary fields, called a \textbf{Belavin–Polyakov–Zamolodchikov (BPZ) equation}.
For example, by using
\eqref{differential-equation}, the correlation function
$\expval{\phi(z)X}$ should be governed by the
following differential equation
\begin{equation}
\left\{\mathcal{L} _{- 2}-\frac{3}{2 (2 h + 1)} \mathcal{L} _{- 1}^{2} \right\} \langle \phi (z) X \rangle=0\, ,
\end{equation}
where  $\phi(z)$ has conformal dimension $h_{1,2}$ or
$h_{2,1}$.
One can write the above expression as a differential equation
\begin{equation}
\left\{\sum _{i=1}^{N} \left[ \frac{1}{z-z _{i}} \frac{\partial}{\partial z _{i}} + \frac{h _{i}}{\left(z-z _{i} \right)^{2}} \right]-\frac{3}{2 (2 h + 1)} \frac{\partial^{2}}{\partial z^{2}} \right\} \langle \phi (z) X \rangle=0\, .
\end{equation}
If $X$ is $\phi(w)$ itself, the differential
equation for a two-point function becomes
\begin{equation}
\left\{\frac{1}{z-w} \partial _{w} + \frac{h}{(z-w)^{2}}-\frac{3}{2 (2 h + 1)} \partial _{z}^{2} \right\} \langle \phi (z) \phi (w) \rangle=0\, ,
\end{equation}
which is satisfied by the general form of a two-point function \eqref{two-point-func-form}.
For a three-point function \eqref{three-point-func-form} with $X=\phi_1(z_1)
\phi_2(z_2)$, we have
\begin{equation}
\left\langle \phi (z) \phi _{1} \left(z _{1} \right) \phi _{2} \left(z _{2} \right) \right\rangle=\frac{C_{h,h_1,h_2}}{\left(z-z _{1} \right)^{h+h _{1}-h _{2}} \left(z _{1}-z _{2} \right)^{h _{1}+h _{2}-h} \left(z-z _{2} \right)^{h+h _{2}-h _{1}}}\, .
\end{equation}
The differential equation gives further constraints to
the highest weights of field operators in the three-point function.
\begin{equation}
h _{2}=\frac{1}{6} + \frac{1}{3} h + h _{1} \pm \frac{2}{3} \sqrt{h^{2} + 3 h h _{1}-\frac{1}{2} h + \frac{3}{2} h _{1} + \frac{1}{16}}\, .
\end{equation}
For $h=h_{1,2}$, and set $h_1=h_{r,s}$ in \eqref{rel-hrs}, we find that $h_2$ is equal to $h_{r,s-1}$ or $h_{r,s+1}$. Similarly for $h=h_{2,1}$, $h_2$ is equal to $h_{r-1,s}$ or $h_{r+1,s}$. Recall that the two-point function vanishes unless the two operators have the same conformal dimension. We thus have the following operator algebra
\bea
\phi _{1,2} \times \phi _{r,s} &=[ \phi _{r,s-1} ]+ [\phi _{r,s + 1}]\notag \\
\phi _{2,1} \times \phi _{r,s} &= [\phi _{r-1 , s}] + [\phi _{r + 1 ,s}]\, ,
\eea
where $\phi_{r,s}$ means the field operator with highest weight $h_{r,s}$. The expression above means that the fields $\phi _{1,2}$ and $\phi_{2,1}$ act as ladder operators in the operator algebra. In addition, the field operators expanded in the right side not only contain the primary fields but their descendant fields as well. The coefficient on the right side may be zero. For instance,
\bea
\phi _{1,2} \times \phi _{2,1} &=[ \phi _{2,0} ]+ [\phi _{2,2}]\, ,\notag \\
\phi _{2,1} \times \phi _{1,2} &= [\phi _{0,2}] + [\phi _{2,2}]\, .
\eea
Since the two OPEs are equivalent, this shows the coefficients before $\phi _{2,0}$ and
$\phi _{0,2}$ must be zero.
Therefore, we have
\begin{equation}
\phi _{1,2} \times \phi _{2,1} =[ \phi _{2,2}]\, .
\end{equation}
The above expression can be generalized to
the following fusion rules
\begin{equation}
\label{fusion-rule-1}
\phi_{r_1, s_1}\times \phi_{r_2,s_2}
= \sum_{
	\begin{subarray}{1}
	\quad k=1+\abs{r_1-r_2}\\
	k+r_1+r_2=1\,\text{mod}\, 2
	\end{subarray}
 }^{k= r_1 + r_2 -1}
\,
\sum_{
	\begin{subarray}{1}
	\quad l=1+\abs{s_1-s_2}\\
	l+s_1+s_2=1\,\text{mod}\, 2
	\end{subarray}
}^{l= s_1 + s_2 -1}
[\phi_{k,l}]\, .
\end{equation}
The summation variables $k$ and $l$ are incremented
by $2$. The expression above implies that
the conformal family $[\phi_{r,s}]$ forms a
closed operator algebra.
\begin{figure}
	\centering
	\includegraphics[width=0.6\linewidth]{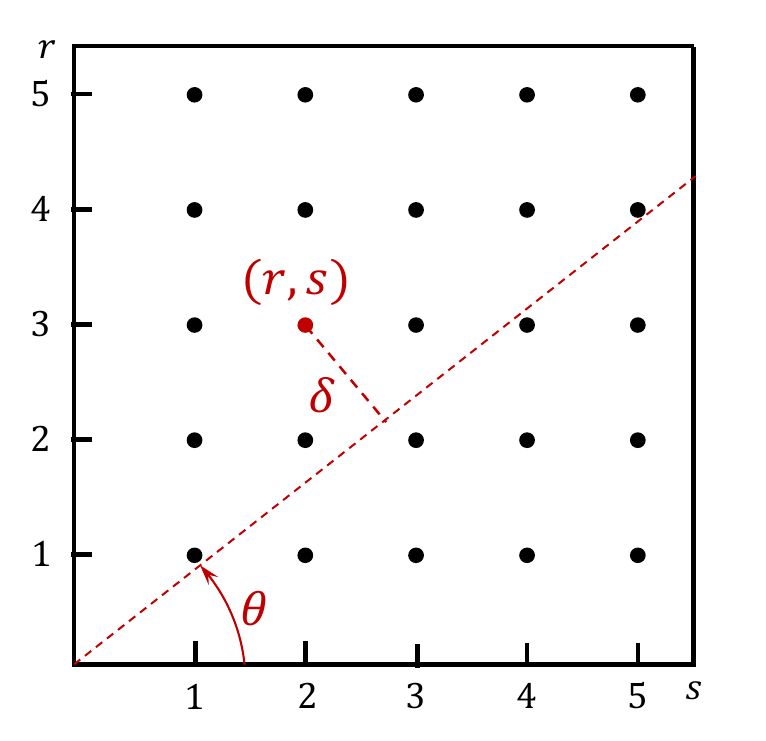}
	\caption{"diagram of dimensions"}
	\label{fig:minimal-model}
\end{figure}

The operator algebra
\eqref{fusion-rule-1}, implies there are infinitely many conformal families present in the theory.
However, we want a finite set of highest weights in
a minimal model. In order to understand the situation
graphically, we consider the "diagram of dimensions" as in Figure
\ref{fig:minimal-model}.
The points $(r,s)$ in the first quadrant label the various conformal dimensions appearing in the Kac formula.
The dotted line has a slope $\tan\theta=-\alpha_-/\alpha_+$, where $\alpha_{\pm}$ are defined
in \eqref{rel-hrs-alpha}. If $\delta$ is the distance
between a point $(r,s)$ and the dotted line. It is
a good exercise to show the following relation,
\begin{equation}
h _{r , s}=h _{0} + \frac{1}{4} \delta^{2} \left(\alpha _{+}^{2} + \alpha _{-}^{2} \right)\, .
\end{equation}
If the slope $\tan\theta$ is irrational, it will never
go through any integer point $(r,s)$, which means
$\delta$ is always non-zero. There is no periodical
property for $h_{r,s}$. However, if the slope
$\tan \theta$ is rational. That is, if there exist
two coprime integers $p$ and $p'$ such that
\begin{equation}
p^{\prime} \alpha _{-} + p \alpha _{+}=0\, .
\end{equation}
The dotted line will go through the point $(p',p)$, we
have the following periodicity property.
\begin{equation}
h _{r, s}=h _{r + p, s + p^{\prime}}\, .
\end{equation}
Therefore, there are finitely many primary fields and we call it the \textbf{minimal model} denoted by $\cM_{p,p'}$.
Using \eqref{rel-hrs-alpha}, the central charge and
the Kac formula become
\bea
\label{minimal_model}
c &= 1-6 \frac{\left(p-p^{\prime} \right)^{2}}{p p^{\prime}}\notag \, ,\\
h _{r , s} &= \frac{\left(p^{\prime} r-p s \right)^{2}-\left(p-p^{\prime} \right)^{2}}{4 p p^{\prime}}\, .
\eea
We can obtain a symmetry property from above:
\begin{equation}\label{inversion-formula}
h _{r , s }=h _{p -r, p^{\prime}-s}\, .
\end{equation}
Now, we get a finite set of conformal families, which
closes under fusion rules. The corresponding finite
set of conformal dimensions $h_{r,s}$ is given by
$1\leq r < p$ and $1 \leq s < p^{\prime}$. The symmetry above
gives
\begin{equation}
\phi _{r , s}=\phi _{p^{\prime}-r , p-s }\, .
\end{equation}
Hence, there remain $(p-1)(p'-1)/2$ distinct fields in the
theory.
And the fusion rule is modified to be
\begin{equation}\label{MM-fusion}
\phi _{r,s}\times\phi _{m,n}=
\sum^{k_{\text{max}}}_{
\begin{subarray}{1}
  \quad  k=1+ \abs{r-m} \\
  k+r+m=1 \, \text{mod} \, 2
\end{subarray}}
\,
\sum^{l_{\text{max}}}_{
	\begin{subarray}{1}
	\quad  l=1+ \abs{s-n} \\
	k+s+n=1 \, \text{mod} \, 2
	\end{subarray}}
\left[\phi _{k,l}\right]\, ,
\end{equation}
where the upper bound of the summation now is control
by the periodical and symmetric condition.
\bea
k_{\max} &= \min \left(r + m-1,2 p -1-r-m \right)\, ,\\
l _{\max} &= \min (s + n-1,2 p^{\prime}-1-s-n)\, .
\eea
We will see several examples of minimal models in the following subsections.

\subsubsection*{Unitarity for $c> 1$}

A representation of the Virasoro algebra is unitary if it contains no negative-norm states.
An essential tool in establishing unitarity conditions is the Kac determinant, which serves as a means to derive these conditions. In fact, the unitary condition gives us a strong constraint of the choice of $h$ and $c$.

We now sketch the proof that the representations are unitary when $(c>1,h>0)$. The proof is separated into three steps as follows:
\begin{itemize}
\item The first step is to prove that for each curve $C_{r,s}: h= h_{r,s}$, it will either be located below, or on the axis $h=0$ if $c>1$. If $1<c<25$, from \eqref{rel-hrs}, $h_{r,s}$ is not real unless $s=r$, where $h_{r,s}\leq0$. If $c\geq 25$ the choice \eqref{m branches} implies that $-1<p<0$. Then $p(p+1)<0$ and $[(p+1)r-ps]\geq 1$. According to \eqref{rel-hrs-m}, this tells us $h_{r,s}(p)$ is
not positive.  Thus we finish our proof for the first step.

\item The second step is to prove that the Kac determinant is positive throughout the region.
For a given level, we can find a $h$ larger than any $h_{r,s}$, and at such a point the Kac determinant is positive. Since none of the curves lies in the region, the determinant sign can not be changed throughout the region by crossing any curves. This proves the second point.

\item  The last step is to prove that the Gram matrix at level $N$, $M_N$ is positive definite in this region, which is equivalent to the theory
in this region is unitary. Since the Kac determinant is positive, we have even number of negative eigenvalues of $M_N$. Since this number can change only across one of the curves $C_{r,s}$, and is consequently fixed to be the same in the region. Therefore, we need to show that the matrix $M_N$ for each level $N$ is positive definite for at least one point $(c,h)$ in the region.

To show this, we define the length $n(\alpha)$ of a basis vector $\ket{\alpha}$ as the number of operators $L_k$ acting on the primary state $\ket{h}$. For example,
$L^{3}_{-1}$ has length $3$ and $L_{-3}\ket{h}$ has length 1. It is possible to show that the dominant behavior in $h$ of inner products is
\bea
  \langle \alpha | \alpha \rangle &= c _{\alpha} h^{n (\alpha)} [ 1 + \cO (1 / h) ] \quad \left(c _{\alpha} > 0 \right)\cr
  \langle \alpha | \beta \rangle &= \cO(h^{(n (\alpha) + n (\beta)) / 2-1}) + \ldots
  \, .
\eea
where $\ket{\alpha}$ and $\ket{\beta}$. The above expression immediately shows that $M_N$ is dominant by its diagonal element when $h$ is sufficiently large, and thus $M_N$ is positive definite. Therefore, in this region, a Verma module $V(c,h)$ is irreducible.
\end{itemize}

\subsubsection*{Unitarity for $0 \le c \le 1$}
This unitarity criterion significantly constrains the admissible values of the parameters $h$ and $c$, particularly within the regime where $0 \le c \le 1$.
In this region, we merely provide a brief summary. 

\begin{itemize}
    \item It is straightforward to show that, in Figure \ref{fig:kac-determinant}, a point in the region $0 \leq c \leq 1$, $h > 0$ that lies between two vanishing curves of the Kac determinant is non-unitary. The Kac determinant changes sign when crossing a vanishing curve, indicating the presence of a negative-norm state at that level. Consequently, no unitary representations of the Virasoro algebra exist in this region, except possibly on the vanishing curves themselves. 
    \item A more careful analysis tells us that there is an additional negative norm state everywhere on the vanishing curves except at certain points where they intersect.
\end{itemize}

Therefore, the unitary representations of the Virasoro algebra only occur at the values of the central charge \cite[ISZ88-No.3]{Cardy:1984bb}:
\begin{equation}
\label{c<1-unitary-c}
c=1-\frac{6}{p (p + 1)} \, ,\quad p=3,4 , \ldots
\end{equation}
with the corresponding highest weight
\begin{equation}
\label{c<1-unitary-hrs}
h _{r,s} (p)=\frac{[ (p + 1) r-p s ]^{2}-1}{4 p (p + 1)}\, .
\end{equation}
where $r,s$ satisfies $1 \leq s \leq r \leq p-1$. The form is exactly the same as \eqref{rel-hrs-m}, but now $p$ can only take discrete integers. This result tells us that the unitary condition is satisfied only at the point $(c,h)$ where the Verma module has null states.

\subsubsection*{Unitary Minimal Models $\cM_p$}
We now consider the choice of $p$ and $p'$ to make the minimal models unitary. Recalling the admissible conformal dimensions in \eqref{minimal_model}, Bezout's lemma states that there exists a couple of integers $(r_0,s_0)$ in the range $1\leq r_0 < p$ and $1 \leq s_0 < p^{\prime}$ such that
\begin{equation}
p^{\prime} r _{0}-p  s _{0}=1\, .
\end{equation}
Thus, the corresponding conformal dimension
\begin{equation}
h _{r _{0} , s _{0}}=\frac{1-\left(p-p^{\prime} \right)^{2}}{4 p p^{\prime}}
\, ,
\end{equation}
which is always negative, expect if $\abs{p-p'} =1$, in which case it vanishes. In fact, the minimal model is unitary only if $\abs{p-p'} =1$, and the primary field $h _{r _{0} , s _{0}} =0$ is indeed the identity operator. Hence, we can set $p'=p+1$ so that the central charge is expressed in \eqref{c<1-unitary-c}. We denote the unitary minimal model by $\cM_p$. We note that the list of unitary representations given in \eqref{c<1-unitary-c} and \eqref{c<1-unitary-hrs}, coincides with the list of highest weight $h_{r,s}$ of unitary minimal models.

\subsection{Characters of minimal models}\label{sec:characters-MM}
The character of an irreducible representation of the Virasoro algebra
\[
\chi _{h} (\tau) \equiv \operatorname{Tr} _{L(c,h)} q^{L _{0} -\frac{c}{24}}=q^{-\frac{c}{24}} \sum _{N=0}^{\infty} d _{h} (N) q^{h + N}~,
\]
where $d_h(N)$ is the number of states at level $N$, and $d_h(N)\le P(N)$. If there is no null state, $d_h(N)=P(N)$. As seen \eqref{character-decomp},
the torus partition function with periodic boundary conditions is
\bea
Z _{\mathrm{PP}} (\tau , \overline{\tau}) &= \operatorname{Tr}_{\cH} \left(q^{L _{0}-\frac{c}{24}} \overline{q}^{\overline{L} _{0}-\frac{c}{24}} \right)\cr
&= \sum _{h , \overline{h}} \mathcal{N} _{h \overline{h}} ~\chi _{h} (\tau) \chi _{\overline{h}} (\overline{\tau})~,
\eea
where $ \mathcal{N} _{h \overline{h}} $ is the number of primary fields of conformal dimension $(h,\overline h)$ \cite[ISZ88-No.25]{cardy1986operator}. Since there is only one vacuum, we set $\cN_{00}=1$. Under the $T$-transformation, a character behaves as
\be\label{T-trans}
\chi _{h} (\tau + 1)=e^{2 \pi i (h-\frac{c}{24})} \chi _{h} (\tau)~.
\ee
The invariance of $\cZ_{\mathrm{PP}}$ under the $T$-transformation requires
\[\mathcal{N} _{h \overline{h}}=0 , \quad h-\overline{h} \notin \mathbb{Z}~.\]
If we write the $S$-transformation of characters
\[\chi _{h} (-1/\tau)=\sum _{h^{\prime}} S _{h h^{\prime}} ~\chi _{h^{\prime}} \left(\tau \right)~,\]
then the modular invariance of $\cZ_{\mathrm{PP}}$ requires
\begin{equation}
\label{relation N S}
\sum _{h , \overline{h}} \mathcal{N} _{h \overline h} S _{h h'} S _{\overline{h} \overline{h}^{\prime}}=\mathcal{N} _{h^{\prime} \overline{h}^{\prime}}~.
\end{equation}
Below we will determine the partition function of diagonal type $\cN_{hh'}=\delta_{hh'}$ in the unitary minimal models. We will see one example of non-diagonal type in the tricritical Ising model in \S\ref{sec:examples}, and the classification of the modular invariant partition functions is discussed at the end of \S\ref{sec:su2k-character}.

From the Kac determinant, we have learned that the Verma module $V_{h_{r,s}}$ has a singular vector at level $rs$. Since $h_{r,s}+rs=h_{-r,s}$, the singular vector generates a submodule $V_{h_{-r,s}}$. Furthermore. since $h_{p-r,p'-s}+ (p-r) (p'-s)=h_{2p-r,s}$, we have another submodule $V_{h_{2p-r,s}}$. Moreover, these submodule $V_{h_{-r,s}}$, $V_{h_{2p-r,s}}$ contain other submodules $V_{h_{-2p-r,s}}$, $V_{h_{2p+r,s}}$, and one can see repeatedly this pattern \cite[ISZ88-No.9]{feigin1983verma}.

\begin{figure}[ht]
\centering
\begin{tikzpicture}[node distance=3.5cm]
\node (A) at (0,0){$(r,s)$};
\node (B) at (2,1){$(-r,s)$};
\node (C) at (2,-1){$(2p-r,s)$};
\node [ right of =B] (D){$(-2p+r,s)$};
\node [ right of =D] (E){$(-2p-r,s)$};
\node [ right of =E] (F){$(-4p+r,s)$};
\node [ right of =F] (F1){$\ $};
\node [ right of =C] (G){$(2p+r,s)$};
\node [ right of =G] (H){$(4p-r,s)$};
\node [ right of =H] (I){$(4p+r,s)$};
\node [ right of =I] (I1){$\ $};
\draw[->] (A) -- (B);
\draw[->] (A) -- (C);
\draw[->] (B) -- (D);
\draw[->] (D) -- (E);
\draw[->] (E) -- (F);
\draw[->] (C) -- (G);
\draw[->] (G) -- (H);
\draw[->] (H) -- (I);
\draw[->] (F) -- (F1);
\draw[->] (I) -- (I1);
\draw[->] (B) -- (G);
\draw[->] (D) -- (H);
\draw[->] (E) -- (I);
\draw[->] (C) -- (D);
\draw[->] (G) -- (E);
\draw[->] (H) -- (F);
\draw[->] (F) -- (I1);
\draw[->] (I) -- (F1);
\end{tikzpicture}
\end{figure}

Therefore, the character of the irreducible representation $L(c,h_{r,s})$ of the Virasoro algebra is
\begin{align} \chi _{r, s} (\tau) &=\operatorname{Tr} _{L(c,h_{r,s})} q^{L _{0} -\frac{c}{24}} \cr
  &=\frac{1}{q^{-\frac1{24}}\eta (\tau)} q^{-\frac{c}{24}} \sum _{k \in \mathbb{Z}}\left[ q^{h _{r + 2 p k , s}}-q^{h _{2 p k-r },  s} \right] \end{align}
which is called the \textbf{Rocha-Caridi formula} \cite[ISZ88-No.10]{rocha1985vacuum}.

Defining the theta functions by
\begin{equation}
\label{Theta function}
\Theta _{m , k} (\tau)=\sum _{n \in \mathbb{Z}} q^{k \left(n + \frac{m}{2 k} \right)^{2}}
\end{equation}
a character of the unitary minimal model $\cM_p$ can be written by
\be\label{Rocha-Caridi-formula}
\chi _{r, s} (\tau)=\frac{1}{\eta (\tau)} \left(\Theta _{r (p + 1)-s p , p (p + 1)} (\tau)-\Theta _{r (p + 1) + s p , p (p + 1)} (\tau) \right)
\ee
The behavior of $\chi_{r,s}(\tau)$ under the $S$-transformation has been obtained in  \cite[ISZ88-No.29]{Cardy:1986ie}:
\bea\label{RC-S}
 \chi _{r, s} (- 1 / \tau)&= \sum _{(r^{\prime},s^{\prime}):\mathrm{primary}} S_{(r,s),(r',s')} ~\chi_{r',s'} (\tau) \cr
S_{(r,s),(r',s')} &=\left(\frac{8}{p (p + 1)} \right)^{\frac12} (- 1)^{(r + s) \left(r^{\prime} + s^{\prime} \right)} \sin\left( \frac{\pi r r^{\prime}}{p} \right)\sin \left(\frac{\pi ss^{\prime}}{p + 1}\right)
\eea
and the $T$-transformation is given in \eqref{T-trans}.
It is shown that the partition function of diagonal type
\be\label{diagonal-PP}
\cZ_{\cM_p}^{\textrm{diag}}=\sum _{(r,s):\mathrm{primary}} \left| \chi _{r, s} (\tau) \right|^{2}
\ee
is invariant under the modular transformations.

\subsection{Ising model \texorpdfstring{$(p',p)=(4,3)$}{(p',p)=(4,3)}}\label{sec:Ising}
\begin{table}[htbp]\centering
\begin{tabular}{c|cc}
		$3$
		& $\frac{1}{2}$
		& $0$
		\\[5pt]
				$2$
		& $\frac{1}{16}$
		& $\frac{1}{16}$
		\\[5pt]
		$1$
		& $0$
		& $\frac{1}{2}$
		\\[5pt]
		\midrule
		& $1$
		& $2$
	\end{tabular}
	\hspace{2cm}
	\begin{tabular}{c|cc}
		$3$
		& $\varepsilon$
		& $\mathbf{1}$
		\\[5pt]
				$2$
		& $\sigma$
		& $\sigma$
		\\[5pt]
		$1$
		& $\mathbf{1}$
		& $\varepsilon$
		\\[5pt]
		\midrule
		& $1$
		& $2$
	\end{tabular}\end{table}
Let us consider the unitary minimal model $\cM_{p=3}$ whose central charge $c=\frac12$. The primary fields and their conformal dimensions are described in the table above. By comparing the conformal dimensions with \eqref{Ising-conf-dim}, $\varepsilon$ is the energy density operator and $\sigma$ is the spin field in the Ising model. Therefore,   the unitary minimal model $\cM_{p=3}$ describes the Ising model.

Since the Ising model is the most important 2d CFT, let us investigate it more in detail. For the Ising model on a lattice of $(N\times M)$ sites, the statistical partition function is
\begin{equation}
\cZ=\sum_{\{\sigma\}}e^{K\sum_{\langle i , j \rangle}\sigma_i\sigma_j}
\end{equation}
where we define $K=J/(k_BT)$. One can use the identity
\[
\exp \left[ x \sigma _{i} \sigma _{l} \right]=\cosh x \left(1 + \sigma _{i} \sigma _{l} \tanh x \right)
\]
so that the partition function is
\begin{equation}\label{high-temp-exp}
    \cZ=\sum _{\{\sigma \}} \prod _{\langle i , j \rangle} \cosh (K) \left(1 + \sigma _{i} \sigma _{j} \tanh (K) \right)~.
\end{equation}
At high temperature, $\tanh K$ is small and one can take expansion. In the expansion, the generic expression of these terms is
\[
\tanh(K)^{r} \sigma _{1}^{n _{1}} \sigma _{2}^{n _{2}} \sigma _{3}^{n _{3}} \ldots
\]
where $r$ is the total number of  lines connecting the adjacent sites $i$ and $j$, and $n_i$ is the number of lines where $i$ is the final site. Since each spin $\sigma_i$ assumes values $\pm1$, we have a null sum unless all $n_1, n_2, \ldots$ are even numbers, which implies all closed loops $P$ of the lattice as in Figure \ref{fig:high-temp}. Hence, the partition function is
\[
\cZ _{\mathrm{high}}=[ 2 \cosh (K) ]^{N M} \sum _{P} [ \tanh (K) ]^{\mathrm{length} (P)}
\]
\begin{figure}[ht]\centering
\includegraphics[width=4cm]{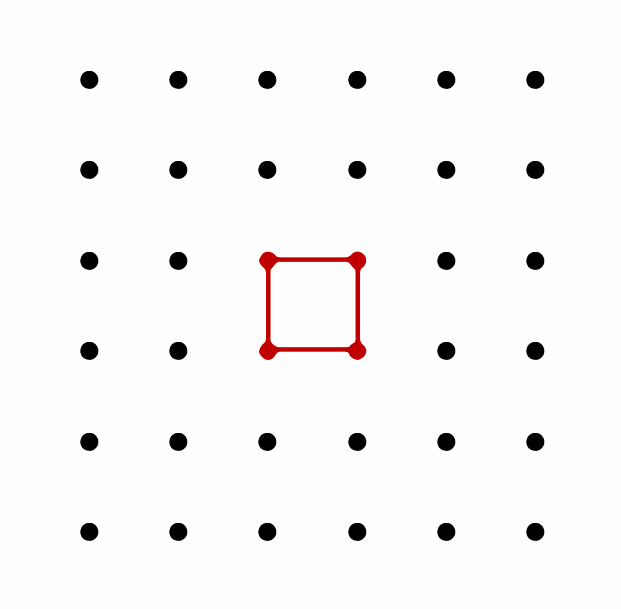}\hspace{2cm}\includegraphics[width=4cm]{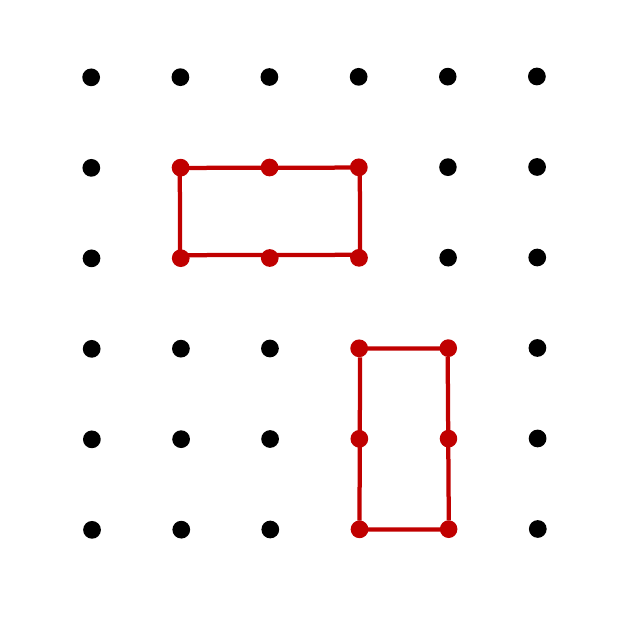}\label{fig:high-temp}
\end{figure}

On the other hand, at low temperatures, the spins tend to align one with another. Taking the dual lattice, the border  between antiparallel spins forms loops $P_D$, and the partition function is given by
\begin{equation}\label{low-temp}
    \cZ _{\mathrm{low}}=2 e^{N M K} \sum _{P_D} e^{- 2 K (\mathrm{length}(P_D))}~.
\end{equation}
where the contribution of all spins down has been factored out of the sum.

\begin{figure}[ht]\centering
\includegraphics[width=5cm]{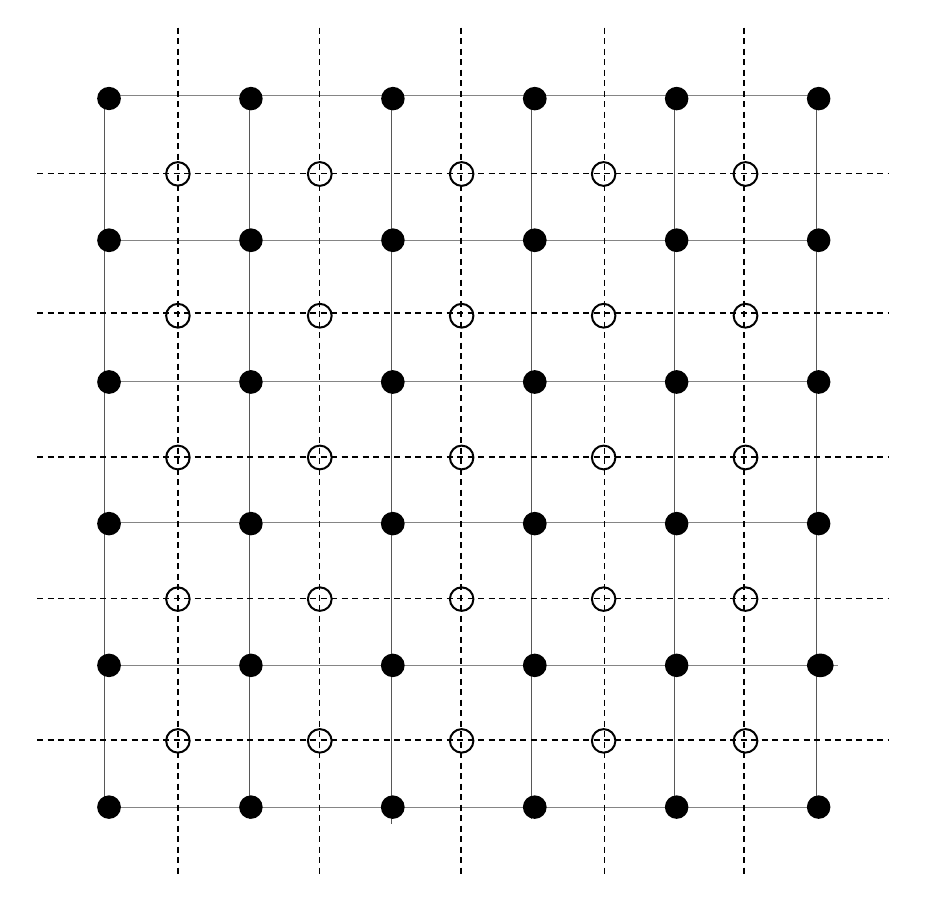}\hspace{2cm}\includegraphics[width=5cm]{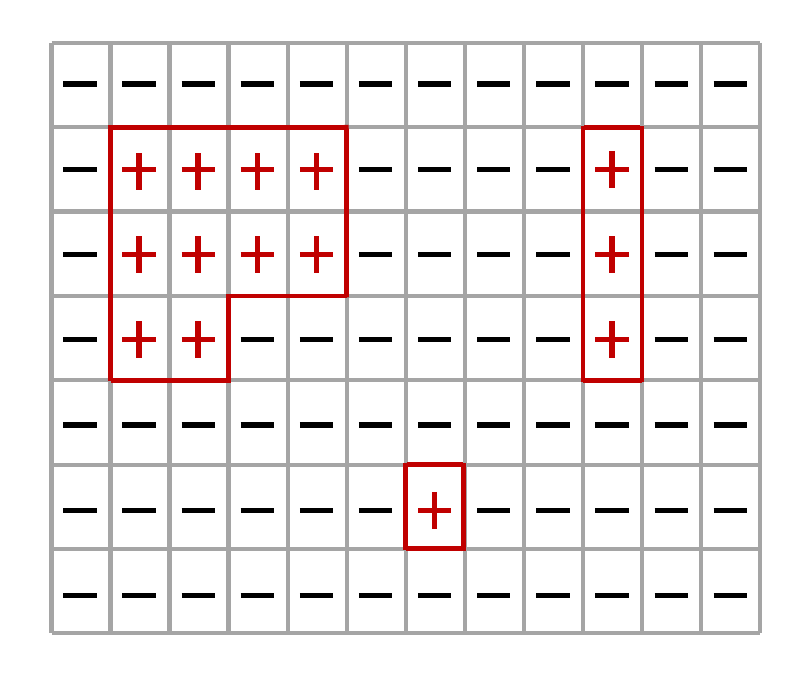}\caption{(Above) In high-temperature regime, the partition function is expressed as a sum over all close loops in the lattice. (Below) In low-temperature regime, the partition function is expressed as a sum over all close loops in the dual lattice.}\label{fig:high-temp}
\end{figure}

The two phases are indeed dual to each other
\begin{equation}
  \cZ _{\mathrm{low}} \left(K^{\prime} \right)=2 (2\sinh 2 \mathrm{K})^{- \frac{N M}2} \cZ _{\mathrm{high}} (K)
\end{equation}
where we identify
\begin{equation}\label{KK'}
e^{- 2 K^{\prime}}=\tanh K~.
\end{equation}
This is called the \textbf{Kramers-Wannier duality}. This can be also written as
\begin{equation}\label{KK'2}
\sinh (2K) \sinh (2K')=1
\end{equation}
so that the critical temperature is at $\sinh 2K=1=\sinh 2K'$ which is
\[K_{c}=\frac{1}{2} \log (1+\sqrt{2})~.\]

\subsubsection*{$(d+1)$-dimensional classical to $d$-dimensional quantum Ising}
Let us first consider the one-dimensional classical Ising model
\begin{equation}
 Z(K)=\sum_{\{s_{i}=\pm1\}}\prod_{i=1}^N   e^{K s_i s_{i+1}}
\end{equation}
with periodic boundary condition $s_1=s_{N+1}$. Locally, the transfer matrix can be written as
\begin{equation}\label{S}
S=e^{K s_i s_{i+1}}=\begin{pmatrix}
e^{K} & e^{-K} \\
e^{-K} & e^{K}
\end{pmatrix}
\end{equation}
Then, the partition function can be written as
\begin{equation}
Z(K)=\Tr S^{N}
\end{equation}
For a quantum statistical system, the quantum partition function is given by
\begin{equation}
\Tilde{Z}(\beta) = \Tr(e^{\beta H}) = \Tr((e^{H\Delta\tau})^{N})
\end{equation}
where $H$ is the Hamiltonian of the system and $\beta=1/k_{B}T$, $N\Delta\tau=\beta$. Therefore, the 1d classical Ising model can be translated into a two-qubit system if we identify $S$ with $e^{H\Delta\tau}$. Moreover, since the eigenvalues of $S$ are $e^K\pm e^{-K}$, the partition function becomes
\begin{equation}
  Z(K)= (e^K+e^{-K})^N+ (e^K-e^{-K})^N=\cosh^N K(1+\tanh^N K)~.
\end{equation}
so that the free energy is
\begin{equation}
  F(K)=\log Z(K)= N \log  \cosh K+\log(1+\tanh^N K)~.
\end{equation}

\begin{figure}[ht]
\centering
\includegraphics[width=16cm]{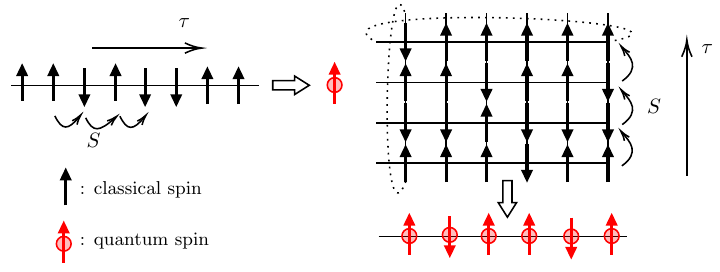}
\caption{Mapping from $(d+1)$-dimensional classical model to $d$-dimensional quantum model}
\label{fig:classical-quantum}
\end{figure}

Now let us move on to the two-dimensional classical Ising model. Analogous to the previous example, it is equivalent to quantum system of $N$ two-qubit sites to which $N$ copies $S^{(1)}\cdots S^{(N)}$ of the transfer matrices \eqref{S} acts on it. Since we have
\begin{equation}
e^{K' \sigma^{x}}=\begin{pmatrix}
\cosh K' & \sinh K' \\
\sinh K' & \cosh K'
\end{pmatrix}
\end{equation}
 \eqref{S} can be expressed as
 \begin{equation}
S=(2 \sinh 2 K)^{\frac12} e^{K' \sigma^{x}}
\end{equation}
where $K$ and $K'$ are related by \eqref{KK'}. The horizontal interaction between two adjacent qubits is diagonalized as $\sum _{i=1}^N \sigma _{i}^z \sigma _{i+1}^z$. Consequently, the partition function can be written as
\begin{equation}
Z\left(K,K^{\prime}\right)=(2 \sinh 2 K)^{\frac{N M}2} \Tr\left(e^{K \sum _{i=1}^N \sigma _{i}^z \sigma _{i+1}^z} e^{K' \sum_{i} \sigma_i^x}\right)^{M}
\end{equation}
The correspondence between classical and quantum statistical systems can be understood visually by Figure \ref{fig:classical-quantum}.

Therefore, the Hamiltonian of quantum one-dimensional Ising model is
\be\label{Ising-Hamiltonian2} \mathcal{H}=-k_BT( K \sum _{i=1}^N \sigma _{i}^z \sigma _{i+1}^z+K' \sum _{i} \sigma _{i}^x)
\ee
This Hamiltonian can be understood as a quantum spin chain with the external magnetic field $K'$ is parallel to the $x$-axis. In the low-temperature $K\gg K'$, we can ignore the interaction with the external magnetic field $K'$. However, in the high-temperature $K\ll K'$, this is not the case. In order to capture the interaction term,  it is natural to consider the \textbf{disorder operator} $\mu_i$ on the dual lattice, which is defined as follows:
\[\mu _{a}^{z} \equiv \prod _{i=- \frac{N}{2}}^{i=a} \sigma _{i}^{x}~,\qquad
\mu _{a}^{x}=\sigma _{a}^{z} \sigma _{a + 1}^{z}
\]
For instance, if we act  $\mu^z_a$ on the parallel-spin state
\[
| \uparrow \rangle \equiv \prod _{b=- \frac{N}{2}}^{\frac{N}{2}} | \uparrow \rangle _{b}
\]
then we have
\begin{equation}
\mu _{a}^{z} | \uparrow \rangle=\prod _{b=- \frac{N}{2}}^{a} | \downarrow \rangle _{b} \prod _{c=a + 1}^{\frac{N}{2}} | \uparrow \rangle _{c}~.
\end{equation}
Thus, the disorder operator $\mu _{a}^{z}$ changes the ordered spin configurations at the dual-site $a$ so that it can be understood as a \textbf{soliton} located at the dual-site $a$.  Moreover, the Hamiltonian \eqref{Ising-Hamiltonian2} can be written as
\[
\mathcal{H}=-k_BT( K' \sum _{\left\langle a,a^{\prime} \right\rangle} \mu _{a}^z \mu _{a^{\prime}}^z+K \sum _{a} \mu _{a}^x)
\]
so that $B$ and $J$ are exchanged. Hence, in the low-temperature
\[
\left\langle \sigma _{a}^{z} \right\rangle \neq 0 , \quad \left\langle \mu _{a}^{z} \right\rangle=0 , \quad T < T _{c}
\]
whereas in the high-temperature
\[
\left\langle \mu _{a}^{z} \right\rangle \neq 0 , \quad \left\langle \sigma _{a}^{z} \right\rangle=0 , \quad T > T _{c}~.
\]

\subsubsection*{Onsager's exact solution}
To solve a one-dimensional quantum Ising chain exactly, we diagonalize the Hamiltonian (\ref{Ising-Hamiltonian2}). This can be done by introducing a set of fermionic operators $\{\psi_{i}|i=1,2,...,2N\}$ satisfying Clifford algebra
\begin{equation}
    \{\psi_{i},\psi_{j}\}=2\delta_{ij}.
\end{equation}
A natural choice of the operators is
\begin{equation}
\begin{aligned}
\psi_{2 k-1} &=\sigma^{z}_{1} \sigma^{z}_{2} \cdots \sigma^{z}_{k-1} \sigma^{y}_{k} =\sigma^{z}_1\otimes\cdots\otimes\sigma^{z}_{k-1}\otimes\sigma^{y}_k\otimes I\otimes\cdots\otimes I\\
\psi_{2 k} &=\sigma^{z}_{1} \sigma^{z}_{2} \cdots \sigma^{z}_{k-1} \sigma^{x}_{k}=\sigma^{z}_1\otimes\cdots\otimes\sigma^{z}_{k-1}\otimes\sigma^{x}_k\otimes I\otimes\cdots\otimes I
\end{aligned}
\end{equation}
for $k=1,2,\cdots,N$. It can be shown that,
\begin{equation}
    \begin{aligned}
    i\psi_{2k-1}\psi_{2k}&=\sigma^{z}_{k}\\
    i\psi_{2k}\psi_{2k+1}&=\sigma^{y}_{k}\sigma^{y}_{k+1}.
    \end{aligned}
\end{equation}

\begin{figure}[ht]
    \centering
    \includegraphics[width=15cm]{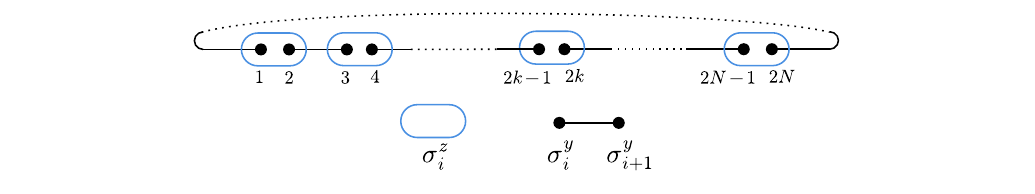}
    \caption{Expressing Ising chain by fermionic operators}
    \label{fig:dimer-chain}
\end{figure}

By rotating the reference frame $x\to y$, $y\to z$, $z\to x$, Hamiltonian (\ref{Ising-Hamiltonian2}) can be rewritten by fermionic operators as
\begin{equation}
    \mathcal{H}=-ik_BT(K\sum_{k=1}^{N}\psi_{2k}\psi_{2k+1}+K'\sum_{k=1}^{N}\psi_{2k-1}\psi_{2k})
\end{equation}
where the periodic boundary condition is imposed by $\psi_{2N+1}=\psi_{1}$. Then, the degree of freedom at each site is split, which could be understood by Figure \ref{fig:dimer-chain}.  This is called the \textbf{Jordan-Wigner transformation}. Set $\Psi=(\psi_{1},\psi_{2},\cdots,\psi_{2N})^{T}$, then $\mathcal{H}=-\frac{ik_BT}{2}\Psi^{T}\mathbf{M}\Psi$,
\begin{equation}
\mathbf{M}=\begin{pmatrix}
0 & K' &  &  &  & -K\\
-K' & 0 & K &  &  & \\
 & -K & 0 & K' &  & \\
 &  & -K' & \ddots  & \ddots  & \\
 &  &  & \ddots  & 0 & K'\\
K &  &  &  & -K' & 0
\end{pmatrix}
\end{equation}
Diagonalizing $H$ is equivalent to performing Fourier transform to a vector
\[
\mathbf{v}=(\cdots,\omega^{-1}a,\omega^{-1}b,a,b,\omega a,\omega b,\cdots)^{T}
\]
\begin{equation}\label{Diagonalize-FT}
\Tilde{\mathbf{v}}=-i\mathbf{M}\mathbf{v}=(\cdots,\omega^{-1}\Tilde{a},\omega^{-1}\Tilde{b},\Tilde{a},\Tilde{b},\omega\Tilde{a},\omega\Tilde{b},\cdots)^{T}
\end{equation}
where $\omega^N=1$. From (\ref{Diagonalize-FT}), $a,b$ and $\Tilde{a},\Tilde{b}$ is related by
\begin{equation}
    \mqty(\Tilde{a}\\\Tilde{b})=-i\mqty(0&K'-K\omega^{-1}\\-K'+K\omega&0)\mqty(a\\b).
\end{equation}
Therefore, the $2N$ eigenvalues of the matrix $-i\mathbf{M}$ is given by
\begin{equation}
\pm  m_{s}=\pm\abs{K'-Ke^{-\frac{2 \pi i s} N}}~,\qquad s=1,2,\cdots,N.
\end{equation}
Hence the partition function of the system is given by
\begin{equation}
Z(K, K')=\sum_{\pm \pm \cdots \pm} e^{\pm m_{1} \pm m_{2} \cdots \pm m_{N}}=\prod_{s=1}^{N}\left(e^{+m_{s}}+e^{-m_{s}}\right)=2^{N} \prod_{s=1}^{N} \cosh \left|K'-K e^{-2 \pi i s / N}\right|.
\end{equation}
So the free energy of the system is
\begin{equation}
F(K, K')=\frac{1}{N}\log Z(K,K')=\log 2+\frac{1}{N} \sum_{s=1}^{N} \log \cosh \left|K'-K e^{-\frac{2 \pi i s} N}\right|.
\end{equation}
In the large-$N$ limit $N\to\infty$, the summation can be replaced by an integral
\begin{equation}
F(K, K')=\log 2+\int_{0}^{2 \pi} \log \cosh \left|K'-K e^{-i \ell}\right| \frac{d \ell}{2 \pi}.
\end{equation}
Moreover, the spectrum of Onsager's solution is given by
\begin{equation}
m_{s}=\sqrt{(K-K')^{2}+ KK'\left(2 \sin\frac{2 \pi s}N\right)^{2}} \geq|K-K'|.
\end{equation}
When $K\neq K'$, the spectrum is gapped. Since a gapped system has the typical length scale, the theory is \emph{not} conformal. On the other hand, when $K= K'=K_c$, we have
\begin{equation}
2 m_{s}=2 K|\sin \frac{2 \pi s}N|
\end{equation}
so that the first excited energy goes to zero if we take $N\to \infty$.
The system is called gapless, signaling that it can be described by CFT. (See Figure \ref{fig:beta-function}.)

\begin{figure}[ht]
\centering
\includegraphics[width=12cm]{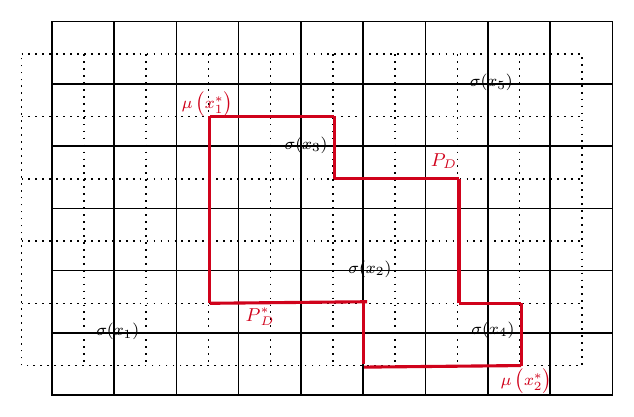}
\caption{Correlation function $\langle\sigma(x_{1}) \cdots \sigma(x_{5}) \mu(x_{1}^{*}) \mu(x_{2}^{*})\rangle$ depends on a choice of a path connecting $x_1^*$ and $x_2^*$ because they are mutually non-local.}
\label{fig:mutually-non-local}
\end{figure}

\subsubsection*{Minimal model $\cM_{p=3}$}
In the high-temperature regime, the partition function admits the expansion \eqref{high-temp-exp}. Hence, the two-point function of spin operators can be obtained by
\begin{equation}
    \langle\sigma(x_1)\sigma(x_2)\rangle_{\textrm{high}}=[ 2 \cosh (K) ]^{N M} \sum _{P} [ \tanh (K) ]^{\mathrm{length} (P)}
\end{equation}
where we sum over all the paths $P$ connecting $x_1$ and $x_2$. On the other hand, the same method cannot be applied in the low-temperature regime because \eqref{low-temp} makes sense only for a closed path $P_D$. However, we can introduce the disordered operator and consider the two-point function as follows. Usually, the Boltzmann weight $e^{ K \sigma_i \sigma_j}$ is assigned to an edge $\langle i,j\rangle$ in the partition function. Choosing a path $P_D$ connecting $x_1^*$ and $x_2^*$, if an edge $\langle i,j\rangle$ intersects $P_D$, the effect of the disordered operators changes the sign of the exponent of the Boltzmann weight $e^{ K \sigma_i \sigma_j}\to e^{- K \sigma_i \sigma_j}$. Namely, we define
\begin{equation}
\left\langle \mu(x_{1}^{*}) \mu(x_{2}^{*})\right\rangle_{P_D} =\sum_{\{\sigma\}}  \prod_{\langle i,j\rangle} e^{\pm K \sigma_i \sigma_j}
\end{equation}
where the sign $\pm$ in the exponent is determined by whether or not the edge crosses $P_D$.
Although there is no unique choice of $P_D$, the two-point function is well-defined
\begin{equation}
    \left\langle \mu(x_{1}^{*}) \mu(x_{2}^{*})\right\rangle_{P_D}=\left\langle \mu(x_{1}^{*}) \mu(x_{2}^{*})\right\rangle_{P_D^*}~.
\end{equation}
Indeed, the Boltzmann weight does not change if we flip all spins inside the closed path $P_D \cdot P_D^*$.

Likewise, we can define a general correlation function as follows:
\begin{equation}
\left\langle\sigma(x_{1})\cdots \sigma(x_{m}) \mu(x_{1}^{*}) \mu(x_{2}^{*})\right\rangle_{P_D} =\sum_{\{\sigma\}}   \sigma(x_{1}) \cdots \sigma(x_{m}) \prod_{\langle i,j\rangle}  e^{\pm K \sigma_i \sigma_j}
\end{equation}
However, this depends on the choice of $P_D$
\begin{equation}
\left\langle\sigma(x_{1}) \cdots \sigma(x_{m}) \mu(x_{1}^{*}) \mu(x_{2}^{*})\right\rangle_{P_{D}}=\pm\left\langle\sigma(x_{1}) \cdots \sigma(x_{m}) \mu(x_{1}^{*}) \mu(x_{2}^{*})\right\rangle_{P_{D}^{*}}
\end{equation}
where the sign $\pm$ is determined by the number of $x_1,\ldots,x_m$ inside the closed path $P_D \cdot P_D^*$. (See Figure \ref{fig:mutually-non-local}.) Put differently, the correlation function with both $\sigma$ and $\mu$ depends on a global structure. Therefore, they are called \textbf{mutually non-local}.
In fact, a disordered operator $\mu$ introduces a branch cut, and a spin operator receives a sign. when it crosses the branch cut:
\[
\sigma (z , \overline{z}) \mu (0,0) \rightarrow \sigma \left(e^{2 \pi i} z , e^{- 2 \pi i }\overline{z} \right) \mu (0,0)=- \sigma (z , \overline{z}) \mu (0,0)
\]
when $\sigma$ is rotated around $\mu$. This can be understood that the fermion shows up in the OPE of $\sigma$ and $\mu$
\be\label{sigma-mu}
\sigma (z , \overline{z}) \mu (0,0)\sim\frac{1}{| z |^{\frac14}} \left(z^{\frac12} \psi (0) + \overline{z}^{\frac12} \overline{\psi} (0) \right)
\ee
Recalling that the central charge of the minimal model $\cM_{p=3}$ is $c=\frac12$, one can deduce that a free fermion describes the continuum limit of the Ising model.

\begin{figure}[ht]\centering
\includegraphics[width=6cm]{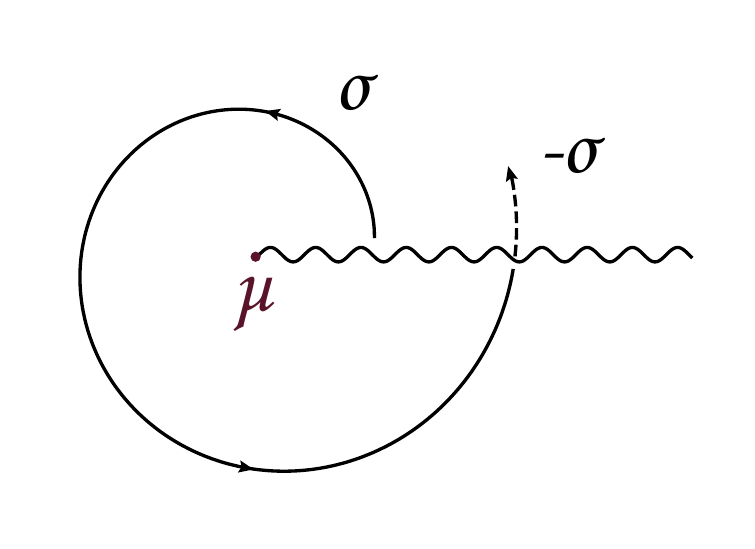}
\end{figure}

In fact, identifying the various operators in Ising model with the primaries in  $\cM_{p=3}$
\bea
\psi (z)& \sim\phi _{2,1} (z) \otimes \phi _{1,1} (\overline{z}) \cr \overline{\psi} (\overline{z})& \sim\phi _{1,1} (z) \otimes \phi _{2,1} (\overline{z}) \cr
\psi \overline{\psi} : =\varepsilon (z , \overline{z})&\sim \phi _{2,1} (z) \otimes \phi _{2,1} (\overline{z})\cr
 \sigma (z , \overline{z}) \ \textrm{or} \ \mu (z , \overline{z}) &\sim\phi _{1,2} (z) \otimes \phi _{1,2} (\overline{z})
 \eea
we obtain desired OPE relations. Note that $\sigma$ and $\mu$ have the same conformal dimension because the Kramers-Wannier duality exchanges the spin and disorder operator $\sigma\leftrightarrow\mu$. The fusion rule in $\cM_{p=3}$ can be read off as
\begin{equation}
  \sigma \times \sigma=\mathbf{1} + \varepsilon ~,\qquad  \varepsilon \times \varepsilon=\mathbf{1} ~,\qquad  \sigma \times \varepsilon=\sigma~.
\end{equation}

Moreover, using the associativity of OPEs, one can derive from \eqref{sigma-mu}
\[
\psi ( z ) \sigma ( w,\overline w ) \sim \frac{1}{( z - w ) ^ { \frac12}} \mu ( w ,\overline w)~.
\]
We can compare this with the mode expansion of free fermion on the Ramond sector
\[
\psi ( z ) \sigma ( 0 ) = \sum _ { r \in \bZ} z ^ { - r - \frac12} b _ { r} \sigma ( 0 )~,
\]
we can deduce
\begin{equation}b _ { 0} \sigma ( 0 ) = \mu ( 0 )~.\end{equation}
In the presence of the spin field $\sigma(0)$ at the origin, the fermion field obeys the anti-periodic boundary condition
\[\psi ( z ) \rightarrow \psi ( e ^ { 2 \pi i} z ) = - \psi ( z )~.\]
The spin field $\sigma$ creates a branch cut, which changes the boundary condition of the fermion field. Hence, $\sigma$ can be regarded as the \textbf{twist operator}.

Finally, let us consider the partition functions of the Ising model. There are three unitary irreducible representations corresponding to the highest weights  $h=0,\frac12,\frac1{16}$, and the corresponding characters can be read off from \eqref{Rocha-Caridi-formula} as
\begin{gather}
\chi_{0}=\frac12 \left(\sqrt{\frac{\vartheta_3}{\eta}}+\sqrt{\frac{\vartheta_4}{\eta}} \right)=\mbox{Tr}_{NS}\left(\frac{1+(-1)^F}{2}q^{L_0-\frac{c}{24}} \right), \nonumber \\
\chi_{\frac12}=\frac12 \left(\sqrt{\frac{\vartheta_3}{\eta}}-\sqrt{\frac{\vartheta_4}{\eta}} \right)=\mbox{Tr}_{NS}\left(\frac{1-(-1)^F}{2}q^{L_0-\frac{c}{24}} \right), \\
\chi_{\frac{1}{16}}=\frac{1}{\sqrt{2}} \sqrt{\frac{\vartheta_2}{\eta}}=\mbox{Tr}_{R}\left(q^{L_0-\frac{c}{24}} \right). \nonumber
\end{gather}
Using these expressions, the partition function of the Ising model of diagonal type is equal to that of the free fermion \eqref{free-fermion-PF}
\begin{equation}\label{free-fermion-minimal-model}
\cZ_{\textrm{F}}(\tau, \overline{\tau})=\chi_0\overline{\chi}_0+\chi_{\frac{1}{2}}\overline{\chi}_{\frac{1}{2}}+\chi_{\frac{1}{16}}\overline{\chi}_{\frac{1}{16}}.
\end{equation}
The structure of this partition function also appears when studying superstrings, and the operator $\frac12 (1+(-1)^F)$ is known as the \textbf{Gliozzi-Scherk-Olive (GSO) projection}.

\subsection{Other examples}\label{sec:examples}

\subsubsection*{Yang-Lee singularity $(p',p)=(2,5)$}
Among the minimal non-unitary models, a simple but particularly significant example is given by the model $\cM_{2,5}$ with central charge is $c=-22/5$.  The primary fields in the theory are identity operator and a field $\phi_{1,2}$ of conformal dimension $h=-1/5$.

The partition function of a statistical model defined on a lattice is an analytic function of its parameters as long as the number $N$ of the fluctuating variables is finite. Let us consider the Ising model at a given temperature $T$  in the presence of an external magnetic field $B$
\begin{equation}\mathcal{H}=- J \sum _{\left\langle i ,\right\rangle} \sigma _{i} \sigma _{j}-B \sum _{i} \sigma _{i}\end{equation}
The Yang-Lee theorem states that the zeroes of the partition function $Z$ are all on the imaginary axis $\Re B=0$. Fisher showed that the points $B=\pm iB_c$ are new critical points of the theory, which are called \textbf{Yang-Lee edge singularities}. Furthermore, he has argued that the effective action is given by the Landau-Ginzburg theory
\[
\mathcal{S}=\int d^{2} z \left[ \frac{1}{2} (\partial \Sigma)^{2} + i \left(B-B _{c} \right) \Sigma + i g \Sigma^{3} \right]
\]
where the non-unitarity of the model manifests itself in the imaginary value of the coupling constant.  This model is described by the non-unitary minimal  model $\cM_{2,5}$ \cite[ISZ88-No.13]{Cardy:1985yy}.

\subsubsection*{Tricritical Ising model $(p',p)=(5,4)$}

\begin{table}[htbp]\centering
\begin{tabular}{c|ccc}
		$4$
		& $\frac32$
		& $\frac7{16}$
		& $0$
		\\ [5pt]
		$3$
		& $\frac35$
		& $\frac3{80}$
		& $\frac1{10}$
		\\[5pt]
		$2$
		& $\frac1{10}$
		& $\frac3{80}$
		& $\frac35$
		\\[5pt]
		$1$
		& $0$
		& $\frac7{16}$
		& $\frac32$
		\\    [5pt]
		\midrule
		$0$
		& $1$
		& $2$
		& $3$
	\end{tabular}\hspace{2cm}
\begin{tabular}{c|ccc}
				$4$
		& $G$
		& $\sigma'$
		& $\mathbf{1}$
		\\ [5pt]
		$3$
		& $t$
		& $\sigma$
		& $\varepsilon$
		\\[5pt]
		$2$
		& $\varepsilon$
		& $\sigma$
		& $t$
		\\[5pt]
		$1$
		& $\mathbf{1}$
		& $\sigma'$
		& $G$
		\\    [5pt]
		\midrule
		$0$
		& $1$
		& $2$
		& $3$
		\end{tabular}\label{Tab-Min4}
\end{table}

Let us consider the variant of the Ising model whose Hamiltonian is given by
\[
\mathcal{H}=- J \sum _{\langle i , j \rangle}^{N} \sigma_{i} \sigma_{j} t _{i} t _{j}- B \sum _{i=1}^{N} \sigma_{i} t _{i}-\mu \sum _{i=1}^{N} t _{i}
\]
where a spin variable $\sigma_k$ takes values $\pm1$, and a vacancy variable $t_k$, with values $0$ and $1$. This variable specifies whether the site
is empty (0) or occupied (1). The chemical potential $\mu$ controls the density of the vacancy.

\begin{figure}[ht]\centering
\includegraphics[width=7cm]{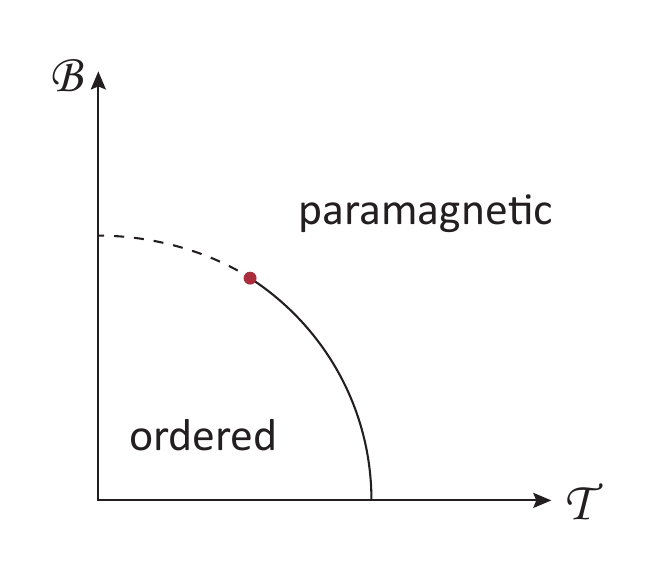}
\caption{Phase diagram of the tricritical Ising model. The (broken) first-order line and the (full) line of Ising critical points meet at the tricritical point $(B_t, T_t)$, shown as a black dot}\label{fig:tricritical}
\end{figure}

A typical phase diagram of this model is sketched in Figure \ref{fig:tricritical}, where the lines of first-order and second-order transitions meet in the tricritical point, located at $B=B_t$ and $T=T_t$. While at a normal critical point the two phases become indistinguishable, at a tricritical point three physically distinct phases merge. For that reason, this model is called the \textbf{tricritical Ising model}. The phenomenology of the tricritical Ising model is described in \cite[\S4]{cardy1996scaling}.

Remarkably, the tricritical model is described by $\cN=1$ superconformal field theory where $G$ is the supercurrent \cite[ISZ88-No.5]{friedan1985superconformal}.

\subsubsection*{3-state Potts model $(p',p)=(6,5)$}
\[\begin{array}{c| c c  c  c } 5& 3 &{\frac{7}{5}} &{\frac{2}{5}} &{0} \\[5pt] 4& \frac{13}{8} &{\frac{21}{40}} &{\frac{1}{40}} &{\frac{1}{8}} \\[5pt] 3&  \frac{2}{3} &{\frac{1}{15}} &{\frac{1}{15}} &{\frac{2}{3}} \\[5pt] 2& \frac{1}{8} &{\frac{1}{40}} &{\frac{21}{40}} &{\frac{13}{8}} \\[5pt] 1& 0 &{\frac{2}{5}} &{\frac{7}{5}} &{3} \\[5pt] \hline &1&2&3&4  \end{array}\]

On a square lattice, the Hamiltonian of the three-state Potts model is given by
\begin{equation}\cH=- \frac{J}{2} \sum _{\langle i, j \rangle} \left(\sigma _{i} \overline{\sigma} _{j} + \overline{\sigma} _{i} \sigma _{j} \right)
\end{equation}
where $\sigma$ takes the values $\omega^i$ (i=1,2,3) with a third root of unity $\omega=e^{2\pi i/3}$. This model undergoes a second-order phase transition at $J_c=\frac{2}{3} \log (\sqrt{3} + 1)$, which is endowed with the $\bZ_3$ symmetry  \cite[ISZ88-No.12]{dotsenko1984critical}. The lattice theory is exactly solvable and consequently all critical exponents are known. For instance,
\[
h_{\sigma}=h_{\overline \sigma}=\frac1{15}~,\qquad h_\varepsilon=\frac{2}{5}~.
\]
It is tempting to identify the primary field of the unitary minimal model $\cM_{p=5}$.
However, the exact solution of the lattice model does not have operators with
conformal dimensions $\frac18$, $\frac1{40}$, $\frac{21}{40}$, and $\frac{13}8$  \cite[ISZ88-No.22]{vonGehlen:1986gk}.

In fact, the partition function of  the 3-state Potts model  is described by the non-diagonal type
\be\label{3-state-Potts}
\cZ _{\text{3-Potts}}=\sum _{r=1,2} \left\{\left| \chi _{r , 1} + \chi _{r , 5} \right|^{2} + 2 \left| \chi _{r , 3} \right|^{2} \right\}~,
\ee
so that the primary fields are
\[
(0,0) , \left(\frac{2}{5} , \frac{2}{5} \right) , \left(\frac{7}{5} , \frac{7}{5} \right)~, 2 \times \left(\frac{1}{15} , \frac{1}{15} \right) , 2 \times \left(\frac{2}{3} , \frac{2}{3} \right)  (0,3) , (3,0) , \left(\frac{2}{5} , \frac{7}{5} \right) , \left(\frac{7}{5} , \frac{2}{5} \right)
\]
It is noteworthy that there are chiral spin-3 currents $W(z)$ and $\overline W(\overline z)$ corresponding to  (0,3), (3,0)  \cite[ISZ88-No.6]{Zamolodchikov:1985wn}. In fact, $W(z)$ with the energy momentum tensor $T(z)$ forms $W_3$-algebra \cite{Fateev:1987vh}.

\subsubsection*{Other models}

It was shown in \cite[ISZ88-No.32]{Huse:1984mn} that a general unitary minimal model $\cM_p$ of diagonal type describes an RSOS model constructed in \cite[ISZ88-No.31]{Andrews:1984af}.
Furthermore, the CFT with $\bZ_k$ symmetry has been constructed in \cite[ISZ88-No.14]{fateev1985parafermionic}, which we will see in \S\ref{sec:coset}, where the central charge is
\be\label{Zk-cc} c = \frac { 2 ( k - 1 )} { k + 2}~.\ee
When $k=4$, the central charge is $c=1$ and the corresponding CFT is called the \textbf{Ashkin-Teller model} where the Hamiltonian is written by the two spin fields $ \sigma _ { i} $ and $ \tau _ { i} $ as
\[\mathcal { H} = - J _ { 2} \sum _ { \langle i , j \rangle} \left( \sigma _ { i} \sigma _ { j} + \tau _ { i} \tau _ { j} \right) - J _ { 4} \sum _ { \langle i , j \rangle} \sigma _ { i} \sigma _ { j} \tau _ { i} \tau _ { j}~.\]
The $\bZ_4$ symmetry is given by
\[
s_j\to e^{\frac{i\pi k}{2}}s_j ~, \qquad \textrm{where}\qquad
s_j\equiv \frac{\sigma _ { j} +i\tau _ { j}}{\sqrt{2}}~.
\]
This Ashkin-Teller model can be described by the free boson on an orbifold space $S^1/\bZ_2$, which will be studied in \S\ref{sec:orbif-part-funct}.
All in all, these models are the main characters in the papers listed in \cite{itzykson1988conformal}.

\subsection{Conformal blocks and bootstrap}\label{sec:bootstrap}
The main object of a field theory is the calculation of correlation functions. We have seen the forms of two- and three-point functions in \eqref{two-point-func-form} and \eqref{three-point-func-form}. As mentioned before, one way to construct all correlation functions is to find the corresponding operator algebra: The complete OPE (including all regular terms) of all primary fields with each other.
So far, we consider the fusion rule of primary fields in a minimal model, which does not tell anything about the coefficients of operator algebra. This section aims to spell out the coefficients of operator algebra and conformal bootstrap approach.

Let us start with a two-point function.
Since the coefficient $C_{12}$ of a two-point function \eqref{two-point-func-form} is symmetric,  we are free to choose a basis of primary fields of conformal dimension $h$ so that $C_{\alpha\beta}=\delta_{\alpha\beta}$. Thus, conformal families associated with different $\phi_\alpha$ are orthogonal in the sense of the two-point function.

As in \eqref{op-algebra}, the general form of OPE of primary fields involves descendants
\begin{equation}
\phi_p^{\{\lambda, \bar{\lambda}\}}(z,\overline{z})
=\mathcal{L}_{-\lambda_1} \cdots \mathcal{L}_{-\lambda_l} \overline{\mathcal{L}}_{-\overline{\lambda}_1} \cdots \overline{\mathcal{L}}_{-\overline{\lambda}_m} \phi_p(z,\overline{z})~.
\end{equation}
In fact, the three-point function \eqref{three-point-func-form} explicitly determines the coefficient of the most singular term of the OPE \eqref{op-algebra} as $C _{12p}^{\{0,0 \}}= C _{12p}$. Moreover,
since the correlations of descendants are built on the correlation of the primaries as in \eqref{descendant-eq}, we expect the coefficient to have the form:
\begin{equation}
C^{\{\lambda,\overline{\lambda}\}}_{12p}=C_{12p}
\beta_{12(p,{\lambda})}
\overline{\beta}_{12(p,\overline{\lambda})}\, .
\end{equation}
As we will see below, $\beta_{ij(p,{\lambda})}$ can be completely determined by the representation of Virasoro algebra as functions of central charge $c$ and of conformal dimensions.
Therefore, once we find all the coefficients $C_{pnm}$ of three-point functions with primary fields, we determine all OPEs. Indeed, the conformal bootstrap is one powerful way to determine  $C_{pnm}$.

\subsubsection*{Conformal Blocks}
To explain the conformal bootstrap, let us consider a four-point:
\begin{equation}
\left\langle \phi _{1} \left(z _{1} , \overline{z} _{1} \right) \phi _{2} \left(z _{2} , \overline{z} _{2} \right) \phi _{3} \left(z _{3} , \overline{z} _{3} \right) \phi _{4} \left(z _{4} , \overline{z} _{4} \right) \right\rangle\,,
\end{equation}
By a suitable conformal transformation, we can set $z_4 =0, z_1=\infty, z_2=1$ and $z_3 =x$ where $x$ is the cross-ratio of $z_{1,2,3,4}$:
\begin{equation}
G _{34}^{21} (x , \overline{x}) =
\lim _{z _{1} , \overline{z} _{1} \rightarrow \infty} z _{1}^{2 h _{1}} \overline{z} _{1}^{2\overline{h} _{1}} \left\langle \phi _{1} \left(z _{1} , \overline{z} _{1} \right) \phi _{2} (1,1) \phi _{3} (x , \overline{x}) \phi _{4} (0,0) \right\rangle\, ,
\end{equation}
Applying the operator algebra \eqref{op-algebra} to
$\phi_3(x,\overline{x})\phi_4(0,0)$ ,
the function $G^{21}_{34}$ becomes
\begin{equation}\label{four-point}
G _{34}^{21} (x , \overline{x})=\sum _{p} C_{12p}C_{34p}  \mathcal{F} _{34}^{21} (p | x) \overline{\mathcal{F}} _{34}^{21} (p | \overline{x})\, ,
\end{equation}
where
\begin{equation}
\label{conformal-block}
\mathcal{F} _{34}^{21} (p | x)=x^{h _{p}-h _{3}-h _{4}} \sum _{\{\lambda \}} \beta _{34(p,\lambda)} x^{|\lambda|} \frac{\left\langle h _{1} \left| \phi _{2} (1) L _{- \lambda_{1}} \cdots L _{- \lambda_l} \right| h _{p} \right\rangle}{\left\langle h _{1} \left| \phi _{2} (1) \right| h _{p} \right\rangle}\, ,
\end{equation}
which are called \textbf{conformal blocks}. Here, we can reformulate this expression using three-point functions. Specifically, we define the vertex function as follows:
\be 
\Gamma_{12(p,\lambda)} := \frac{\left\langle h _{1} \left| \phi _{2} (1) L _{- \lambda_{1}} \cdots L _{- \lambda_l} \right| h _{p} \right\rangle}{\left\langle h _{1} \left| \phi _{2} (1) \right| h _{p} \right\rangle}
\ee 
Additionally, we introduce another vertex function:
\be
\Gamma'_{(p,\lambda')34} := \frac{\langle \phi_p|L _{\lambda'_{N}}\cdots  L _{\lambda'_{1}}\phi_3(1)|\phi_4\rangle}{\langle \phi_p|\phi_3(1)|\phi_4\rangle} = \sum_\lambda \beta_{34(p,\lambda)} \langle \phi_{p}^{-\lambda'} | \phi_{p}^{-\lambda} \rangle
\ee
The last term in the second line of this equation corresponds to the Kac-Shapovalov matrix \(G_{h_p}(\lambda,\lambda')\) in \eqref{KS-matrix}, which enables us to express the structure constants as:
\be 
\beta_{34(p,\lambda)} = \sum_{\lambda} (G_{h_p}(\lambda',\lambda))^{-1} \Gamma'_{(p,\lambda')34}
\ee 
While the Kac-Shapovalov matrix is infinite-dimensional yet block-diagonal, it is possible to compute its inverse at each level. Consequently, the conformal block can be described using vertex functions and the matrix:
\be \label{conf-block}
\mathcal{F} _{34}^{21} (p | x)=x^{h _{p}-h _{3}-h _{4}} \sum _{|\lambda|=|\lambda'|} x^{|\lambda|}\Gamma_{12(p,\lambda)} (G_{h_p}(\lambda,\lambda'))^{-1} \Gamma'_{(p,\lambda')34}
\ee 
 
Therefore, the computation of the vertex functions, the three-point functions with descendants, is crucial for deriving the conformal block. Let us first consider the evaluation of \(\Gamma'\). For simplicity, we assume that the three-point function is properly normalized and omit the denominator in our expressions. Using the operator product expansion (OPE) of the energy-momentum tensor \(T(z)\) with primary fields, we can rewrite the expression as follows:
\begin{align}
\langle L_{-n} \phi_p^{-\lambda'} | \phi_3(1) \phi_4(0) \rangle &= \langle \phi_p^{-\lambda'} | \oint_{C_0+C_1} \frac{dz}{2\pi i} z^{n+1} T(z) \phi_3(1) \phi_4(0) \rangle \\
&= \oint_{C_0} \frac{dz}{2\pi i} z^{n+1} \sum_k \frac{1}{z^{k+2}} \langle \phi_p^{-\lambda'} | \phi_3(1) (L_k \phi_4(0)) \rangle \cr
& \qquad \qquad  + \oint_{C_1} \frac{dz}{2\pi i} z^{n+1} \sum_k \frac{1}{(z-1)^{k+2}} \langle \phi_p^{-\lambda'} | (L_k \phi_3(1)) \phi_4(0) \rangle~.\nonumber
\end{align}
By expanding \(z^{n+1}\) around \(z=1\), we obtain the series expansion:
\be
z^{n+1} = \sum_{l=0}^{n+1} \binom{n+1}{l} (z-1)^l.
\ee
Substituting this series back into the previous expression, we simplify:
\be
\langle L_{-n} \phi_p^{-\lambda'} | \phi_3(1) \phi_4(0) \rangle = \langle \phi_p^{-\lambda'} | \phi_3(1) (L_n \phi_4(0)) \rangle + \sum_{l=0}^{n+1} \binom{n+1}{l} \langle \phi_p^{-\lambda'} | (L_{l-1} \phi_3(1)) \phi_4(0) \rangle.
\ee
For primary fields \(\phi_3\) and \(\phi_4\), terms involving \(L_n\) for \(n \geq 1\) vanish, thereby simplifying the expression to:
\be
\langle L_{-n} \phi_p^{-\lambda'} | \phi_3(1) \phi_4(0) \rangle = \left((n+1)h_3 + \delta_{n,0} h_4\right) \Gamma'_{(p,\lambda')34} + \langle \phi_p^{-\lambda'} | L_{-1} \phi_3(1) \phi_4(0) \rangle.
\ee
For \(n=0\), the second term can be expressed using \(\Gamma'\):
\be
\langle \phi_p^{-\lambda'} | L_{-1} \phi_3(1) \phi_4(0) \rangle = (h_p+|\lambda'| - h_3 - h_4) \Gamma'_{(p,\lambda')34}.
\ee
Substituting this back into our expression, we derive:
\be
\Gamma'_{(p,[n,\lambda']);34} = (h_p+|\lambda'| + n h_3 + h_4) \Gamma'_{(p,\lambda');34}.
\ee
The notation \([n,\lambda']\) refers to the process of adding a row of length \(n\) to the Young diagram \(\lambda' =[\lambda'_1 \geq \ldots \geq \lambda'_{d} \geq 1]\). By repeatedly applying this relation, we obtain
\be\label{gamma'}
\Gamma'_{(p,\lambda');34} = \prod_{i=1}^{l(\lambda')} \Big(h_p + \lambda'_i h_3 - h_4 + \sum_{j>i} \lambda'_j\Big)~.
\ee

Next, let us analyze another component of the correlation function:
\begin{align}
\langle \phi_1(\infty) \phi_2(1) (L_n \phi_p^{-\lambda}(0)) \rangle &= -\oint_{C_0} \frac{dz}{2\pi i} z^{n-1} \langle T(z) \phi_1(\infty) \phi_2(1) \phi_p^{-\lambda}(0) \rangle \\
&= -\oint_{C_1} \frac{dz}{2\pi i} z^{n-1} \sum_k \frac{1}{(z-1)^{k+2}} \langle \phi_1(\infty) (L_k \phi_2(1)) \phi_p^{-\lambda}(0) \rangle \cr
& \qquad - \oint_{C_\infty} \frac{dz}{2\pi i} z^{n-1} \sum_k z^{k-2} \langle (L_k \phi_1(\infty)) \phi_2(1) \phi_p^{-\lambda}(0) \rangle\nonumber 
\end{align}
By deforming the integration contour using \(C_0 = -C_1 - C_{\infty}\) and applying the transformation property of the energy-momentum tensor under \(z = 1 / w\), specifically \(T(z) = w^4 T(w)\), we can perform a Laurent expansion around \(z = \infty\). Upon evaluating the residue, we obtain:
\begin{multline}
    \langle \phi_1(\infty) \phi_2(1) (L_n \phi_p^{-\lambda}(0)) \rangle = \langle (L_n \phi_1)(\infty) \phi_2(1) \phi_p^{-\lambda}(0) \rangle\\
-\sum_{k=-1}^\infty(-1)^{k+1} \binom{k + n - 1}{n - 2}
\langle \phi_1(\infty) (L_k \phi_2)(1) \phi_p^{-\lambda}(0) \rangle ~.
\end{multline}
For primary fields \(\phi_{1,2}\), the same analysis as above leads to
\be
\langle \phi_1(\infty) \phi_2(1) (L_n \phi_p^{-\lambda}(0)) \rangle = -\langle \phi_1(\infty) (L_{-1} \phi_2)(1) \phi_p^{-\lambda}(0) \rangle + \left((n - 1) h_2 + \delta_{n,0} h_1\right) \Gamma_{12(p,\lambda)}.
\ee
Substituting \(n = 0\) into this expression, we arrive at the simplified form
\be
\Gamma_{12, (p,[n,\lambda])} = (h_p+|\lambda| + n h_2 - h_1) \Gamma_{12,(p,\lambda)}~.
\ee
Finally, the correlation function can be written in the form
\be\label{gamma}
\Gamma_{12,(p,\lambda)}= \prod_{i=1}^{l(\lambda)} \Big(h_p + \lambda_i h_2 - h_1 + \sum_{j>i} \lambda_j\Big).
\ee
Comparing with \eqref{gamma'}, we find 
\be 
\Gamma_{12,(p,\lambda)}=\Gamma'_{(p,\lambda)12}~.
\ee 
In this way, we can determine the conformal block \eqref{conf-block} by order-by-order. 

As it stands, the conformal block is derived from the representation theory of the Virasoro algebra and is independent of the specific 2d CFT being considered. In other words, it is a universal construction that can be applied to any 2d CFT by substituting specific values, such as the central charge. In contrast, the structure coefficients in the three-point function \eqref{three-point-func-form} are intricately tied to the particular details of the theory. 

\begin{figure}[ht]
	\centering
	\includegraphics[width=0.6\linewidth]{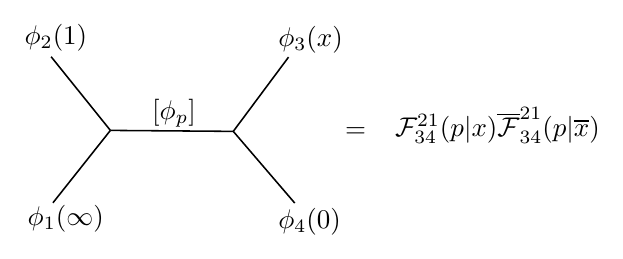}
	\caption{Diagram for the conformal block}
	\label{fig:tree}
\end{figure}

The fusion of fields \(\phi_3\) and \(\phi_4\) generates intermediate operators, denoted as \([\phi_p]\), which subsequently interact with \(\phi_1\) and \(\phi_2\). This interaction process can be visualized as depicted in Figure \ref{fig:tree}.

Let us provide a few examples. By expanding the conformal block as a power of the cross-ratio $x$
series in $x$:
\begin{equation}
\mathcal{F} _{34}^{21} (p | x)=x^{h _{p}-h _{3}-h _{4}} \sum _{K=0}^{\infty} \mathcal{F} _{K} x^{K}\,.
\end{equation}
we can present several leading coefficients
\eqref{conformal-block}.
\begin{equation}
\mathcal{F}_{0}=1\, , \qquad
\mathcal{F} _{1}=\frac{\left(h _{p} + h _{2}-h _{1} \right) \left(h _{p} + h _{3}-h _{4} \right)}{2 h _{p}}
\end{equation}
\bea
\mathcal{F} _{2} &= \frac{\left(h _{p} + h _{2}-h _{1} \right) \left(h _{p} + h _{2}-h _{1} + 1 \right) \left(h + h _{3}-h _{4} \right) \left(h + h _{3}-h _{4} + 1 \right)}{4 h _{p} \left(2 h _{p} + 1 \right)}\notag\\
&+ 2 \left(\frac{h _{1} + h _{2}}{2} + \frac{h _{p} \left(h _{p}-1 \right)}{2 \left(2 h _{p} + 1 \right)}-\frac{3 \left(h _{1}-h _{2} \right)^{2}}{2 \left(2 h _{p} + 1 \right)} \right)^{2}\notag\\
&
\times \left(\frac{h _{3} + h _{4}}{2} + \frac{h _{p} \left(h _{p}-1 \right)}{2 \left(2 h _{p} + 1 \right)}-\frac{3 \left(h _{3}-h _{4} \right)^{2}}{2 \left(2 h _{p} + 1 \right)} \right)^{2} \left(c + \frac{2 h _{p} \left(8 h _{p}-5 \right)}{2 h _{p} + 1} \right)^{- 1}
\eea

\subsubsection*{Crossing Symmetry and the Conformal Bootstrap}
\begin{figure}[ht]
	\centering
	\includegraphics[width=0.8\linewidth]{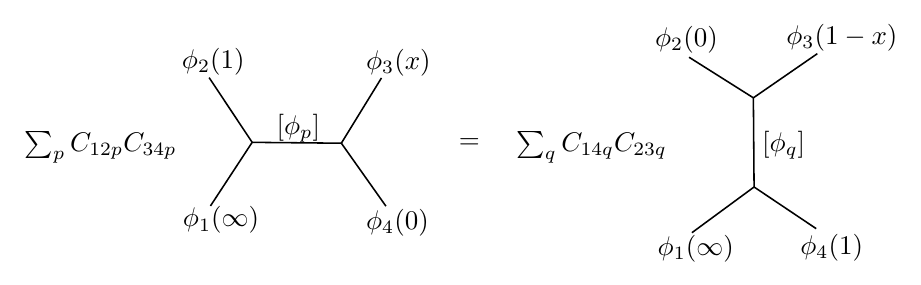}
	\caption{Crossing symmetry}
	\label{fig:crossing-symmetry}
\end{figure}

To obtain a four-point function, we fuse $\phi_3$ and $\phi_4$ in \eqref{four-point}. However, we can also take different pairs for a fusion, say  $\phi_3$ and $\phi_2$. For this, we can send $z_2$ to $0$ and $z_4$ to $1$, and then $z_3$ proves to be $1-x$. Then, the same procedure will give a different basis $\mathcal{F} _{32}^{41} (q | 1-x)$ of conformal blocks. Nonetheless, the final result for the four-point function should be the same
\begin{equation}
\label{Bootstrap method}
G^{21}_{34}(x,\overline{x})=G^{41}_{32}(1-x,1-\overline{x})
\, ,
\end{equation}
which implies
\begin{equation}\label{bootstrap}
\sum _{p} C _{12p} C _{34p} \mathcal{F} _{34}^{21} (p | x) \overline{\mathcal{F}} _{34}^{21} (p | \overline{x})=
\sum _{q} C _{14q} C _{23q} \mathcal{F} _{32}^{41} (q | 1-x) \overline{\mathcal{F}} _{32}^{41} (q | 1-\overline{x})
\end{equation}
This relation is represented graphically in Figure \ref{fig:crossing-symmetry}

If the conformal block $\mathcal{F}$ is known, then \eqref{Bootstrap method} yields a system of equations that can determine $C_{ijk}$'s and $h,\overline{h}$'s. The program of calculating the correlation functions simply by assuming crossing symmetry is known as the \textbf{conformal bootstrap} approach. For minimal models, the bootstrap equations can be solved completely.

\section{Renormalization group flow and \texorpdfstring{$c$}{c}-theorem}

\subsection{Renormalization group flow}

\subsubsection*{Idea}

In the introduction, we have briefly discussed that conformal field theories arise at fixed points of renormalization group flow. The study of renormalization group flows tells us the special role of conformal field theories (See \cite{Zamolodchikov:1989zs} included in \cite{jimbo2014integrable} and \cite{eguchi1989deformations}). In this section, we shall glimpse this in two-dimensional systems.

The name of renormalization group is actually misleading in the following two reasons:
\begin{itemize} \setlength{\itemsep}{-.5pt}
\item the transformations of the renormalization ``group'' are irreversible; therefore, they do not form a group. \item moreover, they do not necessarily concern the renormalization of a theory, i.e. the cure of the divergences of the perturbative series. \end{itemize}
Instead, the concept of renormalization group flow describes how families of physical theories behave under change of the length scale. The effect of averaging over short-distance physics can be summarized in the change of finitely many parameters in the long-distance physics of interest.

There is a wonderful introduction to renormalization group flow in \cite[\S3]{cardy1996scaling}. Following Cardy,
we shall explain the concept of renormalization group flow by coarse-graining for the Ising model.
Let us consider a statistical system defined on a $d$-dimensional regular lattice of spacing $a$, with degrees of freedom placed on its sites. Let $H(\{\sigma_i^{(1)}\},g_k^{(1)})$ be the Hamiltonian of the system, where $g_k^{(1)}$ are the coupling constants of the various interactions among the spins $\sigma_i^{(1)}$. Now we divide the original lattice into blocks of length $ba$, denoted by $\cB_k$, each of them made of $b^d$ spins, and we assign a new spin variable $\sigma_k^{(2)}$ by averaging over all the spins $\sigma_i^{(1)}\in \cB_k$. We are going to do coarse-graining of the original lattice by repeating this procedure. More precisely, we define the new spin variable by
\be\label{average}\sigma_{k}^{(n+1)}= A^{(n)}\sum_{i \in \mathcal{B}_{k}} \sigma_{i}^{(n)}\ee
where $A$ is a suitable normalization constant.

To define the effective hamiltonian for the coarse lattice, we introduce an operator
\[T \left(\sigma_{k}^{(n+1)}, \sigma_{i}^{(n)}\right)=\left\{\begin{array}{l}{1 ~,\quad \text{if}\ \eqref{average}}\\{0 ~,\quad \text{otherwise}} \end{array} \right.\]
so that it satisfies
\[
\sum_{\left\{\sigma_{k}^{(n+1)}\right\}}T \left(\sigma_{k}^{(n+1)}, \sigma_{i}^{(n)}\right)=1
\]
Then, the effective hamiltonian $H^{(n+1)}\left(\left\{\sigma_{k}^{(n+1)}\right\} , g_{k}^{(n+1)}\right)$ can be defined by averaging the spins $\sigma_{i}^{(n)}$ with their Boltzmann factor
\[
\exp \left[-H^{(n+1)}\left(\left\{\sigma_{k}^{(n+1)}\right\} , g_{k}^{(n+1)}\right) \right] =\sum_{\left\{\sigma_{i}^{(n)}\right\}}\prod_{\textrm{blocks}}T \left(\sigma_{k}^{(n+1)}, \sigma_{i}^{(n)}\right) \exp \left[-H^{(n)}\left(\left\{\sigma_{i}^{(n)}\right\} , g_{i}^{(n)}\right) \right]
\]
in this way. Then, ignoring an overall constant, we have the equivalence of the statistical partition functions
\[
\sum_{\left\{\sigma_{k}^{(n+1)}\right\}}\exp \left[-H^{(n+1)}\left(\left\{\sigma_{k}^{(n+1)}\right\} , g_{k}^{(n+1)}\right) \right]=\sum_{\left\{\sigma_{i}^{(n)}\right\}}\exp \left[-H^{(n)}\left(\left\{\sigma_{i}^{(n)}\right\} , g_{i}^{(n)}\right) \right]
\]
This equality also holds for the expectation value of any observable $\cO$.
In this process, the new coupling constants are expressed by the previous one
\[
\left\{g^{(n+1)}\right\}=\mathcal{R}\left(\left\{g^{(n)}\right\} \right)
\]
where $\cR$ is, in general, a complicated nonlinear transformation.Starting from the original coupling constants $\{g_k^{(1)}\}$, the procedure of the coarse-graining provides the sequence  $\{g_k^{(2)}\}$, $\{g_k^{(3)}\}$, $\ldots$, which is called a \textbf{renormalization group flow}.
 In fact, a fixed point is a point in the space of the coupling constants that remains invariant
\[
g^{*}= \mathcal{R}\left(g^{*}\right)
\]
Since the coarse-graining does not affect the theory, it is scale-invariant, implying that it is conformal.  As briefly explained in the introduction, critical phenomena are described by these fixed points.

\subsubsection*{One-dimensional Ising model}

As a concrete example, let us examine the one-dimensional Ising model whose Hamiltonian is
\[
H \left(s_{i}; J \right)=- J \sum_{i}s_{i}s_{i+1}
\]
Each pair of spins has the Boltzmann weight
\begin{equation}
  W \left(s_{i}, s_{i+1}; v \right)=e^{K s_{i}s_{i+1}}=\cosh K \left(1+v s_{i}s_{i+1}\right)
\end{equation}
where $v=\tanh K = \tanh \frac{J}{k_BT}$.

\begin{figure}[ht]\centering
\includegraphics[width=7cm]{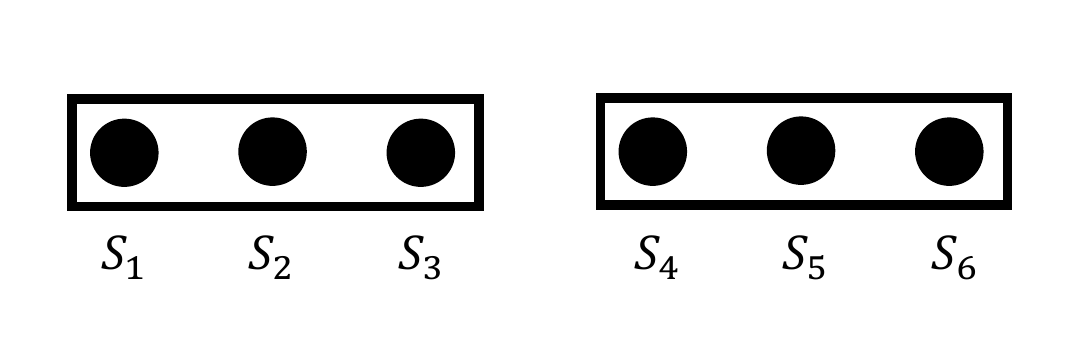}\caption{coarse-graining}\label{fig:coarse-graining}
\end{figure}

As illustrated in Figure \ref{fig:coarse-graining}, we divide the lattice of three spins. Denoting $\sigma_1= s_2$ and $\sigma_2=s_5$, we sum over $s_3$ and $s_4$
\bea
\sum_{s_3,s_4}e^{J \sigma_{1}s_{3}} e^{J s_{3}s_{4}} e^{J s_{4}\sigma_{2}}&=\sum_{s_3,s_4}
(\cosh K)^{3}\left(1+v \sigma_{1}s_{3}\right) \left(1+v s_{3}s_{4}\right) \left(1+v s_{4}\sigma_{2}\right)
\cr
&=
2^{2}(\cosh K)^{3}\left(1+v^{3}\sigma_{1}\sigma_{2}\right)
\eea
In addition to a normalization constant, one can assign the new Boltzmann weight $W (\sigma_{1},\sigma_{2}; \wt v)$ to the new block spins $\sigma_{1},\sigma_{2}$ with the new coupling constant
\[
\wt v= v^{3}~,
\]
which is called the \textbf{renormalization group equation}. It has two fixed points: $v_1^*=0$ and $v_2^*=1$ corresponding to the  high-temperature phase $T\to \infty$, and the low-temperature phase $T=0$.

In addition, the new Hamiltonian of the system is  given by
\[
H \left(\sigma_{i}; \wt J\right)=\wt Np (J)-\wt J\sum_{i}\sigma_{i}\sigma_{i+1}
\]
where $\wt N=N/3$ is the number of new sites, and
\begin{equation}
\wt J  =k_BT\tanh^{- 1}\left[ (\tanh K)^{3}\right]~,\quad p (J)=- \log \left[ \frac{(\cosh K)^{3}}{\cosh \wt K} \right]-2\log 2~.
\end{equation}

\subsubsection*{$\beta$-functions and Callan-Symanzik equation}
In general, the idea described above ends up with mathematical equations describing renormalization group flows in the space of parameters of a physical system. Let us derive these equations by using path integrals. Writing the action of a conformal field theory at a fixed point by $S^*$, the perturbed action can be written as
\be\label{pertubation}
\cS=\cS^{*}+\sum_{a} \int \frac{d^2x}{2\pi}~ g_{a} \cO_{a} (x)
\ee
where the coupling constants $g_{a}$ are sufficiently small. If $\cO_{a}$ is of scaling dimension $\Delta_a$ the coupling constant $g_{a}$ has scaling dimension $2-\Delta_a$. Under an infinitesimal scaling transformation (dilatation) $x\to (1+d\lambda)x$, the perturbed action behaves as
\bea
\cS \rightarrow& \cS^{*}+\sum_{a} (1 +d\lambda)^{2} (1 +d\lambda)^{- \Delta_a} \int \frac{d^2x}{2\pi}~ g_{a} \cO_{a} (x)\cr
&\simeq \cS+d\lambda \sum_{a} \left(2-\Delta_a \right) \int \frac{d^2x}{2\pi}~ g_{a} \cO_{a} (x)
\eea
where we assume that $S^*$ is scale-invariant. On the other hand, we recall the variation under the scaling transformation
\begin{equation}
\cS \rightarrow \cS +d\lambda \int d^2x~ \left(T_{11} (x)+T_{22} (x) \right)=S - d\lambda \int \frac{d^2x}{2\pi}~  \Theta (x)\end{equation}
where
\begin{equation}\label{Theta}
  \Theta :=-2\pi{T^{\mu}}_\mu
\end{equation}
is the trace of the energy-momentum tensor. Therefore, we can write
\be\label{Theta-beta}
 \Theta (x)=-\sum_{a} \beta_{a} (g) \cO_{a} (x)
\ee
where $\beta_{a} $ are first-order approximations of so-called $\beta$-functions
\be\label{1st-approx}
\beta_{a} (g)=\left(2-\Delta_a \right) g_{a}
\ee

The idea of renormalization group flow is that the change of a physical system on the length-scale is absorbed into the change of coupling constants, which is measured by the   $\beta$-functions
\[
\frac{d g_{a}}{d\lambda}=\beta_{a} (g)
\]
Therefore, at a fixed point, we have
\[
\beta_{a} \left(g^{*} \right)=0
\]
 by definition. At the first-order approximation \eqref{1st-approx}, the coupling constants behave as
 \[
 g_{a} \sim e^{(2-\Delta_a)\lambda}
 \]
 around a fixed point $g_{a}^*=0$.

 At the long distant scale $\lambda\gg 1$ that is called \textbf{infra}-\textbf{red} or \textbf{IR},
if $\Delta_a<2$, a coupling constant increases, it is called a \textbf{relevant coupling}. If $\Delta_a>2$, it decreases and is called an \textbf{irrelevant coupling}.  In the space of coupling constants $\{g_{a}\}$, the subspace spanned by irrelevant ones is called the \textbf{critical manifold}, which is the attractive basin for the fixed point $g^*$. Under the RG flow, irrelevant operators are scaled out, and it arrives at the fixed point.
Hence, all the theories belong to the same universality class.
\begin{figure}[ht]\centering
\includegraphics[width=8.5cm]{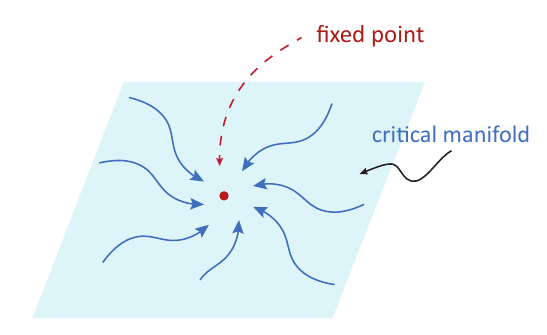}\includegraphics[width=8.5cm]{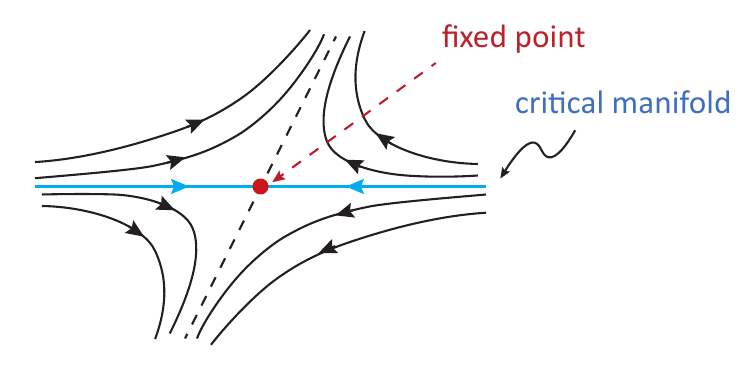}
\end{figure}

On the other hand, the instability nature of a critical point is determined by the number of its relevant couplings. Relevant couplings get amplified as a theory flows to the low-energy region and moves away from the fixed point. When $\Delta_a=2$, an operator $\cO_{a}$ is called \textbf{marginal}, and  its behavior depends on its $\beta$-function, which is determined by taking quantum corrections into account (beyond the first-order approximation). If $\beta_{a}>0$ (resp. $\beta_{a}<0$), then it is called \textbf{marginally relevant} (resp. \textbf{marginally irrelevant}). If $\beta_{a}=0$, then it is called \textbf{exactly marginal}. Our main interest here is the two-dimensional theories, but the same argument can be applied to quantum field theories in any dimension.

Under the perturbation
\be
S = S^{*} + \int \frac{d^2 x}{2\pi}\, g\, \mathcal{O}(x)~,
\ee
the leading contribution to the $\beta$-function is given in \eqref{1st-approx}.
Now we aim to compute the sub-leading (one-loop) correction to the $\beta$-function for the coupling constant $g$. First, let us introduce the dimensionless coupling as $\hat{g} = a^{2-\Delta} g$, where $a$ is a length scale. To determine the $\beta$-function, we analyze how the coupling $g$ must change under an infinitesimal scale transformation $a \to (1 + \delta\lambda) a$ so that the partition function remains invariant.

The partition function admits the perturbative expansion:
\be
\begin{aligned}
\mathcal{Z} &= \int \mathcal{D}\varphi \, e^{-S^* - \hat{g}\int \frac{d^2 x}{2\pi a^{2-\Delta}} \mathcal{O}(x)}\\[6pt]
&= \mathcal{Z}^{*}\biggl[1 - \hat{g}\int \frac{d^2 x}{2\pi a^{2-\Delta}} \langle \mathcal{O}(x) \rangle \\[4pt]
&\quad\quad\quad+ \frac{\hat{g}^2}{2} \int_{|x_1 - x_2|>a}\frac{d^2 x_1}{2\pi a^{2-\Delta}}\frac{d^2 x_2}{2\pi a^{2-\Delta}} \langle \mathcal{O}(x_1)\mathcal{O}(x_2) \rangle + \cdots \biggr]~.
\end{aligned}
\ee
The length scale $a$ appears both explicitly and implicitly through the integration region.

Under the infinitesimal scaling $a \to (1 + \delta\lambda) a$, the coupling $\hat{g}$ transforms as
\be 
\hat{g} \to (1+\delta\lambda)^{2-\Delta}\hat{g} \approx \hat{g} + (2-\Delta)\hat{g}\,\delta\lambda~.
\ee
Additionally, the integral changes according to the rescaling:
\be
\int_{|x_1 - x_2| > a(1+\delta\lambda)}[\cdots] = \int_{|x_1 - x_2| > a}[\cdots] - \int_{a < |x_1 - x_2| < a(1+\delta\lambda)}[\cdots]~.
\ee
The first integral reproduces the original contribution to $\mathcal{Z}$, while the second integral can be evaluated using the operator product expansion (OPE). Suppose the OPE of $\mathcal{O}$ with itself is
\be 
\mathcal{O}(x_1)\mathcal{O}(x_2) = \frac{\mathbf{C}}{|x_{12}|^\Delta}\mathcal{O}(x_2) + \cdots~.
\ee
Using this, we find the second integral becomes
\be
\begin{aligned}
& -\frac{\hat{g}^2}{2}\mathbf{C} a^{-\Delta}\int_{a < |x_1 - x_2| < a(1+\delta\lambda)}\frac{d^2 x_1}{2\pi a^{2-\Delta}}\frac{d^2 x_2}{2\pi a^{2-\Delta}}\langle\mathcal{O}(x_2)\rangle \\
&= -\frac{1}{2}\delta\lambda\, \hat{g}^2\mathbf{C}\int\frac{d^2 x}{2\pi a^{2-\Delta}}\langle\mathcal{O}(x)\rangle~.
\end{aligned}
\ee
Thus, the total infinitesimal scaling effect on the coupling constant is
\be
\hat{g}\to \hat{g} + (2-\Delta)\hat{g}\delta\lambda - \frac{1}{2}\mathbf{C}\hat{g}^2\delta\lambda + \mathcal{O}(\hat{g}^3)~.
\ee
Therefore, we derive the one-loop contribution to the $\beta$-function as
\be\label{1-loop-correction}
\frac{d\hat{g}}{d\lambda} \equiv \beta(\hat{g}) = (2-\Delta)\hat{g} - \frac{1}{2}\mathbf{C}\hat{g}^2 + \mathcal{O}(\hat{g}^3)~.
\ee

Now let us consider the correlation functions of the local fields $\phi_i(x)$, defined as usual by the functional integral
\[
\left\langle \phi_{1} \left(x_{1} \right) \ldots \phi_{n} \left(x_{n} \right) \right\rangle=\int \mathcal {D} \phi ~\phi_{1} \left(x_{1} \right) \ldots \phi_{n} \left(x_{n} \right) e^{- S[ \phi ]}
\]
If the variation of the field $ \phi_{i}$ under the scaling transformation is
\[
\delta \phi_{i}(x)=-\epsilon \left(x^{\mu} \partial_{\mu}+\Delta_{i} \right) \phi_{i}(x)
\]
the corresponding Ward-Takahashi identity  \eqref{conformal-ward-identity-st} is
\be\label{WT-dilaton}
\sum_{i=1}^{n} \left\langle \left(x_i^{\mu} \frac{\partial}{\partial x_i^{\mu}}+\Delta_{i} \right) \phi_{1} \ldots \phi_{n} \right\rangle-\int \frac{d^{2} x}{2\pi} \left\langle \Theta (x) \phi_{1} \left(x_{1} \right) \ldots \phi_{n} \left(x_{n} \right) \right\rangle=0~.
\ee

The fields $\phi_i(x)$ can also depend on the coupling constants and their variation, varying the coupling $g_a$, is expressed in terms of an operator $\widehat {B}_{i}^{(a)}$
\[
\widehat {B}_{i}^{(a)} \phi_{i} (x)=\frac{\partial}{\partial g_{a}} \phi_{i} (x)
\]
Using \eqref{Theta-beta}, we can write
\begin{equation}
  \sum_{a} \beta_{a} \frac{\partial}{\partial g_{a}} \langle \phi_{1}  \ldots \phi_{n} \rangle=\sum_{i=1}^{n} \sum_{a} \beta_{a} \left\langle \widehat {B}_{i}^{(a)}\phi_{1}  \ldots \phi_{n}  \right\rangle+\int \frac{d^{2} y}{2 \pi} \langle \Theta (y) \phi_{1}  \ldots \phi_{n}  \rangle
\end{equation}
Defining the operator
\begin{equation}\label{anomalous}
  \widehat {\Gamma}_{i}= \wh\Delta_{i}+ \sum_{a} \beta_{a} \widehat {B}_{i}^{(a)}
\end{equation}
we can erase $\Theta$ from the \eqref{WT-dilaton}, which is called
the  \textbf{Callan-Symanzik equation}
\begin{equation}\label{CS-eq}
   \sum_{i=1}^{n} \left\langle \left(x_{i}^{\mu} \frac{\partial}{\partial x_{i}^{\mu}}+\widehat {\Gamma}_{i} (g) \right) \phi_{1} \left(x_{1} \right) \ldots \phi_{n} \left(x_{n} \right) \right\rangle  -\sum_{a} \beta^{a} (g) \frac{\partial}{\partial g_{a}} \left\langle \phi_{1} \left(x_{1} \right) \ldots \phi_{n} \left(x_{n} \right) \right\rangle=0~.
\end{equation}
The operator $\widehat {\Gamma}_i$ in \eqref{anomalous} determines the scaling dimension of an operator $\phi_i$ at quantum level where $\wh\Delta_i$ yields the classical scaling dimension and $\sum_{a} \beta_{a} \widehat {B}_{i}^{(a)}$ measures quantum corrections. Thus, it is called the \textbf{anomalous dimension}.
Consequently, the Callan–Symanzik equation is a differential equation describing the evolution of a correlation function under variation of the energy scale.
Since the scaling dimension of $\Theta$ is always two $\widehat {\Gamma}\Theta=2\Theta$ at quantum level, we can derive
\begin{equation}\label{anomalous2}
  \widehat {\Gamma} \cO_{a} (x)=\sum_b \gamma_{a b} \cO_b (x)=\sum_b \left(2 \delta_{a b}-\frac{\partial \beta_b}{\partial g_{a}} \right) \cO_b (x)~.
\end{equation}
Indeed, using \eqref{Theta-beta}, this form yields
\begin{equation}
\begin{aligned}
\widehat{\Gamma} \Theta &=-\left(\widehat{\Delta}+\sum_{a} \beta_{a} \frac{\partial}{\partial g_{a}}\right) \sum_{b} \beta_{b} \cO_b \\
&=-\left(\sum_{b} \beta_{b} \widehat{\Gamma} \cO_b+\sum_{a,b} \beta_{a} \frac{\partial \beta_{b}}{\partial g_{a}} \cO_b\right) \\
&=-\left(\sum_{a,b} \beta_{b}\left(2 \delta_{ba}-\frac{\partial \beta_{a}}{\partial g_{b}}\right) \cO_a+\sum_{a,b} \beta_{a} \frac{\partial \beta_{b}}{\partial g_{a}} \cO_b\right) \\
&=2 \Theta
\end{aligned}
\end{equation}
Also, in the first-order approximation \eqref{1st-approx}, it is easy to see
\[
\widehat {\Gamma} \cO_{a}=\Delta_a \cO_{a}~,
\]
as expected.

\subsection{Zamolodchikov \texorpdfstring{$c$}{c}-theorem} \label{sec:Zc}
In renormalization group flows in 2d, there is an important theorem due to A.B. Zamolodchikov, called the $c$-\textbf{theorem} \cite[ISZ88-No.45]{zamolodchikov1986irreversibility}.
The $c$-Theorem asserts that, under the assumptions of rotation
invariance, unitarity and energy-momentum conservation in a
two-dimensional theory, there exists a function $C$ which is non-increasing along a renormalization group flow. Moreover, at an IR fixed point, it is equal to
the central charge $c$ of the IR CFT. This theorem tells us the global structure of the space of physical theories and the role of conformal field theories.

We begin with energy-momentum conservation,
\begin{equation}
\partial_{\mu} T^{\mu \nu}=0\,,
\end{equation}
where $T^{\mu\nu}$ is symmetric. This can be written in complex coordinates $z$, $\overline{z}$
\begin{equation}
\label{conservation-law-Theta}
\overline\partial T+\frac{1}{4} \partial \Theta=0 \quad,\quad \partial \overline {T}+\frac{1}{4} \overline\partial \Theta=0\, ,
\end{equation}
where $\Theta :=-2\pi{T^{\mu}}_\mu$. (See \eqref{T-barT} and \eqref{holomorphic-EM}.) At an IR fixed point, $\Theta$ vanishes and $T=T(z)$ is holomorphic. However, the energy-momentum tensor is dependent of both holomorphic and anti-holomorphic coordinate $T=T(z,\overline{z})$, and $\Theta(z,\overline{z})\neq0$ for a non-conformal theory.  Since $T$, $\Theta$ and $\overline{T}$
be the components of spin $(2,0)$, $(1,1)$ and $(0,2)$, respectively, of the
energy-momentum tensor, the rotational invariance, spins and dimensions imply for the
correlation functions of $T$ and $\Theta$
\bea\label{T-Theta}
\langle T (z , \overline {z}) T (0,0) \rangle &= \frac{F (z \overline {z})}{z^{4}}\, ,\\
\langle \Theta (z , \overline {z}) T (0,0) \rangle=\langle T (z , \overline {z}) \Theta (0,0) \rangle &= \frac{G (z \overline {z})}{z^{3} \overline {z}}\,,\\
\langle \Theta (z , \overline {z}) \Theta (0,0) \rangle &= \frac{H (z \overline {z})}{z^{2} \overline {z}^{2}}\, .\,
\eea
and similarly for $\overline{T}$. Combining the
first equation of \eqref{conservation-law-Theta} with $T$, we have
\begin{equation}
\left\langle \overline\partial T (z , \overline {z}) T (0,0) \right\rangle+\frac{1}{4} \left\langle \partial \Theta (z , \overline {z}) T (0,0) \right\rangle=0\, ,
\end{equation}
We adopt the notation
$\dot {F} :=z \overline {z} F^{\prime} (z \overline {z})$ where the prime stands for the derivative of $F(x)$ with respect to its argument. One finds
\begin{equation}
\dot {F}+\frac{1}{4} \dot {G}-\frac{3}{4} G=0\, ,
\end{equation}
and similarly by combining with $\Theta$
\begin{equation}
\dot {G}-G+\frac{1}{4} \dot {H}-\frac{1}{2} H=0\, .
\end{equation}
Eliminating $G$ from above two equations,  we define
\begin{equation}\label{c-fn}
C :=2 F-G-\frac{3}{8} H\,.
\end{equation}
Then, we find that it is non-increasing with respect to the scale
\begin{equation}
\label{C-theorem}
\dot {C}=- \frac{3}{4} H \leq 0\, .
\end{equation}
due to unitarity. We see $C$ is a non-increasing function of
$R:=(z\overline{z})^{\frac12}$, for fixed values of couplings $\{g\}$.
The $R$-dependence is related to the dependence on the
$\{g\}$ via the Callan-Symanzik equation \eqref{CS-eq}
\begin{equation}
\label{callan-Symanzik equation}
\left(R \frac{\partial}{\partial R}-\sum_a \beta_a  \left(g\right) \frac{\partial}{\partial g_a} \right) C (R , \{g \})=0\, .
\end{equation}
Using \eqref{Theta-beta} and the Callan-Symanzik equation \eqref{CS-eq} on $\langle \Theta\Theta\rangle$, we obtain
\begin{equation}
\label{c{g}}
\beta^{a} \frac{\partial}{\partial g^{a}} C (g)=- \frac{3}{2} G_{a b} (g) \beta^{a} (g) \beta^{b} (g)\, ,
\end{equation}
where
\begin{equation}\label{Z-metric}
G_{a b} (g)=G_{a b} (1 , g) , \quad G_{a b} (z \overline {z} , g)=( z \overline {z})^{2} \left\langle \cO_{a} (z , \overline {z}) \cO_b (0,0) \right\rangle\, ,
\end{equation}
is positive-definite, since the theory is unitary. In other words, $G_{ab}(g)$ can be regarded as a metric of the space of the coupling constants. Therefore, it is called the \textbf{Zamolodchikov metric}. Moreover, we can write the $\beta$-function as
\be\label{grad}
\beta_{a} (g)=- \frac{2}{3}\sum_{b}G_{ab}^{- 1}\frac{\partial C}{\partial g_{b}}~,
\ee
so that $C$ can be understood as a Morse function on the space of the coupling constants and $\beta_a$ can be understood as the gradient vector fields of the Morse function \cite{milnor1963morse}.

It follows from \eqref{c{g}}  that the function $C(g)$ decreases under  the renormalization
group flow. Since the $\beta$-function vanishes at a fixed point, $C$ is stationary with respect to $R$ at the fixed point.
At the fixed point, the theory is conformal so that \eqref{T-Theta} becomes
\begin{equation}
  F=\frac{c}2~, \qquad G=0=H~.
\end{equation}
Consequently, it is equal to the central charge $C(g^*)=c$ at the fixed point $g=g^*$. This proves the statement of the
$c$-\textbf{theorem}. Consequently,  the central charge of $S$ is less than or equal to that of $S^*$ in \eqref{pertubation}
\[
c_{\textrm{UV}}\ge c_{\textrm{IR}}~.
\]
Hence, one can think that the central charge $c$ encodes degrees of freedom of the theory, and it decreases under a renormalization group flow since some massive particles are integrated out.
There is the corresponding theorem in four dimensions \cite{Komargodski:2011vj}. In six dimensions, the corresponding theorem is proven for supersymmetric theories \cite{Cordova:2015fha}.

\subsection{Landau-Ginzburg effective theory and minimal models}

\subsubsection*{Landau-Ginzburg effective theory}

Zamolodchikov has proposed \cite[ISZ88-No.17]{zamolodchikov1986conformal} that there exists a simple effective Lagrangian description for a special class of minimal models by the following action
\[\cS=\int \mathrm{d}^{2}z \left(\frac{1}{2}(\partial \Sigma)^{2}+V(\Sigma)\right)\]
where the potential
\[V (\Sigma)=g_{1}\Sigma+g_{2}\normord{\Sigma^{2}}+\cdots+g_{2 (p- 2)}\normord{\Sigma^{2 (p- 2)}}+g \normord{\Sigma^{2 (p- 1)}}\]
Note that the terms $\normord{\Sigma^{2 p-3}}$ can always be removed by a shift of the field $\Sigma\to\Sigma +$ const and absorbed in the linear term $g_1\Sigma$. Here we assume $g>0$ so that there are $(p-1)$ local minima, which corresponds to different phases of the model. This model is called \textbf{Landau-Ginzburg effective theory}.
 \begin{table}[htbp]\centering
 \begin{tabular}{c|ccccc}
		$6$
		&
		&
		&
		&
		&	\\ [5pt]
		$5$
		&
		&
		&
		&
		& $\phi_{5,5}$
		\\ [5pt]
		$4$
			&
		&
		&
		& $\phi_{4,4}$
		&$\phi_{5,4}$
		\\[5pt]
		$3$
		&
		&
		& $\phi_{3,3}$
		&$\phi_{4,3}$
     &
		\\[5pt]
			$2$
		&
			& $\phi_{2,2}$
		&$\phi_{3,2}$
		&
     &
		\\    [5pt]
				$1$
			& $1$
		&$\phi_{2,1}$
			&
		&
     &
		\\    [5pt]
		\midrule
		$0$
		& $1$
		& $2$
		& $3$	& $4$	& $5$
			\end{tabular}
\qquad
\begin{tabular}{c|ccccc}
		$6$
		&
		&
		&
		&
		&	\\ [5pt]
		$5$
		&
		&
		&
		&
		& $\Sigma^4$
		\\ [5pt]
		$4$
			&
		&
		&
		& $\Sigma^3$
		&$\Sigma^8$
		\\[5pt]
		$3$
		&
		&
		& $\Sigma^2$
		&$\Sigma^7$
     &
		\\[5pt]
			$2$
		&
			& $\Sigma$
		&$\Sigma^6$
		&
     &
		\\    [5pt]
				$1$
			& $1$
		&$\Sigma^5$
			&
		&
     &
		\\    [5pt]
		\midrule
		$0$
		& $1$
		& $2$
		& $3$	& $4$	& $5$
	\end{tabular}
	\caption{Relevant operators of conformal dimension $h < 1$ in the minimal model $\cM_6$ and the corresponding operators in the Landau-Ginzburg effective theory.}
	\end{table}

The theory consists of $2(p -2)$ scalar relevant fields, associated to the operators $\normord{\Sigma^k}$ for $1\le k\le 2(p-1)$. On the other hand, the unitary minimal model  $\cM_p$ has $2(p -2)$  primary fields  with conformal dimension $h < 1$. Hence, we identify the most relevant field $\phi_{2,2}$ of $\cM_p$ (with the lowest conformal dimension) with the scalar field $\Sigma$ in the Landau-Ginzburg theory. The OPE of $\Sigma$ is
\[
\Sigma (z) \Sigma (0)=\frac{1}{| z |^{2 h_{2,2}}}+\frac{1}{| z |^{2  h_{2,2}- h^{\prime}}}\normord{\Sigma^{2}} (0)+\cdots
\]
where $h'$ is the conformal dimension of $\normord{\Sigma^{2}}$. On the other hand,
recalling the fusion rule of primary fields \eqref{MM-fusion}
\[
\phi_{2,2}\times \phi_{r, s}= \left[ \phi_{r-1 , s-1}\right]+\left[ \phi_{r+1 , s+1}\right]+\left[ \phi_{r+1 , s-1}\right]+\left[ \phi_{r-1 , s+1}\right]~,
\]
so that
\[
\phi_{2,2}\times \phi_{2,2}= [ \mathbf{1}]+\left[ \phi_{3,3}\right]+\left[ \phi_{1,3}\right]+\left[ \phi_{3,1}\right]~.
\]
Apart from the identity operator, the conformal dimension of $\phi_{3,3}$ is lower than the other two  so that it is natural to identify
\[
\phi_{3,3}=\normord{\Sigma^2}~.
\]
The same procedure leads to
\be\label{op-corresp}\begin{array}{llll}\phi_{3,3}=\normord{\Sigma^{2}} ~,& \phi_{4,4}= \normord{\Sigma^{3}} ~,& \cdots &, \phi_{p- 1 , p- 1}= \normord{\Sigma^{p- 2}}  \\  \phi_{2,1}= \normord{\Sigma^{p- 1}} ~, & \phi_{3,2}=\normord{\Sigma^{p}} ~,&\cdots& , \phi_{p- 1 , p- 2}= \normord{\Sigma^{2 p- 4}}  \end{array}\ee
Moreover, $\normord{\Sigma^{2p-3}}$ can be obtained from the OPE of $\phi_{p-1,p-2}$ and $\phi_{2,2}$. The inversion formula \eqref{inversion-formula} identifies $\phi_{1,3}=\phi_{p-1,p-2}$ so that
\[
\phi_{2,2}\times \phi_{1,3}= \left[ \phi_{2,2}\right]+\left[ \phi_{p-2 , p-3}\right]
\]
Therefore, we can identify $\normord{\Sigma^{2p-3}}$ with the descendant $L_{-1}\overline L_{-1}\phi_{2,2}$ of $\phi_{2,2}$
\[
\normord{\Sigma^{2 p-3}}=\partial \overline{\partial} \Sigma
\]
 This is the equation of motion at the multicritical point ($g_i=0$) in Landau-Ginzburg theory.

 If $g_i=0$, the theory is invariant under $\bZ_2$ symmetry $\Sigma\to-\Sigma$ \cite[ISZ88-No.30]{Zuber:1986ng}. When $p=3$, it is the $\Sigma^4$-theory that describes the Ising model. When  $p=4$, it is the $\Sigma^6$-theory that describes the tricritical Ising model. For a generic positive integer $p$, it describes the RSOS model \cite[ISZ88-No.31]{Andrews:1984af}.

The correspondence between the minimal models and the Landau-Ginzburg models can also be seen even with $\cN=2$ superconformal symmetry.  This is related to the ADE classification studied in \S\ref{sec:su2k-character}, which admits a profound interpretation in geometry.

\subsubsection*{Minimal Models and Renormalization Group Flows}

\begin{figure}
	\centering
	\includegraphics[width=0.5\linewidth]{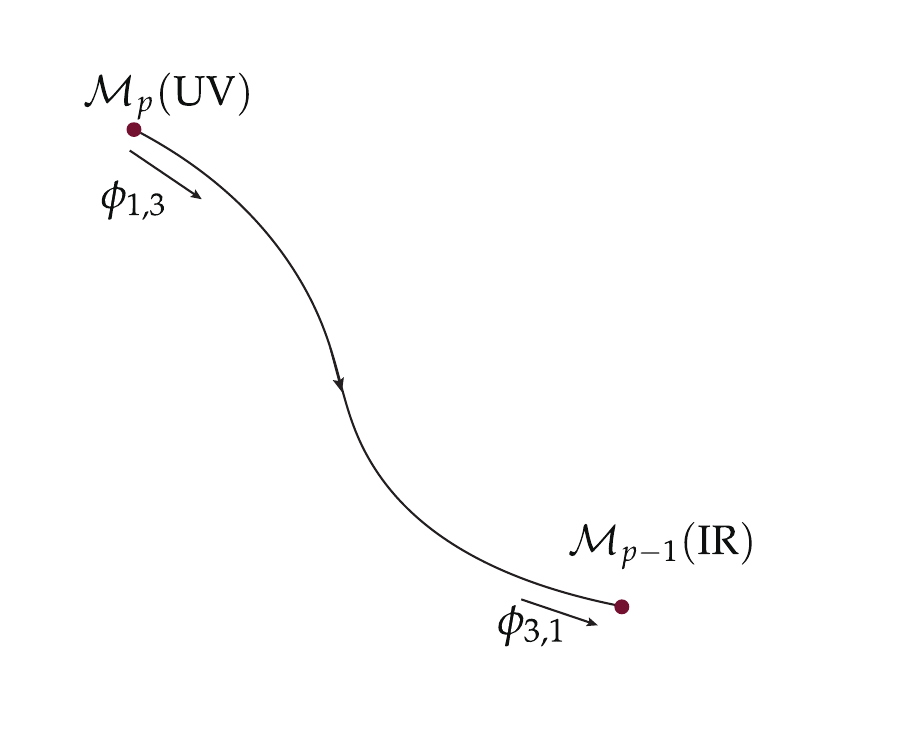}
	\caption{RG-flow between minimal models}
	\label{fig:rg-minimal-model}
\end{figure}

In fact, the minimal models provide a perfect platform to understand renormalization group flows \cite{Zamolodchikov:1987ti}. Let us consider the perturbation of the unitary minimal model $\cM_p$ by the primary field $\phi_{1,3}$
\[\cS=\cS_ {p} + g \int \frac{d^{2} x}{2\pi} \phi_{1,3} ( x )~\]
where we assume $p\gg1$. Then, the primary field $\phi_{1,3}$ can be considered as a quasi-marginal operator since the scaling dimension is
\[\Delta_{1,3}  = 2 h_{1,3} = 2 - \frac{4}{p + 1} \equiv 2 - \epsilon\]
with $\e\ll1$.
Following the fusion rule  \eqref{MM-fusion}, one can read off
\[\phi_{1,3} \times \phi_{1,3} = 1 + \mathbf {C}_{1} \phi_{1,3} + \mathbf {C}_{2} \phi_{1,5}~.\]
Since $ \phi_{1,5}$ is an irrelevant operator, the operator expansion above implies the renormalization of the field  $\phi_{1,3}$, which does not mix with any other fields. In fact, the structure constant is computed in \cite[ISZ88-No.15]{Dotsenko:1984nm}
\[
\begin{aligned} \mathbf {C}_{1} & = \frac{4}{\sqrt {3}} \frac{( 1 - \epsilon )^{2}}{( 1 - \epsilon / 2 ) ( 1 - 3 \epsilon / 4 )} \left[ \frac{\Gamma ( 1 - \epsilon / 4 )}{\Gamma ( 1 + \epsilon / 4 )} \right]^{\frac{3}{2}} \left[ \frac{\Gamma ( 1 + 3 \epsilon / 4 )}{\Gamma ( 1 - 3 \epsilon / 4 )} \right]^{\frac{1}{2}}  \left[ \frac{\Gamma ( 1 + \epsilon / 2 )}{\Gamma ( 1 - \epsilon / 2 )} \right]^{2} \frac{\Gamma ( 1 - \epsilon )}{\Gamma ( 1 + \epsilon )} \\ & = \frac{4}{\sqrt {3}} \left( 1 - \frac{3 \epsilon}{4} + O \left( \epsilon^{2} \right) \right) \end{aligned}
\]
Consequently, \eqref{1-loop-correction} tells us that the $\b$-function can be written as
\[
\frac{d g }{d \lambda} \equiv \beta  ( g ) = \epsilon g - \frac{ \mathbf {C}_{1}}{2}g^{2} +\cO(g^3)~.
\]
Using the gradient flow equation \eqref{grad}, one obtains
\[
C(g)=c_p+\a \left[\frac\epsilon2 g^{2} -\frac{ \mathbf {C}_{1}}{6}g^{3}\right] +\cO(g^4)
\]
where $c_p$ is the central charge \eqref{c<1-unitary-c} of $\cM_p$. Note that $\a$ depends on the Zamolodchikov metric $G_{ab}$ in \eqref{Z-metric}, and $\a$ can be evaluated as follows. The difference between the central charges at UV and IR can be computed by \eqref{C-theorem}
\[
\Delta c = -\frac{3}{4} \int_{0}^{\infty} d ( r^{2} ) r^{2} \langle \Theta ( r ) \Theta ( 0 ) \rangle
\]
Furthermore, \eqref{Theta-beta} tells us that
\bea
\Delta c &= -\frac{3}{4}(\e g)^2 \int_{0}^{\infty} d ( r^{2} ) r^{2} \langle \phi_{1,3} ( r ) \phi_{1,3} ( 0 ) \rangle\cr
&= -\frac{3}{2}(\e g)^2\int_{0}^{\infty} \frac{r^{3}dr}{r^{2 \Delta_{1,3}}}\cr
&=-\frac{3}{4} \e g^2 +\cO({\e^0})~.
\eea
Therefore, we have $\a=-\frac{3}{2}$. At the fixed point $g^{*} =2 \epsilon /\mathbf {C}_{1}$, the Zamolodchikov $C$-function is
\bea
C(g^*)&=c_p- \frac{ \epsilon^{3}}{ \mathbf {C}_{1}^{2}}+\mathcal {O} \left( \epsilon^{4} \right)\cr
& = c_{p} - \frac{3}{4^{2}} \cdot \left( \frac{4}{p} \right)^{3} + \cO \left( p^{- 4} \right) \\ & = c_{p} - \frac{12}{p^{3}} + \cO \left( p^{- 4} \right) \sim c_{p - 1} \eea
At this perturbative order, the new value of the central charge coincides with that of the unitary minimal model $\cM_{p-1}$. Therefore, the perturbation by $\phi_{1,3}$ leads to the renormalization group flow from $\cM_p$ to $\cM_{p-1}$.
Although we have assumed $p\gg1$ in this analysis, the evidence of RG flow from $\cM_p$ to $\cM_{p-1}$ for an arbitrary $p$ has been shown by using factorized $S$-matrices \cite{Zamolodchikov:1991vh}. In fact, from \eqref{anomalous2}, one can deduce the conformal dimension of the operator $\phi_{1,3}$ at the fixed point
\[
2 - \frac { \partial \beta  } { \partial g  }\Big|_{g=g^*}=2+\e+\cO(\e^2)~,
\]
which can be read off as the conformal dimension of $\phi_{3,1}$ in $\cM_{p-1}$ at the fixed point. Thus, the relevant operator at UV becomes the irrelevant operator at IR. (See Figure \ref{fig:rg-minimal-model}.)

In fact, this is consistent with the Landau-Ginzburg effective theory. As we have seen in \eqref{op-corresp}, $\phi_{1,3} = \phi_{p - 1 , p - 2} \sim \normord{\Sigma^{2p-4}}$. Hence, the  Landau-Ginzburg action for  $\cM_{p}$ with the perturbation is
\[\cS=\int \mathrm{d}^{2}z \left(\frac{1}{2}(\partial \Sigma)^{2}+g_{2p-4}\Sigma^{2p-4}+g_{2p-2}\Sigma^{2p-2}\right)~.\]
Then, the operator $\normord{\Sigma^{2p-2}}$ becomes irrelevant and the effective action is
\[\cS=\int \mathrm{d}^{2}z \left(\frac{1}{2}(\partial \Sigma)^{2}+g_{2p-4}\Sigma^{2p-4}\right)~,\]
which is the effective action for $\cM_{p-1}$.

\section{Wess-Zumino-Novikov-Witten Models}
This section is devoted to studying a very important class of exactly solvable 2d conformal field theories called \textbf{Wess-Zumino-Novikov-Witten (WZNW) models} \cite{Wess:1971yu,novikov1982hamiltonian,Witten:1983ar}. The WZNW models have many applications related to condensed matter physics, AdS/CFT correspondence, knot theory, and mathematical physics. We also refer the reader to \cite[\S15,\S16]{francesco2012conformal}, \cite[\S9]{ginsparg1988applied} and \cite[\S3]{Blumenhagen:2009zz}.

\subsection{U(1) symmetry of free boson}
Let us recall the free boson theory in \S\ref{sec:free-boson} whose action is given by
\be\label{free-boson-action}
\cS = \frac{1}{8 \pi} \int d^{2} x \,\partial_{\mu} \varphi \partial^{\mu} \varphi= \frac{1}{8 \pi} \int d^{2} z\, \partial \varphi \overline\partial \varphi
\ee
This action has a symmetry \be\label{U1-symmetry}\varphi\to \varphi+a\ee whose Noether current is $j^{\mu}=- \partial^{\mu} \varphi / 4 \pi$.
The current conservation follows from the equation of motion \eqref{eom-boson}
\[
\partial_\mu j^\mu=0~.
\]
As in \S\ref{sec:free-boson}, one can decompose the current into holomorphic and anti-holomorphic part
\be\label{U1-current}
J (z) \equiv i \partial_z \varphi~,\qquad  \overline J(\overline z)\equiv -i \overline\partial \varphi
\ee
so that the (holomorphic part of) energy-momentum tensor can be written as
\[
T (z)=- \frac{1}{2} : \partial \varphi \partial \varphi : =\frac12:J(z)J(z):~.
\]
As we will see below, this is called \textbf{Sugawara construction} of the energy-momentum tensor. In addition, the action \eqref{free-boson-action}
can be written as
\begin{equation}
 \cS_0=\frac{1}{8\pi} \int_\Sigma \dd^2 x \,
\pd^\mu g^{-1} \pd_\mu g   =\frac{1}{8\pi} \int_\Sigma \dd^2 x \,
\partial g^{-1} \overline\partial g\, ,
\end{equation}
where $g(z,\overline z)$ is the map from \[g:\mathbb{C}\to \U(1);\, z\mapsto g(z,\overline{z})=\exp(i\varphi(z,\overline z))~.\] This action can be regarded as the simplest version of WZNW models.
In fact, the symmetry can be understood as the $\U(1)$ rotation
\[
e^{i\varphi(z,\overline z)} \mapsto e^{i(\varphi(z,\overline z)+a)}~,
\]
so that \eqref{U1-current} is called the $\U(1)$ current.
The $\U(1)$ group is an abelian group and it is natural to ask if we can generalize this simple construction to non-abelian Lie groups. This section answers this question.

\subsection{SU(2) current algebra}\label{sec:SU(2)k}
Before moving to the WZNW models, we shall provide the easiest introduction to the non-abelian generalization. In \eqref{sec:compactified-boson}, we have seen that the compactified boson enjoys the $T$-duality $R\leftrightarrow 2/R$ \eqref{T-duality} and $R=\sqrt{2}$ is the self-dual radius. At the self-radius, one can define the vertex operators
\[
J (z)=\frac{i\partial \varphi(z)}{\sqrt{2}}~,\qquad J^{\pm}(z)=: e^{\pm i \sqrt{2} \varphi(z)} :
\]
where they satisfy the OPE
\be
\begin{aligned} J (z) J^{\pm} (w) & \sim \frac{\pm J^{\pm} (w)}{z-w}  \\ J^{+} (z) J^{-} (w) & \sim \frac{1}{(z-w)^{2}}+\frac{2 J (w)}{z-w}  . \\ J (z) J (w) & \sim \frac{1 / 2}{(z-w)^{2}}  \end{aligned}
\ee
Furthermore, if we write
\be\label{redefinition}
J^{\pm} \equiv J^{1} \pm i J^{2} , \quad J   \equiv J^{3}~,
\ee
then the OPE amounts to
\be \label{level1}
J^{a} (z) J^{b} (w) \sim \frac{\frac{1}{2} \delta_{a b}}{(z-w)^{2}}+\frac{i \epsilon^{a b c} J^{c} (w)}{z-w}
\ee 
where $\epsilon^{abc}$ is the anti-symmetric tensor. It is easy to see the connection to the theory of the angular momentum which is associated to the $\fraksu(2)$ Lie algebra
\be\label{su2-algebra}[t^a,t^b]=i \epsilon^{a b c} t^{c}\ee
where $t^a=\sigma^a/2$ with the Pauli matrices $\sigma^a$ ($a=1,2,3$). In fact, the OPE \eqref{level1} can be generalized to
\be \label{levelk}
J^{a} (z) J^{b} (w) \sim \frac{\frac{k}{2} \delta_{a b}}{(z-w)^{2}}+\frac{i \epsilon^{a b c} J^{c} (w)}{z-w}~.
\ee
We note that
\[
J(z)=\sum_{a=1}^3J^a(z)t^a
\]
is called the $\SU(2)$ current and $k\in \bZ$ is its \textbf{level}. As usual, the mode expansion
\[
J^{a} (z)=\sum_{n \in \mathbb{Z}} z^{- n-1} J_{n}^{a}
\]
leads to the following commutation relations, which is called  is called the $\SU(2)$ \textbf{current algebra},
\[
\left[ J_{m}^{a} , J_{n}^{b} \right]=i \e^{ab c} J_{m+n}^{c}+\frac{k}{2} m \delta^{ab} \delta_{m+n , 0}
\]
where the zero modes satisfy the $\fraksu(2)$ Lie algebra \eqref{su2-algebra}. In mathematics, it is called the \textbf{$\widehat{\fraksu(2)}_k$ affine Lie algebra} with level $k$. This can be further generalized to the so-called \textbf{Kac-Moody algebra}.

Let us briefly study the highest weight representation of the $\SU(2)$ current algebra. From \eqref{redefinition}, we introduce
\[
J_{n}^{\pm} \equiv J_{n}^{1} \pm i J_{n}^{2}
\]
where the Hermitian conjugates are defined as
\[
\left(J_{n}^{\pm} \right)^{\dagger}=J_{- n}^{\mp} , \quad \left(J_{n}^{3} \right)^{\dagger}=J_{- n}^{3}~.
\]
Then, the highest weight state $|h,j\rangle$ is defined as
\be\label{HWS-su2}
 J_{0}^3  |h,j \rangle =j  |h,j \rangle~,\quad  J_{0}^{+}  |h,j \rangle=0 ,\qquad  J_{n}^{a}  |h,j \rangle=0 , \quad n > 0 , \quad a=0 , \pm ~.
\ee
Note that $h$ is the conformal dimension
\[
L_0 |h,j \rangle=
\frac{j (j+1)}{k+2 }|h,j \rangle
\]
which is left as an exercise.
Since the zero modes satisfy the $\fraksu(2)$ Lie algebra, $j$ takes an half-integer value $j=\frac12,1,\frac32,2,\cdots$ associated to the spin-$j$ representation.
In addition, it is straightforward to check that the following generators
\[
J_{(1)}^{+}=J_{- 1}^{+} , \quad J_{(1)}^{-}=J_{+ 1}^{-} , \quad J_{(1)}^3=J_{0}^3-\frac{k}{2}
\]
also form the $\fraksu(2)$ Lie algebra, and we call it $\fraksu(2)_{(1)}$.  Since the eigenvalue of $2 J_{(1)}^3=2 J_{0}^3-k$ is an integer, $k$ has to be an integer. Moreover, the unitarity imposes
\[
\begin{aligned} 0 \leqq \left| J_{(1)}^{+}  |h,j \rangle \right|^{2} &=\langle h,j \left| J_{+ 1}^{-} J_{- 1}^{+} \right |h,j \rangle \\ &=\left\langle h,j \left| \left[ J_{+ 1}^{-} , J_{- 1}^{+} \right] \right| h,j \right\rangle \\ &=- 2 \langle j | (J_{0}^3-k / 2)  |h,j \rangle \\ &=- 2 j+k \end{aligned}
\]
Therefore, for a given level $k\in \bZ_{>0}$, the spin $j$ is allowed to take the values
\[
j=0 , \frac{1}{2} , 1 , \frac{3}{2} , \cdots , \frac{k}{2}~.
\]

\subsection{A crash course on Lie groups and Lie algebras}

It is well-known that the theory of angular momentum in quantum mechanics is described by the representation theory of the $\fraksu(2)$ Lie algebra \eqref{su2-algebra}. We want to generalize the current algebra to arbitrary Lie algebras. Thus, we shall briefly review Lie groups and Lie algebras. For more details, we refer the reader to \cite{kirillov2008introduction}.

\begin{defn}[Lie group]\index{Lie group}
  A \textbf{Lie group} is a manifold $G$ with a group structure such that multiplication $m: G \times G \to G$ and inverse $i: G \to G$ are smooth maps. The \textbf{dimension} of a Lie group $G$ is the dimension of the underlying manifold.
\end{defn}

Some Lie groups will be given by subsets of the space $M_n(\bF)$ of  $n\times n$ matrices where $\bF=\bR$ or $\bC$ specified by certain algebraic equations. For example,
\begin{itemize}
\item  General linear group: $\GL(n, \bF)=\{A \in \mathrm{M}_n(\bF)| \det A \neq 0\}$
\item  Special linear group: $\SL(n, \bF)=\{A \in \GL(n, \bF)| \det A=1\}$
\item Symplectic group $\mathrm { Sp } ( n , \bF )=\{ A \in \GL(2n, \bF)| A^ { \mathrm { T } } J A = J \ \textrm{where} \  J=\begin{pmatrix} 0&I_n\\-I_n&0\end{pmatrix}\}$
\item Unitary group $\U(n)=\{A \in \GL(n, \bC)| AA^\dagger=I\}$
\item Special unitary group $\SU(n) =\{A \in \U(n)| \det A=1\}$
\item  Orthogonal group $\mathrm{O}(n)= \{A \in \GL(n, \bR)| AA^T=I\}$
\item Special orthogonal group $\SO(n)=\{A \in \mathrm{O}(n)| \det A=1\}$
\end{itemize}
In fact, the simple Lie groups are classified by Killing and Cartan, and $\SU(n+1)$, $\SO(2n + 1)$, $\Sp(n,\bR)$ and $\SO(2n)$ are assigned to type $A_n$, $B_n$, $C_n$ and $D_n$, respectively. In addition to these four types, there are five exceptional Lie groups called $E_6$, $E_7$, $E_8$, $F_4$, and $G_2$. The Dynkin diagrams of type $A$, $D$ and $E$ are all simply laced, but the other types are not.

Roughly speaking, an element $X\in \frakg$ of a Lie algebra $\frakg$ describes an element $h\in G$ near the identity element $1\in G$ via
\[
h=1+\e X+\cO(\e^2)~.
\]
More precisely, the tangent space ${T}_1(G)$ of a Lie group $G$ at the identity element $1$ naturally admits a Lie bracket
  \[
    [\cdot, \cdot]: T_1 G \times T_1 G \to T_1 G; (X,Y) \mapsto [X,Y]=XY-YX
  \]
  such that the Lie algebra of
  \[
    \mathfrak{g}=(T_1(G), [\cdot, \cdot])
  \]
  is a Lie algebra.

\begin{defn}[Lie algebra of a Lie group]\index{Lie algebra of a Lie group}
  Let $G$ be a Lie group. The \textbf{Lie algebra} of $G$, written $\mathfrak{g}$, is the tangent space $T_1 G$ under the natural Lie bracket.
\end{defn}

Given a vector $X\in \mathfrak{g} $ in the tangent space of the identity element $1$, the exponential map defines a map $\exp:\mathfrak{g}  \to G$ such that
  \[
    \exp(tX)=\sum_{\ell=0}^\infty \frac{1}{\ell!} (tX)^\ell~.
  \]
Without relying on a Lie group, one can define a Lie algebra as follows:
\begin{defn}[Lie algebra]\index{Lie algebra}
  A \textbf{Lie algebra} $\mathfrak{g}$ is a vector space (over $\bF=\bR$ or $\bC$) with a \textbf{bracket}
  \[
    [\cdot,\cdot] : \mathfrak{g} \times \mathfrak{g} \to \mathfrak{g}
  \]
  satisfying
  \begin{enumerate}
      \item $[\alpha X+\beta Y, Z]=\alpha [X, Z]+\beta [Y, Z]$ for all $X, Y, Z \in \mathfrak{g}$ and $\alpha, \beta \in \bF$ \hfill(bilinearity)
    \item $[X, Y]=-[Y, X]$ for all $X, Y \in \mathfrak{g}$ \hfill(antisymmetry)
    \item $[X, [Y, Z]]+[Y, [Z, X]]+[Z, [X, Y]]=0$ for all $X, Y, Z \in \mathfrak{g}$.\hfill(Jacobi identity\index{Jacobi identity})
  \end{enumerate}
  Note that linearity in the second argument follows from linearity in the first argument and antisymmetry.
\end{defn}

For instance, we have the corresponding Lie algebras
\begin{itemize}
\item $\mathfrak{gl}(n, \bF)=\{A \in \mathrm{M}_n(\bF)\}$
\item $\mathfrak{sl}(n, \bF)=\{A \in \mathrm{M}_n(\bF) |  \Tr A=0\}$
\item  $\mathfrak{ sp } ( n , \mathbb { F } ) = \{ X \in \mathfrak{gl}(2n, \bF)) | X ^ {T} J + J X = 0  \ \textrm{where} \  J=\begin{pmatrix} 0&I_n\\-I_n&0\end{pmatrix} \}$
\item $\mathfrak{su}(n)=\{A \in \mathrm{M}_n(\mathbb{C}) | \Tr A=0, A^\dagger=-A\}$
\item $\mathfrak{so}(n)=\{A \in \mathrm{M}_n(\mathbb{R}) | A^T=-A\}$
\end{itemize}

Given a finite-dimensional Lie algebra $\frakg$, we can pick a basis for $\mathfrak{g}$:
\be\label{Lie-basis}
t^a: a=1, \cdots, \dim \mathfrak{g}~
\ee
with
\[
\Tr(t^at^b)=\frac12 \delta^{ab}~,
\]
which is called the Cartan-Killing form.
Then any $X \in \mathfrak{g}$ can be written as
\[
  X=X^a t^a=i \sum_{a=1}^n X^a t^a,
\]
where $X^a \in \bF$.

By linearity, the bracket of elements $X, Y \in \mathfrak{g}$ can be computed via
\[
  [X, Y] =- X^a Y^b [t^a, t^b].
\]
In other words, the whole structure of the Lie algebra can be given by the bracket of basis vectors. We know that $[t^a, t^b]$ is again an element of $\mathfrak{g}$. So we can write
\[
  [t^a, t^b] =i f_{ab}{}^c t^c,
\]
where $f_{ab}{}^c\in \bF$ are called the \textbf{structure constants}.
By the antisymmetry of the bracket, we know
  \[
    f_{ba}{}^c=-f_{ab}{}^c.
  \]
The Jacobi identity amounts to
  \[
    f_{ab}{}^c f_{cd}{}^e+f_{da}{}^c f_{cb}{}^e+f_{bd}{}^c f_{ca}{}^e=0.
  \]

\subsection*{Affinization}
Hence, one can straightforwardly generalize the current associated to a Lie algebra $\frakg$ as follows. First, the current taking its value on a Lie algebra $\frakg$ can be written
\[
J(z)=\sum_{a}J^a(z)t^a~.
\]
The OPE of the current takes the form
\[
J^{a}(z) J^{b}(w) \sim \frac{\frac k2 \d^{ab}}{(z-w)^{2}}+\frac{i f^{a b }{}_c J^{c}(w)}{z-w }
\]
so that the modes are subject to
\[
\left[ J _{n }^{a}, J _{m }^{b}\right]=i f^{a b}J _{n+m }^{c}+ \frac k2 \d^{a b}n \delta _{n+m , 0 }~.
\]
This is called the $G$ current algebra with level $k$ or the $\wh \frakg_k$ affine Lie algebra with level $k$.

\subsection{WZNW models}
\subsubsection*{Nonlinear Sigma Models}
In searching for an explicit conformal field theory with the $G$ current algebra,
it is natural to first consider the \textbf{nonlinear sigma model}
\begin{equation}
\label{nonlinear-sigma-models}
 \cS_0=\frac{1}{2 a^2} \int_\Sigma \dd^2 x
 \Tr(\pd^\mu g^{-1} \pd_\mu g)\, ,
\end{equation}
where $a^2$ is a positive, dimensionless coupling constant.
The bosonic field $g(x)$ takes its value on the Lie group $G$, namely
\[
g:\Sigma\to G
\]
where $\Sigma$ is generally a two-dimensional manifold, called a Riemann surface.
Under the variation of the action
\begin{equation}
\delta \cS_0=-\frac{1}{a^2}
\int_\Sigma \dd^2 x  \Tr(g^{-1}\delta g \pd^\mu (g^{-1}\pd_\mu g))\, ,
\end{equation}
which results in the following equation of motion
\begin{equation}
\pd^\mu(g^{-1}\pd_\mu g) =0 \, .
\end{equation}
This implies the conservation of the currents
\begin{equation}
  J_\mu=g^{-1}\pd_\mu g \, .
\end{equation}
In the complex coordinate $z=x^0+i x^1$ and
$\overline{z}= x^0 -i x^1$. If we write $J_z= g^{-1}\pd  g$
and $\overline{J}_{\overline{z}}=g^{-1}\overline{\partial} g$, we find
\begin{equation}
\pd  \overline{J}_{\overline{z}}+\overline{\partial}  J_z=0\, .
\end{equation}
So that the holomorphic and anti-holomorphic currents can not
be separately conserved.
\def\tg{\widetilde{g}}

\subsubsection*{WZNW Models}
Hence, a more complicated action must be considered to
enhance the symmetry. We should add a Wess-Zumino term \cite{Wess:1971yu}
\begin{equation}
\label{Wess-Zumino-term}
 \Gamma=\frac{-i}{12\pi} \int_B
 \dd^3 y \epsilon_{\alpha\beta\gamma}
 \Tr(\tg^{-1}\pd^\alpha \tg \tg^{-1} \pd^\beta \tg \tg^{-1}
 \pd^\gamma \tg) \,.
\end{equation}
This is defined on a three-dimensional manifold $B$, whose boundary is the Riemann surface $\partial \Sigma =B$. Here $\tg:B\to G$ is an extension of the original
field $g$. Although the extension is not unique, this is well-defined.
Given another choice $(B',\tg')$, we can glue together the three-manifolds $B$ and $B'$ along $\Sigma$ to get a closed three-manifold $B\cup_\Sigma -B'$. It is easy to check that $\omega\equiv \Tr(\tg^{-1}\pd^\alpha \tg \tg^{-1} \pd^\beta \tg \tg^{-1}
 \pd^\gamma \tg)$ is a closed three-form, and moreover it is pull-back of an element $\Tr(\theta\wedge\theta \wedge\theta)$ of the third homology group $H^3(G;\bZ)$ with integer coefficients \cite[Prop 4.4.5]{pressley1986loop} via $\wt g:M\to G$ where $\theta$ is the Maurer-Cartan form of $G$.
  Hence, the integral of $\omega$ over the closed three-manifold $B\cup_\Sigma -B'$ is an integer.
 Therefore, the Euclidean functional integral, with weight $\exp(-k\Gamma)$ is perfectly well-defined if the level $k$ is
an integer.

\begin{figure}[ht]\centering
\includegraphics[width=10cm]{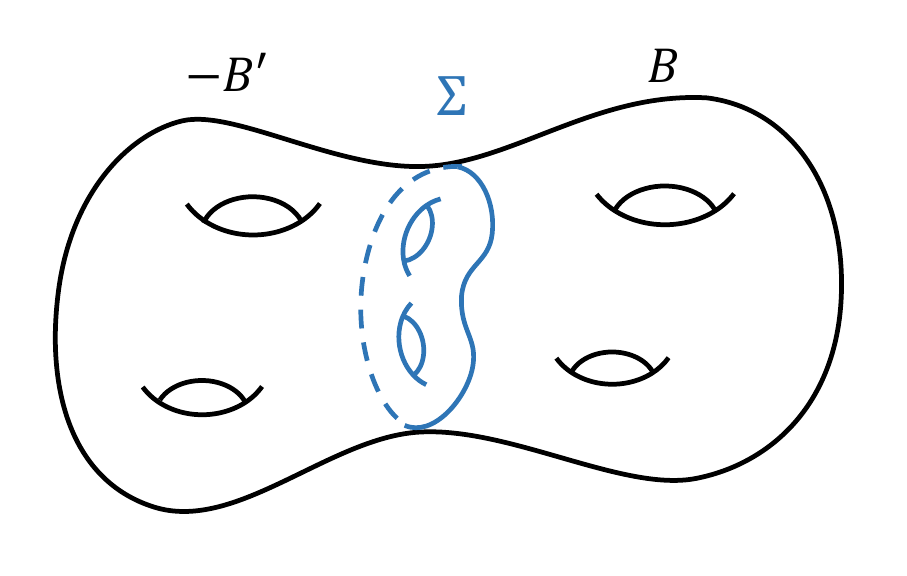}
\end{figure}

We then consider the action
\begin{equation}
\label{wess-zumino-action}
\cS=\cS_0 +k \Gamma\,,
\end{equation}
where $k$ is an integer. Since the variation of the action can be written as a total derivative,
the Wess-Zumino term is a $2$d functional.
The final result is
\begin{equation}
\delta \Gamma=-
\frac{i}{4\pi}
\int_\Sigma \dd^2 x \epsilon_{\mu\nu}
\Tr(g^{-1}\delta g \partial^\mu(g^{-1}\pd^\nu g))\, .
\end{equation}
The equation of motion of the action \eqref{wess-zumino-action}
is then
\begin{equation}
\pd^\mu(g^{-1}\pd_\mu g)+\frac{a^2 i k}{4\pi}
\epsilon_{\mu\nu}\pd^\mu (g^{-1}\pd^\nu g) =0 \, .
\end{equation}
In terms of the complex variable $z$, $\overline{z}$, the equation
of motion becomes
\begin{equation}
\bigg(
1+\frac{a^2 k}{4 \pi}
\bigg)
\pd (g^{-1}\overline{\partial} g)
+
\bigg(
1-\frac{a^2 k}{4 \pi}
\bigg)
\overline{\partial} (g^{-1}\pd  g) =0\, .
\end{equation}
Thus, for
\begin{equation}
a^2=4\pi/k\, ,
\end{equation}
we find the desired conservation law
\begin{equation} \label{current-conservation-WZNW}
\pd (g^{-1}\overline{\partial} g) =0 \, , \qquad \overline{\partial} ((\partial g)g^{-1}) =0 \, .
\end{equation}
The action now becomes
\begin{equation}
 \cS =
 \frac{k}{8\pi} \int_\Sigma \dd^2 x
 \Tr(\pd^\mu g^{-1}\pd_\mu g)+k\Gamma \, ,
\end{equation}
which is called the $\widehat{\frakg}_k$  \textbf{WZNW (WZNW) model}.
Since $a^2$ is positive, $k$ must be a positive integer.
The other solution $a^2=- 4\pi /k$ , which requires
$k<0$, implies the conservation of the dual currents.

Due to the Noether theorem, the conservation of the currents
$J_z$ and $\overline J_{\overline{z}}$ \eqref{current-conservation-WZNW} should imply the invariance of the
action. In fact, \eqref{current-conservation-WZNW} tells us that the field $g$ is factorized as
\[
g(z,\overline{z})=g_L(z)g_R(\overline{z})~.
\]
We will see, under the transformation
\begin{equation}
g(z,\overline{z}) \rightarrow
\Omega(z) g(z,\overline{z}) \overline{\Omega}^{-1}(\overline{z})\, ,
\end{equation}
where $\Omega$ and $\overline{\Omega}$ are two arbitrary matrices
valued in $G$, the action is invariant. Indeed, under the infinitesimal transformation
\begin{equation}
\Omega(z)=1+\omega(z) \, ,
\qquad
\overline{\Omega}(\overline{z})
= 1+\overline{\omega}(\overline{z})\, .
\end{equation}
$g$ transforms as follows
\begin{equation}
\label{transformation-g}
 \delta_\omega g=\omega g \, , \qquad
 \delta_{\overline{\omega}} g=- g\overline{\omega} \, .
\end{equation}
With $a^2=4\pi /k$, the variation of the action for
$g\rightarrow g+\delta_\omega g+\delta_{\overline{\omega}}g$ is
\bea
\label{variation-action}
 \delta \cS &= \frac{k}{\pi}
 \int_\Sigma \dd^2 x \Tr (g^{-1}\delta g [\pd (g^{-1}\overline{\partial} g)])
 \notag \\
 &= \frac{k}{\pi} \int_\Sigma \dd^2 x \Tr
 [\omega(z)\overline{\partial} (\pd  g g^{-1})-
 \overline{\omega}(\overline{z})\pd (g^{-1}\overline{\partial} g)]\,,
\eea
 which vanishes after an integral by parts. Thus the global
 $G \times G$ invariance of the sigma model has thus been
 extended to a local $G(z)\times G(\overline{z})$ invariance.

 By rescaling the conserved currents as
 \bea
 \label{rescaling-current}
  J(z) &\equiv -\frac k2J_z(z)=- \frac k2 \pd  g g^{-1}\, ,\\
  \overline{J}(\overline{z}) &\equiv \frac k2 \overline J_{\overline{z}}(\overline{z})
 =\frac k2 g^{-1}\overline\partial g \, .
  \eea

 We can rewrite \eqref{variation-action} in the form
 \begin{equation}
  \delta \cS=-\frac{2}{\pi}
  \int_\Sigma \dd^2 x \left\{
  \pdbz (\Tr[\omega(z)J(z)])
  +\pdz (\Tr[\obz\jbz])
  \right\}\,.
 \end{equation}
 Replace $d^2 x$ by $(-i/2)\dd z \dd \overline{z}$ and take the
 holomorphic contour to be counterclockwise and the antiholomorphic contour to be clockwise. This lead to
 \begin{equation}
 \delta_{\omega,\overline{\omega}} =
 \frac{i}{\pi}
 \oint \dd z \Tr[\omega(z)J(z)]
 -\frac{i}{\pi}
 \oint \dd \overline{z} \Tr
 [\obz\jbz]\,.
 \end{equation}
 Expanding $J$ and $\omega$ as well as their dual part in terms
 of the basis $t^{\alpha}$ of $G$
 \begin{equation}
  J=\sum_a J^a t^a\, , \qquad
  \omega=\sum_a \omega^a t^a \, ,
 \end{equation}
 The variation of the action now becomes
 \begin{equation}
  \delta_{\omega,\overline{\omega}} S =
  -\frac{1}{2 \pi i}
  \oint \dd z
  \sum_a \omega^a J^a
 +
  \frac{1}{2\pi i}
  \oint \dd \overline{z}
  \sum_a \overline{\omega}^a \overline{J}^a \,.
 \end{equation}
 Let $X$ stand for a list of fields as usual, from \eqref{variation-action-locally} and \eqref{Ward-identity-ver2}, we have
 \begin{equation}
 \delta\expval{X}=\expval{(\delta S) X}\, .
 \end{equation}
 Hence, we immediately obtain the
 corresponding Ward-identity
 \begin{equation}
 \label{WZNW-ward-identity}
 \delta_{\omega,\overline{\omega}}\expval{X}
=-\frac{1}{2\pi i}
 \oint \dd z \sum_a \omega^a \expval{J^a X}
+\frac{1}{2\pi i}
 \oint \dd \overline{z}
 \sum_a \overline{\omega}^a \expval{\overline{J}^a X}\, .
 \end{equation}
 From \eqref{rescaling-current} and \eqref{transformation-g},
 the transformation law for the current is
 \begin{equation}
 \delta_{\omega} J=
 \comm{\omega}{J}-\frac k2\partial_z \omega\, .
 \end{equation}
 Writing in terms of $J^a$ and $\omega^a$, we have
 \begin{equation}
  \delta_\omega J^a=
  \sum_{b,c} i f_{abc} \omega^b J^c- \frac k2 \partial_z \omega^a \, .
 \end{equation}
 The substitution of this transformation into \eqref{WZNW-ward-identity}, leads to so-called
 current algebra.
 \begin{equation}
 J^a (z) J^b(w) \sim
 \frac{\frac k2\delta_{ab}}{(z-w)^2}+ \sum_c i f_{abc}
 \frac{J^c(w)}{(z-w)} \, .
 \end{equation}

 Introducing the modes $J^a_n$ from the Laurent expansion
 \begin{equation}
 \label{mode-expansion-Ja}
   J^a (z)=\sum_{n\in \mathbb{Z}}
   z^{-n-1}J^a_n\,.
 \end{equation}
 We can easily check (as we have done in the Virasoro case),
 the commutation relations of $J^a_m$ become
 \begin{equation}
  \comm{J^a_n}{J^b_m}=\sum_c i f_{abc}J^c_{n+m}
 +\frac k2n\delta_{ab} \delta_{n+m,0}\,,
 \end{equation}
 which is the $\widehat{\frakg}$ affine Lie algebra at level $k$. The dual part $\overline{J}$ is similar.

\subsection{Sugawara construction}
The energy-momentum tensor of the theory is given by Sugawara
construction
\begin{equation}
\label{Sugawara-tensor}
 T(z)=\frac{1}{(k+h^\vee)} \sum_a (J^a J^a)(z)\, .
\end{equation}
The prefactor
$\frac{1}{(k+h^\vee)}$ is determined by the usual form of $TT$ OPE, and $h^\vee$ is called the \textbf{dual Coxeter number} of $\frakg$.

By using the generalized wick theorem, we can derive the $TJ$ OPE
\begin{equation}
\label{TJ-OPE}
 T(z)J^a(w)=
 \frac{J^a(w)}{(z-w)^2}+\frac{\pd J^a (w)}{(z-w)}
 \, .
\end{equation}
Therefore, $J^a$ is a Virasoro primary field with conformal dimension $1$.

$TT$ OPE can be calculated straightforwardly
\begin{equation}
 T(z)T(w)=\frac{c/2}{(z-w)^4}
+\frac{2T(w)}{(z-w)^2}
+\frac{\pd T(w)}{(z-w)} \, ,
\end{equation}
with the central charge $c$
\begin{equation}
 c=\frac{k \,\text{dim} (\frakg)}{k+h^\vee}\, .
\end{equation}

The Sugawara energy-momentum tensor can be written in terms
of mode expansion the same as before
\begin{equation}
 T(z) =\sum_{n\in\mathbb{Z}} z^{-n-2} L_n \, .
\end{equation}
With the OPEs listed above, we can compute the complete
affine Lie and Virasoro algebra
\bea
  \comm{L_n}{L_m} &= (n-m) L_{n+m}+\frac{c}{12}
  (n^3-n) \delta_{n+m,0}\, ,\\
  \comm{L_n}{J^a_m} &= -m J^a_{n+m}\, ,\\
  \comm{J^a_n}{J^b_m} &= \sum_c if_{abc}J^c_{n+m}
 +\frac k2n \delta_{ab}\delta_{n+m,0}\, .
\eea

Finally, for future discussion, we need to write the relation
between $L_n$ and $J^a_n$. From \eqref{Sugawara-tensor} and the mode expansion of
$J^a$ \eqref{mode-expansion-Ja}, the relation can be obtained:
\begin{equation}
\label{rel-Ln-Jan}
L_n=\frac{1}{(k+h^\vee)}
\sum_a \sum_m :J^a_m J^a_{n-m}:\, .
\end{equation}

\subsection{Knizhinik-Zamolodchikov equations}
\subsubsection*{WZNW primary field}
A WZNW primary field is defined as a field that transforms
covariantly with respect to a $G(z)\times G(\overline{z})$ transformation. From \eqref{transformation-g} and \eqref{WZNW-ward-identity}, we reformulate $Jg$ OPE as
following
\bea
J^a(z) g(w, \overline{w}) &\sim \frac{- t^a g(w,\overline{w})}{z-w}\,, \\
\overline{J}^a(z) g(w,\overline{w}) &\sim
\frac{g(w,\overline{w})t^a}{z-w} \, .
\eea
This can be generalized to define a WZNW primary field. Any
WZNW primary field $\phi_{\lambda,\mu}$ with $\lambda$ and $\mu$
specified the representation in the holomorphic and antiholomorphic sector respectively should satisfy the
following OPE
\begin{equation}
\label{WZNW-primary field}
 J^a(z) \phi_{\lambda,\mu}(w,\overline{w}) \sim
 \frac{- t^a_\lambda \phi_{\lambda,\mu}(w,\overline{w})}{z-w}\, ,
\end{equation}
\begin{equation}
 \overline{J}^a(\overline{z}) \phi_{\lambda,\mu}(w,\overline{w})
 \sim
 \frac{\phi_{\lambda,\mu}(w,\overline{w})t^a_\mu}{\overline{z}-\overline{w}}\, ,
\end{equation}
where $t^a_\lambda$ is the matrix $t^a$ in the $\lambda$
representation. By analogy with \eqref{Virasoro-primary}, the above OPE can be extended
\begin{equation}
\label{WZNW-primary-field-extension}
  J^a(z)\phi_1(z_1)\phi_2(z_2)\cdots \phi_n(z_n)
  \sim -\sum_i\frac{t^a_i}{z-z_i} \phi_1(z_1)\phi_2(z_2)
  \cdots \phi_n(z_n)\, ,
\end{equation}
where we consider only the holomorphic sector, with
$t_i^a$ a matrix acting on $\phi_i$.

By expanding the currents in terms of the modes evaluated at
$w$,
\begin{equation}
 J^a(z)=\sum_n (z-w)^{-n-1} J^a_n(w)\, .
\end{equation}
We can write their OPE with an arbitrary field $A$ as we have done in the Virasoro case
\begin{equation}
J^a(z)A(w)=\sum_n (z-w)^{-n-1} (J^a_n A)(w)\, .
\end{equation}
Thus, from \eqref{WZNW-primary field}, for the WZNW holomorphic primary field
$\phi_\lambda$, we have
\bea
 J^a_0 \phi_\lambda &= -t^a_\lambda \phi_\lambda \, ,\\
J^a_n \phi_\lambda &= 0 \qquad \text{for} \qquad n>0 \, .
\eea
Associating the state $\ket{\phi_\lambda}$ to the field
$\phi_\lambda$ as usual
\begin{equation}
\phi_\lambda(0) \ket{0}=\ket{\phi_\lambda}\,.
\end{equation}
Then, the condition for a WZNW primary field translates into
\bea
  J^a_0\ket{\phi_\lambda} &=-t^a_\lambda \ket{\phi_\lambda}\, ,\\
  J^a_n \ket{\phi_\lambda} &= 0 \qquad \text{for} \qquad n>0\, .
\eea
From \eqref{rel-Ln-Jan}, we see that in the expression for $L_n$,
with $n>0$, the rightmost factor $J^a_m$ has $m>0$ which implies that
\begin{equation}
 L_n\ket{\phi_\lambda}=0 \qquad \text{for} \qquad n>0 \, .
\end{equation}
Therefore, the WZNW primary fields are also Virasoro primary fields with the conformal dimension given by
\begin{equation}
 L_0 \ket{\phi_\lambda}=\frac{1}{(k+h^\vee)}
 \sum_a J^a_0 J^a_0 \ket{\phi_\lambda}
=\frac{\sum_a t^a_\lambda t^a_\lambda}{(k+h^\vee)}\ket{\phi_\lambda}
 \, .
\end{equation}
The opposite is indeed not true. For example, $J^a(z)$ is a
Virasoro primary field but not WZNW primary field which can be seen directly from $TJ$ OPE and $JJ$ OPE we have shown before.

Finally, a WZNW descendant state is defined of the form
\begin{equation}
 J^a_{-n_a} J^b_{-n_2}\cdots \ket{\phi_\lambda}\, ,
\end{equation}
with $n_1,n_2 \cdots$ all positive integers.

\subsubsection*{Knizhnik-Zamolodchikov Equations}
From the previous learning, we know that there are many
constraints to the correlation function, such as the
null fields. This section briefly discusses the affine
singular vector at level one, and its constraints.

For $n=-1$, from \eqref{rel-Ln-Jan}, we have
\begin{equation}
  L_{-1}\ket{\phi_i}=\frac{2}{k+h^\vee}
  \sum_a (J^a_{-1}J^a_0) \ket{\phi_i}
 =\frac{-2}{k+h^\vee} \sum_a (J^a_{-1}t_i^a) \ket{\phi_i}\, .
\end{equation}
We consider the insertion of the zero vector
\begin{equation}
 \ket{\chi}=\bigg[
 L_{-1}+ \frac{2}{k+h^\vee}\sum_a(J^a_{-1}t_i^a)
 \bigg]
 \ket{\phi_i}=0\, .
\end{equation}
First, we need to calculate the contribution of the second term.
By using \eqref{mode-expansion-Ja}
, the insertion of the operator $J^a_{-1}$ in the correlator
can be expressed as
\begin{equation}
  \expval{\phi_1(z_1)\cdots (J^a_{-1}\phi_i)(z_i)\cdots\phi_n(z_n)}
 =\frac{1}{2\pi i}
  \oint_{z_i}
  \frac{\dd z}{z-z_i}
  \expval{J^a(z)\phi_1(z_1)\cdots\phi_n(z_n)}\,.
\end{equation}
Now using \eqref{WZNW-primary-field-extension} and
performing the residue calculation, we get
\begin{equation}
\expval{\phi_1(z_1)\cdots (J^a_{-1}\phi_i)(z_i)\cdots\phi_n(z_n)}
= \sum_{j\neq i}
\frac{t^a_j}{z_i-z_j}
\expval{\phi_1(z_1)\cdots \phi_n(z_n)}\,.
\end{equation}
Thus
\begin{equation}
\expval{\phi_1(z_1)\cdots \chi(z_i)\cdots \phi_n(z_n)}=
\bigg[
\partial_{z_i}+ \frac{2}{k+h^\vee}
\sum_{j\neq i}
\frac{\sum_a t^a_i \otimes t^a_j}{z_i-z_j}
\bigg]
\expval{\phi_1(z_1)\cdots\phi_n(z_n)}
\, ,
\end{equation}
where $t^a_j$ and $t^a_i$ act on $\phi_i$ and $\phi_j$
respectively. By construction, it must  vanish so that
we have
\begin{equation}
\bigg[
\partial_{z_i}+ \frac{2}{k+h^\vee}
\sum_{j\neq i}
\frac{\sum_a t^a_i \otimes t^a_j}{z_i-z_j}
\bigg]
\expval{\phi_1(z_1)\cdots\phi_n(z_n)}
=0\, .
\end{equation}
This is called the \textbf{Knizhnik-Zamolodchikov equation} \cite[ISZ88-No.4]{knizhnik1984current}. The
solutions to this equation are the correlation functions
of WZNW primary fields.

\subsection{Coset models}\label{sec:coset}

Using current algebras, one can construct an important class of 2d CFTs called \textbf{coset construction} \cite[ISZ88-No.11]{goddard1986unitary}. This construction has many applications in physics like (supersymmetric) unitary minimal models (see the end of \S\ref{sec:su2k-character}) and Kazama-Suzuki models \cite{Kazama:1988qp}.

We start with the current algebra associated to a Lie algebra $\frakg$ which contains  $\frakh$ as subgroups. The Sugawara energy-momentum tensors of affine Lie algebra $\widehat \frakg_{k_{\frakg}}$ and $\widehat \frakh_{k_{\frakh}}$ are
\bea
T _{\mathfrak{g}} (z) &= \frac{1}{k_{\mathfrak{g}}+h^\vee _{\mathfrak{g}} }\sum _{a=1 }^{\operatorname{dim}\mathfrak{g}} : J _{\mathfrak{g}}^{a}J _{\mathfrak{g}}^{a}: (z) \cr T _{\mathfrak{h}} (z) &= \frac{1}{k_{\mathfrak{h}}+h^\vee _{\mathfrak{h}} }\sum _{b=1 }^{\operatorname{dim}\mathfrak{h}} : J _{\mathfrak{h}}^{b}J _{\mathfrak{h}}^{b}: (z) \eea
We define the energy-momentum tensor of the coset model $\widehat \frakg_{k_{\frakg}}/ \widehat \frakh_{k_{\frakh}}$ as
\[
T _{\mathfrak{g}/ \mathfrak{h}} :=T _{\mathfrak{g}}-T _{\mathfrak{h}}~.
\]
Then, we have
\bea
T _{\mathfrak{g}} (z) J _{\mathfrak{h}}^{b}(w) & =\frac{J _{\mathfrak{h}}^{b}(w)}{(z-w)^{2}}+\frac{\partial _{w}\dot{j}_{\mathfrak{h}}^{b}(w)}{z-w}+ \cdots \cr T _{\mathfrak{h}} (z)  J _{\mathfrak{h}}^{b}(w)&=\frac{J _{\mathfrak{h}}^{b}(w)}{(z-w)^{2}}+\frac{\partial _{w}J _{\mathfrak{h}}^{b}(w)}{z-w}+ \cdots \eea
which yield
\bea
\left(T _{\mathfrak{g}}-T _{\mathfrak{h}} \right) (z) J _{\mathfrak{h}}^{b}(w)&\sim0\cr
 \left(T _{\mathfrak{g}}-T _{\mathfrak{h}} \right) (z) T _{\mathfrak{h}} (w) &\sim0~.
 \eea
Therefore, one can compute the $T _{\mathfrak{g}/ \mathfrak{h}} T _{\mathfrak{g}/ \mathfrak{h}} $ OPE so that the central charge of the coset model is
\be\label{coset-central-charge}
c _{\mathrm{g}/ \mathrm{h}}=c_{\mathfrak{g}}-c _{\mathrm{h}}=\frac{k _{\mathfrak{g}} \operatorname{dim}\mathfrak{g}}{k _{\mathfrak{g}}+h^\vee _{\mathfrak{g}}}- \frac{k _{\mathfrak{h}} \operatorname{dim}\mathfrak{h}}{k _{\mathfrak{h}}+h^\vee _{\mathfrak{h}} }~.
\ee

\subsubsection*{Parafermion}
As the simplest model of the coset model, we consider $\widehat{\mathfrak{su}(2)}_k/\widehat{\mathfrak{u}(1)}_k$. The energy-momentum tensor $T_{\frakg/\frakh}=T _{\frakg }-T _{\frakh }$ where
\bea
T _{\frakg}(z) &= \frac{1}{k+2}\sum _{a=1 }^{3}: J^{a}(z) J^{a}(z) : \cr
T _{\frakh}(z) &= \frac{1}{k}: J^{3}(z) J^{3}(z) :
\eea
If we define $J^{3}$ as
\[
J^{3}(z)=i \sqrt{\frac{k}{2}} \partial \varphi (z)
\]
then $T_H$ is the energy-momentum tensor of the free boson. It is easy to read off central charge
\[
c _{\mathrm{g}/ \mathrm{h}}=\frac{3 k}{k+2}- 1=\frac{2 (k-1)}{k+2 }~,
\]
which is equal to \eqref{Zk-cc}.  Indeed, this coset model describes the 2d CFT with $\bZ_k$ symmetry \cite[ISZ88-No.14]{fateev1985parafermionic}. This model contains a current with fractional spin (conformal dimension), which does obey neither bose nor fermi statistics, and is rather subject to para-statistics. Therefore, it is called  $\bZ_k$ \textbf{parafermion}.
In terms of the free boson $\varphi$ and the $\SU(2)$ current $J^\pm$, the parafermions are defined by
\[
\left(J^{+}\right)^{\ell}= \psi _{\ell}e^{i \ell \sqrt{\frac{2}{k}} X}, \quad \left(J^{-}\right)^{\ell}= \psi _{\ell }^{\dagger}e^{- i \ell \sqrt{\frac{2}{k}} X}, \quad (\ell=0 , \cdots , k)
\]
where $ \psi _{\ell}$ and $\psi _{\ell}^\dagger$ have conformal dimension $h=\ell(k-\ell)/k$.

When $k=1$, the central charge is $c=0$ so that the theory is trivial.
When $k=2$, the central charge is $c=\frac12$ so that the theory is the free fermion.
When $k=3$, the central charge is $c=\frac43$ so that the theory is the 3-state Potts model, which corresponds to the unitary minimal model $\cM_5$. In this case, the primary field with conformal dimension $h=2/3$ corresponds to the parafermion current $\psi_1$. The partition function of the parafermion is evaluated in \cite[ISZ88-No.28]{Gepner:1986hr}.

In fact, the supersymmetric version of the parafermion describes $\cN=2$ unitary minimal models, which plays a crucial role to describe Calabi-Yau sigma-models \cite{Gepner:1987vz,Gepner:1987qi}.

\subsubsection*{Rational conformal field theories}
The minimal models, the WZNW models and the coset models belong to a class of rational conformal field theories (RCFTs). A theory is endowed with a chiral algebra $\cA\otimes \overline \cA$ that contains the Virasoro algebra as a subalgebra. The definition of an RCFT is (roughly) given as follows.
\begin{itemize}
\item It is unitary with a unique vacuum.
\item There are finitely many primary fields $\phi_i$ of the chiral algebra $\cA$. The same statement holds for the anti-holomorphic part.
\item The Hilbert space is spanned by finitely many irreducible representations of $\cA\otimes \overline \cA$
\[
\cH=\bigoplus_{i,j} \cN_{i,j} \mathcal{V}_i\otimes \overline{\mathcal{V}}_j
\]
where the irreducible representation $\mathcal{V}_i$ arises from primary fields $\phi_i$.
\item   Their OPEs of primary fields are closed among themselves
\[
\phi_i(z)\phi_j(w)\sim \sum_{k}\frac{C_{ijk}}{(z-w)^{h_i+h_j-h_k}}\phi_j(w)+\cdots
\]
\end{itemize}
It is called an RCFT because conformal dimensions of primary fields of the chiral algebra $\cA$ as well as the central charge are rational numbers $\in \bQ$ \cite{Vafa:1988ag}. In an RCFT, conformal blocks can be, in principle, determined from structure constants $C_{ijk}$ by bootstrap equations \eqref{bootstrap}, and one can evaluate a correlation function on an arbitrary Riemann surface. For more details, we refer to \cite{Moore:1988qv,Moore:1989vd}.

\section{Modular invariant partition functions}

We have already seen some examples of modular invariant partition functions, such as
torus partition functions of the free boson/fermion \S\ref{sec:free-torus}, and minimal models \S\ref{sec:characters-MM}.
We first review a free boson on a circle and look into its modular invariance in special cases.
In this section, we study modular invariant partition functions more in details. We also refer the reader to \cite[\S10,\S17]{francesco2012conformal} and \cite[\S8]{Blumenhagen:2009zz}.

When we consider the torus partition function, it should always be modular invariant because of torus geometry.
Nonetheless, it is non-trivial to see some cases. Instead, the modular invariance gives us some useful information,
which can be seen in \S\ref{sec:orbif-part-funct}.

\subsection{Free boson on a circle}
\label{sec:free-boson-circle}

Let us first focus on a single free boson on a circle.
The partition function on a circle of radius $R$ is, as we have seen in \eqref{eq:bosonZ}:
\begin{equation}
\mathcal{Z}_{R}(\tau,\overline{\tau})
 = \Tr \bigg( q^{L_0 -\frac{c}{24}} \overline{q}^{\overline{L}_0-\frac{c}{24}} \bigg)
 = \frac{1}{|\eta(\tau)|^2}\sum_{n,w} q^{\frac12 (\frac{n}{R}+\frac{Rw}{2})^2} \overline{q}^{\frac12 (\frac{n}{R}-\frac{Rw}{2})^2}, \label{eq:bosonZ1}
\end{equation}
where we changed conventions, and here, $n$ is for KK-momentum and $w$ for winding number.
Note that $R$ dependence of the partition function means that the theory depends on the radius $R$.
We will see peculiar features of theories (partition functions) at specific values of the radius.
Now, let us study the partition function on a special radius $R = \sqrt{2k}$
with $k \in \mathbb{Z}^{+}$.
Rearranging the summation, the partition function can be written as
\begin{align}
 \mathcal{Z}_{\sqrt{2k}} (\tau,\overline \tau) &= \frac{1}{\abs{\eta(\tau)}^2} \sum_{n,w \in \mathbb Z} q^{\frac{1}{4k} (n+kw)^2} \overline{q}^{\frac{1}{4k} (n-kw)^2}
 = \frac{1}{\abs{\eta(\tau)}^2} \sum_{\substack{m \in \mathbb Z_{2k} \\ n_L,n_R \in \mathbb Z}} q^{\frac{1}{4k} (m+2kn_L)^2} \overline{q}^{\frac{1}{4k} (m-2kn_R)^2}  \nonumber\\
 &= \frac{1}{\abs{\eta(\tau)}^2} \sum_{m \in \mathbb Z_{2k}} \left| \Theta_{m,k} (\tau) \right|^2~,
\end{align}
where the range of $k$ is $-k+1 \le m \le k$, $\Theta$-function was defined in \eqref{Theta function}.
Conformal field theories corresponding to these
partition functions are commonly denoted as
$\widehat{\mathfrak{u}(1)}_k$, and we use this notation for the subscript from now on. Indeed, for $k = 1$, we find
\begin{equation}
\label{partition function k = 1}
  \mathcal{Z}_{\widehat{\mathfrak{u}(1)}_1}(\tau,\overline{\tau})
  = \frac{1}{\abs{\eta(\tau)}^2}
  \bigg(
  \abs{\Theta_{0,1}}^2 + \abs{\Theta_{1,1}}^2
  \bigg)
  = \chi_0^{(1)}\overline{\chi}_0^{(1)}
  +\chi_1^{(1)}\overline{\chi}_1^{(1)}
  \, ,
\end{equation}
where $\chi_j^{(1)}$ is the character of $\widehat{\mathfrak{su}(2)}_1$ as we will see below. As we have seen in \S\ref{sec:SU(2)k}, $k=1$ is actually the self-dual point, and the symmetry $\widehat{\mathfrak{u}(1)}_1$ is enhanced to $\widehat{\mathfrak{su}(2)}_1$,

From the definition of $\Theta$-function, it is easy to see
the T-transformations for $\Theta$- function, we have
\begin{equation}
  \Theta_{m,k}(\tau+1) = e^{\pi i m^2 / 2k}
  \Theta_{m,k}(\tau)\, .
\end{equation}
For the modular S-transformation, we have
\begin{equation}
 \Theta_{m,k} \bigg(
 -\frac{1}{\tau}
 \bigg)
 = \sqrt{-i \tau}
 \sum_{m' = -k + 1}^ k
 S_{m,m'}
 \Theta_{m',k}(\tau)\, ,
\end{equation}
with the modular S-matrix
\begin{equation}
 S_{m,m'} = \frac{1}{\sqrt{2k}}
 \exp(-\pi i \frac{m m'}{k})\, .
\end{equation}
For $k=1$ case, the S-transformation for the characters
in \eqref{partition function k = 1} can be
inferred from the transformation properties of the
$\Theta$- and $\eta$-functions
\begin{equation}
\chi^{(1)}_m\big(
-\frac{1}{\tau}
\big) = \sum_{m'=0}^1
S_{m,m'} \chi_{m'}^{(1)}(\tau)
\end{equation}
Following what we have done when constructing modular invariant
partition function in minimal model, we can write the partition function
\eqref{partition function k = 1} as
\begin{equation}
 \mathcal{Z}_{\widehat{\mathfrak{u}(1)}_1}(\tau,\overline{\tau}) =
 (\chi^{(1)}_0, \chi_1^{(1)})
 \mqty(1&0\\0&1)
 \mqty(\overline{\chi}_0^{(1)}\\ \overline{\chi_1}^{(1)})
 =\vec{\chi}^T M \vec{\overline{\chi}}\, ,
\end{equation}
where $M$ is the identity matrix in this case.
The S-transformation of the partition function now reads
\begin{equation}
  \mathcal{Z}_{\widehat{\mathfrak{u}(1)}_1} \bigg(
  -\frac{1}{\tau},
  -\frac{1}{\overline{\tau}}
  \bigg)
  =\vec{\chi}^T S^T M S^*\vec{\overline{\chi}}\, .
\end{equation}
Since $S^T = S$ from the definition of S matrix, the
condition for invariance under modular S-transformation
for the partition function now reads
\begin{equation}
 S M S^{\dagger} = M \, .
\end{equation}
This is obviously true in this simple case. Because
under T-transformations the characters $\chi_m^{(1)}$
only acquire a phase, we have shown that
$\mathcal{Z}_{\widehat{\mathfrak{u}(1)}_1}(\tau,\overline{\tau})$ is modular-invariant.
This process can be generalized to construct more
complicated modular invariant partition functions.

Let us look into the $k=2$ case as well. The modular invariant partition function is given as follows.
\begin{align}
 \cZ_{\widehat{u(1)}_2} = \sum_{-1 \le m \le 2} \left| \frac{\Theta_{m,2}(\tau)}{\eta(\tau)} \right|^2
 = \frac{1}{2} \left\{\left| \frac{\vartheta_2(\tau)}{\eta(\tau)} \right|^2 +\left| \frac{\vartheta_3(\tau)}{\eta(\tau)} \right|^2
 +\left| \frac{\vartheta_4(\tau)}{\eta(\tau)} \right|^2 \right\}
\end{align}
Notice that the expression above is precisely the same as the torus partition function of a free complex fermion on a circle \eqref{free-fermion-minimal-model},
though the second equality above is non-trivial.
This means that a free boson on a circle at $R=2$ is equivalent to a free complex fermion, and it is called ``Boson-Fermion correspondence''
(or bosonization, fermionization).
This is a simple example of ``duality'', which provides us a pair of theories that describe the same physics
(hence their physical quantities like multi-point function and partition function agree with each other).

\subsection{Orbifold partition function}
\label{sec:orbif-part-funct}
As we saw in a free boson on a circle case, the theory depends on a ``target space''\footnote{
The target space is a space where fields take their values. In the case of a free boson on a circle,
$\varphi$ takes a value on the circle $x \sim x +2\pi R$. Hence, the target space is the circle.}
on which the theory lives.
In particular, the boundary condition plays an important role. For the circle case it is
\begin{align}
 \varphi(z e^{2\pi i}, \overline z e^{-2\pi i}) = \varphi(z, \overline z) + 2\pi w R , \qquad w \in \mathbb Z \quad \textrm{: winding number}.
 \label{eq:bosonS1BoundaryCond}
\end{align}

Now, we want to extend our knowledge to other kinds of CFT. Therefore, we consider different target spaces.
What we consider here is the so-called orbifold.
For a manifold $M$ and a discrete group $\Gamma$, a space defined by a quotient $M/\Gamma$ with an equivalence relation
\begin{align}
 x \sim g x \qquad g \in \Gamma , \quad x \in M ,
\end{align}
could be singular, and it is, in general, called orbifold.

\begin{figure}[ht]
 \centering
 \begin{tikzpicture}[scale=1]
  \draw[thick] (0,0) circle[radius=1];
  \draw (0,1) node{$\times$} ;
  \draw (0,1.5) node{$x=0$} ;
  \draw (0,-1) node{$\times$} ;
  \draw (0,-1.5) node{$x=\pi R$} ;
  \draw[Stealth-Stealth,blue] (-0.7,0.6) -- (0.7,0.6) ;
  \draw[Stealth-Stealth,blue] (-0.9,0.2) -- (0.9,0.2) ;
  \draw[Stealth-Stealth,blue] (-0.9,-0.2) -- (0.9,-0.2) ;
  \draw[Stealth-Stealth,blue] (-0.7,-0.6) -- (0.7,-0.6) ;
  \draw (-2.5,0) node[blue]{Identify} ;
  \draw[-Triangle] (2,0) -- (3,0) ;
  \draw[thick] (4,-1) node{$\times$} -- (4,1) node{$\times$} ;
 \end{tikzpicture}
 \caption{From a circle ($S^1$) with radius $2\pi R$ to a $\mathbb Z_2$-orbifold ($S^1/\mathbb Z_2$) of the circle.
 We identify points $x$ and $-x$ in the circle, which leads to a finite line. There are two singular points, which are illustrated by crosses in the figure.}
 \label{fig:orbifold}
\end{figure}
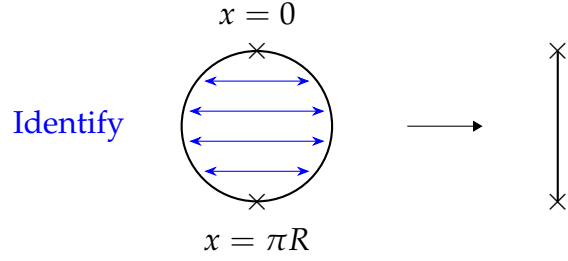

We consider, as the simplest example, the $\mathbb{Z}_2$-orbifold of a circle with radius $R$ \cite[ISZ88-No.40]{Yang:1987mf}.
In this case $M = S^1$ and $\Gamma = \mathbb Z_2$ whose elements are $g = \{1,-1\}$.
The non-trivial equivalence relation is, therefore, $x \sim -x$.
The resultant orbifold is denoted by $S^1/\mathbb Z_2$ (see Fig.~\ref{fig:orbifold}).
Our goal is to calculate a partition function of a free boson on the orbifold.
In order to achieve this we consider a free boson on a circle $\varphi\sim \varphi + 2\pi R$,
and define a $\mathbb{Z}_2$ symmetry (orbifold) operator $\mathcal{R}$:
\begin{equation}
 \mathcal{R}: \varphi(z,\overline{z})\mapsto -\varphi(z,\overline{z})\, , \qquad \mathcal R^2 = \mathrm{Id} \, .
\end{equation}
We calculate a partition function with states
invariant under the orbifold action.
Namely, we
insert an orbifold projection operator $\frac{1}{2}(1+\mathcal{R})$ into the trace in (\ref{eq:bosonZ1}),
and the partition function reads
\begin{align}
\mathcal{Z}_{\textrm{orb}}(\tau,\overline{\tau}) &= \Tr
\bigg(
\frac{1+\mathcal{R}}{2} q^{L_0 -\frac{c}{24}}
\overline{q}^{\overline{L}_0-\frac{c}{24}}
\bigg)\notag\\
& =
\frac{1}{2}\mathcal{Z}_{R}(\tau,\overline{\tau})
+\frac{1}{2}\Tr\big(
\mathcal{R} q^{L_0-\frac{c}{24}}
\overline{q}^{\overline{L}_0-\frac{c}{24}}
\big)\, .
\end{align}
The first term turns out to be the partition function of a
free boson on a circle which we have already computed in
\eqref{eq:bosonZ1}.
Let us focus on the second term.

The action of $\mathcal{R}$ on the Laurent modes $a_n$
of $ i\partial\varphi(z)$ can be found
\def\mR{\mathcal{R}}
\begin{equation}
  \mathcal{R}a_n\mathcal{R}= -a_n \, .
\end{equation}
Hence
\begin{align}
\mathcal{R}\ket{n_1,n_2,n_3,\dots} & =
(\mR a_{-1}\mR)^{n_1}
(\mR a_{-2}\mR)^{n_2}
\dots
\mR \ket{\Omega}\notag \\
& = (-1)^{n_1+n_2+n_3 + \dots}
\ket{n_1, n_2, n_3,\dots}\, .
\end{align}
where $\ket{\Omega}$ is a ground state, and we assumed that it is invariant under $\mR$, i.e $\mR\ket{\Omega} = \ket{\Omega}$.
Now, we consider the action $\mathcal R$ on the ground states in detail. Remember that the ground state is
characterized by KK-momentum and winding number, $\ket{\Omega} = |n,w \rangle$.
Since the zero mode operator $a_0$ gives eigenvalues of the states
\begin{align}
 a_0 \ket{n,w} = \left( \frac{n}{R} +\frac{Rw}{2} \right) \ket{n,w} ,
\end{align}
we see that
\begin{align}
 &a_0 \mathcal{R}\ket{n,w}
 =\mathcal{R}(\mR a_0 \mR)\ket{n,w}
 =\mathcal{R}(-a_0) \ket{n,w}
 =-\bigg(
 \frac{n}{R} + \frac{R w}{2}
  \bigg)
 \mR \ket{n,w}\, ,  \nonumber\\
 &\therefore\quad \mR \ket{m,n} \propto \ket{-m,-n}\, .
\end{align}
Therefore, only states with $\ket{m=0, n= 0}$ will
contribute to the partition function.
Then following the same steps in \eqref{free-boson-process},
we replace the result of the free boson as
\begin{equation}
q^{-\frac{1}{24}}\prod_{n=1}^\infty\frac{1}{1-q^n}
\quad \to \quad
q^{-\frac{1}{24}}\prod_{n=1}^\infty\frac{1}{1-(-q^n)}
= \sqrt{2}\sqrt{\frac{\eta{(\tau)}}{\vartheta_2(\tau)}}\, .
\end{equation}
We thus write the partition function as
\begin{equation}
\label{Orbifold-partition-function-1}
\mathcal{Z}(\tau,\overline{\tau}) = \frac{1}{2}
\mathcal{Z}_R(\tau,\overline{\tau}) +
\abs{\frac{\eta(\tau)}{\vartheta_2(\tau)}}\, .
\end{equation}

Notice that the result \eqref{Orbifold-partition-function-1}
can not be the correct partition function because the second term is not invariant under modular transformation.
Recalling the free fermions partition function result, $\vartheta$-functions
enjoy the modular transformations illustrated in Fig.\ref{fig:modular-theta}.
In order for a free boson on the orbifold to have modular invariance,
we are required to add the so-called twisted sector, whose contributions should be
\begin{equation}
  \mathcal{Z}_{\text{tw}} (\tau,\overline{\tau}) =
  \abs{\frac{\eta(\tau)}{\vartheta_4(\tau)}}
  +
  \abs{\frac{\eta(\tau)}{\vartheta_3(\tau)}}\, .
\end{equation}
Then, the modular invariant partition function of a free boson on the
$\mathrm{Z}_2$-orbifold reads
\begin{equation}
\mathcal{Z}_{\text{orb}}(\tau,\overline{\tau})=
\frac{1}{2}
\mathcal{Z}_R(\tau,\overline{\tau}) +
\abs{\frac{\eta(\tau)}{\vartheta_2(\tau)}}
+
\abs{\frac{\eta(\tau)}{\vartheta_4(\tau)}}
+
\abs{\frac{\eta(\tau)}{\vartheta_3(\tau)}}\, .
\label{eq:orbfioldPF}
\end{equation}
Although we derived the correct result by requiring modular invariance here,
it is instructive to see what is the physical realization of the twisted sector.

\subsubsection*{Twisted sector}

Recall that when we calculate the torus partition function of a free fermion on a circle,
we considered Neveu-Schwarz (NS) sector as well as Ramond (R) sector, which are realized as follows.
\begin{align}
 &\psi\left( e^{2\pi i} z \right) = \psi(z) \qquad\!\quad \textrm{NS} , \\
 &\psi\left( e^{2\pi i} z \right) = -\psi(z) \qquad \textrm{R} ,
\end{align}
where only the holomorphic parts are shown.
We realize that NS sector corresponds to
the boundary condition of a free boson (\ref{eq:bosonS1BoundaryCond}),
and wonder if there is a boundary condition like the R sector.
Indeed, there exists due to the operation $\mR$:
\begin{align}
 \varphi(z e^{2\pi i}, \overline z e^{-2\pi i}) = \mR \varphi(z, \overline z) + 2\pi w R = -\varphi(z, \overline z) + 2\pi w R .
\end{align}
To realize this boundary condition, the mode has to become ``half-integers'' modes:
\begin{align}
 &\varphi(z,\overline z) = \varphi(z) +\overline\varphi(\overline z) , \nonumber\\
 &\quad\varphi(z) = \frac{x}{2} +i \sum_{n \in \mathbb Z +\frac{1}{2}} \frac{1}{n} \frac{a_n}{z^n} .
\end{align}
There are two possible values for the zero mode $x$; one is $x=0 \ (w=0)$ and the other is $x = \pi R \ (w=1)$.
Due to the shift of the modes the zero point energy is also modified as follows:
\begin{align}
 \frac{1}{2} \sum_{n=0}^\infty n = -\frac{1}{24} \quad \to \quad
 \frac{1}{2} \sum_{n=0}^\infty (n+\frac{1}{2}) = -\frac{1}{24} +\frac{1}{16} = \frac{1}{48} .
\end{align}
Therefore, we have the following contribution:
\begin{align}
 &\frac{1}{2} \Tr_\mathrm{tw} \left[ q^{L_0 -\frac{1}{24}} \overline q^{\overline{L_0} -\frac{1}{24}} \right] \quad \textrm{with} \quad
 L_0 = \sum_{n=1}^\infty a_{-n+\frac{1}{2}} a_{n-\frac{1}{2}} +\frac{1}{16} \nonumber\\
 &= \frac{1}{2}\cdot 2 \cdot |q|^{2\cdot \frac{1}{48}} \prod_{n=1}^\infty \left| \frac{1}{1-q^{n-\frac{1}{2}}} \right|^2
 = \left| \frac{\eta(\tau)}{\vartheta_4(\tau)} \right| ,
\end{align}
where the factor of two in the coefficient is coming from the two contributions of zero modes $x=0$ and $x=\pi R$.
Similarly,
\begin{align}
 &\frac{1}{2} \Tr_\mathrm{tw} \left[ \mathcal R q^{L_0 -\frac{1}{24}} \overline q^{\overline{L_0} -\frac{1}{24}} \right] \quad \textrm{with} \quad
 L_0 = \sum_{n=1}^\infty a_{-n+\frac{1}{2}} a_{n-\frac{1}{2}} +\frac{1}{16} \nonumber\\
 &= |q|^{\frac{1}{24}} \prod_{n=1}^\infty \left| \frac{1}{1+q^{n-\frac{1}{2}}} \right|^2
 = \left| \frac{\eta(\tau)}{\vartheta_3(\tau)} \right| .
\end{align}
As we expected, we have correct contributions from the twisted sector.
There are several comments:
\begin{itemize}
 \item $\cZ_\mathrm{orb}$ depends on radius $R$ only through $\cZ_R$, and hence, it enjoys T-duality:
       $\cZ_\mathrm{orb}(R) = \cZ_\mathrm{orb}(2/R)$.
 \item One can see that $\cZ_\mathrm{orb}(R=\sqrt 2) =\cZ_R (R = 2\sqrt 2)$.
       Namely, a free boson on a circle with $R=\sqrt 2$ is equivalent to that on an $S^1/\mathbb Z_2$ with $R=2\sqrt 2$.
       As we stressed before, theories depend on their radii, and two branches of a free boson theory
       intersect at the point of moduli space of the theory (see Fig.~\ref{fig:moduliOfPF}).
      \item  The $\cN=2$ superconformal symmetry arises at $R_{\textrm{circ}}=\sqrt{3}$, and  $\cN=1$ superconformal symmetry arises at $R_{\textrm{orb}}=\sqrt{3}$ \cite[ISZ88-No.39]{Friedan:1989yz} \cite{Yang:1987bj}.
       \item  In fact, a CFT with $c=1$ can be constructed by taking a quotient of the $\SU(2)_1$ current algebra by its finite subgroup. There is a one-to-one correspondence between the finite subgroups of $\SU(2)$ and the ADE Dynkin diagrams, which is called the \textbf{McKay correspondence}. The CFTs corresponding to $T$ $(E_6)$, $O$ $(E_7)$, and $I$ $(E_8)$ are isolated (does not allow deformation) \cite[ISZ88-No.33]{Pasquier:1986jc} \cite[ISZ88-No.41]{Ginsparg:1987eb}.
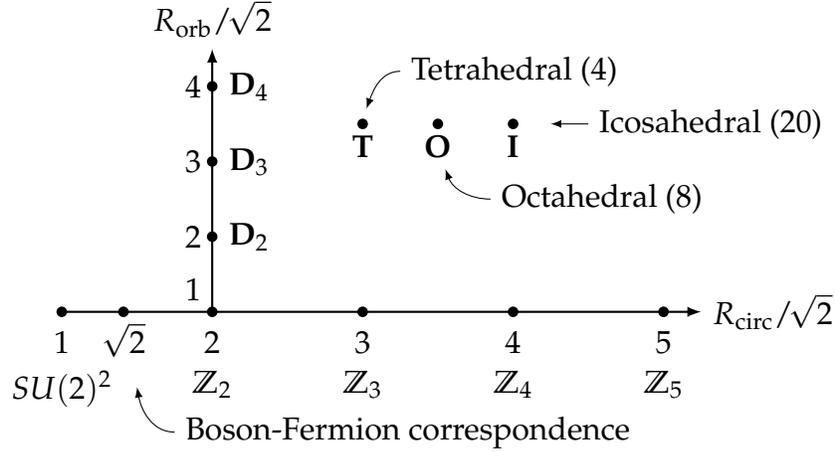
\begin{figure}[ht]
 \centering
 \begin{tikzpicture}
  \draw [-latex,thick] (0,0) -- (8.5,0) node[right]{$R_\mathrm{circ}/\!\sqrt{2}$} ;
  \draw [-latex,thick] (2,0) -- (2,3.5) node[above]{$R_\mathrm{orb}/\!\sqrt{2}$};
  \fill (0,0) circle (2pt) node[below=3pt]{1} node[below=18pt]{$SU(2)^2$};
  \fill (0.82,0) circle (2pt)  node[below]{$\sqrt 2$} ;
  \draw [latex-, out=-80, in=180] (1,-1) to (1.5,-1.6) node[right]{Boson-Fermion correspondence} ;
  \fill (2,0) circle (2pt) node[below=3pt]{2} node[above left]{1} node[below=18pt]{$\mathbb Z_2$} ;
  \fill (4,0) circle (2pt) node[below=3pt]{3} node[below=18pt]{$\mathbb Z_3$} ;
  \fill (6,0) circle (2pt) node[below=3pt]{4} node[below=18pt]{$\mathbb Z_4$} ;
  \fill (8,0) circle (2pt) node[below=3pt]{5} node[below=18pt]{$\mathbb Z_5$} ;
  \fill (2,1) circle (2pt) node[left]{2} node[right=2pt]{$\mathbf{D}_2$} ;
  \fill (2,2) circle (2pt) node[left]{3} node[right=2pt]{$\mathbf{D}_3$} ;
  \fill (2,3) circle (2pt) node[left]{4} node[right=2pt]{$\mathbf{D}_4$} ;
  \fill (4,2.5) circle (2pt) node[below]{\textbf T} ;
  \draw [latex-, out=70, in=180] (4.05,2.7) to (4.5,3.2) node[right]{Tetrahedral (4)} ;
  \fill (5,2.5) circle (2pt) node[below]{\textbf O} ;
  \draw [latex-, out=-70, in=180] (5.1,1.9) to (5.7,1.5) node[right]{Octahedral (8)} ;
  \fill (6,2.5) circle (2pt) node[below]{\textbf I} ;
  \draw [latex-] (6.5,2.5) to (7,2.5) node[right]{Icosahedral (20)} ;
 \end{tikzpicture}
 \caption{Moduli space of 2d CFTs with $c=1$. The horizontal arrow expresses the so-called toroidal branch, and the vertical arrow does the orbifold branch.
 $R_\mathrm{circ}$ is the radius of a circle and $R_\mathrm{orb}$ is that of the orbifold.
 $SU(2)^2$ is the symmetry that the theory has at the self-dual point. $\mathbb Z_k$, $\mathbf D_k$, $T$, $O$, and $I$ are
 the subgroups of the $SU(2)^2$ at the corresponding points illustrated above.. For more details see \cite[ISZ88-No.41]{Ginsparg:1987eb}.}
 \label{fig:moduliOfPF}
\end{figure}
\end{itemize}

So far, only some orbifold examples are illustrated.
For general orbifold $\Gamma$ we can do parallel computations as follows.
Let us consider a field in a torus coordinates $\Phi(\sigma,\tau)$, and for $g, h \in \Gamma$, we have
\begin{align}
 \Phi(\sigma,\tau+2\pi) = g \cdot \Phi(\sigma,\tau) , \nonumber\\
 \Phi(\sigma+2\pi,\tau) = h \cdot \Phi(\sigma,\tau) .
\end{align}
Note that $g$ and $h$ are commutative, i.e $gh=hg$.
As we have seen before, the time-like orbifold condition (the first line above) is realized by operator insertions,
and the space-like orbifold condition (the second line above) is realized by twisted sectors (trace over twisted Hilbert space):
\begin{align}
 \cZ_{(g,h)} (\tau,\overline \tau) = \Tr_{\mathcal H_h} \left[ g q^{L_0 -\frac{c}{24}} \overline q^{\overline{L_0} -\frac{c}{24}} \right] ,
\end{align}
and the full partition function is given by the sum over all possible values of $g$ and $h$:
\begin{align}
 \cZ_{M/\Gamma} (\tau,\overline \tau) = \frac{1}{|\Gamma|} \sum_{ \substack{g,h \in \Gamma \\ gh = hg}} \cZ_{(g,h)} (\tau,\overline \tau) ,
\end{align}
where $|\Gamma|$ is the degree of a discrete group $\Gamma$.
One should check that this formula reproduces our result of $S^1/\mathbb Z_2$.

\subsection{\texorpdfstring{$\widehat{\mathfrak{su}(2)}_k$}{su(2)k} characters}\label{sec:su2k-character}
It has been briefly mentioned in \S\ref{sec:free-boson-circle} that $\widehat{\mathfrak{u}(1)}_1$ theory has actually $\widehat{\mathfrak{su}(2)}_1$ symmetry,
therefore, its partition function can be expressed by so-called $\widehat{\mathfrak{su}(2)}_1$ character.
Here, we derive the $\widehat{\mathfrak{su}(2)}_1$ character by considering a free boson on a circle with $R = \sqrt 2$.

Recall that the theory can be characterized by three operators (for more details see \S\ref{sec:SU(2)k}):
\begin{align}
 J(z) = \frac{i\partial\varphi (z)}{\sqrt 2} , \qquad J^\pm (z) = :\! e^{\pm i \sqrt 2 \varphi(z)} \!: .
\end{align}
OPE calculation leads to the following commutation relations
\begin{align}
 &[J^3_n,J^\pm_m] = \pm J^\pm_{n+m}, \nonumber\\
 &[J^+_n,J^-_m] = 2 J^3_{n+m} +n\delta_{n+m,0}, \nonumber\\
 &[L_0,J^a_n] = -n J^a_{n} \qquad (a=3,\pm) .
\end{align}
Let us consider the highest weight representation of the theory.
Notice that $L_0$ and $J^3_0$ commutes, hence, the highest weight vector(states) should be characterized by their eigenvalues $h$ and $j$,
and the vector is defined as follows.
\begin{align}
 &J^\pm_n \ket{h,j} = J^3_n \ket{h,j} = 0 \qquad (n \ge 1) ,  \nonumber\\
 &J^+_0 \ket{h,j} = \left(J^-_0\right)^{2j+1} \ket{h,j} = 0 ,  \nonumber\\
 &J^3_0 \ket{h,j} = j \ket{h,j} .
\end{align}
Note that from the argument of \S\ref{sec:SU(2)k} $j$ can only be $j=0,\frac12$ for $k=1$.
For $j=0$, the corresponding conformal dimension is $h=0$, and for $j=\frac{1}{2}$, that is $h=\frac{1}{4}$.
We can construct an integrable representation by acting the following operators to the highest weight vectors:
\begin{align}
 J^-_n \quad (n\le 0) , \qquad J^3_m, \ J^+_m \quad (m < 0).
\end{align}
We organize these creation operators in the following way:
\begin{align}
 J^+_{(n)} = J^+_{-n} , \qquad
 J^-_{(n)} = J^-_{+n} , \qquad
 J^3_{(n)} = J^3_{0} -\frac{n}{2} .
\end{align}
Notice that each $J^a_{(n)}$ forms $\mathfrak{su}(2)$ algebra (call them $\mathfrak{su}(2)_{(n)}$).

Let us consider one $\ket{0,0}$ of the highest vectors.
Since $J^3_{(0)} \ket{0,0} = 0$ the vector is singlet under the $\mathfrak{su}(2)_{(0)}$.
On the other hand, $J^3_{(1)} \ket{0,0} = -\frac{1}{2} \ket{0,0}$, $J^-_{(1)} \ket{0,0} = 0$, and hence,
it should be a spin-$\frac{1}{2}$ representation of $\mathfrak{su}(2)_{(1)}$.
Therefore, there is exactly one other states $J^+_{(1)} \ket{0,0}$.
Repeating similar procedures for $J^+_{(1)} \ket{0,0}$, and it turns that
$J^+_{(1)} \ket{0,0}$ is singlet under $\mathfrak{su}(2)_{(2)}$ but doublet under $\mathfrak{su}(2)_{(3)}$.
In summary, we have a tower of $J^+_{(n)}$ for odd $n$ (let us set $n=2p-1$):
\begin{align}
 &J^+_{(1)} J^+_{(3)} J^+_{(5)} \cdots J^+_{(2p-1)} \ket{0,0} := \ket{p; 0,0} ,  \nonumber\\
 &L_0 \ket{p; 0,0} = \sum_{m=1}^{p} (2m-1) \ket{p; 0,0} = p^2 \ket{p; 0,0} ,  \nonumber\\
 &J^3_0 \ket{p; 0,0} = p \ket{p; 0,0} .
\end{align}
We can also consider negative $n=-2p+1$ ($p \in \mathbb Z_{\ge 1}$), and the resultant tower is
\begin{align}
 &J^-_{(-1)} J^-_{(-3)} J^-_{(-5)} \cdots J^-_{(-2p+1)} \ket{0,0} := \ket{-p; 0,0} ,  \nonumber\\
 &L_0 \ket{-p; 0,0} = \sum_{m=1}^{p} (2m-1) \ket{-p; 0,0} = p^2 \ket{-p; 0,0} ,  \nonumber\\
 &J^3_0 \ket{-p; 0,0} = -p \ket{-p; 0,0} .
\end{align}
Notice that the tower corresponds to vertex operators:
\begin{align}
 \ket{p; 0,0} \quad &\leftrightarrow \quad :\! e^{+ i p \sqrt 2 \varphi(z)} \!:  , \nonumber\\
 \ket{-p; 0,0} \quad &\leftrightarrow \quad :\! e^{- i p \sqrt 2 \varphi(z)} \!:  , \nonumber\\
 \ket{0; 0,0} := \ket{0,0} \quad &\leftrightarrow \quad 1  .
\end{align}

\begin{figure}[ht]
 \centering
 \begin{tikzpicture}
 \begin{scope}
  \draw[step=1,very thin, lightgray] (-3,0) grid (3,9) ;
  \draw [-latex] (-3.5,0) -- (3.5,0) node[right]{$J^3_0$} ;
  \draw [-latex] (0,0) -- (0,9.5) node[above]{$L_0$};
  \fill (0,0) circle (2pt) node[below=3pt]{0} ;
  \fill (1,1) circle (2pt) (2,4) circle (2pt) (3,9) circle (2pt) (-1,1) circle (2pt) ;
  \draw[-Stealth, thick] (0,0) -- (1,1) ;
  \draw (1.1,0.4) node{$J^+_{(1)}$} ;
  \draw[-Stealth, thick] (1,1) -- (2,4) ;
  \draw (2,2.2) node{$J^+_{(3)}$} ;
  \draw[-Stealth, thick] (2,4) -- (3,9) ;
  \draw (3,6.2) node{$J^+_{(5)}$} ;
  \draw[-Stealth, thick] (0,0) -- (-1,1) ;
  \draw (-1.1,0.4) node{$J^-_{(-1)}$} ;
  \fill (-2.7,1.5) circle (2pt) (-2.7,3) circle (2pt) (-2.7,5.5) circle (2pt) ;
  \draw[-Stealth, thick] (-2.7,1.5) -- (-2.7,2.5) ;
  \draw (-2,1.9) node{$J^3_{(-1)}$} ;
  \draw[-Stealth, thick] (-2.7,3) -- (-2.7,5) ;
  \draw (-1.8,4) node{$\substack{J^3_{(-2)}, \\ \left(J^3_{(-1)}\right)^2}$} ;
  \draw[-Stealth, thick] (-2.7,5.5) -- (-2.7,8.5) ;
  \draw (-1.3,7) node{$\substack{J^3_{(-3)},\ \left(J^3_{(-1)}\right)^3, \\ J^3_{(-2)} J^3_{(-1)}} $} ;
  \draw [latex-, out=120, in=180] (-2.9,7) to (-2.7,9) node[right]{For each dot} ;
  \draw [latex-, out=120, in=180] (-2.9,4) to (-2.7,9) ;
  \draw [latex-, out=120, in=180] (-2.9,2.2) to (-2.7,9) ;
  \draw (1,0) node[below=3pt]{1};
  \draw (2,0) node[below=3pt]{2};
  \draw (3,0) node[below=3pt]{3};
  \draw (-1,0) node[below=3pt]{$-1$};
  \draw (-2,0) node[below=3pt]{$-2$};
  \draw (-3,0) node[below=3pt]{$-3$};
  \draw (0,-1.3) node{$j=0$} ;
 \end{scope}
 \begin{scope}[shift={(8,0)}]
  \draw[step=1,very thin, lightgray] (-3,0) grid (3,9) ;
  \draw [-latex] (-3.5,0) -- (3.5,0) node[right]{$J^3_0$} ;
  \draw [-latex] (0,0) -- (0,9.5) node[above]{$L_0$};
  \fill (-1.5,9/4) circle (2pt) (0.5,1/4) circle (2pt) (1.5,9/4) circle (2pt) (2.5,25/4) circle (2pt) (-0.5,1/4) circle (2pt) ;
  \draw[Stealth-, thick] (-1/2,1/4) -- (1/2,1/4) ;
  \draw (-0.3,0.75) node{$J^-_{(0)}$} ;
  \draw[-Stealth, thick] (-1/2,1/4) -- (-3/2,9/4) ;
  \draw (-1.6,1.1) node{$J^-_{(-2)}$} ;
  \draw[-Stealth, thick] (1/2,1/4) -- (3/2,9/4) ;
  \draw (1.6,1.1) node{$J^+_{(2)}$} ;
  \draw[-Stealth, thick] (3/2,9/4) -- (5/2,25/4) ;
  \draw (2.6,4.1) node{$J^+_{(4)}$} ;
  \draw[thick] (5/2,25/4) -- (3,37/4) ;
  \draw (0,0) node[below=3pt]{0};
  \draw (1,0) node[below=3pt]{1};
  \draw (2,0) node[below=3pt]{2};
  \draw (3,0) node[below=3pt]{3};
  \draw (-1,0) node[below=3pt]{$-1$};
  \draw (-2,0) node[below=3pt]{$-2$};
  \draw (-3,0) node[below=3pt]{$-3$};
  \draw (0,-1.3) node{$j=\frac{1}{2}$} ;
 \end{scope}
 \end{tikzpicture}
 \caption{Highest weight representation of $\widehat{\mathfrak{su}(2)}_{1}$.
 For each dot, there must be degenerate vertical arrows with integer length, which express descendant fields.}
 \label{fig:su2_1states}
\end{figure}
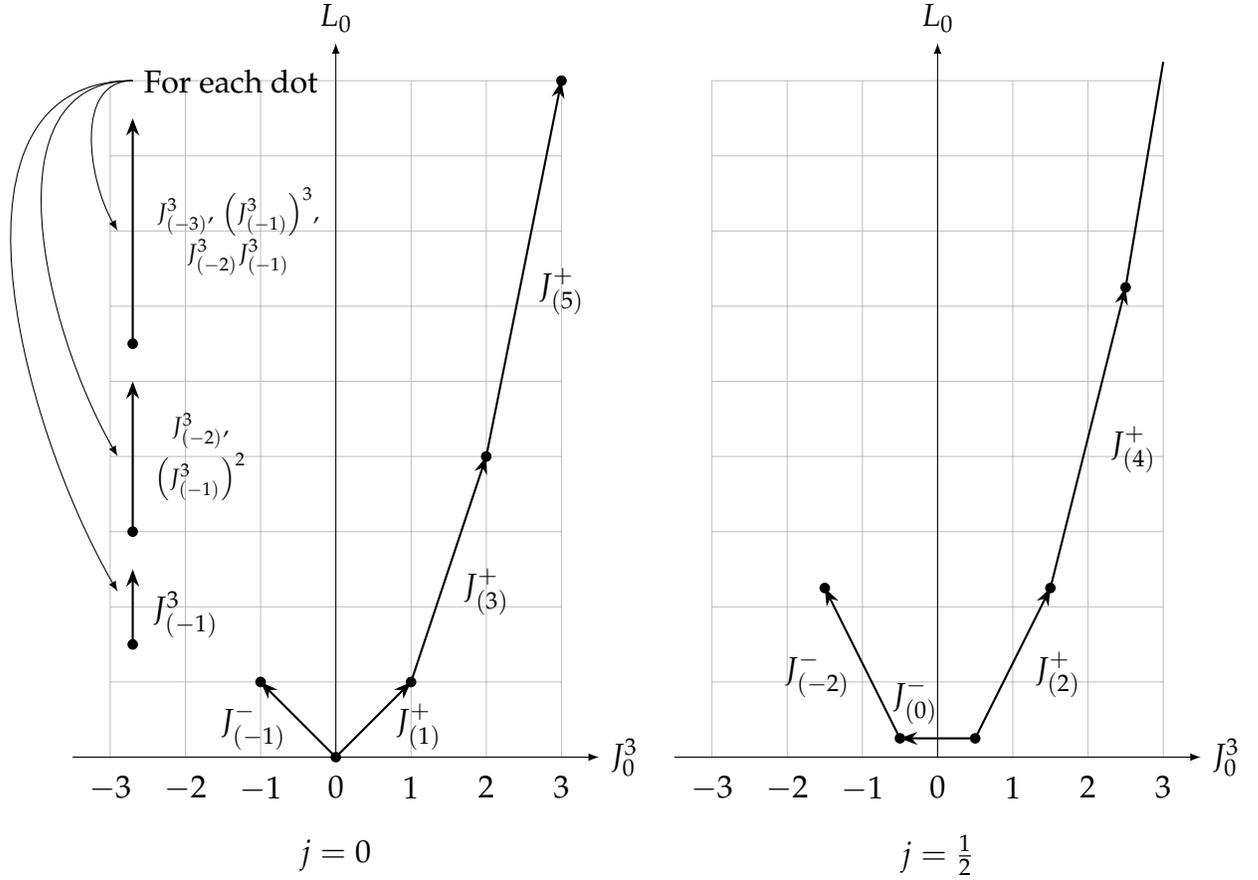

The other highest weight vector $\ket{\frac{1}{4},\frac{1}{2}}$ is doublet under $\mathfrak{su}(2)_{(0)}$.
$\ket{\frac{1}{4},\frac{1}{2}}$ is singlet under $\mathfrak{su}(2)_{(1)}$ and doublet under $\mathfrak{su}(2)_{(2)}$. On the other hand, $J^-_{(0)} \ket{\frac{1}{4},\frac{1}{2}}$ is singlet under $\mathfrak{su}(2)_{(-1)}$ and doublet under $\mathfrak{su}(2)_{(-2)}$.
Therefore, similar tower for $\ket{\frac{1}{4},\frac{1}{2}}$ is ($n=2q \ge 2$)
\begin{align}
 &J^+_{(2)} J^+_{(4)} \cdots J^+_{(2q)} \ket{\frac{1}{4},\frac{1}{2}} := \ket{q; \frac{1}{4},\frac{1}{2}} ,  \nonumber\\
 &L_0 \ket{q; \frac{1}{4},\frac{1}{2}} = \left(\frac{1}{4} +\sum_{m=1}^{q} 2m \right) \ket{q; \frac{1}{4},\frac{1}{2}}
 = \left(q+\frac{1}{2}\right)^2 \ket{q; \frac{1}{4},\frac{1}{2}} ,  \nonumber\\
 &J^3_0 \ket{q; \frac{1}{4},\frac{1}{2}} = \left(q+\frac{1}{2} \right) \ket{p; 0,0} ,
\end{align}
and for $J^-_{(0)} \ket{\frac{1}{4},\frac{1}{2}}$ ($n=-2q \le 0$)
\begin{align}
 &J^-_{(0)}  J^-_{(-2)} J^-_{(-4)} \cdots J^-_{(-2q)} \ket{\frac{1}{4},\frac{1}{2}} := \ket{-q; \frac{1}{4},\frac{1}{2}} ,  \nonumber\\
 &L_0 \ket{-q; \frac{1}{4},\frac{1}{2}} = \left(\frac{1}{4} +\sum_{m=1}^{q} 2m \right) \ket{q; \frac{1}{4},\frac{1}{2}}
 = \left(q+\frac{1}{2}\right)^2 \ket{q; \frac{1}{4},\frac{1}{2}} ,  \nonumber\\
 &J^3_0 \ket{q; \frac{1}{4},\frac{1}{2}} = -\left(q+\frac{1}{2} \right) \ket{p; 0,0} .
\end{align}
Again, we can see corresponding vertex operators
\begin{align}
 \ket{q; \frac14,\frac12} \quad &\leftrightarrow \quad :\! e^{+ i \frac{2q-1}{\sqrt 2} \varphi(z)} \!:  , \nonumber\\
 \ket{-q;\frac14,\frac12} \quad &\leftrightarrow \quad :\! e^{- i \frac{2q+1}{\sqrt 2} \varphi(z)} \!:  .
\end{align}
Recall that we still have $J^3_{-n}\ (n\ge 1) \sim \varphi^{(n)} = \frac{\partial^n}{\partial z^n} \varphi(z)$,
which can freely act on each tower.
Consequently,  the character of the spin-$j$ highest weight representation of $\widehat{\mathfrak{su}(2)}_{1}$
\begin{align}
 \chi_{2j}^{(1)} (\tau,z) = \Tr_{j} \left[ q^{L_0 -\frac{1}{24}} y^{J^3_0} \right] , \quad (y=e^{2\pi i z})
\end{align}
can be read off
\begin{align}
 &\chi_{0}^{(1)} (\tau,z) = \frac{\Theta_{0,1}(\tau,z)}{\eta(\tau)} ,  \nonumber\\
 &\chi_{1}^{(1)} (\tau,z) = \frac{\Theta_{1,1}(\tau,z)}{\eta(\tau)} ,   \nonumber
\end{align}
where
\[\Theta_{l,k}(\tau,z) := \sum_{n\in\mathbb{Z}+\frac{l}{2k}}
 q^{k n^2}y^{k n}\]
 reduces to the one
introduced in \eqref{Theta function} at $z=0$.
It is now obvious that the $\eta$-function is coming from the action of $\varphi^{(n)}$ ($n\ge 1$)
and the rest is from $:\! e^{i n\sqrt 2 \varphi(z)} \!:$ ($n \in \mathbb Z +j$).
See Fig.~\ref{fig:su2_1states} for the summary.

In general, irreducible representations of $\widehat{\mathfrak{su}(2)}_k$  require more careful treatment. Given a  highest weight state $|h,j\rangle$ in \eqref{HWS-su2},  one can obtain the Verma module $V(h,j)$ by acting $J_0^-$ and $J_n^a$ for $n<0$. However, like the Virasoro modules considered in \S\ref{sec:MM}, the Verma module is generally not irreducible  because it contains null states $|\chi\rangle$ which satisfy
\[
 J_{0}^{+}  |\chi \rangle=0 ,\qquad  J_{n}^{a}  |\chi \rangle=0 , \quad n > 0 , \quad a=0 , \pm ~.
\]
The module $J(h,j)$ generated by the null states $|\chi\rangle$ becomes a (maximal proper) submodule of $V(h,j)$. Subsequently,  the quotient module
\[L(h,j):=V(h,j)/J(h,j)\]
becomes irreducible, which is called the \textbf{integrable representation} of highest weight $|h,j\rangle$. For a general affine Lie algebra $\wh \frakg_k$, one can similarly construct an irreducible module and its character is called the \textbf{Weyl-Kac character formula}. Referring the reader to \cite[\S14.4]{francesco2012conformal} for the derivation, we just write the Weyl-Kac character formula of the integrable representation of highest weight $|h,\frac l2\rangle$ for  $\widehat{\mathfrak{su}(2)}_k$  for $0\leq l\leq k$
\begin{equation}
\label{su2k-character}
\chi_l^{(k)} (\tau,z)
= \frac{\Theta_{l+1,k+2}(\tau,z) - \Theta_{-l-1,k+2}(\tau,z)}
{\Theta_{1,2}(\tau,z)-\Theta_{-1,2}(\tau,z)}\, .
\end{equation}
We can calculate the modular $S$-matrix for the character
\eqref{su2k-character} as
\begin{equation}
\label{S-matrix character}
 S^{(k)}_{l l'} =
 \sqrt{\frac{2}{k+2}}
 \sin \bigg(
 \frac{\pi}{k+2}
 (l+1)(l'+1)
 \bigg)
 \qquad
 \text{with}\quad
 l,l' = 0, \dots , k\, .
\end{equation}

We now want to construct modular invariant partition
functions out of the characters \eqref{su2k-character}.
This amounts to determining all matrices
$M_{ll'}$ such that
\begin{equation}
\label{su2k-partition function}
 \mathcal{Z}_{\widehat{\mathfrak{su}(2)}_k}(\tau,\overline{\tau}) =
 \sum_{l,l'}
 \chi_l^{(k)}(\tau)
 M^{(k)}_{ll'}
 \overline{\chi_{l'}}^{(k)}(\overline{\tau})\, .
\end{equation}
is modular-invariant. The invariance under the $S$-transformation
requires
\begin{equation}
 S^{(k)T} M^{(k)} S^{(k)*} = M^{(k)}\, .
\end{equation}
$S^{(k)}$ in \eqref{S-matrix character} is symmetric
and real, thus we find the condition
\begin{equation}
 \comm{M^{(k)}}{S^{(k)}} = 0\, .
\end{equation}
We have already encountered this relation in the minimal
model modular invariant formula \eqref{relation N S}.
There are further constraints for entries $M^{(k)}_{ll'}$.
Since $M^{(k)}_{ll'}$ admits the interpretation
as the number of degeneracies of states in the Hilbert space,
they must be non-negative integers.
In addition, for the vacuum to only appear once, we have to
require $M^{(k)}_{00}=1$.
Furthermore, it turns out that in order for \eqref{su2k-partition function} to be invariant under
$T$-transformations, one has to satisfy the level-matching
condition $h_l - \overline{h}_{l'}\in \mathbb{Z}$.

\begin{table}[htbp]
	\centering
	\begin{tabular}{lll}
		\toprule
		Level
		& Partition function ($\mathcal{Z}_{\widehat{\mathfrak{su}(2)}_k}$)
		& Type
		\\
		\midrule
		$k=n$
		&
		$ \sum_{l=0}^{n}\abs{\chi_l}^2$
		& $A_{n+1}\, , \, n\geq 1 $
		\\
		&&\\
		$k = 4n$
		&
		$\sum_{l=0}^{n-1}
		\abs{\chi_{2l} + \chi_{k-2l}}^2
		+2|\chi_{\frac{k}{2}}|^2
		$
		& $
		D_{2n+2}\, , \, n\geq 1
		$
		\\
		&&\\
		$k=4n-2$
		& $
		 \sum_{l=0}^{k/2}
		|\chi_{2l}|^2
		+ \sum_{l=0}^{2n-2}
		\chi_{2l+1}\overline{\chi}_{k-2l-1}
		$
		& $
		D_{2n+1}\, , \, n\geq 2
		$
		\\
		&&\\
		$k=10$
		& $
		 \abs{\chi_0 + \chi_6}^2
		+ \abs{\chi_3 + \chi_7}^2
		+ \abs{\chi_4 + \chi_{10}}^2
		$
		& $
		E_6
		$
		\\
		&&\\
		$k= 16$
		&
		$
		 \abs{\chi_0+ \chi_{16}}^2
		+ \abs{\chi_4 + \chi_{12}}^2
		+ \abs{\chi_6 + \chi_{10}}^2
		$
		&
		$E_7$\\
		&
		$\quad\quad
		+(\chi_2 + \chi_{14})\overline{\chi}_8
		+\chi_8(\overline{\chi}_2 + \overline{\chi}_{14})
		+ \abs{\chi_8}^2
		$
		&
		\\
		&&\\
		$k=28$
		&
		$
		\abs{\chi_0 + \chi_{10}
			+\chi_{18}+\chi_{28}}^2
		$
		&
		$E_8$
		\\
		&
		$\quad\quad
		+ \abs{\chi_6 +\chi_{12}+ \chi_{16} + \chi_{22}}^2
		$
		&
		\\
		\bottomrule
	\end{tabular}
		\caption{All $\widehat{\mathfrak{su}(2)}_{k}$ modular
		invariant partition functions}
	\label{Tab:su2k-modular-invariant}
\end{table}

\subsubsection*{ADE classification of modular invariant partition functions}

Cappelli, Itzykson and Zuber have conjectured \cite[ISZ88-No.26]{cappelli1987modular} that all matrices $M$ with properties above for $\widehat{\mathfrak{su}(2)}_k$ characters are in one-to-one correspondence with the Dynkin diagrams of ADE types, which has been proven by themselves  \cite{cappelli1987ade} and Kato \cite{kato1987classification}. The corresponding partition functions are
listed in Table \ref{Tab:su2k-modular-invariant}, which is
known as the ADE classification. The ADE classification appears in the minimal $\cN=2$ superconformal field theories ($c<3$), which admits a geometric interpretation as Arnold's singularities \cite{Lerche:1989uy}. This interpretation plays an important role in the discovery of \textbf{mirror symmetry}. (See Itzykson's review in \cite{jimbo2014integrable}, and \cite{cappelli2009ade} and the references therein.)

In fact, the classification of modular invariant partition functions of the unitary minimal model $\cM_p$ attributes to the ADE classification of the modular invariant combinations of the $\widehat{\mathfrak{su}(2)}_k$ characters.  This stems from the fact that the coset model can describe the unitary minimal model $\cM_p$
\[
\frac{\widehat{\mathfrak{su}(2)}_{p-2}\oplus \widehat{\mathfrak{su}(2)}_{1}}{\widehat{\mathfrak{su}(2)}_{p-1}}~.
\]
As an easy check, the central charge of the coset model can be computed by \eqref{coset-central-charge}, yielding
\[
c=1-\frac{6}{p(p+1)}~,
\]
which is equal to \eqref{c<1-unitary-c}.
We refer the reader to \cite[\S18.3]{francesco2012conformal} for the details, and we just state the result in the following.
As seen in \S\ref{sec:characters-MM}, the modular invariant partition functions of $\cM_p$
\[
\cZ_{\cM_p}=\sum_{\substack{1\le s\le r\le p-1 \\ 1\le  \wt s\le \wt r\le p-1}}\cN_{(r,s),(\wt r,\wt s)}^{(p)}\chi_{(r,s)}\overline{\chi_{(\wt r,\wt s)}}
\]
are combinations of the Rocha-Caridi character $\chi_{r,s}$ in \eqref{Rocha-Caridi-formula}. This partition function turns out to be written in terms of two modular invariant partition combinations of the $\widehat{\mathfrak{su}(2)}_k$ characters
\bea \nonumber
\cZ_{\widehat{\mathfrak{su}(2)}_{p-1}}& = \sum_{r , \wt{r} = 1}^{p-1} M_{r , \wt{r}}^{( p-1 )} \chi_{r - 1}^{( p-1 )}  \overline {\chi_{\wt{r} - 1}^{( p-1 )} }  \cr \cZ_ {\widehat{\mathfrak{su}(2)}_{p-2}} &= \sum_{s , \wt{s} = 1}^{p-2} M_{s , \wt{s}}^{( p-2 )} \chi_{s - 1}^{( p-2 )}  \overline {\chi_{\wt{s} - 1}^{( p-2 )} } ~,
\eea
via
\[
\cN_{(r,s),(\wt r,\wt s)}^{(p)}=M_{r , \wt{r}}^{( p-1 )}M_{s , \wt{s}}^{( p-2 )}~.
\]
Hence, the modular invariant partition functions of $\cM_p$ can be specified by a pair of ADE Dynkin diagrams with adjacent levels as in Table \ref{Tab:ADE-minimal model}. Note that an $A$-series is always present in a pair. In particular, the partition function  \eqref{diagonal-PP} of diagonal type is of $(A_{p-1},A_p)$ type whereas the non-diagonal  partition function \eqref{3-state-Potts} of the 3-state Potts model is of $(A_4,D_4)$ type.

\begin{table}[htbp]
	\centering
	\begin{tabular}{lll}
		\toprule
		$p$
		& Partition function ($\cZ_{\cM_p}$)
		& Type
		\\
		\midrule
		&&\\
		$\geq 3$
		&
		$ \frac{1}{2}
		\sum_{r=1}^{p-1}\sum_{s=1}^{p}
		\abs{\chi_{r,s}}^2
		$
		& $ (A_{p-1},A_p) $
		\\
		&&\\
		\midrule
		&&\\
		$4\ell + 1$
		&
		$
		\sum_{r=1}^{p-1}\sum_{a=0}^\ell
		\abs{\chi_{r,2a+1}}^2
		+
		\sum_{r=1}^{p-1}\sum_{a=0}^{\ell-1}
		\chi_{r,2a+1}
		\overline{\chi}_{p-r,2a+1}
		$
		& $
		(A_{p-1},D_{2\ell+2})
		$
		\\
		&&\\
		$4\ell+2$
		& $
		\sum_{a=0}^{\ell}\sum_{s=1}^p
		\abs{\chi_{2a+1,s}}^2
		+
		\sum_{a=0}^{\ell-1}\sum_{s=1}^{p} \chi_{2a+1,s}
		\overline{\chi}_{2a+1,m+1-s}
		$
		& $
		(D_{2\ell+2},A_p)
		$
		\\
		&&\\
		$4\ell +3$
		& $
		\sum_{r=1}^{p-1}
		\sum_{a=0}^{\ell}
		\abs{\chi_r,2a+1}^2
		+
		\sum_{r=1}^{2\ell+1} \abs{\chi_{r,2\ell+2}}^2
		+\sum_{r=1}^{p-1}\sum_{a=1}^\ell
		\chi_{r,2a} \overline{\chi}_{p-r,2a}
		$
		& $
		(A_{p-1},D_{2\ell+3})
		$
		\\
		&&\\
		$4\ell+4$
		&
		$
		\sum_{a=0}^\ell\sum_{s=1}^p
		\abs{\chi_{2a+1,s}}^2 +
		\sum_{s=1}^{2\ell+2}
		\abs{\chi_{2\ell+2,s}}^2
		+\sum_{a=1}^\ell\sum_{s=1}^p
		\chi_{2a,s}\overline{\chi}_{2a,m+1-s}
		$
		&
		$(D_{2\ell+4, A_p})$\\
		&&\\
		\midrule
		&&\\
		$11$
		&
		$
		\frac{1}{2}
		\sum_{r=1}^{10}\qty[
		\abs{\chi_{r,1}+\chi_{r,7}}^2
		+\abs{\chi_{r,4}+\chi_{r,8}}^2
		+ \abs{\chi_{r,5}+\chi_{r,11}}^2
		]
		$
		&
		$(A_{10},E_6)$
		\\
		&&\\
		$12$
		&
		$
		\frac{1}{2}
		\sum_{s=1}^{12} \qty[
		\abs{\chi_{1,s}+\chi_{7,s}}^2
		+ \abs{\chi_{4,s}+\chi_{8,s}}^2
		+ \abs{\chi_{5,s}+\chi_{11,s}}^2
		]
		$
		&$(E_6 , A_{12})$
		\\
		&&\\
		\midrule
		&&\\
		$17$
		&
		$\frac{1}{2}
		\sum_{r=1}^{16} \Big[
		\abs{\chi_{r,1}+\chi_{r,17}}^2
		+\abs{\chi_{r,5}+\chi_{r,13}}^2
		+\abs{\chi_{r,7}+\chi_{r,11}}^2
		$
		&$(A_{16},E_7)$
		\\
		&
		$+\abs{\chi_{r,9}}^2 + (\chi_{r,3}+\chi_{r,15})^{*}\chi_{r,9}
		+\chi_{r,9}^*(\chi_{r,3}+\chi_{r,15})\Big]$
		&
		\\
		&&\\
		$18$
		&
		$\frac{1}{2}
		\sum_{s=1}^{18} \Big[
		\abs{\chi_{1,s}+\chi_{17,s}}^2
		+\abs{\chi_{5,s}+\chi_{13,s}}^2
		+\abs{\chi_{7,s}+\chi_{11,s}}^2
		$
		&
		$(E_7,A_{18})$
		\\
		&
		$+\abs{\chi_{9,s}}^2
		+(\chi_{3,s}+\chi_{15,s})^*\chi_{9,s}
		+\chi^*_{9,s}(\chi_{3,s} + \chi_{15,s})
		\Big]
		$
		&\\
		&&\\
		\midrule
		&&\\
		$29$
		&
		$
		\frac{1}{2}\sum_{r=1}^28\Big[
		\abs{\chi_{r,1}+\chi_{r,11}+\chi_{r,19}+\chi_{r,29}}^2
		+\abs{\chi_{r,7}+\chi_{r,13}+\chi_{r,17}+\chi_{r,23}}^2\Big]
		$
		&
		$(A_{28},E_8)$
		\\
		&&\\
		$30$
		&
		$ \frac{1}{2}
		\sum^{30}_{s=1}\Big[
		\abs{\chi_{1,s}+\chi_{11,s}+\chi_{19,s}+\chi_{29,s}}^2
		+ \abs{\chi_{7,s}+\chi_{13,s}+\chi_{17,s}+\chi_{23,s}}^2\Big]
		$
		&
		$(E_8,A_{30})$
		\\
		&&\\
		\bottomrule
	\end{tabular}	\caption{Modular invariant partition functions of unitary minimal models $\cM_p$. }
	\label{Tab:ADE-minimal model}
\end{table}

\section{Entanglement entropy}

In this section, we will learn how 2d CFTs are related to entanglement entropy \cite{Calabrese:2004eu} and holography.  There are nice reviews \cite{Calabrese:2009qy,Nishioka:2009un,Takayanagi:2012kg,Solodukhin:2011gn,Harlow:2014yka,Rangamani:2016dms}. At the end of this section, we will introduce the celebrated Ryu-Takayanagi entanglement entropy formula \cite{Ryu:2006bv,Ryu:2006ef}.

\subsection{Introduction to entanglement entropy}\label{sec:EE}
In this subsection, we briefly introduce some basic concepts
and properties in entanglement entropy.

In quantum mechanics, the state of a quantum system is represented by a state vector, denoted $| \psi \rangle$, whose time evolution is governed by the Schr\"odinger equation
\[
i \frac{\partial}{\partial t} | \Psi \rangle = H | \Psi \rangle
\]
A quantum system with a state vector $|\psi\rangle$ is referred to as a \textbf{pure state}. However, a system can also exist in a statistical ensemble of different state vectors. For example, there might be a 50\% probability that the state vector is $|\psi_1\rangle$ and a 50\% probability that it is $|\psi_2\rangle$. Such a system is in a \textbf{mixed state}. The density matrix $\rho_{\text{tot}}$ is particularly useful for describing mixed states, as it characterizes any state, whether pure or mixed, using a single matrix.

The density matrix $\rho_{\text{tot}}$ describes the statistical state of a system in quantum mechanics. For instance, in the canonical ensemble for a system with Hamiltonian $H$ and temperature $T$, the density matrix is given by
\[
\rho_{\text{tot}}= \frac{e^{- \beta H}}{\cZ}~.
\]
Note that we always normalize $\Tr_{\cH_{\text{tot}}} \rho_{\text{tot}}$. Then, the entropy can be expressed in terms of the density matrix
\be\label{VN-entropy}
S=\frac{E-F}{T}=\frac{- 1}{\cZ} \left( \operatorname {Tr}_{\mathcal {H}_{\textrm{tot}}} \left[ \beta H e^{- \beta H} \right] + Z \log Z \right)=-\rho_{\text{tot}} \log \rho_{\text{tot}}
\ee
which is called \textbf{von Neumann entropy}.

If we express a pure state by the density operator
\begin{equation}
  \rho=\ket{\phi}\bra{\phi} \, ,
\end{equation}
the expectation value of operator $\cO$ can be expressed by
density operator
\begin{equation}
 \expval{\cO}_\rho := \Tr(\rho \cO)\, .
\end{equation}

\begin{figure}[ht]\centering
\includegraphics[width=8cm]{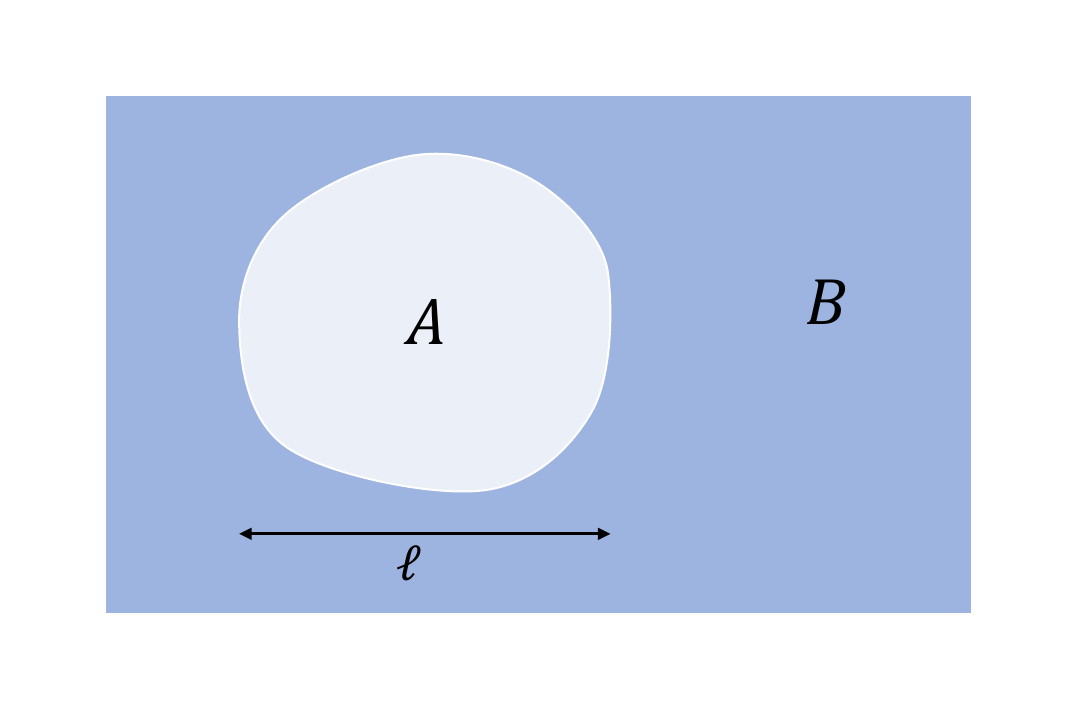}
	\caption{The subsystem $A$ on which we consider entanglement entropy and its complement $B$.}
	\label{fig:ee-illustration}
\end{figure}

Let us consider the situation in which a system is divided into two subsystems $A$ and $B$ where  the Hilbert space is simply a direct product
\[
\mathcal {H}_ {\textrm{tot}} = \mathcal {H}_{A} \otimes \mathcal {H}_{B}
\]
Then, we can consider the density operator $\rho_A$  restricted to the subsystem $A$ by taking the trace over $\cH_B$
\begin{equation}
 \rho_A := \Tr_{\cH_B}(\rho_{\text{tot}})\, .
\end{equation}
Then, the \textbf{entanglement entropy} in the subsystem $A$  is defined as
\begin{equation}
  S_A := -\Tr_{\cH_A}(\rho_A \log \rho_A)\, ,
\end{equation}
analogous to the von Neumann entropy \eqref{VN-entropy}.

For example, consider a situation where $\mathcal{H}_A$ and $\mathcal{H}_B$ are two Hilbert spaces spanned by the states $\{ |0\rangle_A, |1\rangle_A \}$ and $\{ |0\rangle_B, |1\rangle_B \}$, respectively. Suppose we have a state:
\begin{equation}
\label{entanglement-state}
  \ket{\phi}=\frac{1}{\sqrt{2}}
  (\ket{1}_A \otimes \ket{0}_B+\ket{0}_A\otimes\ket{1}_B)
\end{equation}
No matter how far apart the subsystems are, if a measurement in subsystem $A$ yields $ |1\rangle_A $, then the measurement in subsystem $B$ will definitely yield $ |0\rangle_B $, and vice versa. This indicates that subsystem $ A $ is entangled with subsystem $ B $, a phenomenon known as \textbf{quantum entanglement}.

In fact, the density matrix for the state \eqref{entanglement-state},
\bea
\rho_{\text{tot}} &= |\phi\rangle \langle \phi|\cr
&= \frac{1}{2}
(\ket{0}_A\bra{0}_A\otimes\ket{1}_B\bra{1}_B
+\ket{0}_A\bra{1}_A\otimes\ket{1}_B\bra{0}_B\notag\\
&+\ket{1}_A\bra{0}_A\otimes\ket{0}_B\bra{1}_B
+\ket{1}_A\bra{1}_A\otimes\ket{0}_B\bra{0}_B)\,,
\eea
By tracing out the degrees of freedom associated with subsystem $ B $, we obtain the reduced density matrix for subsystem $ A $
\begin{equation}
 \rho_A:=\Tr_{\cH_B}(\rho_{\text{tot}})=\frac{1}{2}
 (\ket{0}_A\bra{0}_A+ \ket{1}_A\bra{1}_A)\,~.
\end{equation}
Thus the entanglement entropy can be evaluated as
\begin{equation}
  S_A=\log 2\,.
\end{equation}
This result is related to the fact that both Hilbert spaces $ \mathcal{H}_A $ and $ \mathcal{H}_B $ are two-dimensional. In fact, the maximum entanglement entropy is given by $ \min(\log \dim \mathcal{H}_A, \log \dim \mathcal{H}_B) $. Therefore, the state \eqref{entanglement-state} is maximally entangled.

On the other hand,  a state $\ket{\psi}$ is called \textbf{separate} if it can be written as
the direct product of two pure states, i.e
\begin{equation}
 \ket{\psi}=\ket{\phi}_A \otimes \ket{\phi}_B \, ,
 \qquad
 \ket{\phi}_A \in \cH_A\, \&
 \ket{\phi}_B \in \cH_B\, .
\end{equation}
In this case, it is easy to check that the entanglement entropy vanishes, meaning there is no entanglement between the two subsystems.

In conclusion, the information lost by tracing over the Hilbert space $ \mathcal{H}_B $ is somehow recovered in $ S_A $ through quantum entanglement. Thus, entanglement entropy plays a crucial role in situations where we cannot observe subsystem $ B $.

There are several equivalent formulas for entanglement entropy that will be useful in subsequent calculations:
\bea
S_A=\lim_{n\to 1} \frac{\Tr_{\cH_A} \rho^n_A -1}{1-n}
=-\pdv{}{n}\Tr_{\cH_A} \rho^n_A|_{n=1}
=- \pdv{}{n}\log \Tr_{\cH_A} \rho_A^n|_{n=1}\,.
\eea

\subsubsection*{Properties of entanglement entropy}
Entanglement entropy has several general properties that are particularly useful. Some of these properties are summarized as follows:
\begin{itemize}
\item
 Suppose $B$ is the complement of $A$. If the density matrix $\rho_{\text{tot}}$ is constructed from a pure state, such as in a zero-temperature system, then the following relation holds
\begin{equation}
\label{complement-prop}
 S_A=S_B \, .
\end{equation}
    Thus, entanglement entropy is not an extensive quantity. This relation, however, does not hold for mixed states.

\item
For three
non-intersecting subsystems $A$, $B$, and $C$, the following two inequalities
hold
\bea
\label{strong-subadditivity-property}
S_{A+B+C}+S_B &\leq S_{A+B}+S_{B+C}\,,\cr
S_A+S_C &\leq S_{A+B}+S_{B+C}\, ,
\eea
which are called \textbf{strong subadditivity} \cite{lieb1973fundamental}.

\item
If the subsystem $B$ is empty in the above setting, they reduce to the subadditivity relation
\begin{equation}
\label{subadditivity relation}
S_{A+B} \leq S_A+S_B \, .
\end{equation}
\item
We thus are allowed to define a 
non-negative quantity called \textbf{mutual
information} $I(A,B)$ by
\begin{equation}
  I(A,B)=S_A+S_B-S_{A+B}\geq 0 \, .
\end{equation}
\end{itemize}

\subsection{Conformal field theory on \texorpdfstring{$\bZ_n$}{Zn} orbifold}

Before exploring entanglement entropy in conformal field theory, we shall generalize the twist operator discussed in \S\ref{sec:Ising} to the $\mathbb{Z}_n$ orbifold. In \S\ref{sec:Ising}, the continuum limit of the Ising model is described by the free fermion $\psi$, where the presence of a spin field $\sigma$ introduces the $\mathbb{Z}_2$ twist condition to the free fermion in the complex plane:
\begin{equation}
  \psi(e^{2\pi i}z)=- \psi(z) \, ,
\end{equation}
This means that the field operator acquires a minus sign when taken around the origin, a global effect that renders the field operator double-valued in space-time. The $\mathbb{Z}_2$ twist operator $\sigma$ can be generalized to any finite group. Specifically, we are interested in the $\mathbb{Z}_n$ twist, defined as
\begin{equation}
\label{ZN-orbifold-definition}
 X(e^{2\pi i}z, e^{-2\pi i}\overline{z}) =
 e^{2\pi k i /N} X(z,\overline{z})\, ,
\end{equation}
where $k$ is an integer called the monodromy of the field $X$. For the definition of \eqref{ZN-orbifold-definition} to make sense, $X$ must be complex. Therefore, we combine two copies of free real scalar field $\varphi_1$ and $\varphi_2$ to introduce the complex free boson:
\begin{equation}
X=\varphi_1(z,\overline{z})+i \varphi_2(z,\overline{z})\, ,
\qquad
\overline{X}=\varphi_1(z,\overline{z})-i \varphi_2(z,\overline{z})\, .
\end{equation}
Since we have studied free scalar fields in \S\ref{sec:free}, the properties of complex free
boson are easy to derive. Let us list some of them in brief.
First, the holomorphic fields for complex free bosons are as follows:
\def\pd{\partial}
\begin{equation}
  \pd X(z)=\pd \varphi_1(z)+i \pd \varphi_2(z) \, .
\end{equation}
The holomorphic energy-momentum tensor is
\begin{equation}
 T(z)=\frac{1}{2} \big(
 :\pd\varphi_1 \pd\varphi_1:
 +
 :\pd\varphi_2\pd\varphi_2:
 \big)
=-\frac{1}{2}
 :\pd X \pd \overline{X} : \, ,
\end{equation}
which is just the sum of the energy-momentum tensor
of two real scalar fields. The central charge of this theory
becomes $2$. The OPE of $\pd X$ and $\pd\overline{X}$ is
\begin{equation}
  \pd X(z) \pd X(w) \sim
  -\frac{2}{(z-w)^2} \, .
\end{equation}
In a similar fashion to \eqref{def-EM-tensor 2}, the vacuum expectation value of energy-momentum is
\begin{equation}
 \expval{T(z)} := \bigg[
 \expval{
 -\frac{1}{2} \pd X(z) \pd \overline{X}(w)
}
-\frac{1}{(z-w)^2}
 \bigg]_{z=w}\,.
\end{equation}
To generalize the definition, we need to replace the original vacuum state with the twist
ground state. The twist ground state
for the $\bZ_n$ orbifold is defined by acting a twist operator on the vacuum state at the zero point.
\begin{equation}
  \sigma_{k/n}(0)\ket{0}=\ket{\sigma_{k/n}}\, .
\end{equation}
$\sigma_{k/n}$ is defined to twist the field $X$ by
$e^{2\pi i k/n}$ when $X$ turns around it counterclockwise.
The conjugate state then is given by inserting anti-twist
operator $\sigma_{-k/n}$  at infinity which is denoted by
$\bra{\sigma_{-k/n}}$. By introducing twist operators, the global ground effect can be illustrated by inserting a local operator \cite[ISZ88-No.47]{Dixon:1986qv}.

\def\sm{\bra{\sigma_{-k/n}}}
\def\sp{\ket{\sigma_{k/n}}}

The expectation value of energy-momentum in the $\bZ_n$ twist
ground state now can be evaluated as
\begin{equation}
\label{def-vacuum-expectation-e-m-tensor-complex-boson}
  \frac{\bra{\sigma_{-k/n}} T(z) \ket{\sigma_{k/n}}}
 {\bra{\sigma_{-k/n}}\ket{\sigma_{k/n}}} :=
  \bigg[
  \frac{\sm -\frac{1}{2}\pd X(z)\pd \overline{X}(w) \sp}
 {\bra{\sigma_{-k/n}}\ket{\sigma_{k/n}}}
 -\frac{1}{(z-w)^2}
  \bigg]_{z=w}\, ,
\end{equation}
where the normalized factor in the denominator can be adjusted
to $1$.
We have calculated this quantity in the R sector of free fermion. To perform the calculation, we need to study the mode expansion of $X$ first.

From \eqref{ZN-orbifold-definition}, we immediately obtain that
\bea
 \pd X(e^{2\pi i} z , e^{-2\pi i}\overline{z})
 &= e^{2\pi i(k/n-1)} \pd X(z,\overline{z})\, ,\\
 \pd \overline{X} (e^{-2\pi i}\overline{z},e^{2\pi i} z)
 &= e^{-2\pi i(k/n+1)} \pd \overline{X}(\overline{z},z)\, .
\eea
Therefore, the Laurent expansions of $\pd X$ and $\pd \overline{X}$
must have the form
\bea
\label{ZN-twist-mode}
 i\pd_z X &= \sum_{m\in\mathbb{Z}}
 \alpha_{m-k/n} z^{-m-1+k/n} \, , \notag\\
 i\pd_z \overline{X} &=\sum_{m\in\mathbb{Z}}
 \overline{\alpha}_{m+k/n} z^{-m-1-k/n}\, .
\eea
The mode operators have the following canonical commutation
relations
\begin{equation}
 \comm{\overline{\alpha}_{m+k/n}}{\alpha_{n-k/n}}
 =2(m+k/n)\delta_{m,-n}\, ,
\end{equation}
where $2$ comes from the fact that a complex boson field
consists of two real scalar fields. And the twist ground
state $\sp$ is annihilated by all the positive frequency
mode operators
\bea
  \alpha_{m-k/n} \sp &= 0\, , \qquad m>0 \,,\\
  \overline{\alpha}_{m+k/n}\sp &=0\, , \qquad m\geq 0\, .
\eea
Using the mode expansion \eqref{ZN-twist-mode} and their
properties we mentioned above, we can calculate the expectation
value in the twist ground state
\begin{equation}
 -\frac{1}{2}\sm\pd X(z)\overline{\pd}X(w)\sp
=z^{-(1-k/n)}w^{-k/n}
 \bigg[
 \frac{(1-k/n)z+k w/n}{(z-w)^2}
 \bigg] \, .
\end{equation}
From \eqref{def-vacuum-expectation-e-m-tensor-complex-boson},
we can compute
\begin{equation}
\label{T-expval-twist-ground}
 \sm T(z) \sp =\frac{1}{z^2}\cdot \frac{1}{2}\frac{k}{n}\bigg(1-\frac{k}{n}\bigg)\, .
\end{equation}
From our previous studies, we know that as a local operator, the OPE of $T(z)$ with $\sigma_{k/n}(0)$ should have the following form
\begin{equation}
 T(z)\sigma_{k/n}(0) =\cdots \frac{h \sigma_{k/n}(0)}{z^2}+\cdots\, ,
\end{equation}
where $h$ is the conformal dimension of the twist field $\sigma_{k/n}$. Plugging into \eqref{T-expval-twist-ground}, we immediately find that the conformal dimension of $\sigma_{k/n}$ is
\be \label{twist-conf-dim}
h_{k/n} = \frac{1}{2} \cdot \frac{k}{n} \left(1 - \frac{k}{n} \right)
\ee 
The same analysis can be applied for $\sigma_{-k/n}$ and the anti-holomorphic sector.

\subsection{Entanglement entropy in 2d CFTs}
Next, we study the entanglement entropy in 2d CFT. We define a subsystem $A$ at a fixed time $t = t_0$, with its complement denoted as $B$. The boundary of $A$ is denoted by $\partial A$. Given that we normalize the reduced density matrix such that $\text{Tr}(\rho_A) = 1$, the entanglement entropy $S_A$ is defined as:
\be\label{SA-cal-formula}
\begin{aligned}
S_{A} &=\lim _{n \rightarrow 1} \frac{\operatorname{Tr}_{\mathcal{H}_{A}} \rho_{A}^{n}-1}{1-n} \\
&=-\left.\frac{\partial}{\partial n} \operatorname{Tr}_{\mathcal{H}_{A}} \rho_{A}^{n}\right|_{n=1}=-\left.\frac{\partial}{\partial n} \log \operatorname{Tr}_{\mathcal{H}_{A}} \rho_{A}^{n}\right|_{n=1} .
\end{aligned}
\ee
This is called the \textbf{replica trick}. Therefore, what we need to do is to evaluate $\Tr_{\cH_A}\rho^n_A$ in our 2d system.

To this end, let us recall how a wave function of the ground state can be expressed in the path-integral. Here, we consider the Euclidean 2d QFT, and we write the coordinates as $(x_0,x_1)$.
The probability that the field configuration becomes $\phi\left(x_{1}\right)$ at $x_0=0$ is given by the following path-integral expression
\begin{equation}
\Psi\left[\phi\left(x_{1}\right)\right] =\frac{1}{\sqrt{\cZ_1}} \int_{x_0=-\infty}^{x_0=0} \mathcal{D} \phi~ e^{-\mathcal{S}(\phi)} \delta\left[\phi\left(0, x_{1}\right)-\phi\left(x_{1}\right)\right]~.
\end{equation}
In fact, the integration from $x_0=0$ to $x_0=-\infty$ corresponds to the operation $\lim_{T\to \infty} e^{-TH}$ so that this realizes the ground state wave function. Likewise, we can define its complex conjugate
\begin{equation}
\Psi^{*}\left[\phi\left(x_{1}\right)\right] =\frac{1}{\sqrt{\cZ_1}} \int_{x_0=0}^{x_0=\infty} \mathcal{D} \phi~e^{-\mathcal{S}(\phi)} \delta\left[\phi\left(0, x_{1}\right)-\phi\left(x_{1}\right)\right]~.
\end{equation}

\begin{figure}[ht]
	\centering
	\includegraphics[width=0.3\linewidth]{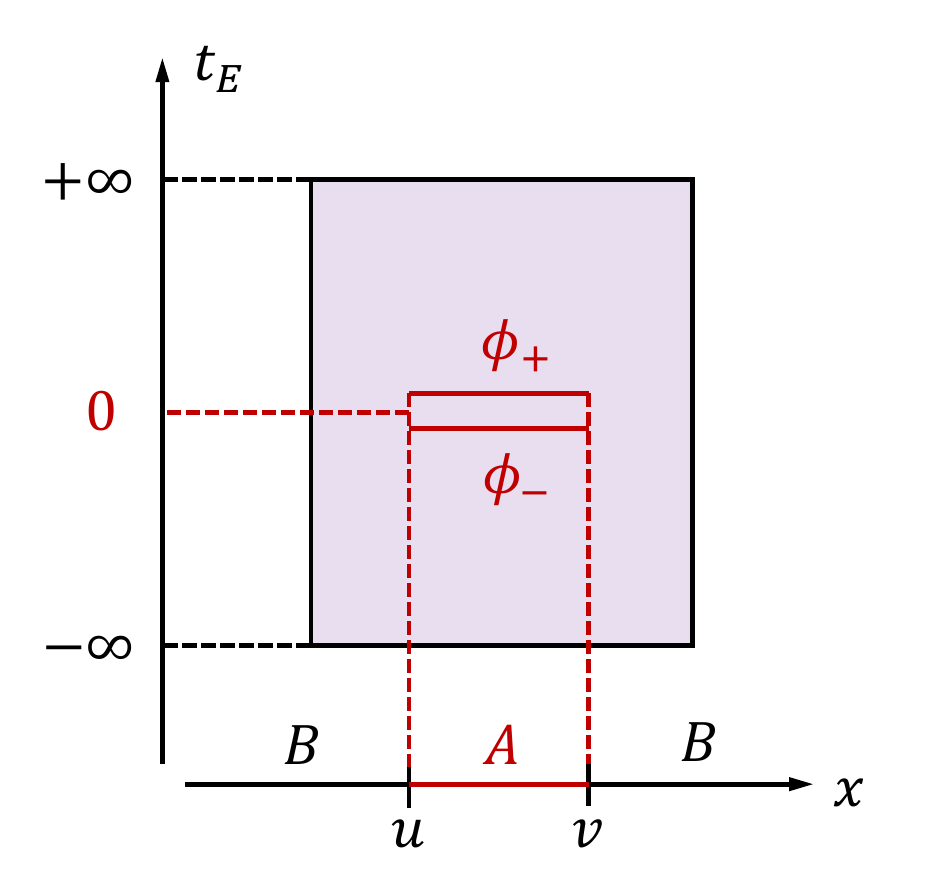}
\qquad
		\includegraphics[width=0.3\linewidth]{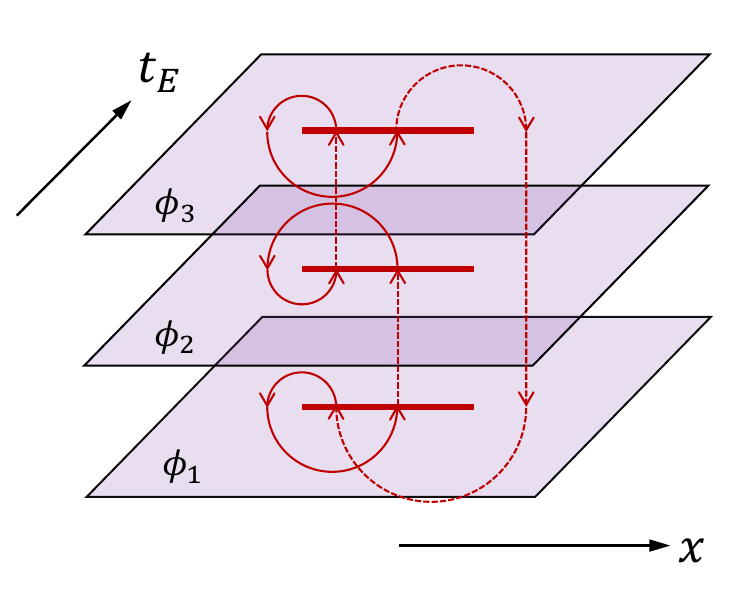}
	\caption{Left: the reduced density matrix $[\rho_A]_{\phi_-\phi_+}$ admits an interpretation of the path-integral over the slit with boundary condition $(\phi_-,\phi_+)$. Right: $\Tr_{\cH_A} \rho^n_A$ can be understood as the path-integral over $n$-sheeted Riemann surface $\mathcal{R}_n$}
	\label{fig:9-1}
\end{figure}

Now, the total density matrix $\rho_{\textrm{tot}}$ is given by the product of two wave functions
\be \left[\rho_{\textrm{tot}}\right]_{\phi_{-}\left(x_{1}\right), \phi_{+}\left(x_{1}^{\prime}\right)}=\Psi\left[\phi_{-}\left(x_{1}\right)\right] \Psi^{*}\left[\phi_{+}\left(x_{1}^{\prime}\right)\right] ~.\ee
Using the above path-integral expressions, the density matrix restricted to $A$ can be written as
\begin{equation}
\label{reduce-density-matrice}
\left[\rho_{A}\right]_{\phi_{-} \phi_{+}}=\frac{1}{\cZ_1} \int_{x_0=-\infty}^{x_0=\infty} \mathcal{D} \phi~ e^{-\mathcal{S}(\phi)} \prod_{x \in A} \delta\left(\phi(-0, x)-\phi_{-}(x_{1})\right) \cdot{\delta}\left(\phi(+0, x_{1})-\phi_{+}(x_{1})\right)~,
\end{equation}
where $\phi_-$ and $\phi_+$ are both boundary wave functions at $A$. Note that taking the trace over the complement $B$ amounts to the integration over the subsystem $B$ at $x_0=0$. (See Figure \ref{fig:9-1}.)

To evaluate $\Tr_{\cH_A} \rho^n_A$, we just need to multiply the $n$ copies of
\eqref{reduce-density-matrice} with the same boundary conditions for the adjacent matrices
\begin{equation}
\left(\prod_{j=1}^{n}\left[D \phi_{j}\right]\right)\left[\rho_{A}\right]_{\phi_{1} \phi_{2}}\left[\rho_{A}\right]_{\phi_{2} \phi_{3}} \cdots\left[\rho_{A}\right]_{\phi_{n} \phi_{1}} .
\end{equation}
This corresponds to gluing $n$-sheets along the subsystem $A$ successively as in Figure \ref{fig:9-1}. Writing the resultant $n$-sheeted Riemann
surface $\mathcal{R}_n$, the final path-integral expression can be written as
\begin{equation}
 \Tr_{\cH_A} \rho^n_A=(\cZ_1)^{-n} \int_{(x_0,x_1)\in\mathcal{R}_n}
 \mathcal{D}\phi~e^{-\cS(\phi)} \equiv \frac{\cZ_n}{(\cZ_1)^n}\,.
\end{equation}

\begin{figure}[ht]
	\centering
	\includegraphics[width=0.5\linewidth]{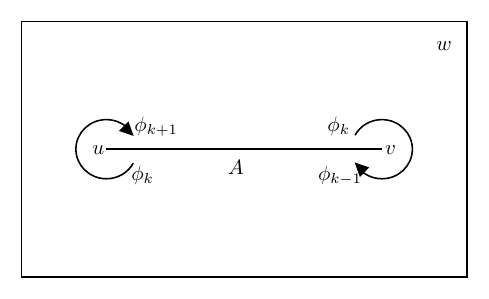}
	\caption{Twisted boundary condidions on replica fields}
	\label{fig:replica}
\end{figure}

To evaluate this path-integral, we consider that there is a distinct field at each sheet, called \textbf{replica fields},
and we denote them by $\phi_k(x_0,x_1) (k=1, 2,\cdots n)$. To realize the field on the $n$-sheeted Riemann
surface $\mathcal{R}_n$, we impose the twisted boundary conditions
\begin{equation}
\phi_k (e^{2\pi i}(w-u))=\phi_{k+1}(w-u)\,, \qquad
\phi_k(e^{2\pi i}(w-v))=\phi_{k-1}(w-v) \, ,
\end{equation}
where we write the complex coordinate $w=x_0+ix_1$, and $u$ and $v$ are the endpoints of the subsystem $A$. (See Figure \ref{fig:replica}.)
Assuming that $\phi$ is a complex scalar field with central charge $c=2$, then we
can introduce $n$ new fields
$\wt{\phi}_k=\frac{1}{n}\sum_{l=1}^n e^{2\pi ilk/n} \phi_l$. They obey the boundary condition
\begin{equation}
 \wt{\phi}_k(e^{2\pi i}(w-u))=e^{2\pi ik/n}
 \wt{\phi}_k(w-u)\,,
 \qquad
 \wt{\phi}_k(e^{2\pi i}(w-v))=e^{-2\pi i k/n}
 \wt{\phi}_k(w-v)\, .
\end{equation}
As we have learned in the last section, the system is equivalent to $n$-disconnected sheets with two twist operators $\sigma_{k/n}$ and $\sigma_{-k/n}$ inserted in the $k$-th sheet
for each value of $k$. In the end, we find
\begin{equation}
\label{trrhonA}
 \Tr_{\cH_A} \rho^n_A=\prod_{k=0}^{n-1}
 \expval{\sigma_{k/n}(u)\sigma_{-k/n}(v)}\sim
 (u-v)^{-4\sum_{k=0}^{n-1}h_{k/n}}
 =(u-v)^{-\frac{1}{3}(n-1/n)}\,,
\end{equation}
where $h_{k/n}$ is the
conformal dimension of $\sigma_{k/n}$ given in \eqref{twist-conf-dim}. When we have $m$ such
complex scalar fields we simply obtain
\begin{equation}
\label{trace-rho-A-n}
 \Tr_{\cH_A} \rho^n_A \sim (u-v)^{-\frac{c}{6}(n-1/n)}\, ,
\end{equation}
setting the central charge $c= 2m$.
Applying the formula \eqref{SA-cal-formula} to \eqref{trace-rho-A-n}, we find that
\begin{equation}
 S_A \sim \frac{c}{3}\log \ell\,
\end{equation}
where we set $\ell\equiv u-v$. Introducing the UV cut-off $\e$ into the expression above we get
\begin{equation}
\label{EE-infinite}
 S_A=\frac{c}{3}\log \frac{\ell}{\e}\, .
\end{equation}

\subsubsection*{Entanglement entropy of finite size}
In the previous subsection, we assume the space direction $x$ is
infinite, while in this section we consider the subsystem
$A$ is defined in a finite space region.
\begin{equation}
 A=\{
 x| x\in [r, s]
 \}\, ,
\end{equation}
where we assume $-\frac{L}{2} < r < s \leq \frac{L}{2}$.
The system is related to the previous system via the
conformal map
\begin{equation}
 w=\tan\bigg(
 \frac{\pi \w '}{L}
 \bigg)\, .
\end{equation}
And $u=\tan \big(
\frac{\pi r}{L}\big)$ and
$v=\tan\big(\frac{\pi s}{L}\big)$.
From \eqref{trrhonA}, we can calculate $\Tr_{\cH_A} \rho^n_{A\omega'} $
by applying conformal transformation of the local twist field.
\bea
\Tr_{\cH_A} \rho^n_{A\omega'}
&= \prod_{k=0}^{n-1}
\bigg(
\dv{w}{w'}
\bigg)^{2h_{k/n}}_r
\bigg(
\dv{w}{w'}
\bigg)^{2h_{k/n}}_s
\expval{\sigma_{k/n}(u) \sigma_{-k/n}(v)}\notag\\
& \sim \bigg[
\frac{L}{\pi}\cos\bigg(\frac{\pi}{L}r\bigg)
\cos\bigg(\frac{\pi}{L}s\bigg) (u-v)
\bigg]^{-\frac{c}{6}(n-1/n)}\notag\\
& \sim \bigg[
\frac{L}{\pi} \sin\frac{\pi}{L}(r-s)
\bigg]^{-\frac{c}{6}(n-1/n)}\,.
\eea
Following the same calculational step, we
obtain the entanglement entropy in a finite space region
\begin{equation}\label{EE-circle}
S_A=\frac{c}{3} \cdot \log\bigg(
\frac{L}{\pi \e} \sin\bigg(
\frac{\pi \ell}{L}
\bigg)
\bigg)\, ,
\end{equation}
where $\ell= r-s$. This result is invariant under the $\ell\to L-\ell$, satisfying the property \eqref{complement-prop}.

\subsubsection*{Entanglement entropy at finite temperature}
In the Euclidean 2-dimension theory, the Euclidean
time is equivalent to the inverse of temperature. As in statistical mechanics \eqref{Boltzmann}, we can compactify the Euclidean
time as $t_E \sim t_E+\beta$ at finite temperature $T=\beta^{-1}$. We can map this system
to the infinite system via the conformal map
\begin{equation}
 w=e^{\frac{2\pi}{\beta} w'}\, .
\end{equation}
And $u=e^{\frac{2 \pi r}{\beta}}$ ,
$v=e^{\frac{2\pi s}{\beta}}$. $t_E$ is the imaginary part of
$w'$. This conformal map will lead to the extra factor
\begin{equation}
\bigg[
  \frac{\beta}{2 \pi} e^{-\frac{\pi}{\beta}}
\bigg]^{-\frac{c}{6}(n-1/n)}
\end{equation}
Following the similar calculation steps, we obtain
\begin{equation}\label{entropytemp}
S_A=\frac{c}{3} \log
\bigg(
\frac{\beta}{\pi \e}
\sinh\bigg(
\frac{\pi \ell}{\beta}
\bigg)
\bigg)\, ,
\end{equation}
where $\ell= r-s$. In the zero temperature limit $T\to 0$,
this reduces to the previous result \eqref{EE-infinite}.
In the high-temperature limit  $T \to \infty$, it
approaches
\begin{equation}\label{EE-finite}
S_A \simeq \frac{\pi c}{3}\ell T+\frac{c}{3} \log \frac{\beta}{2 \pi \epsilon} \,.
\end{equation}
The leading term is the thermodynamic entropy, which is proportional to the volume of the subsystem $A$ (extensive property)
as expected.

\subsection{Zamolodchikov \texorpdfstring{$c$}{c}-theorem revisited}

In \S\ref{sec:Zc}, we discussed Zamolodchikov $c$-theorem stating that in a unitary, rotational (Lorentz) invariant 2d theory, there exists a monotonically decreasing function $C$ with respect to the length scale $R$
\[
\frac{d C (R )}{d R} \leq 0
\]
where   $C(R=0)$ is the central charge at ultra-violet and $C(R=\infty)$ is the one at infra-red. This theorem can be shown easily by using the strong subadditivity \eqref{strong-subadditivity-property} of entanglement entropy.

\begin{figure}[ht]\centering
\includegraphics[width=9cm]{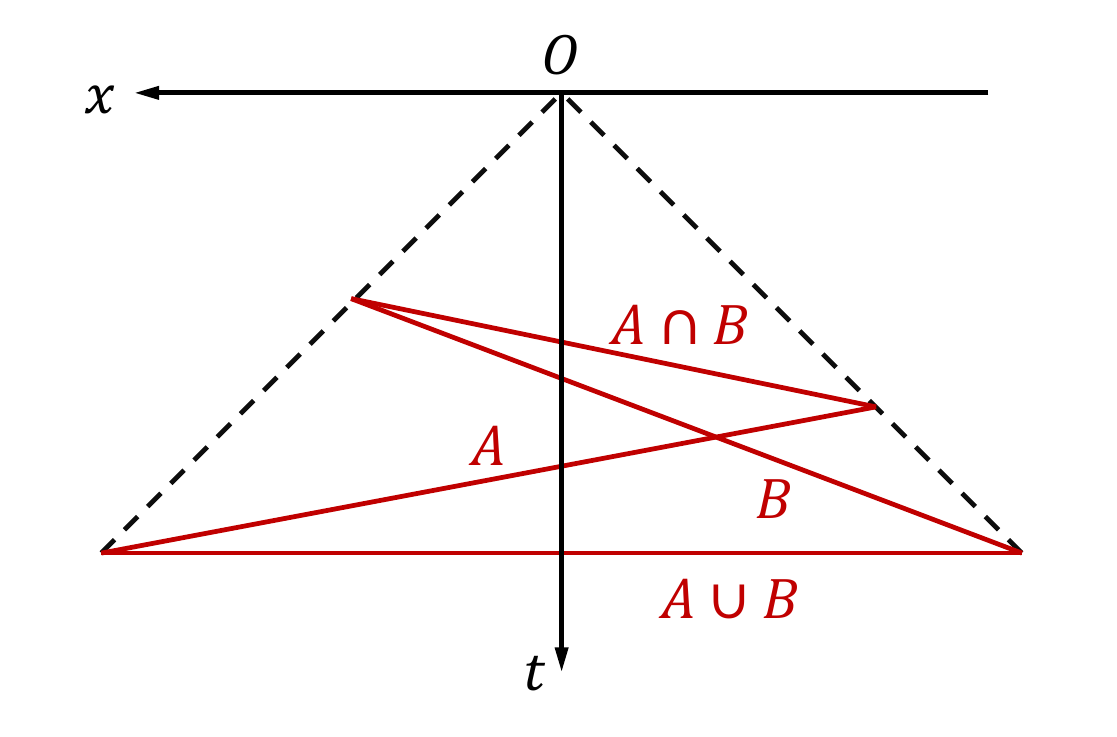}
\caption{The black lines represent subsystems and the blue lines express the light cone}\label{fig:subsystem}
\end{figure}

We consider the subsystems $A$ and $B$ as in Figure \ref{fig:subsystem}. Although $A$ and $B$ do not lie at the constant time slice in Figure \ref{fig:subsystem}, an appropriate Lorentz boost will bring them to the constant time slice so that we can define the Hilbert space $\mathcal{H}_{A} $ and $ \mathcal{H}_{B}$.  The Hilbert spaces $\mathcal{H}_{A} \cup \mathcal{H}_{B}$ and  $\mathcal{H}_{A} \cap \mathcal{H}_{B}$ are defined on $A\cup B$ and $A\cap B$.
Let $\ell(A)$ denote the Lorentz-invariant length of subsystem $A$. Then, a simple computation yields
\be
\ell(A ) \cdot \ell(B )=\ell(A \cup B ) \cdot \ell(A \cap B )~.
\ee
In particular, when
\[
\ell(A )=\ell(B )=e^{\frac{a+b}{2}} , \quad \ell(A \cup B )=e^{a} , \quad \ell(A \cap B )=e^{b}~,
\]
the strong subadditivity \eqref{strong-subadditivity-property} implies that
\[
2 S \left(\frac{a+b}{2} \right) \geq S (a )+S (b )~.
\]
Thus, $S (R)$ is concave and
\[
\frac{d^{2}}{d R^{2}} S (R) \leq 0
\]
where $\ell =e^R$.
If we define the function
\[
C(R )=3 \frac{d S (R)}{d R}~,
\]
then it is monotonically decreasing
\[
\frac{dC}{d R}  \leq 0~.
\]
Compared with \eqref{EE-infinite}, it is equal to the central charge at the fixed point.  This corresponds to the Zamolodchikov $C$-function \eqref{c-fn}.

\subsection{Black hole thermodynamics and Bekenstein-Hawking entropy}

The aforementioned Ryu-Takayanagi formula evaluates entangle entropy of a CFT from the holographic viewpoint. First, we shall provide a brief introduction to black hole thermodynamics and AdS/CFT correspondence based on which the Ryu-Takayanagi formula is constructed. However, we glimpse only the tip of the iceberg and the subject would actually deserve the entire semester. If you are interested in this fertile subject, we refer to the standard references \cite{wald2010general,Townsend:1997ku,Aharony:1999ti,DHoker:2002nbb,Maldacena:2003nj,Dabholkar:2012zz,Ramallo:2013bua,nuastase2015introduction}.

Let us recall the Schwarzschild black hole of mass $M$ with metric
\[
d s ^{2}=-\left( 1-\frac{r_{H}}{r}\right) c^2d t ^{2}+ \frac{d r ^{2}}{1-\frac{r_{H}}{r}}+r ^{2}d \Omega ^{2}
\]
where the radius of the horizon is given by
\[
r_{H}= \frac{2 GM}{c^2}~.
\]
To see the thermodynamic property of the Schwarzschild black hole, we use the naive trick.
It turns out that much of the exciting physics having to
do with the quantum properties of black holes comes from the
region near the event horizon.
To examine the region \emph{near the horizon $r_H$}, we analytically continued to the Euclidean metric $t= -it_E$, and we set
\[
r-\frac{2GM}{c^2}=\frac{x^2c^2}{8GM}~.
\]
Then, the metric near the event horizon $x\ll1$
\[
ds^2_{\textrm{E}} \approx  (\kappa  x)^2dt_E^2+dx^2
+\frac{1}{4\kappa^2}d\Omega^2~,
\]
where $\kappa=\frac{c^4}{4GM}$ is called the \textbf{surface gravity} because it is indeed the acceleration of a static
particle near the horizon as measured at spatial infinity.
The first part of the metric is just $\bR^2$ with polar coordinates if we make the
{periodic identification}
\[
t_E \sim t_E +\frac{2\pi}{\kappa}~.
\]
Using the relation between Euclidean periodicity and temperature,
we can deduce \textbf{Hawking temperature} of the Schwarzschild black hole
\begin{equation}\label{hawktemp}
k_BT_H = \frac{\hbar\kappa}{ 2 \pi c}=\frac{\hbar  c^3}{8\pi GM}~,
\end{equation}
where $k_B$ is Boltzmann constant. This is a very heuristic way to introduce the Hawking temperature, which was not found originally in this way.

If the black hole has temperature, then it should obey the thermodynamics law. Classically, a stationary black hole is characterized by its mass $M$, angular momentum $J$, and charge $Q$. This is called a black hole no hair theorem. However, in \cite{bekenstein1972black}, Bekenstein proposed that a black hole has entropy proportional to the area of the black hole.
Soon after that, Bardeen, Carter and Hawking point out similarities between the laws of black hole mechanics and the laws of thermodynamics in \cite{Bardeen:1973gs}. More concretely, they find the laws of corresponding to the three laws of thermodynamics.
\begin{enumerate}
\item[{(0)}]
Zeroth Law: In thermodynamics, the zeroth law states that the temperature $T$ of a thermal equilibrium object is constant throughout the body. Correspondingly, for a stationary black hole, its surface gravity $\kappa=1/4GM$ is constant over the event horizon.

\item[{(1)}]
First Law: The first law of thermodynamics states that energy is conserved, and the variation of energy is given by
\be dE = TdS + \mu dQ + \Omega dJ \ee
where $E$ is the energy, $Q$ is the charge with chemical potential $\mu$ and $J$ is the angular momentum with chemical potential
$\Omega$ in the system.
Correspondingly, for a black hole, the variation of its mass is given by
\be dM = \frac{\kappa}{ 8\pi G} dA + \mu dQ + \Omega dJ \ee
where $A$ is the area of the horizon,  and $\kappa$ is the surface gravity, $\mu$ is the chemical potential conjugate to $Q$, and $\Omega$ is the angular velocity conjugate to $J$.

\item[{(2)}]
Second Law: The second law of thermodynamics states that the total entropy $S$ never decreases, $\delta S \geq 0$.  Correspondingly, for a black hole, the area theorem states that the total area of a black hole in any process never decreases, $\delta A \geq 0$.
\end{enumerate}

\begin{table}[ht]
\centering
\begin{tabular}{c|c}
\hline
\textbf{Laws of thermodynamics} & \textbf{Laws of black hole mechanics}\\
 \hline
 Temperature is constant & Surface gravity is constant \\
 throughout a body at equilibrium. &  on the event horizon.\\
 $T$=constant. & $\kappa$ =constant.\\
 \hline
Energy is conserved.& Energy is conserved. \\
$dE = T dS + \mu dQ + \Omega dJ. $& $dM = \frac{\kappa c^2}{8\pi G} dA + \mu dQ + \Omega dJ . $\\
 \hline
 Entropy never decreases.  & Area never decreases.\\
 $\delta S \geq 0$. & $ \delta A \geq 0 $. \\
 \hline
\end{tabular}
\caption{\small{Laws of black hole thermodynamics}}
\label{blackholelaws}
\end{table}

This result can be understood as one of the highlights of general relativity.
Moreover, Hawking has shown this is indeed more than an analogy \cite{Hawking:1974sw}.
Hawking has applied techniques of quantum field theories on a curved background to the near-horizon region of a black hole and showed that a black hole indeed radiates \cite{Hawking:1974sw}.
This can be intuitively understood as follows: in a quantum theory, particle-antiparticle creations constantly occur in the vacuum. Around the horizon, after pairs are created, antiparticles fall into a black hole due to the gravitational attraction whereas particles escape to infinity.  Although we do not deal with Hawking's calculation unfortunately (see \cite{Townsend:1997ku}), it indeed justifies this picture. Moreover, it revealed that the spectrum emitted by the black hole is precisely subject to the thermal radiation with temperature \eqref{hawktemp}. Indeed, a black hole is not black at quantum level.
Hence, we can treat a black hole as a thermal object, and the analogy of the laws in Table \ref{blackholelaws} can be understood as the natural consequence of the laws of thermodynamics.

In fact, because the energy of the Schwarzschild black hole is equal to its mass, the first law of
thermodynamics can be written as
\[
T_{H}d S=c^2d M
\]
Using the formula
for the Hawking temperature \eqref{hawktemp}, the Schwarzschild black hole has the entropy
\[
S=4 \pi \frac{k_BGM ^{2}}{c\hbar}~.
\]
The entropy is proportional to the area of the horizon
\be\label{BH-entropy}
S=\frac{k_B c^3 \operatorname{Area} }{4G\hbar} ~.
\ee
This is a universal result for any black hole, and this remarkable relation between the thermodynamic properties
of a black hole and its geometric properties is called the celebrated \textbf{Bekenstein-Hawking entropy formula} \cite{bekenstein1972black,Bekenstein:1973ur,Hawking:1974sw}.
This formula involves all four fundamental constants of nature; $(G,c,k_B,\hbar)$. Also, this is the first place where the Newton constant $G$ meets with the Planck constant $\hbar$.
Thus, this formula shows a deep connection between black hole geometry, thermodynamics and quantum mechanics.

In thermodynamics, the energy and entropy are examples of \textbf{extensive properties} \textit{i.e.} they are proportional to the size of the system. However, the Bekenstein-Hawking entropy is proportional to the area. Moreover, in general relativity, the maximal entropy in a certain subsystem $R$ is proportional to the area of $\partial R$
\[
S\le \frac{\operatorname{Area}(\partial R)}{4G}~,
\]
which is called \textbf{entropy bound}. Subsequently, people come up with the idea that a gravity theory in $(d+1)$-dimension is equivalent to a quantum field theory without gravity in $d$-dimension, which is called the \textbf{holographic principle}.

\subsection{AdS/CFT correspondence}

The AdS/CFT correspondence \cite{Maldacena:1997re} is a special case of the holographic principle, which states that a quantum gravity in an $\AdS_d$ space is equivalent to a CFT in $d$-dimensions. Even though the correspondence was originally proposed in the context of string theory, it is now generalized in broader contexts. This subsection briefly introduces the basics of the  AdS/CFT correspondence to pave the way to the Ryu-Takayanagi formula.

\subsubsection*{AdS space}
To begin with, let us investigate the geometry of AdS spaces.
An anti-de Sitter (AdS) space is a maximally symmetric manifold with constant negative scalar curvature.
It is a solution to Einstein's equations for an empty universe with negative cosmological constant.
The easiest way to understand it is as follows.

A Lorentzian AdS$_{d+1}$ space can be illustrated by the hyperboloid in $(2,d)$ Minkowski space:
\be
 X_0^2 +X_{d+1}^2 -\sum_{i=1}^{d} X_i^2 = R^2 \ .
 \label{embeding}
\ee
The metric can be naturally induced from the Minkowski space
\be
 ds^2 = -dX_0^2 -dX_{d+1}^2 +\sum_{i=1}^{d} dX_i^2 \ .
\ee
By construction, it has $\SO(2,d)$ isometry, which is the first connection to
the conformal group in $d$-dimensions studied in \S\ref{sec:conf-gen}.

\subsubsection*{Global coordinate}

A simple solution to (\ref{embeding}) is given as follows.
\bea\nonumber
 &X_0^2 +X_{d+1}^2=R^2 \cosh^2 \rho \ , \cr
 &\sum_{i=1}^{d} X_i^2=R^2 \sinh^2 \rho \ .
\eea
Or, more concretely,
\bea\nonumber
 X_{0} &= R \cosh \rho\ \cos \tau \ , \qquad
 X_{d+1}=R \cosh \rho\ \sin \tau \ ,  \nonumber \cr
 X_i &= R \sinh \rho\ \Omega_{i} \quad (i=1,\cdots,d,
 \text{and} \sum_i \Omega_i^2=1).
\eea
These are $S^{1}$ and $S^{d-1}$ with radii $R\cosh\rho$ and $R\sinh\rho$, respectively.
The metric is
\bea\nonumber
 ds^2 &= R^2 \left(-\cosh^2 \rho \ d\tau^2 +d\rho^2 +\sinh^2 \rho \ d\Omega_{(d-1)}^2 \right) \ .
\eea
Note that $\tau$ is a periodic variable and if we take $0 \le \tau <2\pi$
the coordinate wraps the hyperboloid precisely once.
This is why this coordinate is called a \textbf{global coordinate}.
The manifest sub-isometries are $\SO(2)$ and $\SO(d)$ of $\SO(2,d)$.
To obtain a causal space-time, we simply unwrap the circle $S^1$,
namely, we take the region $-\infty < \tau < \infty$ with no identification,
which is called the \textbf{universal cover} of the hyperboloid.

In literature, another global coordinate is also used,
which can be derived by redefinitions
$r \equiv R \sinh \rho$ and $dt \equiv R d\tau$:
\bea\nonumber
 ds^2 &=-f(r) dt^2 +\frac{1}{f(r)} dr^2+r^2 d\Omega_{(d-1)}^2 \ , \qquad
 f(r)=1+\frac{r^2}{R^2} \ .
\eea

\subsubsection*{Poincar\'e coordinates}

There is yet another coordinate, called \textbf{Poincar\'e coordinates}.
As opposed to the global coordinate, this coordinate covers only half of the hyperboloid.
It is most easily (but naively) seen in $d=1$ case:
\bea\nonumber
 x^2 -y^2=R^2 \ ,
\eea
which is the hyperbolic curve.
The curve consists of two isolated parts in regions $x>R$ and $x<-R$.
We simply use one of them to construct the coordinate.

Let us get back to general $d$-dim.
We define the coordinate as follows.
\bea
 X_{0} &= \frac{1}{2u} \left(1+u^2 \left(R^2 +x_i^2-t^2 \right) \right) \ , \cr
 X_{i} &= R u x_i \qquad (i=1,\cdots,d-1) \ , \cr
 X_{d} &= \frac{1}{2u} \left(1-u^2 \left(R^2 -x_i^2+t^2 \right) \right) \ , \cr
 X_{d+1} &= R u t \ ,
\eea
where $u > 0$.
As it is stated
the coordinate covers half of the hyperboloid; in the region, $X_0 > X_{d}$.
The metric is
\bea\label{ads-poincare}
 ds^2 &= R^2 \left(\frac{du^2}{u^2} +u^2 (-dt^2 +dx_i^2) \right)
=R^2 \left(\frac{du^2}{u^2} +u^2 dx_\mu^2 \right) \ .
\eea
The coordinates $(u,t,x_i)$ are called \textbf{Poincar\'e coordinates}.
This metric has manifest $ISO(1,d-1)$ and $\SO(1,1)$ sub-isometries of $\SO(2,d)$;
the former is the Poincar\'e transformation and the latter corresponds to the dilatation
\bea\label{scaling}
 (u,t,x_i) \to (\lambda^{-1}u,\lambda t,\lambda x_i) \ .
\eea

If we further define $z=1/u$ ($z>0$), then,
\bea\label{ads-poincare2}
 ds^2 &= \frac{R^2}{z^2} \left(dz^2 +dx_\mu^2 \right) \ .
\eea
This is called \textbf{the upper (Poincar\'e) half-plane model}.
The hypersurface given by $z=0$ is called the \textbf{(asymptotic) boundary} of the AdS space,
which corresponds to $u \sim r \sim \rho=\infty$.

\subsubsection*{AdS/CFT correspondence}
The AdS/CFT correspondence \cite{Gubser:1998bc,Witten:1998qj} is an exact duality between a quantum gravity in an asymptotically $\AdS_{d+1}$ space-time and a $\CFT_d$ without
gravity.  It is \textbf{holographic} since the gravitational
theory lives in one extra dimension. It is often useful to think that the CFT lives at the \textbf{boundary} of the \textbf{bulk} AdS space.
Indeed, the CFT lives in a space-time parameterized by $\vec{x}$ while gravity
fields are functions of $\vec{x}$ and the radial coordinate $z$.
For instance,  the scaling transformation of the CFT can be translated into the transformation of the AdS coordinate \eqref{ads-poincare2}
\[
x_{\mu} \rightarrow \lambda x_{\mu} , \quad z \rightarrow \lambda z~.
\]
Hence, the coordinate $z$ parametrizes the energy scale, and the ultra-violet limit corresponds to the boundary at $z=0$. As $z$ increases, the energy scale decreases.

\begin{figure}[ht]\centering
\includegraphics[width=6cm]{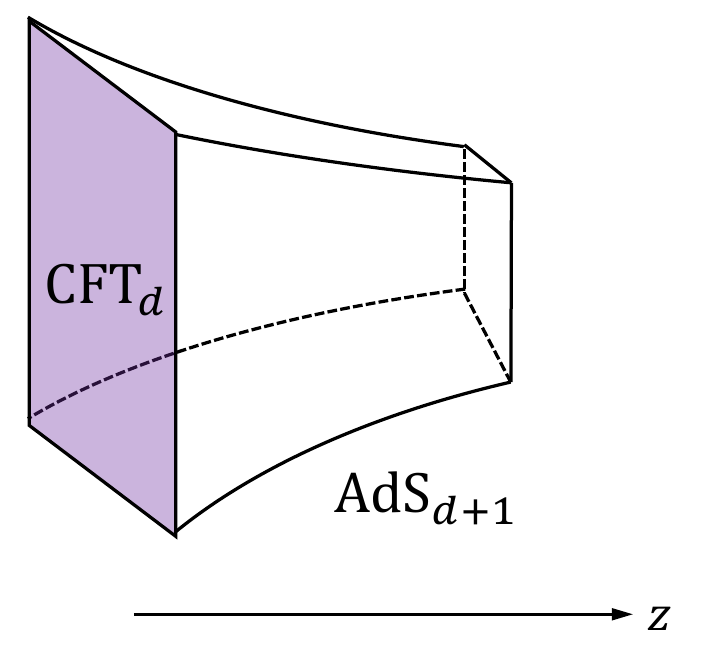}\qquad
\includegraphics[width=7.5cm]{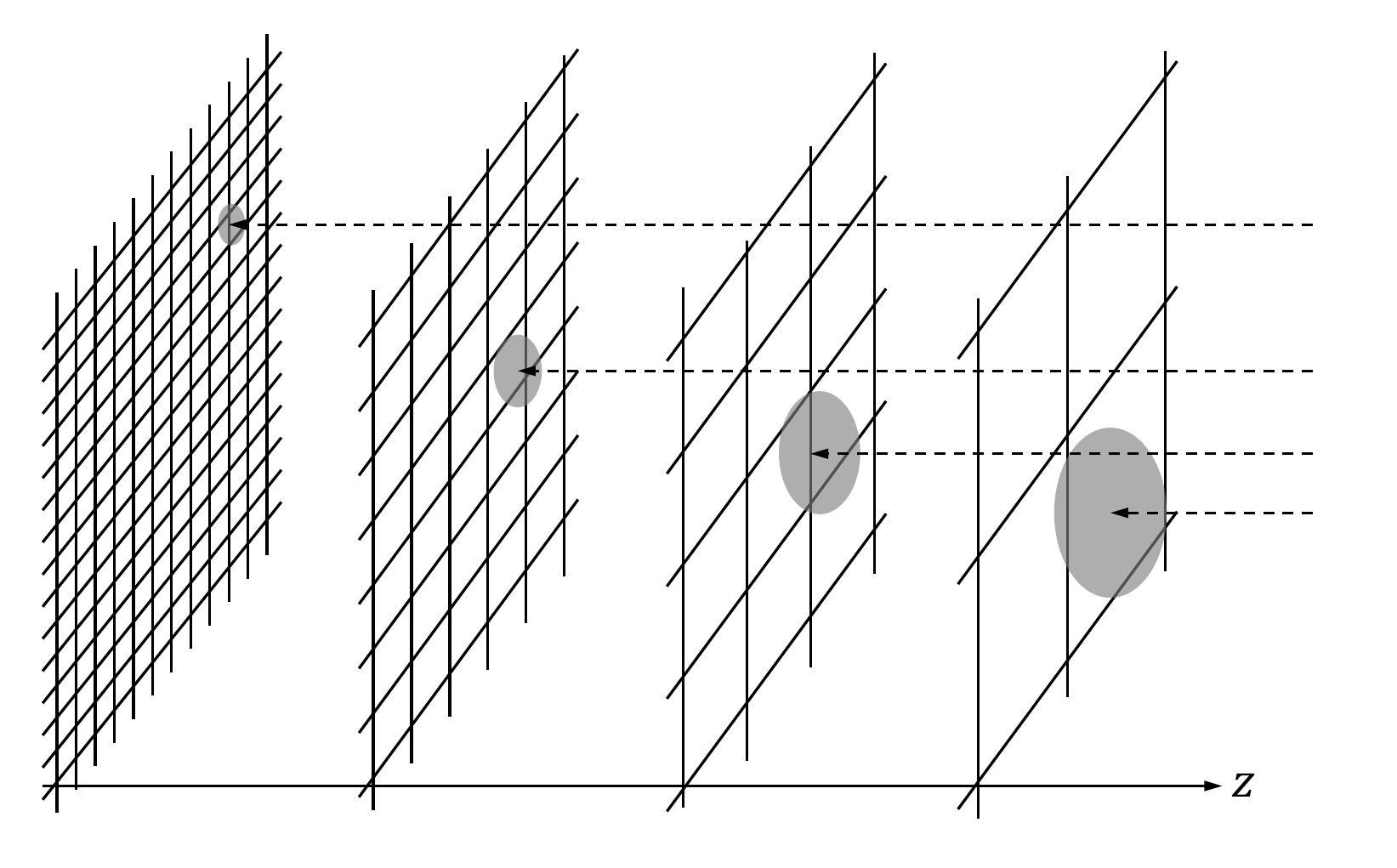}
\caption{}\label{fig:AdS-CFT}
\end{figure}
%

The theories are believed to be entirely
equivalent: any physical (gauge-invariant) quantity that can be computed in one theory
can also be computed in the dual. As a typical feature of a duality, computation of physical quantity becomes much easier on one side than on the other side.
More precisely, the gravitational partition function on asymptotically AdS space is equal to the generating function of correlation functions of the corresponding CFT:
\bea\label{GKPW}
Z_{\textrm{grav}}[\phi\to \phi_0]
=
\Bigg\langle \exp \biggl (\int _{\partial AdS} \overline \phi_0 \cO \biggr)
\Bigg\rangle_{\textrm{CFT}}
\eea
that is called the \textbf{GKPW} relation.
For any bulk field $\phi$ in gravity theory on AdS, there exists the corresponding operator
$\cO$ in the CFT. The gravitational partition function can be schematically written as
\[
Z_{\textrm{grav}}[\phi\to \phi_0]=\int_{\phi\to \phi_0}\cD\phi ~e^{-S_{\textrm{string}}[\phi]}~.
\]

\subsubsection*{Bulk field/boundary operator}
Each field propagating on AdS space
is in a one-to-one correspondence with
an operator in the field theory. The spin of the
bulk field is equal to the spin of the CFT operator; the mass of the bulk field fixes the
scaling dimension of the CFT operator. Here are some examples:

\begin{itemize}
\item  Every theory of gravity has a massless spin-2 particle, the graviton $g_{\m\n}$. This
is dual to the stress tensor $T_{\m\n}$ in CFT. This makes sense since every CFT has a stress
tensor. The fact that the graviton is massless corresponds to the fact that the CFT
stress tensor is conserved.
\item  If our theory of gravity has a spin-1 vector field $A_\m$, then the dual CFT has a
spin-1 operator $J_\m$. If $A_\m$ is gauge field (massless), then $J_\m$
 is a conserved current so that gauge symmetries in the
bulk correspond to global symmetries in the CFT.
\item A bulk scalar field is dual to a scalar operator in  the CFT. The boundary
value of the bulk scalar field
acts as a source in the CFT.
\end{itemize}
For more details, we refer the reader to \cite{Aharony:1999ti}.

\subsection{Ryu-Takayanagi formula}
\begin{figure}[ht]
	\centering
	\includegraphics[width=8cm]{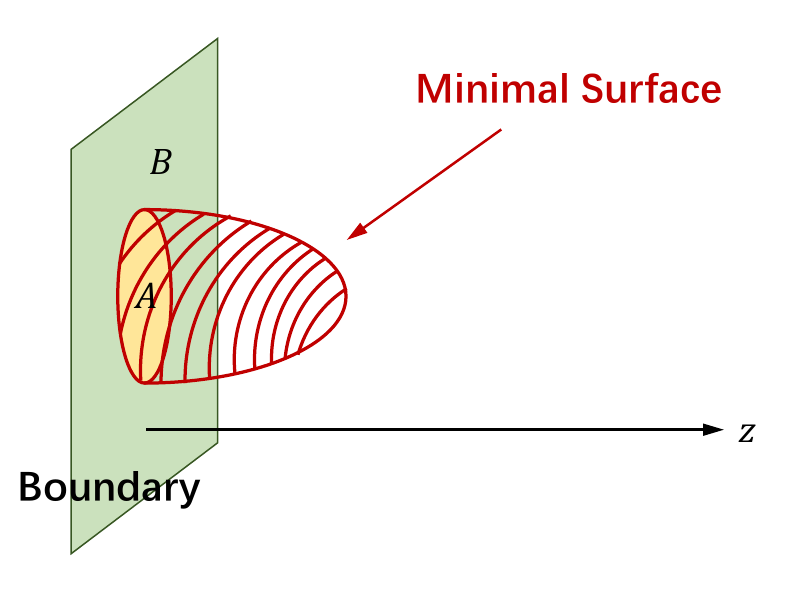}
	\caption{Holographic entanglement entropy}
	\label{fig:holographicEE}
\end{figure}

The Bekenstein-Hawking entropy encodes the information restored in the interior of the horizon. As explained in \S\ref{sec:EE}, the entanglement entropy can be used when the observer cannot access a certain subsystem like the interior of the horizon. Hence, it is natural to seek the relation between the Bekenstein-Hawking entropy and the entanglement entropy. As the first step to understanding the relation, one can also ask how the $\AdS_{d+1}$ space-time encodes the entanglement entropy in the subsystem $A$ of a $\CFT_d$. The answer to this question is the Ryu-Takayanagi (RT) formula \cite{Ryu:2006bv,Ryu:2006ef}:
\begin{equation}
  \label{R-T formula}
S_{A}=\frac{\operatorname{Area} \left(\gamma_{A} \right)}{4 G^{(d+1 )}}.
\end{equation}
The manifold $\gamma_A$ is the $d$-dimensional submanifold with minimal area  in $\mathrm{AdS}_{d+1}$ whose boundary is given by $\pd A$ as shown in Figure \ref{fig:holographicEE}. This minimal submanifold is often called Ryu-Takayanagi (RT) surface. Its area is denoted by $\operatorname{Area}(\gamma_A)$. $G^{(d+1)}_N$ is the $d+1$ dimensional Newton constant. Since $\sqrt{g}$ diverges at the boundary of AdS space $z=0$,
the area $\gamma_A$ is divergent there. Thus, the result is still divergent of order $\e^{-(d-2)}$
\[\begin{aligned} S_{A}& \sim \frac{R^{d-1}}{G } \operatorname{Area} \left(\partial{A} \right) \int_{\epsilon} \frac{d z}{z^{d-1}}\\ & \sim \frac{R^{d-1}}{G}\frac{\operatorname{Area} \left(\partial{A} \right)}{\e^{d-2}} \end{aligned}~.\]

As we will see at the end, when the region $A$ is large enough, the minimal surface (submanifold) wraps the horizon of AdS black hole, and the  RT formula can explain the Bekenstein-Hawking entropy as a special case. However, minimal surfaces (submanifolds) are more general so that the RT formula can be interpreted as a generalization of the Bekenstein-Hawking entropy formula. Moreover, the entanglement entropy depends on a CFT and it also encodes the quantum entanglement. Therefore, the entanglement entropy in a CFT can also be understood as the quantum correction to the Bekenstein-Hawking entropy in the context of the AdS/CFT correspondence.

\subsubsection*{Holographic derivation of strong subadditivity}

The RT formula provides a geometric viewpoint to the entanglement entropy. As one of its advantages, the RT formula leads to the straightforward geometric derivation of strong subadditivity  \eqref{strong-subadditivity-property}.

\begin{figure}[ht]
	\centering
	\includegraphics[width=8cm]{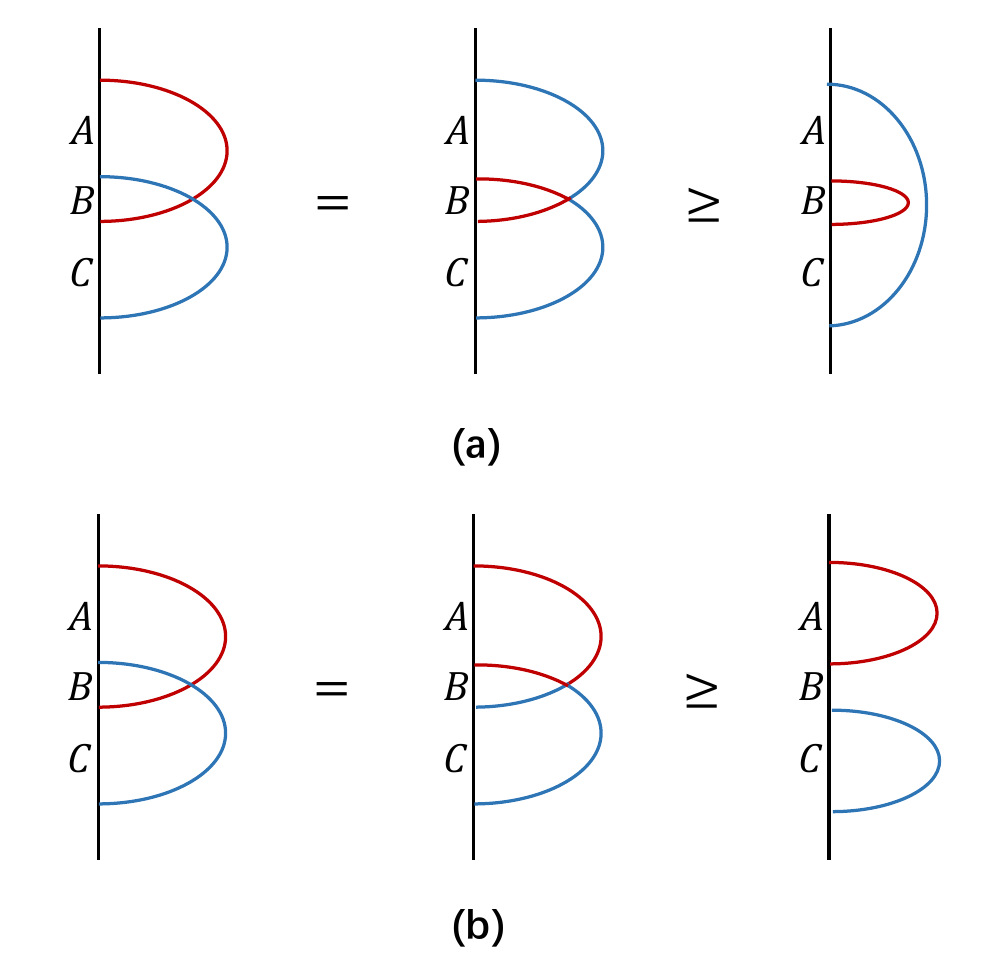}
	\caption{Holographic proof of strong subadditivity}
	\label{fig: holographic proof}
\end{figure}

To derive the strong subadditivity, we start with three subsystems $A$, $B$ and $C$ on a spatial slice as in Figure \ref{fig: holographic proof}. The entanglement entropy $S_{A+B}$ and $S_{B+C}$ are given by the minimal surfaces $\gamma_{A+B}$ and $\gamma_{B+C}$. (left panel of (a).) We decompose the total surface of $\gamma_{A+B}$ and $\gamma_{B+C}$ into the surfaces $\gamma'_{B}$ and $\gamma'_{A+B+C}$. (middle panel of (a).) The total area is clearly greater than or equal to that of the minimal surfaces $\gamma_{B}$ and $\gamma_{A+B+C}$. (right panel of (a).)  This gives the first equation of \eqref{strong-subadditivity-property}. For the second equation, we decompose the total surface of $\gamma_{A+B}$ and $\gamma_{B+C}$ into the surfaces $\gamma'_{A}$ and $\gamma'_{B}$.  (middle panel of (b).) The total area is again greater than or equal to that of the minimal surfaces $\gamma_{A}$ and $\gamma_{B}$. (right panel of (b).)  This provides the second equation of \eqref{strong-subadditivity-property}.
\begin{align*}
 \textrm{Area}(\gamma_{A+B}) + \textrm{Area}(\gamma_{B+C})
 &= \textrm{Area}(\gamma'_{B}) + \textrm{Area}(\gamma'_{A+B+C})
 \geq \textrm{Area}(\gamma_B) + \textrm{Area}(\gamma_{A+B+C})\, ,\\
 \textrm{Area}(\gamma_{A+B}) + \textrm{Area}(\gamma_{B+C})
 &= \textrm{Area}(\gamma'_{A}) + \textrm{Area}(\gamma'_{C})
 \geq \textrm{Area}(\gamma_A) + \textrm{Area}(\gamma_{C})\, .
\end{align*}

\subsubsection*{Holographic entanglement entropy}
We focus our discussion in $\mathrm{AdS}_3/\mathrm{CFT}_2$ case, and the $\mathrm{AdS}_3$ space is
defined in the Poincare coordinates.
\[
d s^{2}=R^{2} \left(\frac{d z^{2}+d x^{2}-d t^{2}}{z^{2}} \right).
\]
At the fixed time $t_0$, the whole spatial region of $\mathrm{CFT}_2$ is an infinite line in the $x$ direction.
We pick up subsystem $A$ along $x$ direction: $-l/2 \leq x \leq l/2$ with its boundary coordinates given by
\[
\begin{aligned} P : (t , x , z ) &=\left(t_{0} ,-\frac{\ell}{2} , \epsilon \right), \\
 Q : (t , x , z ) &=\left(t_{0} , \frac{\ell}{2} , \epsilon \right). \end{aligned}
\]

Now $\operatorname{Area}(\gamma_A)$ in \eqref{R-T formula} is the length of geodesic line in the AdS space,
with two fixed points $P$ and $Q$ . Therefore, we can apply the RT formula to compute the entanglement entropy
for subsystem $A$
\begin{equation}
S_{A}=\frac{2 R}{4 G} \int_{\epsilon}^{\ell/ 2} \frac{d z}{z \sqrt{1-4 z^{2} /\ell^{2}}}
=\frac{R}{2 G} \log \left(\frac{\ell}{\epsilon} \right),
\end{equation}
where we have introduced $\epsilon$ as the UV cutoff.
Since the AdS${}_3$/CFT${}_2$ correspondence identifies the central charge with the Newton constant
\[
c=\frac{3 R}{2 G}~,
\]
it is equal to \eqref{EE-infinite}.

\begin{figure}
	\centering
	\includegraphics[width=0.6\linewidth]{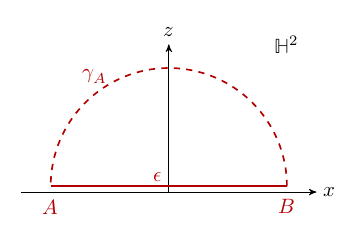}
	\caption{Holographic entanglement entropy in the Poincare coordinates}
	\label{fig:poincare-disk}
\end{figure}

\subsubsection*{Holographic entanglement entropy on a circle}
In the global coordinate of $\mathrm{AdS}_3$, the metric of space-time is written as
\be\label{ads3-metric}
d s^{2}=R^{2} \left(- \cosh^{2} \rho d t^{2}+d \rho^{2}+\sinh^{2} \rho d \theta^{2} \right)
\ee
The $\mathrm{CFT}_2$ is identified with the cylinder $(t, \theta)$ at the (regularized)
boundary $\rho = \rho_\infty$. The whole spatial region at a fixed time is a circle.
The subsystem $A$ corresponds to $0\leq \theta \leq 2\pi \ell/L$.
In this case, the minimal surface $\gamma_A$ is the geodesic line which connects two
boundary points at $\theta = 0$ and $\theta = 2\pi l /L$ with $t$ fixed
The explicit form of the geodesic $\mathrm{AdS}_3$, expressed in the ambient $\vec{X}\in \mathbb{R}^{2,2}$ space, is
\[
\vec{X}=\frac{R}{\sqrt{\alpha^{2}-1}} \sinh (\lambda / R ) \cdot \vec{x}+R \left[ \cosh (\lambda / R )-\frac{\alpha}{\sqrt{\alpha^{2}-1}} \sinh (\lambda / R ) \right] \cdot \vec{y},
\]
where  $\alpha = 1+ 2 \sinh^2 \rho_0 \sin^2(\pi \ell/L)$ and $x$ and $y$ are defined by

\begin{align*} \vec{x} &=\left(\cosh \rho_{\infty} \cos t , \cosh \rho_{\infty} \sin t , \sinh \rho_{\infty} , 0 \right) \\ \vec{y} &=\left(\cosh \rho_{\infty} \cos t , \cosh \rho_{\infty} \sin t , \sinh \rho_{\infty} \cos (2 \pi\ell/L) , \sinh \rho_{\infty} \sin (2 \pi \ell/L) \right), \end{align*}

The length of geodesic can be found as
\[ \cosh\left(\frac{\mbox{Length}(\gamma_A)}{R}\right)=1+2 \sinh^{2} \rho_{\infty} \sin^{2} \frac{\pi\ell}{L}\]
Assuming that the UV cutoff energy is large $e^{\rho_\infty} \geqq 1$,  we can obtain the entropy as follows
\[
S_{A} \simeq \frac{R}{4 G} \log \left(e^{2 \rho_{\infty}} \sin^{2} \frac{\pi\ell}{L} \right)=\frac{c}{3} \log \left(\frac{L}{\epsilon} \sin \frac{\pi\ell}{L} \right).
\]
This coincides with entanglement entropy in a finite size $L$ \eqref{EE-circle}.

\begin{figure}
	\centering
	\includegraphics[width=0.6\linewidth]{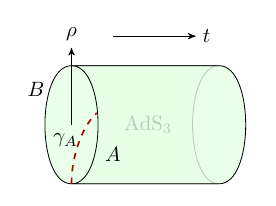}
	\caption{Holographic entanglement entropy on a circle}
	\label{fig:ads3-cft2-holographic}
\end{figure}

\subsubsection*{BTZ black hole}
We have seen that the AdS space has the scaling symmetry \eqref{scaling}. However, one can also consider an asymptotically AdS space whose dual is a QFT which flows to a CFT at IR $z\to 0$. A typical example of asymptotically AdS spaces is an AdS black hole whose metric can be written as
\bea\nonumber
d s^{2} = R^{2}\left(- \frac{f ( z )}{z^{2}} d t^{2} +  \frac{d z^{2}}{z^{2} f ( z )} + \frac{d x^{2}}{z^{2}}  \right) ~,\qquad f ( z ) = 1 - \left( \frac{z}{z_0} \right)^{2}~,\eea
where the horizon is located at $z=z_0$. It is easy to see that the metric asymptotes to the $\AdS_3$ space as $z\to 0$. By the change $r=R/z$ of the coordinates, the metric is written as
\[
d s^{2} = -  (r^{2} - r_{0}^{2})  d t^{2} + \frac{R^{2}}{r^{2} - r_{0}^{2}} d r^{2} + r^{2} d x^{2}~.
\]
We can identify the spatial coordinate $x\sim x+2\pi$, which is called the \textbf{BTZ black hole} and it has temperature
\[
T=1/\beta=\frac{r_0}{2\pi R}~.
\]
Hence, if we use the Euclidean time $\tau=-it$, the time direction is compactified as $\tau\sim \tau+\beta$.
Therefore, the dual $\CFT_2$ is living on a torus boundary parametrized by $(\tau,x)$ with periodicities $\tau\sim\tau+\beta$ and $x\sim x+2\pi$.

Let us consider the RT entropy  for the subsystem $A$
given by $0\leq x\leq 2\pi \ell/R$ at the boundary. To evaluate the length of the geodesic from $x=0$ to $x=2\pi \ell/R$, one can use the fact that the BTZ metric can be changed to the $\AdS_3$ metric \eqref{ads3-metric} by
\[
r\to r_0\cosh\rho~, \quad x\to i R  t~,\qquad \tau \to R \theta~.
\]
Hence, the same analysis above can be applied to compute the length by exchanging $L\leftrightarrow \beta$
\[\cosh\left(\frac{\mbox{Length}(\gamma_A)}{R}\right)
=
1 + 2 \cosh^{2} \rho_{\infty} \sinh^{2} \left( \frac{\pi \ell}{\beta} \right)
\]
The relation between the cut-off $\e$ in CFT and
the one $ \rho_{\infty}$ of AdS is given by $e^{\rho_{\infty}} =\frac{\beta}{
\e}$ so that it reproduces the  CFT result \eqref{entropytemp}.

\begin{figure}[ht]
\begin{center}
\includegraphics[width=\textwidth]{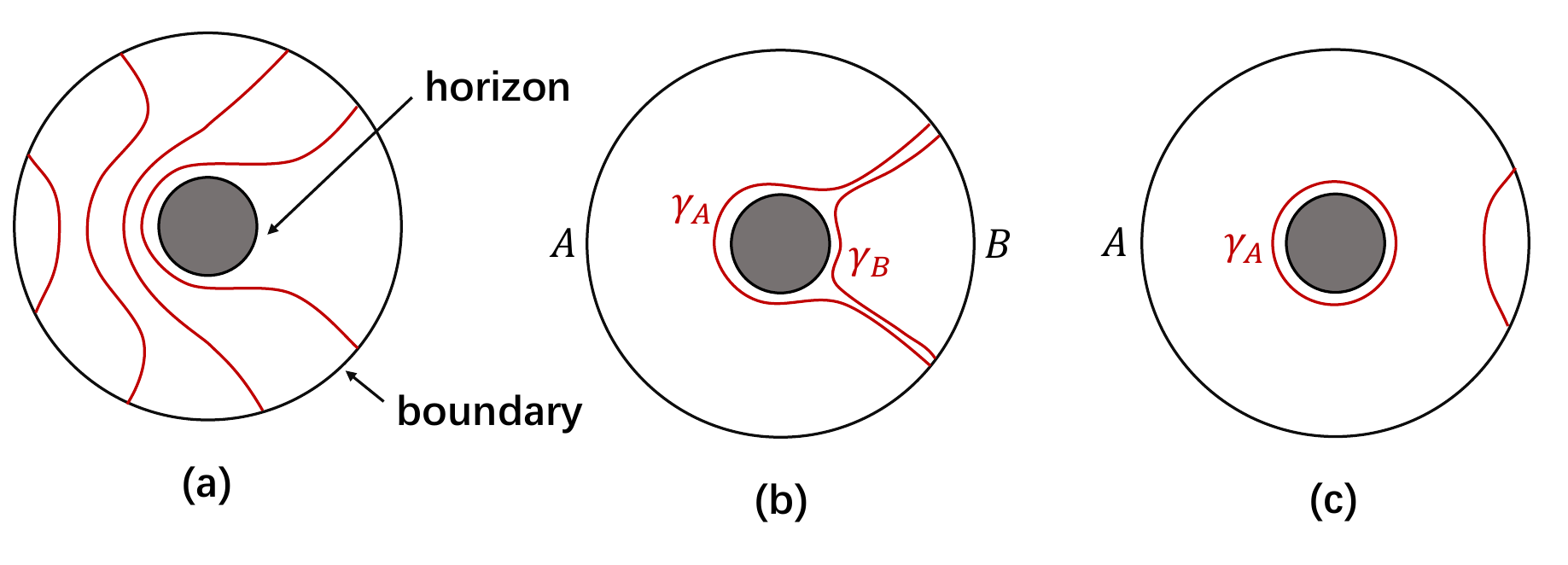}
\end{center}
\caption{(a) As the subsystem $A$ gets larger, the minimal surface starts bent due to the presence of the black hole. (b) Due to the horizon, the minimal surfaces of the subsystems $A$ and its complement $B$ are different. (c) When the subsystems $A$ occupies most of the boundary region, the minimal surface consists of two separate parts: the entire horizon and the minimal surface of its complement.} \label{fig:ads_blackhole}
\end{figure}

Now let us come back to the Bekenstein-Hawking entropy \eqref{BH-entropy} of the BTZ black hole from the viewpoint of the AdS/CFT correspondence and the RT formula.  As in Figure \ref{fig:ads_blackhole} (a), when a subsystem $A$ in the dual CFT is small enough, the RT surface is similar to that of the pure AdS space. As the subsystem $A$ becomes larger, the RT surface starts touching the event horizon of the BTZ black hole. Therefore, the RT surfaces of the subsystems $A$ and its complement $B$ are different due to the presence of the event horizon as in Figure \ref{fig:ads_blackhole} (b). When the subsystem $A$ covers the major part of the boundary CFT region, the RT surface splits into the entire horizon and the minimal surface of its complement as in Figure \ref{fig:ads_blackhole} (c).
Therefore, the  Bekenstein-Hawking entropy of the BTZ black hole is given by
\[S_{A} - S_{B} = S_{\textrm{BH}}~.\]
Consequently, when $A$ is the entire boundary, the RT formula \eqref{R-T formula} reproduces the Bekenstein-Hawking entropy \eqref{BH-entropy}.
As we see in \eqref{EE-finite}, the leading term of the entangle entropy in the CFT is given by thermodynamic entropy in the high-temperature limit.
Therefore, the viewpoint of the AdS/CFT correspondence and the RT formula provides a natural interpretation to the area law of the black hole entropy \eqref{BH-entropy}.

\bibliography{conformal-ref}
\bibliographystyle{hyperamsalpha}

\bigskip

\noindent
Satoshi Nawata, {\sf Department of Physics and Center for Field Theory and Particle Physics, Fudan University, 
Songhu Road 2005, 200438 Shanghai, China}, snawata@gmail.com

\

\noindent
Runkai Tao, {\sf Department of Physics and Astronomy, Rutgers University,
126 Frelinghuysen Road, Piscataway NJ 08854, USA}, runkaitao@gmail.com

\

\noindent
Daisuke Yokoyama, {\sf Department of Physics, Meiji University, Kanagawa 214-8571, Japan},  ddyokoyama@meiji.ac.jp

\end{document}